\documentclass[fleqn,usenatbib]{mnras}

\usepackage{amssymb}	% Extra maths symbols
\usepackage{newtxtext,newtxmath}
\usepackage{amsfonts}
\usepackage[T1]{fontenc}
\usepackage{ae,aecompl}
\usepackage[position=top]{subfig}
\usepackage{float}
\usepackage{comment}
\usepackage[normalem]{ulem}

\usepackage{booktabs}
\usepackage[position=top]{subfig}
\usepackage{grffile}% A pacakge which changes the algorithm to check for known extensions, so that we can insert pdf figures into this text.
\usepackage{etoolbox}
\makeatletter % Workaround to only color the year for cite commands
  \patchcmd{\NAT@citex}
    {\@citea\NAT@hyper@{%
      \NAT@nmfmt{\NAT@nm}%
      \hyper@natlinkbreak{\NAT@aysep\NAT@spacechar}{\@citeb\@extra@b@citeb}%
      \NAT@date}}
    {\@citea\NAT@nmfmt{\NAT@nm}%
    \NAT@aysep\NAT@spacechar\NAT@hyper@{\NAT@date}}{}{}

  \patchcmd{\NAT@citex}
    {\@citea\NAT@hyper@{%
      \NAT@nmfmt{\NAT@nm}%
      \hyper@natlinkbreak{\NAT@spacechar\NAT@@open\if*#1*\else#1\NAT@spacechar\fi}%
        {\@citeb\@extra@b@citeb}%
      \NAT@date}}
    {\@citea\NAT@nmfmt{\NAT@nm}%
    \NAT@spacechar\NAT@@open\if*#1*\else#1\NAT@spacechar\fi\NAT@hyper@{\NAT@date}}
    {}{}
\makeatother

\defcitealias{Brown2019}{B19}

 \makeatother
\usepackage{graphicx}	% Including figure files
\usepackage{amsmath}	% Advanced maths commands
\usepackage{pifont}
\usepackage{xcolor}
\hypersetup{allcolors=cyan}

\title[X-ray outputs from Be-XRBs]{Population synthesis of Be X-ray binaries: %and their impact at Cosmic Dawn: 
metallicity dependence of total X-ray outputs}

\author[B. Liu et al.]{Boyuan Liu\textsuperscript{\href{https://orcid.org/0000-0002-4966-7450}{\includegraphics[width=2.5mm]{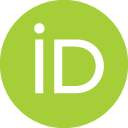}}\,}\thanks{E-mail: bl527@cam.ac.uk}$^{1}$, Nina S. Sartorio\textsuperscript{\href{https://orcid.org/0000-0003-2138-5192}{\includegraphics[width=2.5mm]{orcid.png}}\,}$^{2}$, Robert G. Izzard\textsuperscript{\href{https://orcid.org/0000-0003-0378-4843}{\includegraphics[width=2.5mm]{orcid.png}}\,}$^{3}$, Anastasia Fialkov$^{1,4}$
\\
$^{1}$Institute of Astronomy, University of Cambridge, Madingley Road, Cambridge, CB3 0HA, UK\\
$^{2}$Department of Physics and Astronomy, University of Ghent, Technologiepark 903, Ghent, 9052 Zwijnaarde, Belgium\\
$^{3}$Astrophysics Research Group, University of Surrey, Guildford, Surrey, GU2 7XH, UK\\
$^{4}$Kavli Institute for Cosmology, Madingley Road, Cambridge, CB3 0HA, UK}
\date{Accepted XXX. Received YYY; in original form ZZZ}

\pubyear{2023}

\begin{document}
\label{firstpage}
\pagerange{\pageref{firstpage}--\pageref{lastpage}}
\maketitle

% Abstract of the paper
\begin{abstract}
X-ray binaries (XRBs) are thought to regulate cosmic thermal and ionization histories during the Epoch of Reionization and Cosmic Dawn ($z\sim 5-30$). 
Theoretical predictions of the X-ray emission from XRBs are important for modelling such early cosmic evolution. 
Nevertheless, the contribution from Be-XRBs, powered by accretion of compact objects from decretion disks around rapidly rotating O/B stars, has not been investigated systematically. Be-XRBs are the largest class of high-mass XRBs (HMXBs) identified in local observations and are expected to play even more important roles in metal-poor environments at high redshifts. 
In light of this, we build a physically motivated model for Be-XRBs based on recent hydrodynamic simulations and observations of decretion disks. Our model is able to reproduce the observed population of Be-XRBs in the Small Magellanic Cloud with appropriate initial conditions and binary stellar evolution parameters. 
We derive the X-ray output from Be-XRBs as a function of metallicity in the (absolute) metallicity range $Z\in [10^{-4},0.03]$ {with a large suite of binary population synthesis (BPS) simulations. 
The simulated Be-XRBs can explain a non-negligible fraction ($\gtrsim 30\%$) of the total X-ray output from HMXBs observed in nearby galaxies for $Z\sim 0.0003-0.02$. 
The X-ray luminosity per unit star formation rate from Be-XRBs in our fiducial model increases by a factor of $\sim 8$ from $Z=0.02$ to $Z=0.0003$, which is similar to the trend seen in observations of \textit{all} types of HMXBs. We conclude that Be-XRBs are potentially important X-ray sources that deserve greater attention in BPS of XRBs.}
%significant fraction ($\sim 60\%$) 
%A similar fraction of observed ultra-luminous ($\gtrsim 10^{39}\ \rm erg\ s^{-1}$) X-ray sources can also be explained by Be-XRBs. 
%We also estimate the X-ray output from Be-XRBs in the extremely metal-poor regime ($Z\lesssim 10^{-6}$) relevant for the first stars, which turns out to be comparable to the values predicted for other types of XRBs. 
%We conclude that Be-XRBs are important X-ray sources that should be taken into account in BPS of XRBs.  
\end{abstract}
% Select between one and six entries from the list of approved keywords.
% Don't make up new ones.

%%%%%%%%%%%%%%%%%%%%%%%%%%%%%%%%%%%%%%%%%%%%%%%%%%

%%%%%%%%%%%%%%%%% BODY OF PAPER %%%%%%%%%%%%%%%%%%
\section{Introduction}\label{sec:intro}
%The baryonic component of the early Universe is strongly influenced by X-ray photons that can easily escape the host galaxies/haloes of their sources to heat and ionise the interstellar medium (IGM) at long distances. 
During the Epoch of Reionization and Cosmic Dawn ($z\sim 5-30$), X-ray binaries (XRBs) are expected to be the dominant sources of X-rays that regulate the thermal and ionization evolution and small-scale structure of the interstellar medium \citep[IGM, e.g.,][]{Fragos2013,Fialkov2014,Fialkov2014rich,Pacucci2014,Madau2017,Eide2018}, as well as early star formation \citep[e.g.,][]{Jeon2012,Artale2015,Hummel2015,Ricotti2016,Park2021I,Park2021II,Park2022}\footnote{Other agents, such as cosmic rays and Lyman-band photons, can also have competitive effects as X-rays \citep[e.g.,][]{Stacy2007,Safranek-Shrader2012,Fialkov2013,Hummel2016,Kulkarni2021,Reis2021,Schauer2021,Bera2023,Gessey-Jones2023}.}. They can leave unique signatures in the 21-cm signal from neutral hydrogen, which is one of the most promising probes of early structure/galaxy/star formation and cosmology \citep[e.g.,][]{Fialkov2017,Bowman2018,Ewall-Wice2018,Ma2018,Madau2018,Mirocha2019,Schauer2019,Chatterjee2020,Qin2020,Gessey-Jones2022,Kamran2022,Kaur2022,Kovlakas2022,Magg2022tr,Munoz2022,Acharya2023,Bevins2023,Hassan2023,Lewis2023,Ma202321,Mondal2023,Shao2023,Ventura2023,Yang2023}, and in the Lyman-$\alpha$ forest from long-lasting relics of reionization \citep{Montero-Camacho2023}. To fully unleash the power of the 21-cm probe and break potential degeneracy between astrophysics, dark matter physics and cosmology {\citep[e.g.,][]{Barkana2018,Liu2019,Yang2021,Yang2023,Ghara2022,Acharya2023,Mondal2023dm,Shao2023}}, it is necessary to model the X-ray emission from XRBs accurately. 

The metallicity dependence of X-ray outputs from high-mass XRBs \citep[HMXBs, reviewed by, e.g.,][]{Walter2015,Kretschmar2019,Fornasini2023}, which dominate the cosmic XRB luminosity density at $z\gtrsim 3$ \citep{Fragos2013xrb}, is particularly important in the early Universe when the metal content of XRB host galaxies evolves rapidly \citep[e.g.,][]{Wise2012,Johnson2013,Xu2013,Pallottini2014,Liu2020did,Ucci2023}, and the X-rays from active galactic nuclei are subdominant \citep[see, e.g.,][]{Fragos2013}. 
In fact, X-ray observations of nearby and distant (up to $z\sim 2$) galaxies {\citep[e.g.,][]{Antoniou2010,Antoniou2019,Basu-Zych2013,Prestwich2013,Douna2015,Antoniou2016,Brorby2016,Lehmer2016,Lehmer2019,Lehmer2021,Lehmer2022,Aird2017,Fornasini2019,Fornasini2020,Riccio2023}} and theoretical predictions by binary population synthesis (BPS) of XRBs \citep[e.g.,][]{Linden2010,Fragos2013,Sartorio2023} all suggest a strong metallicity dependence of the X-ray output from HMXBs. This drives the redshift evolution of the scaling relation between X-ray luminosity and star formation rate (SFR), and can have significant impact on the 21-cm signal \citep[e.g.,][]{Kaur2022}. For instance, \citet{Fragos2013} provides fitting formulae for the X-ray luminosity of HMXBs per unit SFR as a function of metallicity from BPS simulations \citep{Fragos2013xrb}. In their case, the X-ray luminosity increases by a factor of $\sim 6$ from solar metallicity to $\sim$1\% solar, consistent with the trend seen in observations. \citet{Sartorio2023} derives the X-ray outputs of XRBs from metal-free stars, i.e., the so-called Population III (Pop~III). They find that in optimistic cases the X-ray emission of Pop~III XRBs can be significantly stronger (up to a factor of 40) compared with that of XRBs from metal-enriched stars predicted by \citet{Fragos2013}. %Note that these BPS studies do not consider Be-XRBs, it is crucial to compare their results with those of Be-XRBs and evaluate the role played by Be-XRBs in the metallicity dependence of X-ray outputs from XRBs. 

However, the aforementioned BPS studies only consider the XRBs powered by Roche lobe overflow (RLO) and (spherical) stellar winds but ignore an important type of HMXBs, Be-XRBs, likely due to their transient nature. A Be-XRB is made of a compact object and a rapidly-rotating, massive ($\gtrsim 6\ \rm M_{\odot}$), main sequence (MS) star \citep[reviewed by, e.g.,][]{Reig2011,Rivinius2013,Rivinius2019}. Here the massive star is typically of spectral type B (and O) with an rotation velocity above $\sim70$ percent of the equatorial Keplerian limit and shows Balmer emission lines, which can be well explained by a viscous decretion disk\footnote{The disks are very light structures compared with the stars, with typical masses $M_{\rm d}\sim 10^{-11}-10^{-8}\ \rm M_{\odot}$ \citep{Granada2013,Klement2017,Rivinius2019}. Formation of VDDs can be regarded as simply the means of losing angular momentum for a massive star getting close to the critical limit of rotation \citep{Rivinius2019}.} (VDD) around the star. This VDD is formed by materials ejected from the star due to redistribution of angular momentum caused by fast rotation\footnote{The detailed mechanisms for the formation of VDDs around O/B stars are still in debate. Possible mechanisms include mechanical mass loss at critical rotation \citep[e.g.,][]{Granada2013,Hastings2020,Zhao2020}, pulsations \citep[e.g.,][]{Cranmer2009,Rogers2013,Lee2014} and small-scale magnetic fields \citep[e.g.,][]{Ressler2021}. Nevertheless, in all these scenarios, rapid rotation is required \citep{Rivinius2013}.} \citep{Rivinius2013}. In most observed Be-XRBs, neutron stars (NSs) are identified as the compact companion, %{
although there %is one binary, AS 386, 
are two binaries, MWC 656\footnote{With new spectroscopic data of MWC 656, it is found by \citet{Janssens2023} that the compact companion in this system has a mass $M_{\rm X}\sim 0.6-2.4\ \rm M_{\odot}$, which disfavours the black hole interpretation that is based on previous estimates $M_{\rm X}\sim 4-7\ \rm M_{\odot}$ \citep{Casares2014}.} and AS 386, 
that contain a {Be star and a black hole (BH) candidate} but show very faint X-ray emission %\citep{Khokhlov2018}, 
\citep{Casares2014,Munar-Adrover2014,Grudzinska2015,Khokhlov2018,Zamanov2022}, 
and one Be-XRB that contains a white dwarf \citep[Swift J011511.0-725611, ][]{Kennea2021}. The X-ray output of Be-XRBs is dominated by X-ray outbursts produced by strong accretion of the compact object from the VDD, which typically occurs close to periastron \citep{Okazaki2001,Okazaki2001bexrb} and/or when the compact object crosses a tidally warped (eccentric) VDD \citep{Okazaki2013,Franchini2021}. %$v\nu$%"" ``''

Be-XRBs make up the largest class of HMXBs identified in observations \citep{Fornasini2023}, especially in metal-poor environments. In the Milky Way (MW), 74 Be-XRBs have been found among the total 152 known HMXBs \citep{Fortin2023}. In the Large Magellanic Cloud (LMC) at approximately half solar metallicity, there are 33 Be-XRBs among the 40 confirmed HMXBs \citep{Antoniou2016}, while in the Small Magellanic Cloud (SMC) at about one quarter solar metallicity, %all but one (SMC X-1) of the $70$ classified 
69 X-ray pulsars are identified as Be-XRBs among the 121 HMXB candidates \citep{Coe2015,Haberl2016}. The HMXB population of M33 is also dominated by Be-XRBs \citep{Lazzarini2023}. Besides, theoretical models find that Be-XRBs can be an important component of the X-ray luminosity function of HMXBs in the MW \citep{Zuo2014,Misra2022}. However, previous BPS studies of Be-XRBs \citep[e.g.,][]{Zhang2004,Belczynski2009,Linden2009,Shao2014,Shao2020,Zuo2014,Vinciguerra2020,Xing2021,Misra2022} focus on the cases of solar and SMC metallicities in which they either do not model the X-ray emission or use rough estimates and empirical scaling laws to characterize the VDDs and X-ray outbursts of Be-XRBs \citep[e.g.,][]{Dai2006,Coe2015,Klement2017}. The overall X-ray outputs from Be-XRB populations \textit{as a function of metallicity} has not been investigated quantitatively, while a strong metallicity dependence is expected from the reduced mass loss at low metallicities \citep{Linden2010,Fragos2013xrb,Sartorio2023}. 
Therefore, it is crucial to include Be-XRBs in BPS models to evaluate the metallicity dependence of X-ray outputs from the entire population of HMXBs. %across cosmic time. 
%It is expected that there could be a significant metallicity dependence, as implied by 

In light of the potential importance of Be-XRBs for the cosmic thermal and ionization history, %and the 21-cm signal, 
we build a physically motivated Be-XRB model to predict the X-ray output from Be-XRBs as a function of metallicity with BPS. Inspired by the recent advancements in hydrodynamic simulations of VDDs in Be-XRBs %that are able to explain the key features of X-ray outbursts in observed Be-XRBs 
\citep{Okazaki2013,Panoglou2016,Cyr2017,Brown2018,Brown2019,Suffak2022}, our model fully captures for the first time the dependence of X-ray outburst properties (i.e., strength and duty cycle) on stellar and orbital parameters of Be-XRBs by combining simulation results \citep{Brown2019} with VDD properties inferred from observations of Be stars \citep{Vieira2017,Rimulo2018}. In this paper we focus on the \textit{absolute} metallicity\footnote{Throughout this paper, we use the absolute metallicity (i.e., mass fraction of metals). %to avoid confusions in the definition of solar metallicity.
When comparing our results with observations, we convert $\rm \log[O/H]+12$ to absolute metallicity using a solar oxygen abundance of $\rm \log[O/H]_{\odot}+12=8.69$ and a bulk solar metallicity of $\rm Z_{\odot}=0.0142$ \citep{AllendePrieto2001,Asplund2004,Asplund2009}.} range $Z\in [10^{-4},0.03]$ where observational constraints for HMXBs are available. Our model can also be applied to more metal-poor regimes (e.g., $Z\lesssim 10^{-6}$ for Pop~III stars) that are likely more important at Cosmic Dawn \citep{Sartorio2023}. %under certain assumptions motivated by observations (see Sec.~\ref{sec:xray}).

The paper is structured as follows. In Section~\ref{sec:bps}, we provide an overview of our method and discuss the setup of BPS parameters, binary sample and initial conditions. In Section~\ref{sec:bexrb}, we explain our physically motivated model for the identification and characterization of Be-XRBs. In Section~\ref{sec:spec}, we build a empirical model for the X-ray spectra of Be-XRBs. 
In Section~\ref{sec:res} we present our predictions on the formation efficiency (Sec.~\ref{sec:eps}), mass and orbital parameter distributions (Sec.~\ref{sec:properties}), and X-ray outputs of Be-XRBs (Sec.~\ref{sec:lxsfr}), focusing on how they evolve with metallicity. %{\color{blue}(optional) We also look into the special situation of Be-XRBs from Pop~III stars in Sec.~\ref{sec:pop3}.} 
Finally, we summarize our main findings in Section~\ref{sec:sumy}, 
and discuss their caveats and our outlook to future work in Section~\ref{sec:dis}. 
The key physical quantities used in this paper are summarized in Table~\ref{tab:symbol}.

\begin{table}
    \centering
    \caption{Key physical quantities.}
    \begin{tabular}{l|l}
    \hline
        $Z$ & absolute metallicity (mass fraction of metals) \\
        $M_{1}$ & initial primary mass \\
        $M_{2}$ & initial secondary mass\\
        %$W$ & ratio of equatorial rotation velocity and Keplerian velocity\\
        $a$ & orbital separation (semi-major axis)\\
        $e$ & orbital eccentricity\\
        $P_{\rm orb}$ & orbital period\\
        $P_{\rm s}$ & spin period of the NS\\
        $M_{\rm X}$ & compact object (NS/BH) mass\\
        $M_{\star}$ & mass of the donor star\\
        $R_{\star}$ & equatorial radius of the donor star\\
        $v_{\rm Kep}$ & equatorial Keplerian velocity of the donor star\\
        $v_{\rm rot}$ & equatorial rotation velocity of the donor star\\
        $W$ & $\equiv v_{\rm rot}/v_{\rm Kep}$ with the initial value denoted by $W_{0}$\\
        %$T_{\rm eff}$ & effective temperature of the donor star\\
        $R_{\rm L1}$ & Roche lobe size of the donor star at periastron (Eq.~\ref{rl1})\\
        $R_{\rm trunc}$ & $\equiv f_{\rm trunc}a$, average tidal truncation radius (Eq.~\ref{rtrunc})\\
        $R_{\rm crit}$ & VDD boundary beyond which gas flows are subsonic (Eq.~\ref{rcrit})\\
        $c_{s}$ & sound speed in the ionised isothermal VDD \\%$=\sqrt{2k_{\rm B}T_{\rm d}/m_{\rm p}}$\\
        $\Sigma_{0}$ & base surface density of the VDD\\
        $\alpha$ & viscosity parameter of the VDD\\
        $\dot{M}_{\rm ej}$ & mass ejection rate of the O/B star\\
        $\dot{M}_{\rm acc}$ & peak/outburst accretion rate in the Be-XRB\\
        $L_{\rm bol}$ & bolometric luminosity during outbursts\\
        $\epsilon$ & radiative efficiency\\
        $\dot{M}_{\rm Edd}$ & Eddington limit of accretion rate (Eq.~\ref{edd})\\
        $\eta$ & $\equiv \dot{M}_{\rm acc}/\dot{M}_{\rm Edd}$, Eddington ratio during outbursts\\
        $f_{\rm duty}$ & effective fraction of time the Be-XRB spends in outbursts\\
        $L_{\rm X}$ & outburst X-ray luminosity for a certain band\\
        $\psi_{\rm X}$ & calibration parameter for the observed $L_{\rm X}-P_{\rm orb}$ relation\\
        $f_{\rm corr}$ & correction factor for outburst luminosity (Sec.~\ref{sec:xray})\\
        $\tau$ & lifetime of the Be-XRB\\
        $M_{\rm tot}$ & total stellar mass underlying the Be-XRB population\\
        $\mathcal{N}_{\rm X}$ & number of Be-XRBs in the outburst phase per unit SFR\\
        $\mathcal{L}_{\nu}$ & specific X-ray
luminosity per unit SFR\\
        $\mathcal{L}_{\rm X}$ & X-ray
luminosity per unit SFR for a certain band\\
    \hline
    \end{tabular}
    \label{tab:symbol}
\end{table}

%\section{Methodology}\label{sec:method}

\section{Binary population synthesis}\label{sec:bps}

We add a new module for the identification and characterization of Be-XRBs to the BPS %binary population synthesis (BPS) 
code \href{https://binary_c.gitlab.io/}{\textsc{binary\_c}} {\citep{Izzard2004,Izzard2006,Izzard2009,Izzard2017,Izzard2018,Izzard2023,Mirouh2023,Hendriks2023gw,Hendriks2023ms,Yates2023}}, which simulates the evolution of stars in each binary and the binary orbit governed by binary interactions (e.g., mass transfer and tidal effects) and stellar evolution processes such as winds and supernovae (SNe). We evolve large populations of binaries from zero-age main-sequence (ZAMS) for $15$~Gyr in the (absolute) metallicity range $Z\in [10^{-4},0.03]$ with randomly sampled initial binary properties %, using the standard Binary Star Evolution (BSE) models \citep{Hurley2002} in \textsc{binary\_c} (with additional updates, see below) 
through the \texttt{python} interface \href{https://gitlab.com/binary_c/binary_c-python}{\textsc{binary\_c-python}} \citep{Hendriks2023} of \textsc{binary\_c}. The Be-XRB model is explained in detail in the next Section~\ref{sec:bexrb}. 
An X-ray spectral model based on observations (Sec.~\ref{sec:spec}) is applied to the Be-XRB populations generated by \textsc{binary\_c} in post-processing to calculate their X-ray outputs. 
%DeMarco2017

As shown in previous BPS studies \citep[e.g.,][]{Vinciguerra2020,Xing2021}, the main channel of Be-XRB formation is expected to be stable mass transfer during the main sequence (MS) and Hertzsprung gap (HG) phases, where the initial secondary star grows by accretion from the initial primary star and is meanwhile spun up to become an O/Be star. Thereafter, if the initial primary star collapses into a compact object when the O/Be star is still on MS, and the system remains bound, we can obtain a Be-XRB. Therefore, formation of Be-XRBs is sensitive to binary stellar evolution parameters governing the stability and efficiency of mass transfer, angular momentum loss, as well as natal kicks of SNe \citep[see, e.g.,][]{Shao2014,Vinciguerra2020,Xing2021}. %\citep[see][for a comprehensive discussion of the impact of these parameters on Be-XRB formation]{Vinciguerra2020}. 
The initial binary properties may also play an important role. 

In this work, we use the standard BSE models \citep{Hurley2002} with the default setup of \textsc{binary\_c}\footnote{\textsc{binary\_c} has been updated recently with a new treatment of pair-instability SNe {\citep{Farmer2019,Hendriks2023gw}}, an improved stellar wind prescription \citep{Schneider2018,Sander2020} and stellar evolution of zero-metallicity stars based on \textsc{mesa} data \citep{Paxton2018,Paxton2019}. It is also used to study XRBs from zero-metallicity stars \citep{Sartorio2023}, but only considering XRBs powered by RLO and stellar winds (without Be-XRBs).} with an updated stellar wind model from \citet{Schneider2018} and \citet{Sander2020} as well as a special treatment of mass-transfer efficiency (Sec.~\ref{sec:acc}). In addition to the mass-transfer efficiency, we briefly explain our choices of select BSE parameters that are important for Be-XRBs in Sec.~\ref{sec:others} according to the default setup of \textsc{binary\_c} detailed in \citet{Izzard2017}. 
It is shown in Appendix~\ref{apx:smc} that our choices of BSE parameters, combined with standard initial conditions of binary stars (Sec.~\ref{sec:ic}), can reproduce the population of observed Be-XRBs in the SMC at the metallicity $Z_{\rm SMC}=0.0035$ \citep{Davies2015}. For simplicity, we assume that the BSE parameters and initial conditions do not evolve with metallicity for $Z\in[10^{-4},0.03]$. %and defer a thorough investigation into the 

{The BSE models used here keep track of the spin evolution of each star regulated by mass loss, accretion and tidal interactions, as detailed in \citet{Hurley2002}. In particular, during stable mass transfer via RLO, the accretor gains angular momentum from the accreted material, which is assumed to come from the inner edge of an accretion disc with the specific angular momentum of the circular orbit on the surface of the accretor. %, which is a conservative estimate of the accretion spin-up efficiency. 
However, }the effects of rotation on stellar evolution are not considered, which can be significant (particularly for initially fast-rotating stars) and complex, covering various aspects (e.g., mass loss, timescales of evolution phases, stellar structure, nucleosynthesis and remnant masses), especially at low $Z$, as shown in detailed stellar-evolution simulations \citep[e.g.,][]{Ekstrom2012,Georgy2013,Choi2017,Groh2019,Murphy2021}. Such effects may also be important in the modelling of Be-XRBs, as fast-rotating O/B stars are involved by definition. However, it is beyond the scope of this work to take into account these effects because the detailed mechanisms that connect the formation and properties of VDDs (i.e., the so-called `Be phenomenon') with stellar evolution processes are still unresolved \citep{Rivinius2013}.

{ For simplicity, we ignore the mass growth of compact objects via accretion in Be-XRBs, so that our Be-XRB module does not affect binary stellar evolution. This approximation is justified by the fact that VDDs are very light structures \citep{Rivinius2019} that cannot supply much mass to compact objects. We have verified that among all compact objects in the Be-XRBs simulated in this work, the mass accreted from the VDD is less then a few percent of the initial mass in most ($\gtrsim 99\%$) cases, and remains below $50\%$ of the initial mass in the most extreme systems with massive VDDs.}

%\subsection{Binary evolution parameters}\label{sec:param}
\subsection{Mass-transfer efficiency}\label{sec:acc}
It is found by \citet{Vinciguerra2020} using the \textsc{compas} code \citep{Riley2022} that efficient accretion during stable mass transfer is required to reproduce the observed orbital period distribution of Be-XRBs in the SMC\footnote{Otherwise mass and angular momentum loss during mass transfer shrink binary orbits too efficiently leading to over-prediction of low-period systems.}. Since \textsc{compas} is also based on the BSE models, we expect that an enhancement of mass-transfer efficiency with respect to the default prescription of \textsc{binary\_c} is necessary in our case. Therefore, we set the mass-transfer efficiency parameter, i.e., the ratio of the accreted mass to mass lost by the donor, as
\begin{align}
    \beta=\min\left(1,\beta_{\rm thermal}\times\frac{\dot{M}_{\rm acc,\max}}{\dot{M}_{\rm donor}}\right)\ .\label{beta}
\end{align}
Here $\dot{M}_{\rm donor}$ is the mass loss rate of the donor, $\beta_{\rm thermal}$ is the maximal mass-transfer efficiency defined with respect to the maximal steady-state mass acceptance rate $\dot{M}_{\rm acc,\max}\sim(\epsilon_{\rm g,acc}/L_{\rm acc})^{-1}\sim L_{\rm acc}R_{\rm acc}/(GM_{\rm acc})\sim M_{\rm acc}/t_{\rm KH,acc}$, which is limited by the thermal (Kelvin-Helmholtz) timescale $t_{\rm KH}\sim GM_{\rm acc}^{2}/(R_{\rm acc}L_{\rm acc})$ of the accretor, given $\epsilon_{\rm g,acc}\sim GM_{\rm acc}/R_{\rm acc}$ the specific energy carried by the in-falling matter that needs to be radiated away, the luminosity $L_{\rm acc}$, mass $M_{\rm acc}$ and radius $R_{\rm acc}$ of the accretor. \textsc{binary\_c} adopts a conservative choice $\beta_{\rm thermal}=1$ by default, while here we use $\beta_{\rm thermal}=30$, which is approximately the value required to match observations inferred by \citet{Vinciguerra2020}. This rather large value of $\beta_{\rm thermal}$ captures the variation of $t_{\rm KH}$ during mass transfer by expansion and increase of luminosity \citep{Paczynski1972,Hurley2002,Vinciguerra2020}. Similarly, we also increase the upper limit on $\beta$ from the dynamical timescale of the accretor as well as the thermal and dynamical timescales of the donor by a factor of 30. Although such efficient mass transfer is required to reproduce the observed Be-XRBs in the SMC (see Appendix~\ref{apx:smc}), we find by numerical experiments that the total X-ray output from Be-XRBs is insensitive to $\beta_{\rm thermal}$. The reason is that the total X-ray output is dominated by luminous Be-XRBs mainly on eccentric orbits ($e\gtrsim 0.1$) whose progenitor primary stars only undergo weak mass loss (i.e., $\beta=1$ for $\dot{M}_{\rm donor}\ll \dot{M}_{\rm acc,\max}$, independent of $\beta_{\rm thermal}$), as discussed in Sec.~\ref{sec:res}. 
%time required for the accretor to radiate away .

%\citep{Hurley2000,Hurley2002,Izzard2004,Izzard2006,Izzard2009,Izzard2017,Izzard2023,Schneider2018,Sander2020,Sartorio2023}
%,Paxton2018,Paxton2019}

\subsection{Other key BSE parameters}\label{sec:others}
%Default \textsc{binary\_c} models...
In addition to mass-transfer efficiency, the properties of Be-XRBs are also expected to depend on the prescriptions for mass transfer stability, angular momentum loss, remnant masses and SN natal kicks \citep{Vinciguerra2020}. We plan to explore their effects in the future {(see Sec.~\ref{sec:dis})}. Here we briefly describe the default choices for these parameters adopted in our work for \textsc{binary\_c}. 

{Mass transfer is stable when $M_{\rm accretor}/M_{\rm donor}>q_{\rm crit}$ at the onset of mass transfer given the critical mass ratio $q_{\rm crit}$.} %In our case, only mass transfer to stellar accretors (not compact objects) are relevant for the formation of Be-XRBs. Here 
We adopt $q_{\rm crit}=5/8$, 1/3 and 1/4 in the hydrogen MS, helium MS and HG phases for both hydrogen and helium burning of the donor, respectively. In the giant phase, we use the prescription in \citet[see their sec.~2.6]{Hurley2002}. Mass transfer beyond what is allowed by the mass-transfer efficiency (Eq.~\ref{beta}) is lost from the system. {We adopt the %isotropic re-emission model \citep{Soberman1997} 
fast (also called Jeans) model \citep{Huang1963} to calculate the angular momentum loss in this process, i.e., the specific angular momentum carried by the lost mass is equal to the specific angular momentum of the donor. We have verified by numerical experiments using the isotropic re-emission model \citep{Soberman1997}, in which the lost material carries the specific angular momentum of the accretor, that the prescription of angular momentum loss has minor effects on our results (with $\lesssim 20\%$ and $\lesssim 50\%$ changes in the total X-ray output for $Z\lesssim 0.01$ and $Z\sim 0.01-0.03$, respectively). The reason is that under the high mass-transfer efficiency described in Sec.~\ref{sec:acc}, the mass/angular momentum loss during stable mass transfer has little impact on the orbital parameters (and luminosities) of Be-XRBs, which are more sensitive to stellar winds and SN natal kicks (see below and Sec.~\ref{sec:comp_bps}).}

The masses of compact object remnants are determined by the CO core masses of progenitors using the original BSE models in \citet{Hurley2000} and \citet{Hurley2002}. We apply natal kicks to Type~II and Ib/c SNe that follow a Maxwellian distribution with a dispersion of $\sigma_{\rm kick}=190\ \rm km\ s^{-1}$ \citep{Hansen1997}, while electron-capture SNe have no natal kicks. The latter typically happen to highly stripped stars in progenitor binaries of Be-XRBs (see Sec.~\ref{sec:properties}), which are expected to have weak natal kicks \citep[$\lesssim 30\ \rm km\ s^{-1}$, see the discussion in sec.~3.1 of][]{Vinciguerra2020}. Therefore, we use zero natal kicks for simplicity.

\subsection{Binary sample and initial conditions}\label{sec:ic}
To construct the input catalog of binary stars, we sample $N_{\rm B}=3\times 10^{5}$ binaries randomly from widely used distributions of mass and orbital parameters. %\subsubsection{Mass distribution}
To be specific, the primary stellar mass $M_{1}$ is drawn from the \citet{Kroupa2001} initial mass function (IMF) in the mass range of $[5,100]\ \rm M_{\odot}$ and the mass ratio $q\equiv M_{2}/M_{1}$ is generated from a uniform distribution in the range $q\in [0.1\ {\rm M_{\odot}}/M_{1},1]$. Here we only consider binaries with $M_{1}\ge 5\ \rm M_{\odot}$ because only massive primary stars can form the compact objects considered in our Be-XRB model\footnote{We have verified by numerical experiments, including systems with less-massive primary stars, that all Be-XRB progenitors must have $M_{1}>5\ \rm M_{\odot}$. } (see Sec.~\ref{sec:id}). In the way, \textit{the binary stars in our catalog} only make up a small fraction of the whole underlying stellar population. Following \citet{Misra2022}, we assume that the whole stellar population is made of $70\%$ binary stars \citep{Sana2012} and $30\%$ single stars. For the whole stellar population, the single stars and the primary stars in binaries also follow the \citet{Kroupa2001} IMF but in the range $M_{1}\in [0.01,100]\ \rm M_{\odot}$, and the mass ratio distribution for the entire binary star population is uniform in $q\in[0.01\ {\rm M_{\odot}}/M_{1},1]$. Under these assumptions\footnote{In our case, the mass distribution of all stars, including single stars, primary and secondary stars in binaries, do not strictly follow the \citet{Kroupa2001} IMF. Nevertheless, for stars above $1\ \rm M_{\odot}$ that are relevant for Be-XRBs, the mass distribution is very close to the \citet{Kroupa2001} IMF with small ($\lesssim 20$\%) deviations above $\sim 60\ \rm M_{\odot}$ and a minute Wasserstein distance of 0.04. Therefore, we expect this imperfect sampling of the \citet{Kroupa2001} IMF to have negligible effects on our results.}, we estimate that the total mass of stars in our binary sample accounts for $f_{\rm sample}=0.2$ of the total mass of the whole stellar population, using the method in appendix A of \citet{Bavera2020}. Here $f_{\rm sample}$ serves as a normalization factor for the calculation of X-ray outputs per unit stellar mass or SFR. The mass of the whole stellar population corresponding to our binary sample is $M_{\rm tot}=f_{\rm sample}^{-1}\sum_{i}^{N_{\rm B}}(M_{1,i}+M_{2,i})\sim 3\times 10^{7}\ \rm M_{\odot}$. 

For orbital parameters, by default we follow \citet{Izzard2017} to draw the initial semi-major axis $a$ from a log-flat distribution for $a\in[3,10^{4}]\ \rm R_{\odot}$ and the initial eccentricity $e$ from a thermal distribution for $e\in [0,1)$. We assume no correlations between $a$, $e$ and masses of stars, while evidence of such correlations has been found in observations \citep[e.g.,][]{Moe2017}. %{\color{blue}(optional)}
We also consider an alternative model in which we draw the initial orbital period $P_{\rm orb}$ (in the unit of day) from a hybrid distribution $f_{P_{\rm orb}}\equiv {dN}/{d\log P_{\rm orb}}$ based on observations of low-mass \citep{Kroupa1995} and massive \citep{Sana2012} stars, motivated by the ideas in \citet[see their appendix B1]{Izzard2017} and \citet[see their sec. 3.2.2]{Sartorio2023}:
\begin{align}
    f_{P_{\rm orb}}=\begin{cases}
        f_{\rm Kroupa}(1-\tilde{m})+f_{\rm Sana}\tilde{m}\ , \quad &\tilde{m}<1\ ,\\
        f_{\rm Sana}\ ,\quad &\tilde{m}\ge 1\ ,
    \end{cases}\label{fporb}
\end{align}
where $\tilde{m}\equiv M_{1}/M_{\rm O}$, given $M_{\rm O}=16\ \rm M_{\odot}$ as the minimum mass of O stars, and \citep{Kroupa1995,Sana2012}
\begin{align}
    f_{\rm Kroupa}=2.5(\log P_{\rm orb}-1)/[45+(\log P_{\rm orb}-1)^{2}]\ ,\label{fporb_kroupa}\\
    f_{\rm Sana}\propto (\log P_{\rm orb})^{-0.55}\ .\label{fporb_sana}
\end{align}
It turns out that that the results in this case are very similar to those of the default model (see Appendix~\ref{apx:model}). Therefore, we only show the results of the default model in our main text. We defer a more detailed investigation of initial binary parameters to future work.

Finally, another initial condition parameter that can be important for Be-XRBs is the initial stellar rotation velocity $v_{\rm rot,0}$, which can be characterized by the parameter $W_{0}\equiv v_{\rm rot,0}/v_{\rm Kep,0}$ given the initial Keplerian velocity $v_{\rm Kep,0}$ at the stellar equator. The reason is that rapid rotation is required to make O/Be stars, which is also used in our model to identify Be-XRBs (see Sec.~\ref{sec:id}). The chance of forming O/Be stars is expected to be higher for stars with faster initial rotation. Here we consider two models for $W_{0}$. {In our slowly-rotating (SR) model, we adopt the fit formula for $v_{\rm rot,0}$ from \citet[see their sec.~7.2]{Hurley2000} based on the MS data in \citet{Lang1992}:
\begin{align}
    v_{\rm rot,0}=330\ {\rm km\ s^{-1}}(M/{\rm M_\odot})^{3.3}/[15+(M/{\rm M_\odot})^{3.45}]
\end{align}
given the initial stellar mass $M$. In this case, $W_{0}$ is a function of $M$ and metallicity (which determines the initial stellar radius and $v_{\rm Kep,0}$ given $M$).
%as in \citet[see their sec.~7.2]{Hurley2000} and \citet{Hurley2002}. %, such that $W_{0}\sim 0.17-0.58$ for $M_{\star}\sim 2-100\ \rm M_{\odot}$. %and the peak value $W_{0}\sim 0.43$ is reached at $M_{\star}\sim 4\ \rm M_{\odot}$. %(see Fig.~\ref{fig:w0}). 
}
In our fast-rotating (FR) model, we set $W_{0}=0.9$ for all stars to obtain an upper limit on the formation efficiency as well as X-ray output of Be-XRBs. %as an extremely optimistic choice. 
The FR model can also be regarded as the asymptotic situation when we decrease metallicity, %($Z\lesssim 0.5\ \rm Z_{\odot}$), 
since more metal-poor stars are more likely to be fast-rotating \citep[e.g.,][]{Ekstrom2008,Bastian2017,Schootemeijer2022}. %, and zero-metallicity stars are expected to \citep{Stacy2011,Stacy2013,Hirano2018}.

\begin{comment}
\begin{figure}
    \centering
    \includegraphics[width=1\columnwidth]{W0_ms1.pdf}
    \vspace{-25pt}
    \caption{Initial rotation rate parameter as a function of initial stellar mass in \citet{Hurley2000} and \citet{Hurley2002}.}
    \label{fig:w0}
\end{figure}
\end{comment}

\section{Be-XRB model}\label{sec:bexrb}

\subsection{Identification of Be-XRBs}\label{sec:id}

Inspired by previous BPS studies on Be-XRBs \citep[e.g.,][]{Zhang2004,Belczynski2009,Linden2009,Shao2014,Shao2020,Zuo2014,Vinciguerra2020,Xing2021,Misra2022}, we identify a binary as in the Be-XRB phase with the following criteria:
\begin{enumerate}
    \item {The binary is made of a massive MS (O/B) donor star with $M_{\star}>6\ \rm M_{\odot}$, and a compact object (NS/BH) companion with a mass $M_{\rm X}> 1.29\ \rm M_{\odot}$.} We ignore Be-XRBs with white dwarfs for simplicity considering their faintness and rareness \citep{Kennea2021}. We set the mass threshold for donor stars at $6~\rm M_{\odot}$ as a conservative estimate of the minimum mass of Be stars in Be-XRBs \citep{Hohle2010}, which is larger than the minimum mass of single Be stars \citep[$\sim 3\ \rm M_{\odot}$,][]{Vieira2017}. This choice is supported by the fact that only early spectral types of Be stars (e.g., no later than B5 in the \citealt{Coe2015} SMC catalogue) that are expected to be massive have been found in observations of Be-XRBs \citep{Antoniou2009,Reig2011,Maravelias2014,Shao2014}.
    %($q_{\rm X}\equiv M_{\rm X}/M_{\star}$, $q_{\star}\equiv M_{\star}/M_{\rm X}$) %\citep{Belczynski2009,Zuo2014} or $M_{\star}>6\ \rm M_{\odot}$ \citep{Hohle2010,Misra2022} or $M_{\star}\sim 3-20\ \rm M_{\odot}$ \citep{Xing2021} corresponding to early A to late O stars involved in observed Be-XRBs \citep{Rivinius2013}
    \item {There is a VDD around the donor star}, which we assume to be present when the following conditions are satisfied:
    \begin{enumerate}
        \item The donor star is fast-rotating 
        %\footnote{For initial conditions, we can assume that all Pop~III stars are born fast-rotating, e.g., with $W\sim 0.8$ \citep{Stacy2011,Stacy2013,Hirano2018} and also consider the pessimistic case with $W=0$ initially. Since it is likely that most O/Be stars are spun up by mass transfer \citep{Klement2019,DorigoJones2020,Vinciguerra2020,Dallas2022}, the initial rotation may be unimportant.} 
        with $W\equiv v_{\rm rot}/v_{\rm Kep}> 0.7$ to eject mass that can potentially settle into a VDD, where $v_{\rm rot}$ is the rotation velocity and $v_{\rm Kep}$ is the Keplerian velocity at the equator of the donor star. Here we adopt the minimum rotation rate $W_{\rm crit}=0.7$ for decrection disk formation suggested by \citet{Rivinius2013} based on observations.
        \item The orbital period is not too small, i.e., $P_{\rm orb}> 7$ days, otherwise the VDD cannot form due to tidal forces from the companion \citep{Panoglou2016,Panoglou2018,Rivinius2019}. All of the $\sim 170$ Be-XRBs detected so far have $P_{\rm orb}>10$~days \citep{Raguzova2005,Coe2015,Antoniou2016} except for one object in the SMC, [MA93] 798, with $P_{\rm orb}\sim 0.7-2.7$~days \citep{Schmidtke2013}.
        %{\color{cyan} Alternatively, for simplicity, we can assume that a fraction $f_{\rm Be}\sim 0.2-0.3$ of B stars will be O/Be stars (in B star compact object binaries).}
    \end{enumerate}

    \item {The VDD overfills the Roche lobe of the O/B star at periastron to allow accretion by the compact object from the VDD}: $R_{\rm out}>R_{\rm L1}(a,e,q_{\star})$, where \citep{Eggleton1983} 
    \begin{align}
            R_{\rm L1}(a,e,q_{\star})= \frac{0.49q_{\star}^{2/3}(1-e)a}{0.6q_{\star}^{2/3}+\ln(1+q_{\star}^{1/3})}\label{rl1}
    \end{align}
    is the Roche lobe radius given the semi-major axis $a$, eccentricity $e$ and mass ratio $q_{\star}\equiv M_{\star}/M_{\rm X}$, and $R_{\rm out}$ is the effective disk boundary. 
    
    In previous studies, $R_{\rm out}$ is either fixed to a typical value \citep[e.g., $100\ \rm R_{\odot}$ in][]{Misra2022}, or set to the average tidal truncation radius \citep{Zhang2004,Xing2021}
    \begin{align}
        R_{\rm trunc}=f_{\rm trunc}a\ ,\quad f_{\rm trunc}=N_{\rm trunc}^{-2/3}(1+q_{\rm X})^{-1/3}\ ,\label{rtrunc}
    \end{align}
    given $q_{\rm X}\equiv M_{\rm X}/M_{\star}$, and $N_{\rm trunc}=3$ assuming a typical viscosity parameter $=0.63$ \citep{Rimulo2018}\footnote{In the calculation of $f_{\rm trunc}$ for each Be-XRB, we further introduce a relative scatter with respect to Eq.~\ref{rtrunc} following a Gaussian distribution of $\sigma=10\%$ to capture the variations of $\alpha$ (and $N_{\rm trunc}$) from system to system. }. The latter definition excludes Be-XRBs with low eccentricities $e\lesssim 0.1$ \citep[see, e.g., fig.~6 in][]{Xing2021}, even though such systems have been found in observations \citep[e.g., CPD-29 2176,][]{Richardson2023}. The reason is that tidal truncation at $R_{\rm trunc}$ is not an absolute cutoff. What happens is that materials accumulate within $R_{\rm trunc}$ and the disk density profile becomes much steeper beyond $R_{\rm trunc}$ than within $R_{\rm trunc}$ \citep{Okazaki2002,Panoglou2016}, such that accretion is still possible, although weaker, when $R_{\rm trunc}<R_{\rm L1}(a,e,q_{\star})$. Such weak accretion beyond the truncation radius can also explain the presence of persistent low-luminosity Be-XRBs in observations \citep[e.g.,][]{Sguera2023}.
    
    In light of this, we consider an optimistic and physically motivated definition of VDD boundary as the radius beyond which gas flows in the disk become subsonic \citep{Krticka2011}: 
    \begin{align}
        R_{\rm crit}=0.3(v_{\rm Kep}/c_{s})^{2}R_{\star}\ \label{rcrit}    
    \end{align}
    according to the equatorial radius $R_{\star}\equiv R_{\rm eq}$ of the O/B star, the sound speed $c_{s}=\sqrt{2k_{\rm B}T_{\rm d}/m_{\rm p}}$ in the ionised isothermal VDD of a temperature $T_{\rm d}=0.6T_{\rm eff}$ \citep{Carciofi2006} given the stellar effective temperature $T_{\rm eff}$, where $k_{\rm B}$ is the Boltzmann constant and $m_{\rm p}$ is proton mass. In this case, the tidal truncation effect is considered in the calculation of peak accretion rate (see Sec.~\ref{sec:lx}). Given this optimistic definition of VDD boundary, we are able to reproduce the nearly circular ($e=0.06\pm 0.06$) Be-XRB, CPD-29 2176, from progenitor binaries of stars in the mass range $M_{1,2}\sim 10-12\ \rm M_{\odot}$ with weak SN natal kicks, similar to the progenitors identified in the BPASS models \citep[see their table~2]{Richardson2023}. %\citep[3:1 resonant truncation;][]{Zhang2004}, $n=3$ for $\alpha\sim 0.17-0.7$ and $n=4$ for $\alpha\sim 0.037-0.17$ \citep{Okazaki2002}\\
            %{\color{cyan}Alternatively, we can derive $R_{\rm trunc}$ (and $f_{\rm trunc}=R_{\rm trunc}/a$) from the empirical relation $a=(7\times 10^{-12})R_{\rm trunc}^{2}+0.4524R_{\rm trunc}+4.3\times 10^{10}$ (with both $a$ and $R_{\rm trunc}$ in metres) based on observations of Be-XRBs in the SMC \citep{Coe2015}. We further consider the optimistic case with $R_{\rm out}=R_{\rm crit}$, assuming that accretion onto the NS is non-negligible even when the inner, dense part of the disk within $R_{\rm trunc}$ does not fill the Roche lobe (at pericenter). In this case ($R_{\rm trunc}<R_{\rm L1}$), we reduce the accretion rate by a factor (see below) to take into account the steepening of disk density profile beyond $R_{\rm trunc}$. }
    \item {The O/B star itself does not fill the Roche lobe at the periastron}: $R_{\star}<R_{\rm L1}(a,e,q_{\star})$. Otherwise, the system will be classified as a RLO XRB \citep{Reig2011}.
    
    %\item X-ray luminosity: see Eq.~5 in \citet{Misra2022}, which is a fit by \citet{Dai2006} to the observational data from \citet{Raguzova2005}
    %\item Duty cycle: $0.1-0.3$ in observations \citep{Reig2011,Sidoli2019}%\\ (Can we design a physically motivated model?)
\end{enumerate}

\subsection{X-ray outbursts of Be-XRBs}\label{sec:xray}
%During the Be-XRB phase, X-ray outbursts are triggered by strong accretion of the compact object from the VDD. 
To the first order (ignoring the contributions from quiescent phases), X-ray emission of a Be-XRB can be described by (1) the bolometric luminosity of accretion flows around the compact object $L_{\rm bol}=\epsilon\dot{M}_{\rm acc}c^{2}$ during outbursts where $\epsilon$ is the radiative efficiency and $\dot{M}_{\rm acc}$ is the peak accretion rate, and (2) the duty cycle $f_{\rm duty}$, i.e., the effective fraction of time the binary spends in X-ray outbursts during which the average luminosity is $L_{\rm bol}$. Previous studies \citep[e.g.,][]{Zuo2014,Misra2022} usually adopt empirical scaling laws or typical values for $L_{\rm bol}$ and $f_{\rm duty}$, which do not fully take into account the dependence of X-ray emission on stellar and orbital properties (e.g., eccentricity) of Be-XRBs. %, making their models difficult to generalize for the low-metallicity regime ($Z\lesssim 0.2\ \rm Z_{\odot}$) where the properties of Be-XRBs can be different. 
Here we fully capture such dependence\footnote{This is an essential step in our Be-XRB modelling since the stellar and orbital properties of Be-XRBs can vary with metallicity and contribute to the metallcicty evolution of the total X-ray output.} with a physically motivated X-ray outburst model. To be specific, our model adopts simulation results calibrated to observational data to calculate $L_{\rm bol}$ (Sec.~\ref{sec:lx}), and also considers the classification of X-ray outbursts which, combined with observational constraints, is used to estimate $f_{\rm duty}$ (Sec.~\ref{sec:fduty}).  %, to capture that X-ray output to fully capture the of any metallicity evolution of the Be-XRB.

\subsubsection{Peak accretion rate \& luminosity}\label{sec:lx} 
\begin{figure}
    \centering
    \includegraphics[width=1\columnwidth]{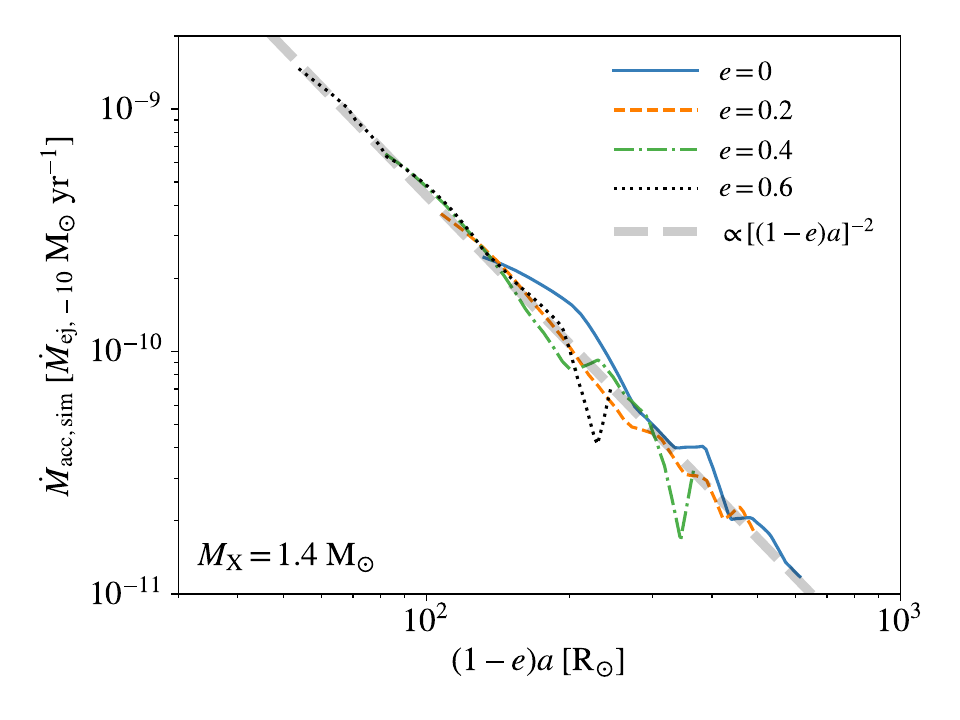}
    \vspace{-25pt}
    \caption{Relation between the peak accretion rate and pericenter distance based on the simulation results in \citet[see their fig.~7]{Brown2019} for eccentricity $e=0$ (solid), 0.2 (\textit{thin} dashed), 0.4 (dash-dotted) and 0.6 (dotted), in the unit of $\dot{M}_{\rm ej,-10}\ \rm M_{\odot}\ yr^{-1}$ given $\dot{M}_{\rm ej,-10}\equiv \dot{M}_{\rm ej}/(10^{-10}\ \rm M_{\odot}\ yr^{-1})$. The relation can be well described by a power-law $\dot{M}_{\rm acc,sim}\propto [(1-e)a]^{-2}$ as shown by the \textit{thick} dashed line. These simulations consider a NS with $M_{\rm X}=1.4\ \rm M_{\odot}$ around a Be star of $M_{\star}=18\ \rm M_{\odot}$ and $R_{\star}=7\ \rm R_{\odot}$ with $\alpha=0.63$ for $P_{\rm orb}\sim 40-400\ \rm days$ and derive the median peak accretion rate from 5 orbits after the system settles to steady state. }%, which is further calibrated to reproduce the X-ray properties of observed Be-XRBs \citep[see Eq.~\ref{lxorig} and Fig.~\ref{fig:lx_dis}]{Raguzova2005,Dai2006,Cheng2014}. }
    \label{fig:macc}
\end{figure}

We start with the steady-state peak accretion rate (assumed to be identical to the gas capture rate) predicted by the hydrodynamic simulations in %\citetalias{Brown2019} 
\citet{Brown2019}, which satisfies a simple power-law $\dot{M}_{\rm acc,sim}\propto [(1-e)a]^{-2}\dot{M}_{\rm ej}$ as shown in Fig.~\ref{fig:macc}, where $\dot{M}_{\rm ej}\propto \Sigma_{0}/\alpha$ \citep{Carciofi2008} is the steady-state mass ejection rate %\footnote{For classical O/Be stars, the steady-state mass ejection rate is typically larger by two orders of magnitude than the decretion rate, i.e., the rate at which materials settle into the VDD \citep{Vieira2017,Rimulo2018}, as most ejected mass will normally fall back to the star \citep{Okazaki2002,Haubois2012}. } 
given the base surface density $\Sigma_{0}$ and viscosity parameter $\alpha$ of the VDD. For simplicity, we fix $\alpha=0.63$ throughout our calculation based on the measurements by \citet{Rimulo2018}, so that the simulation results can be well described by
\begin{align}
    \dot{M}_{\rm acc,sim}(a,e,\Sigma_{0})&\simeq 4.4\times 10^{-10}{\ \rm M_{\odot}\ yr^{-1}}\notag\\
	&\times \left[\frac{(1-e)a}{100\ \rm R_{\odot}}\right]^{-2}\left(\frac{\Sigma_{0}}{0.015\ \rm g\ cm^{-2}}\right)%\left(\frac{0.63}{\alpha}\right)
 \ .\label{macc0}
\end{align}
This relation is obtained from a series of simulations for a Be-XRB made of a NS with $M_{\rm X}=1.4\ \rm M_{\odot}$ and a Be star of $M_{\star}=18\ \rm M_{\odot}$ and $R_{\star}=7\ \rm R_{\odot}$ with constant mass ejection rates, covering $e\in [0,0.6]$ and $P_{\rm orb}\sim 40-400$~days, 
%$P_{\rm orb}\equiv2\pi\sqrt{a^{3}/(GM)}=24.3\ {\rm day}\times\sqrt{(a/6.5\times 10^{12}\ \rm cm)^{3}/(M/18.4\ {\rm M_{\odot}})} \sim 40-400\ \rm days$ \citep{Martin2011} given $M=M_{\star}+M_{\rm X}$
which will be extrapolated to broader ranges of $e$ and $a$ in our model. %This may lead to overestimation of $\dot{M}_{\rm acc}$ considering that tidal truncation is stronger for higher $e$ and lower $a$. Besides, i
For such a Be star, %(with $M_{\star}=18\ \rm M_{\odot}$ and $R_{\star}=7\ \rm R_{\odot}$), 
we estimate the stellar luminosity as $L_{\star}\sim 3.2\times 10^{4}\ \rm L_{\odot}$ and the disk temperature as $T_{\rm d}\sim2\times 10^{4}\ \rm K$, from which we derive the disk surface density as $\Sigma_{0,\rm ref}\sim 0.015\ \rm g\ cm^{-2}$ for $\dot{M}_{\rm ej}=10^{-10}\ \rm M_{\odot}\ yr^{-1}$ (given the volume density $\rho_{0}\sim 5\times 10^{-13}\ \rm g\ cm^{-3}$ of the disk at the stellar surface for $\alpha=0.63$), which sets the normalization of Eq.~\ref{macc0}. 
%A linear relation between $\dot{M}_{\rm acc}$ and $\dot{M}_{\rm ej}$ is also found in \citet{Brown2019}. Since $\dot{M}_{\rm ej}\propto\Sigma_{0}/\alpha$ in steady state \citep{Carciofi2008}, the linear relation with $\dot{M}_{\rm ej}$ translates to a linear relation with $\Sigma_{0}$ for a fixed viscosity parameter ($\alpha=0.63$). 
%In the case of $\dot{M}_{\rm ej,-10}=1$, the base disk surface density is $\Sigma_{0,\rm ref}\sim 0.015\ \rm g\ cm^{-2}$ given the base volume density $\rho_{0}\sim 5\times 10^{-13}\ \rm g\ cm^{-3}$, luminosity $L\sim 3.2\times 10^{4}\ \rm L_{\odot}$ and radius $R_{\star}\sim 7\ \rm M_{\odot}$ of such a O/Be star with a disk temperature $T_{\rm d}\sim2\times 10^{4}\ \rm K$, which sets the normalization of Eq.~\ref{macc0}. 

In reality, mass ejection can be highly variable, even leading to disk dissipation/formation at timescales of a few years \citep{Reig2011}, and the viscosity parameter $\alpha$ can vary from system to system \citep{Vieira2017,Rimulo2018}, so that VDDs are more complex in reality than simulated by \citet{Brown2019} at steady state. Therefore, the accretion rate predicted by Eq.~\ref{macc0} should be regarded as an order-of-magnitude estimate that captures the increasing trend with decreasing pericenter distance $(1-e)a$. Finally, we multiply the peak accretion rate from Eq.~\ref{macc0} by a factor of $(R_{\rm trunc}/R_{\rm L1})^{8}$ for systems with $R_{\rm trunc}<R_{\rm L1}$ to capture the steepening of disk density profile beyond $R_{\rm trunc}$ \citep{Okazaki2002}.

\begin{figure}
    \centering
    \includegraphics[width=1\columnwidth]{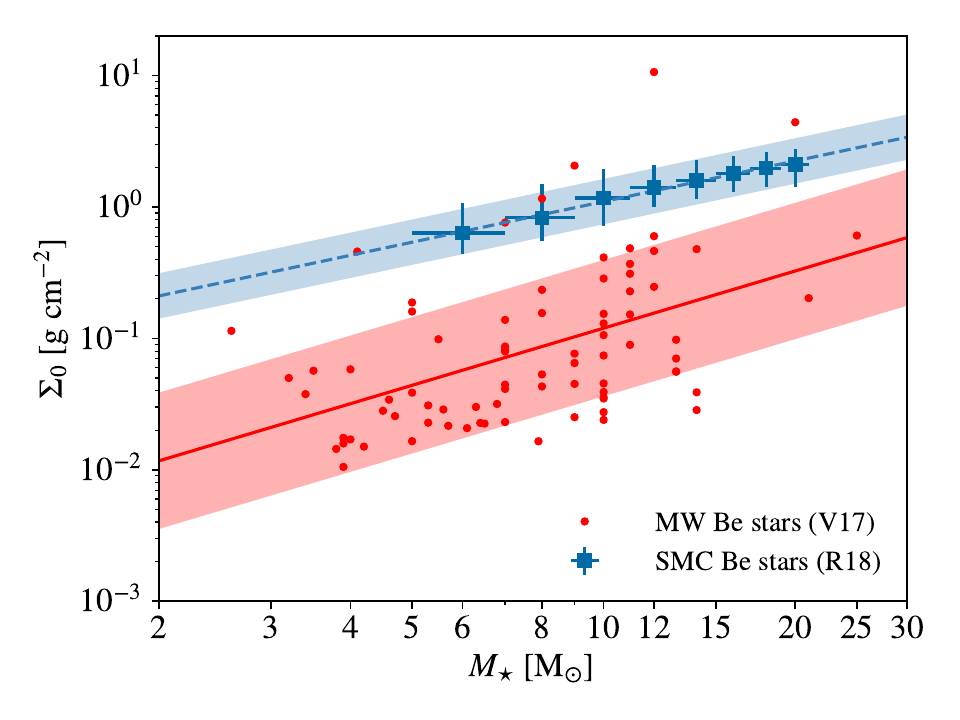}
    \vspace{-25pt}
    \caption{Relation between base disk surface density and Be star mass. The dots denote the 80 Be stars observed in the MW from \citet[V17]{Vieira2017}, which can be fit with a power-law relation $\log(\Sigma_{0}\ [{\rm g\ cm^{-2}}])\simeq 1.44\log(M_{\star}\ [{\rm M_{\odot}}])-2.37$ of $\simeq 0.52$~dex errors (shaded region around the solid line). The squares show the median base disk surface densities in 8 mass bins measured from 54 Be stars in the SMC \citep[R18]{Rimulo2018}, where the errorbars show the bin size and 25-75\% percentiles of $\Sigma_{0}$ in each bin. These results can also be described with a power-law $\log(\Sigma_{0}\ [{\rm g\ cm^{-2}}])\simeq 1.03\log(M_{\star}\ [{\rm M_{\odot}}])-0.99$ of $\simeq 0.17$~dex errors (shaded region around the dashed line). }%Note that the sample in \citet{Rimulo2018} is likely biased towards more massive O/Be stars with denser disks due to source selection, which can be used to estimate the upper limit of $\Sigma_{0}(M_{\star})$.}
    \label{fig:sigma_m}
\end{figure}

%To evaluate Eq.~\ref{macc0}, we derive $\Sigma_{0}$ from empirical scaling laws 
Next, we associate the VDD base density $\Sigma_{0}$ with the donor star mass $M_{\star}$ by fitting observational data \citep{Vieira2017,Rimulo2018}. The obtained empirical scaling laws capture the increasing trend of $\Sigma_{0}$ with $M_{\star}$ \citep{Arcos2017,Klement2017,Vieira2017,Rimulo2018}, as shown in Fig.~\ref{fig:sigma_m}. To be specific, we have
\begin{align}
    \log(\tilde{\Sigma}_{0,\rm MW}\ [{\rm g\ cm^{-2}}])\simeq 1.44\log(M_{\star}\ [{\rm M_{\odot}}])-2.37 \label{sig_mw}
\end{align}
with $\simeq 0.52$~dex scatter by fitting the data of 80 Be stars observed in the MW \citep{Vieira2017}, and 
\begin{align}
    \log(\tilde{\Sigma}_{0,\rm SMC}\ [{\rm g\ cm^{-2}}])\simeq 1.03\log(M_{\star}\ [{\rm M_{\odot}}])-0.99   \label{sig_smc} 
\end{align}
of $\simeq 0.17$~dex scatter for 54 Be stars observed in the SMC \citep{Rimulo2018}. The observations by \citet{Rimulo2018} are likely biased towards dense disks, such that Eq.~\ref{sig_smc} should be regarded as an upper limit. Besides, to consider the large scatter in $\Sigma_{0}$ at similar $M_{\star}$, for each Be-XRB, we draw a random number $\chi$ from a Gaussian distribution of a standard deviation $\sigma=0.52\ (0.17)$~dex, and set $\Sigma_{0}=10^{\chi}\tilde{\Sigma}_{0,\rm MW\ (SMC)}(M_{\star})$, given the prediction of the best-fit model $\tilde{\Sigma}_{0,\rm MW\ (SMC)}(M_{\star})$ for the MW (SMC). In addition to the donor mass dependence, we also consider the metallicity $Z$ dependence of $\Sigma_{0}$ with two cases. In the conservative (CS) case, we always use the MW model independent of $Z$, motivated by the finding that the X-ray luminosity per luminous HMXB is insensitive to metallicity for $Z\sim 0.0004-0.03$ in nearby galaxies \citep[see their fig.~5]{Douna2015}, while in the optimistic (OP) case, we assume that $\Sigma_{0}$ increases with decreasing metallicity with a linear relation between $\log(\Sigma_{0})$ and $Z$ from solar to SMC metallicities, i.e., $Z\sim 0.0035-0.0142$, and adopt the MW model for $Z>\rm Z_{\odot}=0.0142$ \citep{Asplund2009} and the SMC model for $Z<Z_{\rm SMC}=0.0035$ \citep{Davies2015}. 
 
%$\Sigma_{0}(M_{\star})\propto M_{\star}^{x}$ {given $x\sim 1.4$ with $\simeq0.52$~dex scatters in the MW \citep[V17]{Vieira2017} as the fiducial model, and $x\sim 1$ with $\simeq0.17$~dex scatters for dense disks around massive O/Be stars in the SMC \citep[R18]{Rimulo2018}} as the upper limit.

To model the peak accretion rate and luminosity more precisely, we calibrate our model with observations of Be-XRBs in the MW. To do so, we apply the above formalism (at solar metallicity) to a randomly generated sample of 10000 NS-Be star binaries with a log-flat distribution of $a$ in the range of $[10-1000]\ \rm R_{\odot}$, a uniform distribution of eccentricity for $e\in[0,0.6]$ and a log-flat distribution of $M_{\star}$ for $M_{\star}\in [6,20]\ \rm M_{\odot}$, given fixed $M_{\rm X}=1.4\ \rm M_{\odot}$. From this sample we select a mock population of Be-XRBs with $P_{\rm orb}\sim 10-300\ \rm days$ and $L_{\rm X}> 10^{34}\ \rm erg\ s^{-1}$ to be compared with observations. These conditions are chosen to mimic the statistics of most ($\sim 90$\%) well observed Be-XRBs in the MW \citep{Raguzova2005,Cheng2014,Brown2018}. The calibration target is the relation between the (outburst) X-ray luminosity $L_{\rm X}$ and orbital periods $P_{\rm orb}$, derived by \citet[D06]{Dai2006} based on 36 observed Be-XRBs from \citet{Raguzova2005}: %\footnote{We can reproduce this fit from the (38) Be-XRBs with measurements of both $P_{\rm orb}$ and $L_{\rm X}$ in \citet{Raguzova2005} excluding peculiar objects V615 Cas and 2206+543 whose $P_{\rm orb}$ measurements are based on radio and X-ray signals and may not be reliable.}:
	    \begin{align}
	    &\log\left(\frac{L_{\rm X}}{10^{35}\rm\ erg\ s^{-1}}\right)\notag\\
	    &\quad\quad =4.53\pm 0.66-(1.50\pm 0.33)\log\left(\frac{P_{\rm orb}}{1\rm day}\right)\ .\label{lxorig}    
	    \end{align}
The best-fit model indicates that the typical X-ray luminosity follows $L_{\rm X}\propto P_{\rm orb}^{-3/2} \propto a^{-9/4}$, which is similar to the $a$ dependence in $\dot{M}_{\rm acc,sim}\propto a^{-2}$. In light of this, we assume that $L_{\rm X}$ is proportional to the bolometric luminosity predicted by simulations with a calibration parameter $\psi_{\rm X}$: $L_{\rm X}=%\psi_{\rm X} L_{\rm bol,sim}=
\psi_{\rm X}\epsilon\dot{M}_{\rm acc,sim}c^{2}$, given $\epsilon=0.2$ the typical radiative efficiency for NSs. This calibration factor captures the difference between X-ray luminosity and bolometric luminosity in observations as well as the difference between the peak accretion rates predicted by simulations and those in reality. % and use the empirical law (Eq.~\ref{lxorig}) as a calibration target to derive the proportional factor $\psi_{\rm X}$. %As shown in Fig.~\ref{fig:lx_p} and \ref{fig:lx_dis}, with the fiducial model of X-ray luminosity (Eq.~\ref{lxns}), the original scaling law (Eq.~\ref{lxorig}), its scatters, and the observed $L_{\rm X}$ distribution \citep[see, e.g., fig.~3 in][C14]{Cheng2014} can be well recovered 

%For each Be-XRB in the mock population, we estimate the peak X-ray luminosity as $L_{\rm X}\sim \psi_{\rm X}\epsilon \dot{M}_{\rm acc,sim}(a,e,\Sigma_{0})$, where $\Sigma_{0}=10^{\chi}\tilde{\Sigma}_{0}(M_{\star})$, given the best-fit model $\tilde{\Sigma}_{0}(M_{\star})$ in Fig.~\ref{fig:sigma_m} for MW O/Be stars and a corresponding random variable $\chi$ that follows a Gaussian distribution of $\sigma=0.52$~dex to capture the scatters in the observed scaling relation between $\Sigma_{0}$ and $M_{\star}$. 

\begin{figure}
    \centering
    \includegraphics[width=1\columnwidth]{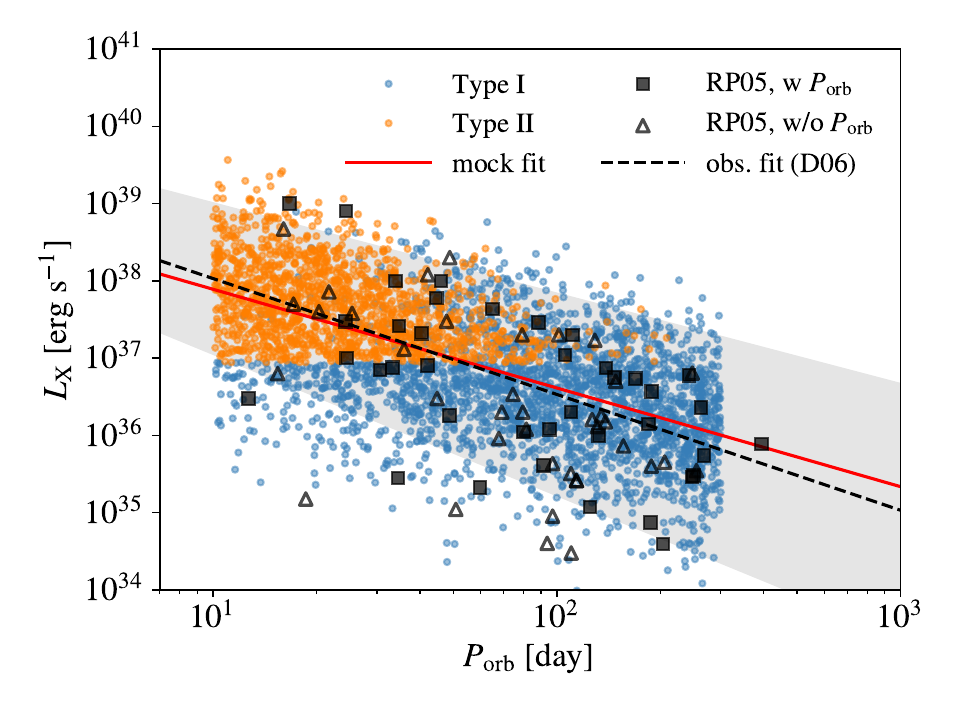}
    \vspace{-25pt}
    \caption{Relation between outburst X-ray luminosity and orbital period for the mock population of Be-XRBs with $P_{\rm orb}\sim 10-300\ \rm days$, $e\sim 0-0.6$ and $L_{\rm X}> 10^{34}\ \rm erg\ s^{-1}$, given the calibration paremeter $\psi_{\rm X}=0.25$. Type I/II Be-XRBs are shown in blue/orange. These mock Be-XRBs are identified from 10000 randomly generated NS-Be star binaries with a log-flat distribution of separations for $a\in [10-1000]\ \rm R_{\odot}$, a uniform distribution of eccentricities for $e\in[0,0.6]$ and a log-flat distribution of Be star masses for $M_{\star}\in[6,20]\ \rm M_{\odot}$ given a fixed NS mass $M_{\rm X}=1.4\ \rm M_{\odot}$. %, which are classified with the accretion rate and tidal warp criteria (Eqs.~\ref{a_tidal}-\ref{mdotc}). To evaluate Eq.~\ref{a_tidal}, we derive the (equatorial) radius $R_{\star}$ and luminosity $L$ of a O/Be star with a mass $M_{\star}$ from empirical fitting formulas based on the observed O/Be stars in the MW \citep{Vieira2017}: $\log(R_{\star}\ [{\rm R_{\odot}}])=0.41\log(M_{\star}\ [{\rm M_{\odot}}])+0.46$ of 0.13~dex scatters, $\log(L\ [{\rm L_{\odot}}])=2.74\log(M_{\star}\ [{\rm M_{\odot}}])+1.10$ of 0.16~dex scatters. A mock population of Be-XRBs, highlighted with the thick dots, is selected with $P_{\rm orb}\sim 10-300\ \rm days$, $e\sim 0-0.6$ and $L_{\rm X}> 10^{34}\ \rm erg\ s^{-1}$ to be compared with observations. 
    We show the observed Be-XRBs compiled by \citet[RP05, excluding V615 Cas and 2206+543]{Raguzova2005} with filled squares and unfilled triangles. The filled squares have $P_{\rm orb}$ measured by observations, while for the unfilled triangles without $P_{\rm orb}$ measurements, we use the empirical scaling law $\log(P_{\rm orb}\ [{\rm day}])=0.4329\log(P_{\rm s}\ [{\rm s}])+ 1.043$ \citep{Vinciguerra2020} to estimate $P_{\rm orb}$ given the spin period $P_{\rm s}$. The dashed line represents the best fit to observations from \citet{Dai2006} with the shaded region denoting the $1\sigma$ uncertainties in the fitting parameters (see Eq.~\ref{lxorig}). The solid line shows the best-fit power-law model for our mock population. It turns out that the mock population reproduces well observational data with a slightly shallower best fit and a similar scatter given $\psi_{\rm X}=0.25$.}
    %The relation between $L_{\rm X}$ and $P_{\rm orb}$ (shaded region around the dashed line, see Eq.~\ref{lxorig}) derived by \citet[D06]{Dai2006} from observed Be-XRBs \citep{Raguzova2005} is well reproduced by the mock population, which results in a slightly shallower best-fit power-law model (solid line) and a similar scatter. The observed Be-XRBs compiled by \citet[excluding V615 Cas and 2206+54]{Raguzova2005} are shown with the filled squares and unfilled triangles. The filled squares have $P_{\rm orb}$ measured by observations, while for the unfilled triangles $P_{\rm orb}$ is inferred from the empirical scaling law $\log(P_{\rm orb}\ [{\rm day}])=0.4329\log(P_{\rm s}\ [{\rm s}])+ 1.043$ \citep{Vinciguerra2020} given the spin period $P_{\rm s}$.}
    \label{fig:lx_p}
\end{figure}

If the observed X-ray luminosity completely dominates the bolometric luminosity and the simulations are realitic, we should have $\psi_{\rm X}\simeq 1$. However, we find that the empirical $L_{\rm X}$-$P_{\rm orb}$ scaling relation (Eq.~\ref{lxorig}) can be reproduced by the mock population with $\psi_{\rm X}= 0.25\ll 1$, as shown in Fig.~\ref{fig:lx_p}. There are two possible reasons for this low $\psi_{\rm X}$ value: (1) The accretion rates predicted by the simulations in \citet{Brown2019} are overestimated and/or the mass ejection rates of Be stars in Be-XRBs are lower than those of Be stars at large in the MW \citep{Vieira2017}. (2) The X-ray luminosity derived from observations in fact only represents a (small) fraction of the bolometric luminosity. To capture these uncertainties, we introduce a correction factor $f_{\rm corr}\in [\psi_{\rm X},1]$ (fixing $\psi_{\rm X}=0.25$) for the peak accretion rate $\dot{M}_{\rm acc}$. We also include a power-law term of $M_{\rm X}$ with index $\xi$ to account for the mass dependence, so that
\begin{align}
    \dot{M}_{\rm acc}=f_{\rm corr}\dot{M}_{\rm acc,sim}(a,e,\Sigma_{0})\left(\frac{M_{\rm X}}{1.4\ \rm M_{\odot}}\right)^{\xi}&\ ,\label{macc}
\end{align}
where %$\Sigma_{0}$ can be inferred from the scaling relation with the O/Be star mass $M_{\star}$ shown in Fig.~\ref{fig:sigma_m} (including the uncertainties), based on either MW Be-XRBs (fiducial) or {\color{cyan}SMC Be-XRBs (upper limit, likely more suitable for metal-poor stars)}, and 
we adopt $\xi= 2$ for Bondi-like accretion. 
Then the bolometric luminosity during outbursts of a Be-XRB can be written as
\begin{align}
	%&(f_{\rm corr}/\psi_{\rm X})L_{\rm X}\simeq L_{\rm bol}=\epsilon\dot{M}_{\rm acc}c^{2}\notag\\
        L_{\rm bol}&=\epsilon\dot{M}_{\rm acc}c^{2}=f_{\rm corr}\times 5\times 10^{36}\ {\rm erg\ s^{-1}}\notag\\
	&\times \left(\frac{\epsilon}{0.2}\right)\left[\frac{(1-e)a}{100\ \rm R_{\odot}}\right]^{-2}\left(\frac{\Sigma_{0}}{0.015\ \rm g\ cm^{-2}}\right)\left(\frac{M_{\rm X}}{1.4\ \rm M_{\odot}}\right)^{\xi}\ .\label{lbol}
\end{align}
Here $f_{\rm corr}=1$ corresponds to the optimistic case in which the discrepancy is only caused by observational effects, while $f_{\rm corr}=0.25$ is the opposite end where the overestimation in simulations needs to be corrected the most. By default, we adopt $f_{\rm corr}=\psi_{\rm X}/f_{\rm BC,0}=0.5$, assuming a typical bolometric correction (BC) factor $f_{\rm BC,0}=L_{\rm X}/L_{\rm bol}=0.5$. The motivation is that most observations of Be-XRBs in \citet{Raguzova2005} come from the $2-10$~keV band, which typically counts for $\sim 50$\% of the bolometric luminosity (see Sec.~\ref{sec:spec} below). 
%as an optimistic choice, in which case $L_{\rm bol}$ can be overestimated by up to a factor of $1/\psi_{\rm X}\sim 4$. 

Besides, we do not cape $\dot{M}_{\rm acc}$ at the Eddington rate 
\begin{align}
    \dot{M}_{\mathrm{Edd}}%=\frac{L_{\rm Edd}}{c^{2}}&\simeq \frac{1.8\times 10^{38}\ \rm erg\ s^{-1}}{c^{2}}\left(\frac{M_{\mathrm{X}}}{1.4\ \mathrm{M_{\odot}}}\right)\notag\\
    &\simeq 1.55\times 10^{-8}\ \mathrm{M_{\odot}\ yr^{-1}}\ \left(\frac{\epsilon}{0.2}\right)^{-1}\left(\frac{M_{\mathrm{X}}}{1.4\ \mathrm{M_{\odot}}}\right)\ ,\label{edd}
\end{align}
considering that Be-XRBs are promising candidates of ultra-luminous X-ray sources %(ULXs)
\citep[ULXs, with $L_{\rm X}\gtrsim 10^{39}\ \rm erg\ s^{-1}$, reviewed by, e.g.,][]{Kaaret2017,Fabrika2021,King2023} 
and a few Be-XRBs with outburst luminosities above the Eddington luminosity $L_{\rm Edd}=\epsilon\dot{M}_{\rm Edd}c^{2}\sim 2\times 10^{38}\ \rm erg\ s^{-1}$ for typical NSs with $\epsilon=0.2$ and $M_{\rm X}=1.4\ \rm M_{\odot}$ have been observed (see table~1 of \citealt{Karino2022} and Fig.~\ref{fig:lx_p}). Moreover, BPS studies show that XRBs (with both BH and NS accretors) can undergo episodes of highly super-Eddington (up to $\eta\sim 10^{3}$) mass transfer and potentially become ULXs \citep[e.g.,][]{Marchant2017,Wiktorowicz2017,Wiktorowicz2019,Wiktorowicz2021,Shao2019,Shao2020,Abdusalam2020,Kuranov2020,Misra2020}. It is discussed below that ULXs are important in our Be-XRB populations.

{We still use $L_{\rm bol}=\epsilon \dot{M}_{\rm acc}c^{2}$ when $\dot{M}_{\rm acc}>\dot{M}_{\rm Edd}$, ignoring any possible suppression of $L_{\rm bol}$ in the super-Eddington regime by, e.g., radiation-driven winds from the accretion disk\footnote{If such winds keep the accretion rate at the local Eddington rate everywhere in the disk, the total accretion luminosity is $L_{\rm bol}\simeq [(1+\ln\eta)/\eta] \epsilon\dot{M}_{\rm acc}c^{2}$ given $\eta\equiv \dot{M}_{\rm acc}/\dot{M}_{\rm Edd}>1$ \citep{Shakura1973}. We find by numerical experiments that applying this correction to $L_{\rm bol}$ reduces the X-ray outputs and number counts of ULXs from our Be-XRB populations by up to $\sim 60$\% and a factor of $\sim 10$, respectively.} \citep{Shakura1973}, so that our results should be regarded as optimistic estimates. We also ignore the beaming effects of accretion disk geometry that can boost the apparent luminosities (and reduce the \textit{observed} duty cycles or number counts) of ULXs with $\eta\equiv\dot{M}_{\rm acc}/\dot{M}_{\rm Edd}\gtrsim 8.5$ \citep{King2001,King2009,Lasota2023}. Our simple approach is motivated by the lack of features around the Eddington limits of NSs and BHs in the observed luminosity function of HMXBs, which implies that super-Eddington systems are most likely `normal' XRBs similar to their sub-Eddington counterparts \citep{Gilfanov2023}.}
%Motivated by the lack of features around the Eddington limits of NSs and BHs in the observed luminosity function of HMXBs \citep{Gilfanov2023} , which is smaller than the normal $\epsilon\dot{M}_{\rm acc}c^{2}$ by a factor of $(1+\ln\eta)/\eta$ , which is suppressed by a factor of $(1+\ln\eta)/\eta$ with respect to the normal value $L_{\rm bol}=\epsilon\dot{M}_{\rm acc}c^{2}$

\subsubsection{Classification of X-ray outbursts \& duty cycle}\label{sec:fduty} 
Now we can calculate the outburst strength by Eqs.~\ref{macc} and \ref{lbol}. We further classify the outbursts into two categories following the convention in observations to estimate the duty cycle \citep{Reig2011,Rivinius2013}: 
%\begin{itemize}
\begin{enumerate}
    \item Type~I outbursts are regular, (quasi-)periodic and short-lived ($\sim 0.1-0.3\ P_{\rm orb}$) increases of X-ray flux by a factor of $\sim 10-100$, peaking at or close to the pericentric passage of the compact object with X-ray luminosities $L_{\rm X}\lesssim 10^{37}\ \rm erg\ s^{-1}$. The duty cycle is typically $f_{\rm duty,I}\sim 0.1-0.3$ \citep{Reig2011,Sidoli2018}.
    \item Type~II outbursts are major enhancements of X-ray flux, by a factor of $10^{3}-10^{4}$, even reaching the Eddington limit. They do not have preferred orbital phases and last longer than Type~I outbursts (up to a few orbital periods). During a Type~II outburst, a radiatively efficient thin accretion disk is expected to form around the compact object, and the VDD structure can be significantly disrupted. The duty cycle is usually lower than the Type~I case: $f_{\rm duty,II}\sim 10^{-3}-0.1$ \citep{Sidoli2018,Xu2019}.
\end{enumerate}
%\end{itemize}
These two types of outbursts generally correlate with the two peaks in the observed bi-model spin period distribution of NSs in Be-XRBs, which can be divided by a critical spin period $P_{\rm s,crit}=40\ \rm s$ \citep{Cheng2014,Haberl2016,Xu2019}. To explain this correlation, it is proposed by \citet{Okazaki2013} with considerations of accretion timescale and spin-up efficiency that the two types of outbursts experience different modes of accretion: During a Type II outburst, the NS accretes \textit{at a high rate} via a radiatively efficient thin accretion disk and is spun up efficiently to have spin periods $P_{\rm s}\lesssim 40\ \rm s$. In Type I outbursts, accretion is in the form of advection dominated accretion flow (ADAF) resulting in low spins with $P_{\rm s}\gtrsim 40\ \rm s$. It is further shown by \citet{Cheng2014} that disk warping plays an important role in the spin evolution of NSs, such that Type~II outbursts tend to occur when NSs interact with tidally warped VDDs. Motivated by these results (and generalizing them to BHs), we assume that a Be-XRB will have Type~II outbursts when two criteria for (1) tidal warping and (2) accretion rate are satisfied, as defined below. 

%Motivated by the observed bi-modal spin period distribution of Be-XRBs that correlate with the two types of outbursts and the lack of periodicity in Type~II outbursts, as well as theoretical considerations of accretion timescales and spin-up efficiency \citep{Okazaki2013,Cheng2014,Haberl2016,Xu2019,Franchini2021}, we assume that a Type II outburst only occurs when the compact object accretes (via a radiatively efficient thin accretion disk) from a (misaligned) \textit{tidally warped} (eccentric) VDD at a \textit{high enough rate}. Otherwise, accretion is in the form of advection dominated accretion flow (ADAF) in Type I outbursts. 

\begin{figure}
    \centering
    \includegraphics[width=1\columnwidth]{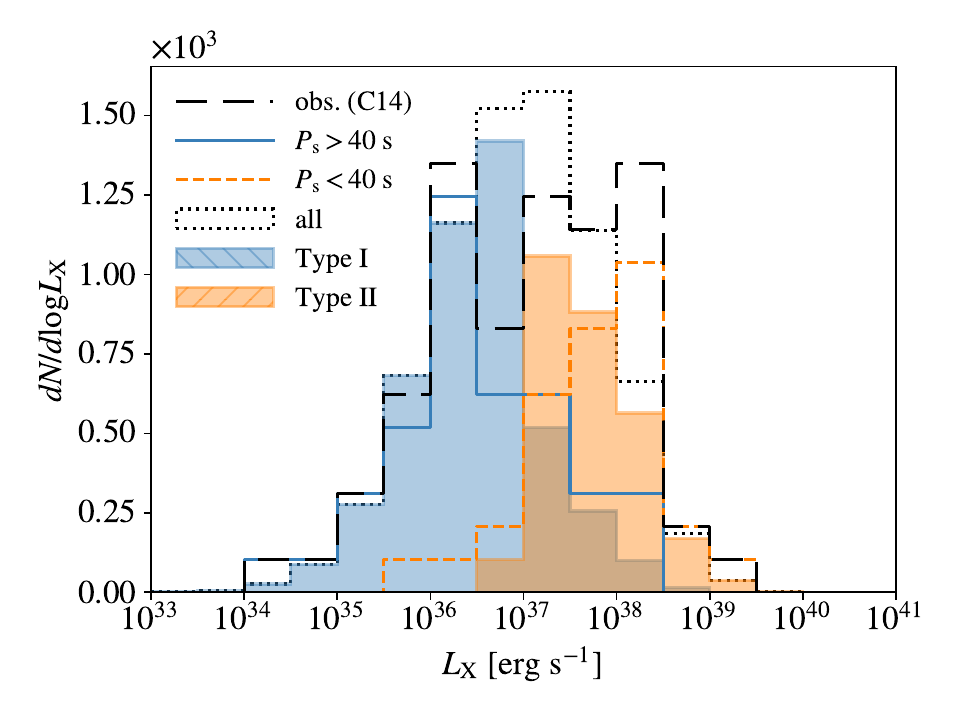}
    \vspace{-25pt}
    \caption{Distribution of outburst X-ray luminosity for the mock population of Be-XRBs (dotted contour, see Fig.~\ref{fig:lx_p}) in comparison with that of the 71 observed Be-XRBs (long-dashed contour) complied by \citet[C14, see their fig.~3]{Cheng2014}. The observed distribution is re-normalized to have the same total number of Be-XRBs. For the mock population, contributions from Type I and II outbursts are plotted with the right and left shaded histograms in blue and orange, respectively, while for the observed population, the sub groups of X-ray pulsars with spin periods $P_{\rm s}$ above and below 40 seconds are shown with the solid and dashed contours. The $L_{\rm X}$ distribution and fraction of Type I (II) events from our mock population are generally consistent with those of the observed Be-XRBs with $P_{\rm s}>\ (<)\ 40\ \rm s$.}%, which justifies our simple method of accretion mode (outburst type) distinction based on the tidal warping and accretion rate criteria.}
    \label{fig:lx_dis}
\end{figure}

\begin{enumerate}
\item[(1)] Following the analysis in \citet{Cheng2014} for the power-law+Gaussian VDD model \citep[see sec. 2.2 of][]{Martin2011}, {the tidal warping criterion is satisfied when the disk truncation radius at periastron $R_{\rm trunc,p}=f_{\rm trunc}(1-e)a$ is larger than the tidal warping radius} \citep[eq.~30 in][]{Martin2011}:
%For BH Be-XRBs, only the accretion rate criterion is considered 
%{\color{cyan}Alternatively, we can only use one of the accretion rate criterion and tidal warp criterion described below. The accretion rate criterion alone is more physically motivated, although in this case the presence of luminous ($L_{\rm X}\gtrsim 10^{37}\ \rm erg\ s^{-1}$) Be-XRBs with long spin periods ($P_{\rm s}>40\ \rm s$) may need further explanations.} 
\begin{align}
    R_{\rm tid}&=\left[\frac{2\nu_{\star}(G M_{\star})^{1/2}\bar{R}_{\rm b}^{3}}{3GM_{\rm X}R_{\star}^{n-2}}\right]^{\gamma}\notag\\
    &=a^{3\gamma}(1-e^{2})^{3\gamma/2}\left[\frac{2\alpha H_{\star}^{2}M_{\star}}{3M_{\rm X}R_{\star}^{n-1/2}}\right]^{\gamma}\ ,\label{r_tid}
\end{align}
where $\nu_{\star}=\alpha H_{\star}^{2}R_{\star}^{3/2}(GM_{\star})^{1/2}$ is the (base) disk viscosity at the stellar surface, $\bar{R}_{\rm d}=a(1-e^{2})^{1/2}$ is the average separation, $H_{\star}=\sqrt{2}c_{s}v_{\rm Kep}^{-1}R_{\star}$ \citep[$\simeq 0.04R_{\star}$,][]{Wood1997} is the (base) scale height of the disk at the stellar surface \citep{Klement2017}, and the power-law index $\gamma=2/(11-2n)\sim 0.5$ is given by $n\sim 3.5$ which is the slope\footnote{For simplicity, we adopt $n=3.5$ throughout this work assuming that the part of the disk that interacts with the compact object can be well approximated with the steady-state solution (with constant $\dot{M}_{\rm ej}$). In fact, the inner disk structure can vary significantly (with $n\sim 2-5$) in response to the variations of $\dot{M}_{\rm ej}$, magnetorotational instability and/or the presence of a companion object \citep{Carciofi2008,Haubois2012,Krticka2015,Panoglou2016,Vieira2017,Rimulo2018}.} of the disk (mid-plane) density profile \citep[see eqs.~12, 16 and 29 in][]{Martin2011}. Substituting the formula of $R_{\rm tid}$ (Eq.~\ref{r_tid}) to the iniquity $f_{\rm trunc}a(1-e)>R_{\rm tid}$, we have
\begin{align}
    & a<a_{\rm crit}\equiv \left[\frac{f_{\rm trunc}(1-e)}{(1-e^{2})^{\frac{3\gamma}{2}}}\left(\frac{3}{2\alpha}\frac{M_{\rm X}}{M_{\star}}\frac{R_{\star}^{n-1/2}}{ H_{\star}^{2}}\right)^{\gamma}\right]^{\frac{1}{{(3\gamma-1)}}}\ ,\label{a_tidal}
\end{align}
%in which $R_{\rm tid}$ is the tidal warp radius, and 
in which $f_{\rm trunc}$ is given by Eq.~\ref{rtrunc}. %, $c_{s}=\sqrt{2k_{\rm B}T_{\rm d}/m_{\rm p}}$ is the sound speed in the ionised isothermal decretion disk with a temperature $T_{\rm d}=0.6T_{\rm eff}$ \citep{Klement2017} given the effective temperature $T_{\rm eff}$ at the O/Be star equator, where $k_{\rm B}$ is the Boltzmann constant and $m_{\rm p}$ is proton mass. 
Here we use $\alpha=0.63$ to evaluate the critical seperation $a_{\rm crit}\propto\alpha^{-\gamma/(3\gamma-1)}\sim \alpha^{-1}$ in Eq.~\ref{a_tidal}, motivated by the finding in \citet{Cheng2014} that the observed populations of Be-XRBs with low ($P_{\rm s}>40$~s) and high ($P_{\rm s}<40$~s) spins, roughly corresponding to Type I and II outbursts, can be well divided by the tidal warping criterion with $\alpha\sim 0.5-1$. However, the adopted value of $\alpha$ here is much lower than the viscosity parameter for \textit{vertical} shear $\alpha_{2}=2.66$ considered in \citet{Martin2011}. It is shown below that our choice of $\alpha$ is justified by comparing the mock population of Be-XRBs with observations (Fig.~\ref{fig:lx_dis}). The discrepancy here may be caused by the fact that the VDD flares (reaching $H/R\gtrsim 0.1$ at $R\gtrsim 10 R_{\star}\sim 100\ \rm R_{\odot}$) %with $H/R\sim 0.1-0.3$ at the outer region ($R\sim 10-50R_{\star}\sim$ a few hundred $\rm R_{\odot}$) that is typically subject to tidal warping, 
while the value in \citet{Martin2011} is derived for thin ($H/R\ll 1$), flat disks \citep[]{Ogilvie1999,Lodato2010}. The disk flaring may reduce the viscosity for vertical shear and enhance vertical diffusion, making the disk more vulnerable for tidal warping (with $R_{\rm tid}$ smaller by a factor of $\sim 2$). %However, 

\item[(2)] {The accretion rate criterion can be written as} %following \citet[see also Eq.~\ref{edd}]{Okazaki2013}:
\begin{align}
    \eta\equiv \dot{M}_{\rm acc}/\dot{M}_{\rm Edd}>\eta_{\rm crit}\ .         \label{mdotc}
\end{align}
Here we adopt the typical radiative efficiencies $\epsilon=0.2$ for NSs and $\epsilon=0.1$ for BHs to calculate the Eddington rate $\dot{M}_{\rm Edd}$ (Eq.~\ref{edd}). We expect the transition Eddington ratio to be in the range $\eta_{\rm crit}\sim 0.05-0.2$, where the upper limit is adopted in \citet{Okazaki2013} to explain the outburst strengths of Type~I and II in simulations, while the lower limit is consistent with the theoretical thin disk formation criterion $\eta>0.07\alpha$ adopted in \citet{Takhistov2022} based on \citet{Pringle1987}, given the viscosity parameter $\alpha=0.63$ in our case. We set $\eta_{\rm crit}=0.2f_{\rm corr}$, because with this choice the $L_{\rm X}$ distributions of Type I and II outbursts from the mock population are generally consistent with those of observed Be-XRBs corresponding to the two peaks of spin period distribution at $P_{\rm s}>40$~s and $P_{\rm s}<40$~s \citep[see, e.g., fig.~3 in][]{Cheng2014}, as shown in Fig.~\ref{fig:lx_dis}. Since our mock population of Be-XRBs is not meant to fully capture the statistics of observed Be-XRBs complied by \citet{Cheng2014}, it does not reproduce the quasi-bi-modal feature of the observed $L_{\rm X}$ distribution.

     %, and set the transition Eddington ratio as %$\eta\equiv \dot{M}_{\rm acc}/\dot{M}_{\rm Edd}=0.07\alpha=0.044$ following the theoretical thin disk formation criterion $\eta>0.07\alpha$ adopted in \citet{Takhistov2022} based on \citet{Pringle1987}, given $\alpha=0.63$ in our case. Note that with this choice the $L_{\rm X}$ distributions of Type I and II outbursts from the mock population are generally consistent with those of observed Be-XRBs corresponding to the two peaks of spin period distribution at $P_{\rm s}>40$~s and $P_{\rm s}<40$~s \citep[see, e.g., fig.~3 in][C14]{Cheng2014}, as shown in Fig.~\ref{fig:lx_dis}. 

\end{enumerate}

%Given the typical duty cycles in observations for the two types of outbursts, 
Given the above classification, we combine the 
\textit{optimistic} duty cycles $\hat{f}_{\rm duty}$ in observations of the two types of outbursts, $\hat{f}_{\rm duty,I}=0.3$ and $\hat{f}_{\rm duty,II}=0.1$, with a physical limit $f_{\rm duty,\max}=\dot{M}_{\rm ej}/\dot{M}_{\rm acc}$ to estimate the (average) duty cycle as $f_{\rm duty}=\min(\hat{f}_{\rm duty},f_{\rm duty,max})$ where
\begin{align}
    \dot{M}_{\rm ej}&=f_{\rm corr}\times 10^{-10}\ {\rm M_{\odot}\ yr^{-1}}(\Sigma_{0}/0.015\ \rm g\ cm^{-2})\ ,\label{mej}
\end{align}
is the mass ejection rate for $\alpha=0.63$ based on the results from \citet{Brown2019}\footnote{For conservative estimates of $\dot{M}_{\rm ej}$, the correction factor $f_{\rm corr}$ is also included, assuming that the potential overestimation of accretion rates in simulations is fully caused by overestimated ejection rates.}. %\footnote{One should be cautious with the terms `mass ejection rate' ($\dot{M}_{\rm ej}$) and `decretion rate' ($f_{\rm de}\dot{M}_{\rm ej}$) in the literature, since they can be easily confused with each other but typically differ by 2 orders of magnitude. For instance, we found that the fiducial model in \citet{Brown2019} actually has a \textit{decretion rate} of $10^{-10}\ \rm M_{\odot}\ yr^{-1}$, although it is called `mass ejection rate' in several places of their paper.}. 
The limit $f_{\rm duty,\max}$ captures the simple requirement that the compact object does not accrete more than what is ejected from the O/B star. When $f_{\rm duty,\max}<\hat{f}_{\rm duty}$, we assume that the compact object is able to accrete \textit{all} materials ejected by the O/B star during X-ray outbursts, despite the fact that for classical O/Be stars only a small fraction ($\sim 0.01$) of the ejected materials is expected to settle into the disk according to the standard steady-state VDD model \citep[e.g.,][]{Haubois2012,Rimulo2018}, while the majority will fall back to the star. This optimistic assumption is required to explain the observed high duty cycles \citep[up to $0.3$,][]{Reig2011,Sidoli2018}. %when the Be-XRB phase lasts as long as the remaining lifetime of the O/Be star. according to the ejection rate $\dot{M}_{\rm ej}$ from the empirical scaling relation
It means that mass replenishment of VDDs is much more efficient\footnote{This is likely caused by { enhanced mass loss (under the same angular momentum loss rate) for a O/B star in a binary system due to tidal truncation of the VDD by the companion \citep{Krticka2011,Rivinius2013} and/or} stronger (episodic) mass ejection with non-zero central torques \citep{Nixon2020} than expected from the steady-state rate based on observations of classical Be stars (Eqs.~\ref{sig_mw}, \ref{sig_smc} and \ref{mej}).} for O/Be stars in Be-XRBs than predicted by the standard VDD model (for O/Be stars in isolation), which is supported by the shorter disk timescales of Be stars in Be-XRBs compared with single Be stars in observations \citep{Reig2011}. 
%in the VDD.

With this model, we find that most Be-XRBs in the mock population do satisfy $f_{\rm duty,I}\sim 0.1-0.3$ and $f_{\rm duty,II}\sim 0.01-0.1$, which is generally consistent with observations \citep{Reig2011,Sidoli2018,Xu2019}. %(see Fig.~\ref{fig:fduty} for details). 
In the mock population, evolution of binary and stellar parameters during the Be-XRB phase is ignored, whereas in the BPS runs such evolution can change the outburst type of a Be-XRB. Be-XRBs with both types of outbursts also exist in observations. We classify the systems that experience both types of outbursts as Type~I/II. %They are also counted into the general Type~II category, which refers to all Be-XRBs that once undergo major outbursts with high accretion rates, disk warp and/or disruption. 
Besides, when $\dot{M}_{\rm acc}$ is comparable to $\hat{f}_{\rm duty}^{-1}\dot{M}_{\rm ej}$ (such that $f_{\rm duty}\lesssim \hat{f}_{\rm duty}$), significant disruption of the VDD by the compact object is expected to happen, such that the system will show Type~II features especially with low $f_{\rm duty}$. Therefore, we also regard the Be-XRBs classified as Type~I according to Eqs.~\ref{a_tidal} and \ref{mdotc} with $f_{\rm duty}<0.1$ as Type~I/II in post-processing. 
Finally, we count Type~I/II systems into the general Type~II category, which refers to all Be-XRBs that once undergo major outbursts with high accretion rates, disk warping and/or disruption. 
Recent observations find that the outburst behaviors of Be-XRBs are likely more diverse and complex than the conventional two types \citep{Sidoli2018}. Nevertheless, our consideration of the outburst type only affects the final X-ray output indirectly by the optimistic duty cycle $\hat{f}_{\rm duty}$, and we have verified by numerical experiments that our results are insensitive to outburst classification. The reason is that the majority ($\sim 60-70$\%) of X-ray emission comes from systems with $f_{\rm duty}\simeq f_{\rm duty,max}\lesssim\hat{f}_{\rm duty}$ in all Be-XRB populations considered here.%in this work (see Sec.~\ref{sec:res}).

\section{X-ray spectral model}\label{sec:spec}

Once $L_{\rm bol}$ is known, we only need to determine the spectral shape to obtain the full (intrinsic) spectral energy distribution (SED) of X-ray outbursts. For simplicity, we consider three regimes of accretion rates: low-hard (LH, $\eta<0.05$), high-soft (HS, $\eta\sim 0.05-2$) and super-Eddington (SE, $\eta>2$), for both NSs and BHs. Motivated by the ideas in \citet{Fragos2013xrb}, we find the typical spectral shape in each regime by fitting simple spectral models to the BC factors measured in observations \citep[e.g.,][]{McClintock2006,Wu2010,Anastasopoulou2022} for select energy bands\footnote{Theoretical calculation of the X-ray spectra from accreting compact objects \citep[see, e.g.,][]{yang_comprehensive_2017,chashkina_super-eddington_2019,qiao_systematic_2020,sokolova-lapa_x-ray_2021,pradhan_comprehensive_2021,Mushtukov2023} is beyond the scope of our phenomenological model for Be-XRBs in BPS. We therefore adopt simple observation-based models (i.e., black body or thin disk + power law) to capture the general trends.}. Here we consider the photon energy range $E\sim 10^{-4}-10^{4}$~keV assuming that this contains most of the power. %\footnote{We can also consider galactic composite X-ray spectra of XRBs from local observations and simulations \citep[e.g.,][]{Islam2019,Garofali2020,Lehmer2022,Vladutescu-Zopp2023}.}

\begin{figure}
    \centering
    \includegraphics[width=1\columnwidth]{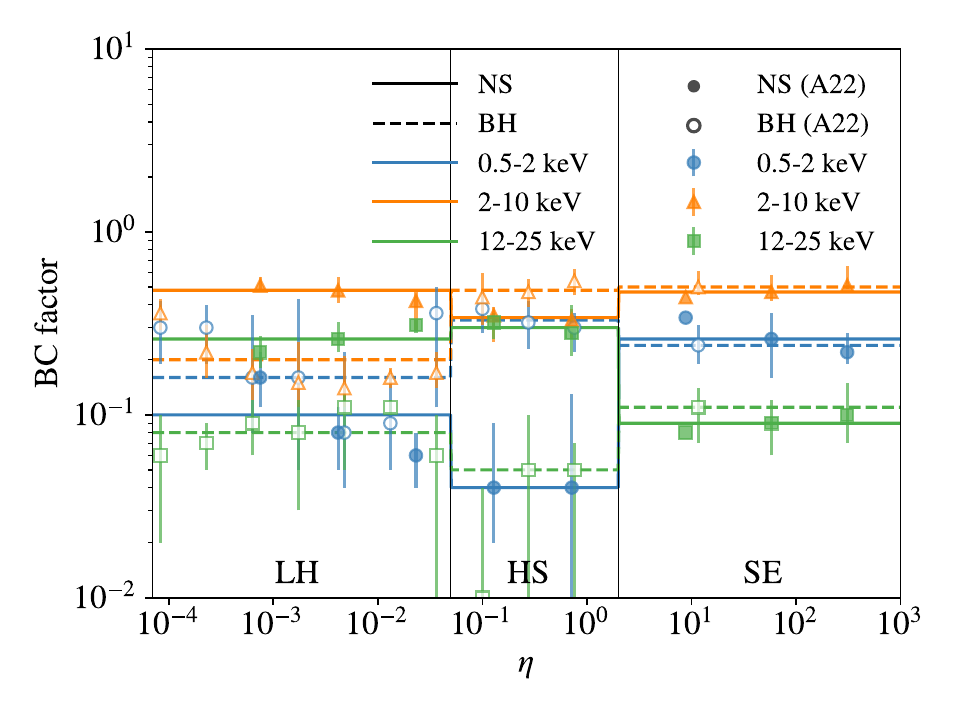}
    \vspace{-25pt}
    \caption{BC factor as a function of Eddington ratio $\eta$. The solid and dashed lines show the values adopted in our spectral fitting process for NSs ($M_{\rm X}\le 2.2\ \rm M_{\odot}$) and BHs ($M_{\rm X}> 2.2\ \rm M_{\odot}$), respectively, considering the energy bands $E\sim 0.5-2$~keV (blue), $2-10$~keV (orange), and $12-25$~keV (green), based on the measurements in \citet[A22, see their table~4, A1, A6, A9 and A10]{Anastasopoulou2022}. For comparison, we also show their original data for $E\sim 0.5-2$~keV (circles), $2-10$~keV (triangles), and $12-25$~keV (squares), where the results for NSs and BHs are denoted by filled and unfilled data points, respectively. }
    \label{fig:fbc}
\end{figure}

We take the recent measurements of BC factors of the energy bands $E\sim 0.5-2$, $2-10$ and $12-15$~keV by \citet[see their table~4, A1, A6, A9 and A10]{Anastasopoulou2022} for both NS and BH HMXBs\footnote{The NS HMXB sample in \citep{Anastasopoulou2022} is purely made of Be-XRBs.} as references. We estimate the typical value (and uncertainty) of the BC factor for each band in each regime, as shown in Fig.~\ref{fig:fbc}. These estimates are used to construct reference spectral energy distributions (SEDs) that serve as the targets of spectral fitting. Since soft X-ray photons in the $E\sim 0.5-2$~keV band are mostly responsible for X-ray heating of the IGM \citep{Das2017}, we increase the weight of this band by a factor of 4 in the fitting process to better reproduce the corresponding BC factor.

For simplicity, we assume that for both NSs and BHs the final spectrum is made of two components: an input spectrum and a spectrum of inverse Compton scattered photons (by corona electrons) with a power-law tail in the high-energy end. Following \citet{Sartorio2023}, we use the SIMPL-1 Comptonizon model in \citet[see their eqs.~1 and 4]{Steiner2009} to connect the up-scattered component with the input spectrum $L_{\nu,\rm in}$ via two parameters: the fraction $f_{\rm scatter}$ of photons in the input (photon number) spectrum $n_{\rm in}(E)\equiv dN_{\rm in}/dE=(hE)^{-1}L_{\nu,\rm in}$ that are up scattered, and the power-law index $\Gamma$ in the spectrum of up-scattered photons: 
\begin{align}
    n_{\rm out}(E)&=(1-f_{\rm scatter})n_{\rm in}(E)\notag\\
    &+f_{\rm scatter}\int_{E_{\min}}^{E}n_{\rm in}(E_{0})G(E,E_{0})dE_{0}\ ,\label{compton}%\\
    %&G(E,E_{0})=(\Gamma-1)(E/E_{0})^{-\Gamma}/E_{0}\ ,\label{green}
\end{align}
where $G(E,E_{0})=(\Gamma-1)(E/E_{0})^{-\Gamma}/E_{0}$ is the Green's function of inverse Compton scattering. The final specific luminosity can then be derived from the output photon number spectrum with $n_{\rm out}(E)\equiv dN_{\rm in}/dE=(hE)^{-1}L_{\nu,\rm out}$.

\begin{figure*}
    \centering
    \subfloat{\includegraphics[width=0.7\columnwidth]{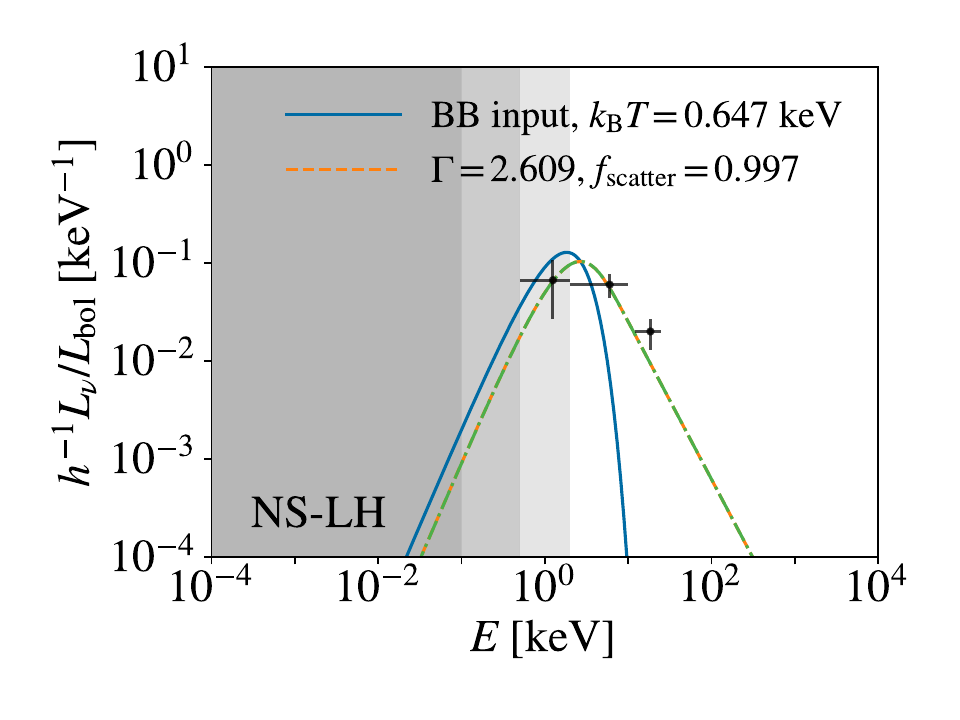}}
    \subfloat{\includegraphics[width=0.7\columnwidth]{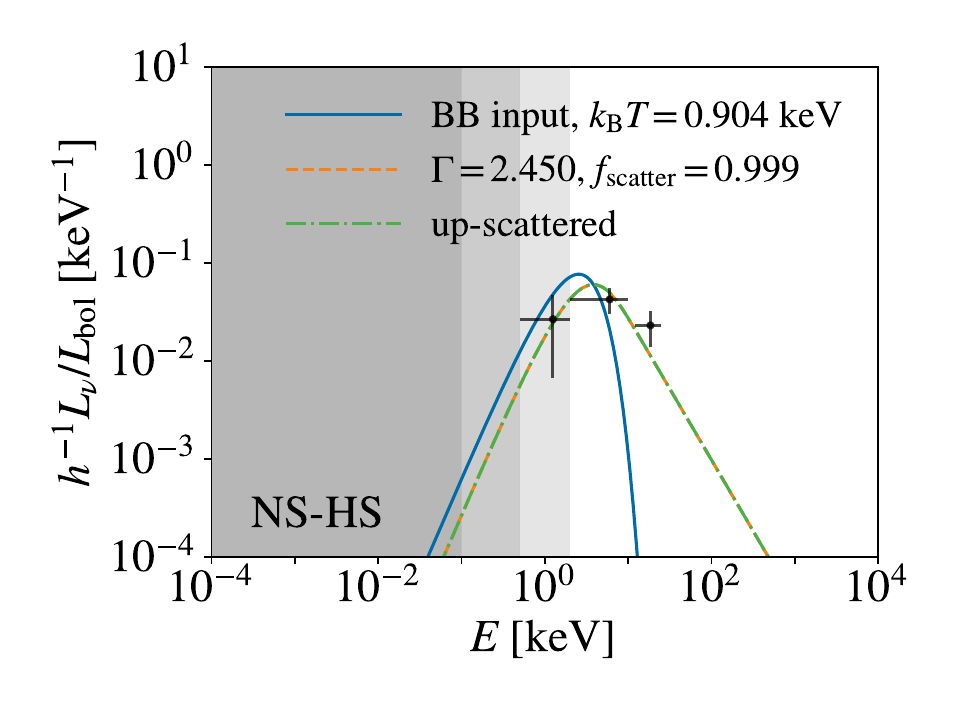}}
    \subfloat{\includegraphics[width=0.7\columnwidth]{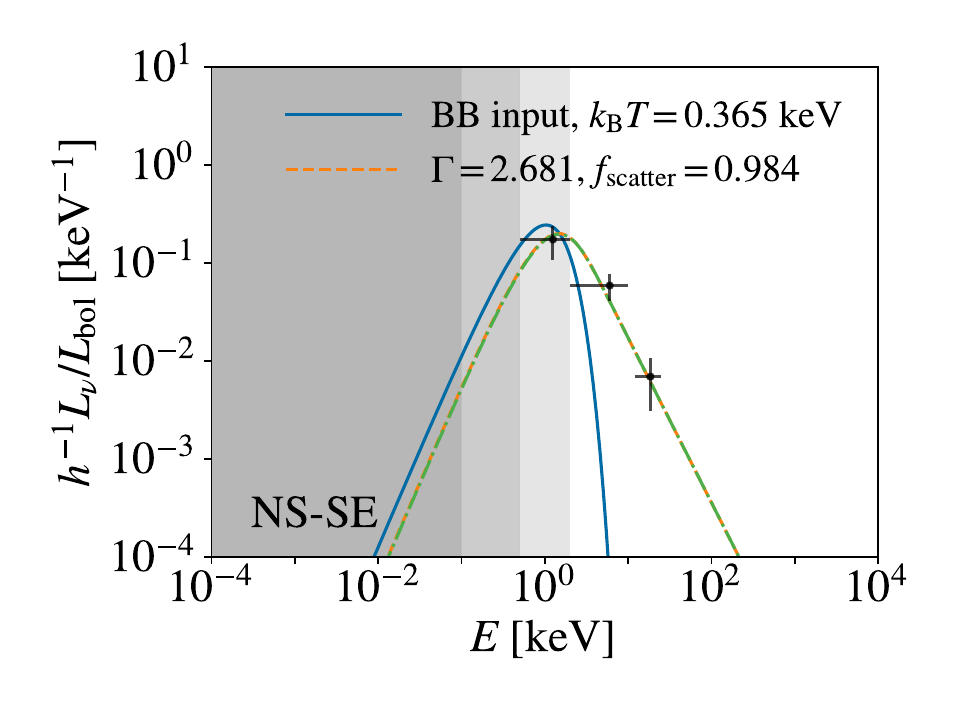}}\\
    \vspace{-29pt}
    \subfloat{\includegraphics[width=0.7\columnwidth]{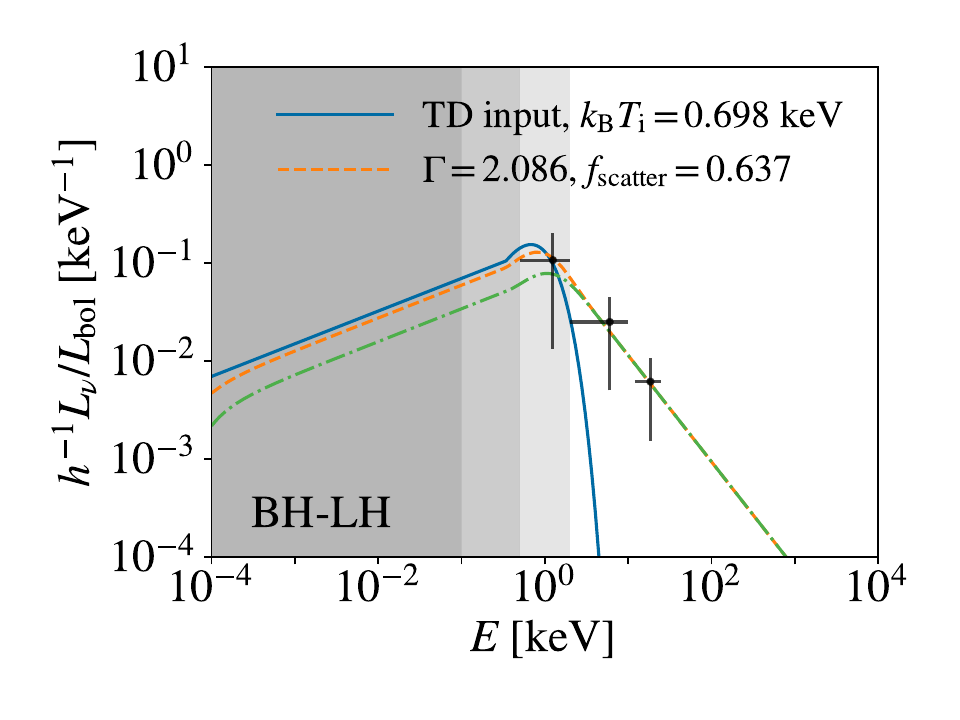}}
    \subfloat{\includegraphics[width=0.7\columnwidth]{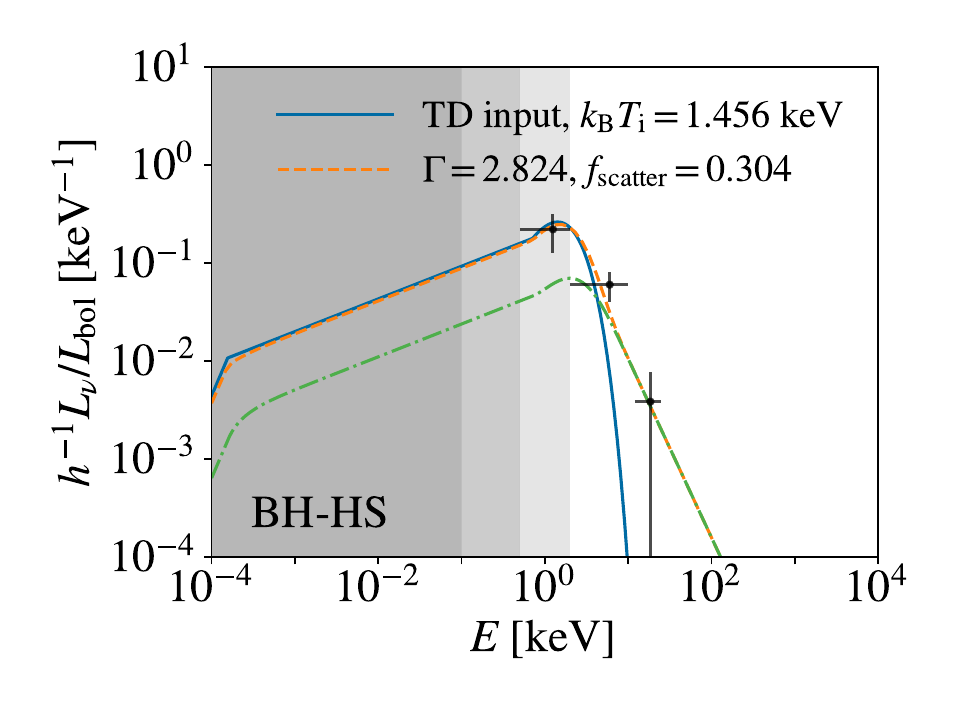}}
    \subfloat{\includegraphics[width=0.7\columnwidth]{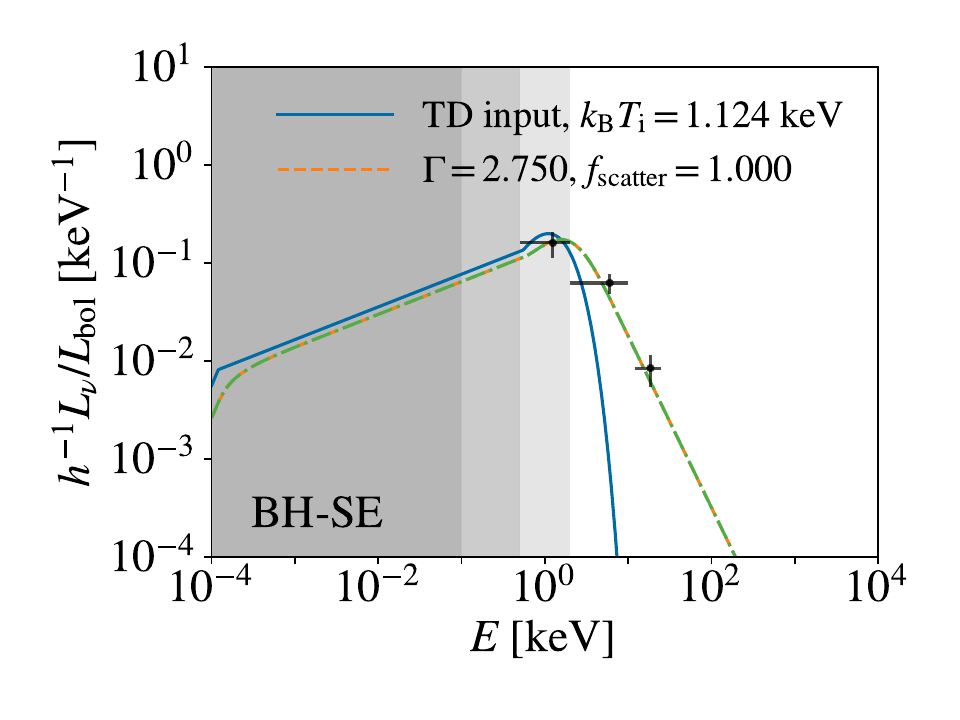}}
    \vspace{-10pt}
    \caption{Best-fit spectral models for NSs (top) and BHs (bottom) in the low-hard (LH, $\eta<0.05$, left), high-soft (HS, $\eta\sim0.05-2$, middle) and super-Eddington (SE, $\eta>2$, right) regimes, in terms of $L_{\nu}/(hL_{\rm bol})$. %in the unit of $L_{38}\rm\ erg\ s^{-1}\ keV^{-1}$ given $L_{38}\equiv L_{\rm bol}/(10^{38}\ \rm erg\ ^{-1})$.
    Here we assume $M_{\rm X}=4\ \rm M_{\odot}$ and $R_{\rm o}=10\ \rm R_{\odot}$ for the BH spectra. In each case the input spectrum is shown with the solid curve, the final spectrum with the dashed curve and the up-scattered component with the dash-dotted curve, respectively. The data points with error bars show the reference SEDs inferred from the BC factor measurements in \citet{Anastasopoulou2022}. The dark ($E\lesssim 0.1\ \rm keV$), intermediate dark ($E\lesssim 0.5\ \rm keV$) and light ($E\sim 0.5-2\ \rm keV$) shaded regions denote the parts of the spectrum that are expected to be absorbed by the ISM in typical high-$z$ HMXB-hosting minihaloes \citep{Sartorio2023}, atomic cooling haloes \citep{Das2017}, and the IGM at $z\sim 15$ \citep{Pritchard2007}.}% For NSs, the $E\sim 0.5\ (0.1)-2\ \rm keV$ band counts for 0.093 (0.096), 0.036 (0.037) and 0.258 (0.274) of the bolemnetric luminosity, in the LH, HS and ULX states, respectively. While for BHs}
    \label{fig:specmodel}
\end{figure*}

For the input spectrum, we adopt the black body (BB) spectrum for NSs %\footnote{Even if an (multi-color) accretion disk forms around the NS that contributes most the X-rays, the disk is usually much smaller than those of BHs, so that the disk radiation can also be approximated with a BB spectrum.}
%\begin{align}
%    L_{\nu,\rm BB}\propto \nu^{3}/\{\exp[h\nu/(k_{\rm B}T)]-1\}\ ,\label{bb}
%\end{align}
assuming that the majority of X-ray radiation is produced at hot spots on the NS surface from inflows channeled by magnetic fields. The top row of Fig.~\ref{fig:specmodel} shows the resulting best-fit spectral models in the three regimes with $k_{\rm B}T\sim 0.4-0.9$~keV, $f_{\rm scatter}\sim 1$ and $\Gamma\sim 2.5-2.7$.
	    
While for BHs, we use the thin disk (TD) model\footnote{In principle, the thin disk solution is only valid at high accretion rates \citep[e.g., $\eta\gtrsim 0.07\alpha$,][]{Pringle1987,Takhistov2022}, which are expected to cover most cases. Besides, we find that the contribution of BHs to the total X-ray output from Be-XRBs is no more than a few percent in all cases explored. %, consistent with the rareness of strong BH accretion from O/Be star disks predicted by previous BPS studies \citep{Zhang2004,Belczynski2009,Brown2018} and in observations \citep{Casares2014,Grudzinska2015,Khokhlov2018,Zamanov2022}.
Therefore, we do not consider the ADAF solution for BHs with lower accretion rates.}  %assuming that the most important X-ray properties will be captured by the power-law tail of the inverse Compton spectrum that is insensitive to the input spectrum.} 
\citep[][]{Pringle1987,Takhistov2022} 
\begin{align}
    &L_{\nu,\rm TD}\propto\begin{cases}
    \left(\frac{T_{\max}}{T_{\rm o}}\right)^{5/3}\left(\frac{\nu}{\nu_{\max}}\right)^{2}\ ,\quad h\nu\le k_{\rm B}T_{\rm o}\ ,\\
    (\nu/\nu_{\max})^{1/3}\ ,\quad k_{\rm B}T_{\rm o}<h\nu<k_{\rm B}T_{\max}\ ,\\
    \left(\frac{\nu}{\nu_{\max}}\right)^{2}\exp\left(1-\frac{\nu}{\nu_{\max}}\right)\ ,\ h\nu>k_{\rm B}T_{\max}\ ,\label{td}
\end{cases}%\notag\\
\end{align}
where $T_{\max} = 0.488 T_{i}$, $T_{\rm o}\simeq T_{\rm i}(R_{\rm i}/R_{\rm o})^{3/4}$ and $\nu_{\max}=k_{\rm B}T_{\max}/h$, given $T_{\rm i}$ as the temperature at the inner edge ($R_{\rm i}=6GM_{\rm X}/c^{2}$) of the TD, and $T_{\rm o}$ as the temperature at the outer disk boundary $R_{\rm o}\sim R_{\rm L1}(a,e,q_{\rm X})$. Since very close binaries ($P_{\rm orb}\lesssim 7$~days) and RLO are forbidden for Be-XRBs \citep{Panoglou2016,Panoglou2018,Rivinius2019} by tidal forces that can slow down the rotation of donor stars, we have $R_{\rm o}\gtrsim 2\ \rm R_{\odot}$ and $k_{\rm B}T_{\rm o}\lesssim 10^{-3}$~keV in most cases. Therefore $R_{\rm o}$ is unimportant for X-rays that we are concerned with ($E>0.1$~keV) and the input TD spectrum is controlled by a single parameter $T_{\rm i}$ during the fitting process. The best-fit models for $M_{\rm X}=4\ \rm M_{\odot}$ and $R_{\rm o}=10\ \rm R_{\odot}$ are shown in the bottom row of Fig.~\ref{fig:specmodel} with $k_{\rm B}T_{\rm i}\sim 0.7-1.5$~keV, $f_{\rm scatter}\sim 0.3-1$ and $\Gamma\sim 2.1-2.8$.
%\end{itemize}

%{Columns of the output Be-XRB catalog:\\
%\footnote{We can also output these quantities at a few timesteps (e.g., the beginning and end of the Be-XRB phase) to verify the assumption of little evolution or output the time-averaged values.}):\\ 
%($t_{\rm ini}$, $t_{\rm fin}$, $a$, $e$, $M_{\star}$, $M_{\rm X}$, $R_{\star}$, $T_{\rm eff}$, $W$, $\Sigma_{0}$, $\eta$, $L_{\rm bol}$, $f_{\rm duty}$, COB type, outburst type, ID)\\ 
%Here COB type shows whether a BH or NS is involved, and ID is the index of the progenitor binary in the input catalog. 

\section{Results}\label{sec:res}

Combining the two choices for initial rotation parameter $W_{0}$: SR or FR (Sec.~\ref{sec:ic}), and the two models for VDD density $\Sigma_{0}$: CS or OP (Sec.~\ref{sec:xray}), we have 4 models to explore, which are summarized in Table~\ref{tab:model}. For each model, we consider ten cases at $Z=10^{-4}$, $3\times 10^{-4}$ $10^{-3}$, 0.0035, 0.005, 0.007, 0.01, 0.0142, 0.02 and 0.03, where $Z=0.0035$, 0.07 and 0.0142 correspond to the situations of the SMC, LMC and MW, respectively. Since the CS and OP models are identical at $Z\ge 0.0142$, we only need to run 34 BPS simulations in total. 
We record the \textit{time-averaged} values of the Be-XRB properties including properties of donors and accretors, orbital parameters as well as X-ray outburst properties\footnote{In the calculation of time-averaged outburst luminosity, $L_{\rm bol}$ at each time step $\delta t$ is weighted by $f_{\rm duty}\delta t$.}, which are then used to calculate the total X-ray output. 
Considering the short lifetimes of Be-XRBs, %(a few to a few tens Myr, see Sec.~\ref{sec:xray}), 
for simplicity, when calculating the total X-ray emission of a Be-XRB population, we assume no variation of X-ray outburst properties during the Be-XRB phase\footnote{In our BPS runs, %the Be-XRB phase typically only lasts for a few to a few tens Myr (see Sec.~\ref{sec:xray}) during which period of time 
the stellar and orbital parameters do not vary much during the Be-XRB phase in most cases, which leads to little evolution of X-ray outburst properties under the assumption of steady-state mass ejection.}, which is a simplification of the reality that VDDs can be highly variable structures with disk dissipation/formation at timescales of a few years \citep{Reig2011,Rivinius2013}. Since our purpose is to evaluate the overall X-ray output from a population of Be-XRBs, higher-order effects are expected to be unimportant. In this section we mainly show the results from the SR\_CS model with $f_{\rm corr}=0.5$, defined as the fiducial case, because it achieves the best agreement with observations and the key trends in the metallicity dependence of Be-XRB properties are similar in the other cases. Select results for the other 3 models in Table~\ref{tab:model} and different values of $f_{\rm corr}$ are included in Appendix~\ref{apx:model}. 
%All results shown below assume $f_{\rm corr}=0.5$ unless annotated specially.

%}

\begin{table}
    \centering
    \caption{Summary of models. The second column shows the choices of the initial rotation parameter $W_{0}\equiv v_{\rm rot,0}/v_{\rm Kep,0}$ given the initial rotation velocity $v_{\rm rot,0}$ and Keplerian velocity $v_{\rm Kep,0}$ at the stellar equator (see Sec.~\ref{sec:ic}), where the slowly-rotating (SR) model $W_{0}(M_{\star},Z)$ with mass and metallicity dependence is based on \citet[see their sec.~7.2]{Hurley2000} and \citet{Hurley2002}, while the fast-rotating (FR) model uses a constant high value. The third column shows the choices of the (base) surface density $\Sigma_{0}$ of decretion disk (see Sec.~\ref{sec:xray}), where the metallicity-dependent optimistic (OP) model uses a linear interpolation for $\log(\Sigma_{0})$ between the MW fit for at $Z=0.0142$ (Eq.~\ref{sig_mw}) and the SMC fit at $Z=0.0035$ (Eq.~\ref{sig_smc}), while the conservative model adopts the MW fit at all metallicities.}
    \begin{tabular}{ccc}
    \hline
        Model & $W_{0}$ & $\Sigma_{0}$ \\
    \hline
        SR\_CS & $W_{0}(M_{\star},Z)$  & $\Sigma_{0}(M_{\star})$ (Eq.~\ref{sig_mw})\\
        FR\_CS & 0.9 & $\Sigma_{0}(M_{\star})$ (Eq.~\ref{sig_mw})\\
        SR\_OP & $W_{0}(M_{\star},Z)$  & $\Sigma_{0}(M_{\star},Z)$ (Eqs.~\ref{sig_mw} and \ref{sig_smc})\\
        FR\_OP & 0.9 & $\Sigma_{0}(M_{\star},Z)$ (Eqs.~\ref{sig_mw} and \ref{sig_smc})\\
    \hline
    \end{tabular}
    \label{tab:model}
\end{table}

\subsection{Formation efficiency}\label{sec:eps}

We first look into the formation efficiency of \textit{active} Be-XRBs, $\mathcal{N}_{\rm X}$, as a function of $Z$ which, considering the short lifetimes (a few to a few tens Myr) of Be-XRBs, is defined as the number of Be-XRBs in the outburst phase per unit SFR for a long enough star formation timescale ($\tau_{\rm SF}\gtrsim 100$~Myr). Given $N$ Be-XRBs predicted by a BPS run for a single-age stellar population of a total mass $M_{\rm tot}$, the formation efficiency can be written as
\begin{align}
    \mathcal{N}_{\rm X}&\equiv \frac{\langle N_{\rm X}\rangle}{{\rm SFR}}=\frac{1}{{\rm SFR}}\sum_{i}^{N}f_{{\rm duty},i}\left(\frac{\tau_{i}{\rm SFR}}{M_{\rm tot}}\right)%\notag\\
    %\left(\frac{M_{\rm tot}}{\tau_{\rm SF}}\right)^{-1}\notag\\
    =\sum_{i}^{N}\frac{f_{{\rm duty},i}\tau_{i}}{M_{\rm tot}}\ ,\label{nxsfr}
\end{align}
where $\tau_{i}$ is the duration of the Be-XRB phase for binary $i$. 

\begin{figure}
    \centering
    \includegraphics[width=\columnwidth]{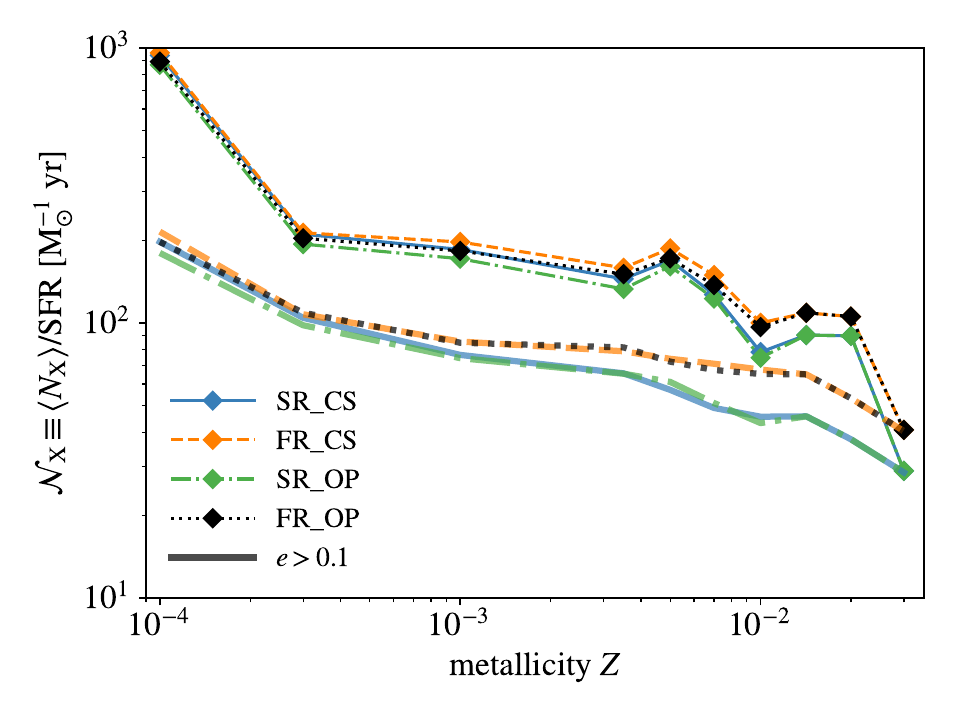}
    \vspace{-25pt}
    \caption{Number of Be-XRBs (in the outburst phase) per unit SFR as a function of metallicity, for the SR\_CS (solid), FR\_CS (dashed), SR\_OP (dash-dotted) and FR\_OP (dotted) models. The results for all Be-XRBs are shown with the thin curves marked by {diamonds}, while the unmarked thick curves show the results for Be-XRBs with $e>0.1$. The thin vertical lines label the metallicities of the MW, LMC and SMC (from right to left).}
    \label{fig:nxsfr}
\end{figure}

The results for all the 4 models in Table~\ref{tab:model} are shown in Fig.~\ref{fig:nxsfr}, where we count both the number of all Be-XRBs (thin curves with markers) and that of Be-XRBs with $e>0.1$ (thick curves without markers). The latter case is meant to capture the situation assumed in previous BPS studies of Be-XRBs \citep[e.g.,][]{Zhang2004,Xing2021} that tidal truncation of the VDD is sharp, leading to a smaller disk boundary at $R_{\rm trunc}$ (Eq.~\ref{rtrunc}) than the one adopted in our model ($R_{\rm crit}$, see Eq.~\ref{rcrit}). In both cases, $\mathcal{N}_{\rm X}$ generally increases with decreasing $Z$, but the evolution is not fully monotonic for all Be-XRBs, which has a small peak at $Z=0.005$ and a small dip at $Z=0.01$. The general trend is driven by the stronger stellar winds at higher $Z$ that increasingly widen binary orbits and reduce the number of stars that can become NSs and BHs. The non-monotonic features are likely caused by the complex interplay between stellar winds and mass transfer rate (see Sec.~\ref{sec:properties}). 
We also find that nearly-circular ($e\le 0.1$) systems make up a significant fraction ($\sim 40-80\%$) of active Be-XRBs at $Z\lesssim 0.02$. Naïvely, this seems in tension with the rareness ($\lesssim 10\%$) of such Be-XRBs in observations \citep{Cheng2014,Sidoli2018}. However, observations are very likely incomplete at the low eccentricity end because the measurement of $e$ is difficult for nearly-circular binaries, and only a small fraction of observed Be-XRBs have eccentricity measurements. Besides, these binaries are typically faint with $L_{\rm bol}\sim 10^{33}-10^{37}\ \rm erg\ s^{-1}$ due to the suppressed accretion rate from disk truncation, and therefore, difficult to detect. In fact, they produce much less X-rays compared with their more eccentric counterparts such that ignoring them has little (up to a few percent) impact on the overall X-ray output. 
%As will be discussed in Sec.~\ref{sec:properties}, the majority of Be-XRBs on nearly-circular orbits form via electron-capture SNe with no kicks, and it is necessary to consider them in order to explain the large population of Be-XRBs observed in the SMC (see Appendix~\ref{apx:smc}). consistent with the findings in \citet{Linden2009}. 

Last but not the least, $\mathcal{N}_{\rm X}$ is higher in the FR models with higher initial rotation rates than in the SR models as expected. 
The difference between the FR and SR models is larger at higher $Z$ but remains below $\sim 40\%$ and becomes even comparable to the uncertainties\footnote{Since our Be-XRB routine (see Sec.~\ref{sec:bexrb}) does not affect binary stellar evolution, the CS and OP models under the same assumption of initial rotation produce almost the same values of $\mathcal{N}_{\rm X}$, and the small difference between their predictions by a few percent reflects the {scatter caused by stochastic VDD densities and SN kicks with} the limited sample size.} in $\mathcal{N}_{\rm X}$ at $Z\lesssim 0.0035$. %This difference is smaller at lower $Z$ is due to the fact the stars are more compact and can be more easily spun up 
The overall small difference indicates that in most binaries that can potentially become Be-XRBs %-like systems (i.e., a MS star with a compact object companion) 
the (initial) secondary star will be spun up to become an O/Be star (via stable mass transfer during the MS and HG phases) regardless of its initial rotation rate. %, implying that binary stellar evolution is an important pathway of making O/Be stars. 
This is consistent with the scenario that all or most (young) O/Be stars are produced by mass and angular momentum transfer from companion stars {\citep{Shao2014,Hastings2020,Hastings2021,Dodd2023,Wang2023}}, which is supported by observations that find a large fraction of classical Be stars with disk truncation \citep[i.e., SED turndown;][]{Klement2019}, the lack of close Be binaries with MS companions \citep{Bodensteiner2020}, and a higher run-away/field frequency of O/Be stars (or fast rotators) compared with normal O/B stars \citep{DorigoJones2020,Dallas2022}.
On the other hand, the difference between the FR and SR models is generally larger when we focus on Be-XRBs with $e>0.1$ at $Z\gtrsim 10^{-3}$. The reason is that mass transfer is weaker in these systems leading to less efficient spin-up of the secondary star, as discussed below (Sec.~\ref{sec:properties}). Besides, the difference is smaller at lower $Z$, where the secondary stars are more compact and are more easily spun up to become O/Be stars. 
%and less significant compared with the $Z$ dependence. 

%\footnote{Since our Be-XRB routine (see Sec.~\ref{sec:bexrb}) does not affect binary stellar evolution, the CS and OP models under the same assumption of initial rotation produce almost the same values of $\mathcal{N}_{\rm X}$, and the small difference between their predictions reflects the uncertainty caused by the limited sample size. }.

%The large fraction of classical O/Be stars with disk truncation \citep[i.e., SED turndown;][]{Klement2019}, lack of close Be binaries with main-sequence companions \citep{Bodensteiner2020} and higher run-away/field frequency of OO/Be stars (or fast rotators) compared with normal OB stars \citep{DorigoJones2020,Dallas2022} in observations imply that all or most O/Be stars can be produced by mass and angular momentum transfer from companion stars \citep{Shao2014,Hastings2021}.

\begin{figure}
    \centering
    \includegraphics[width=\columnwidth]{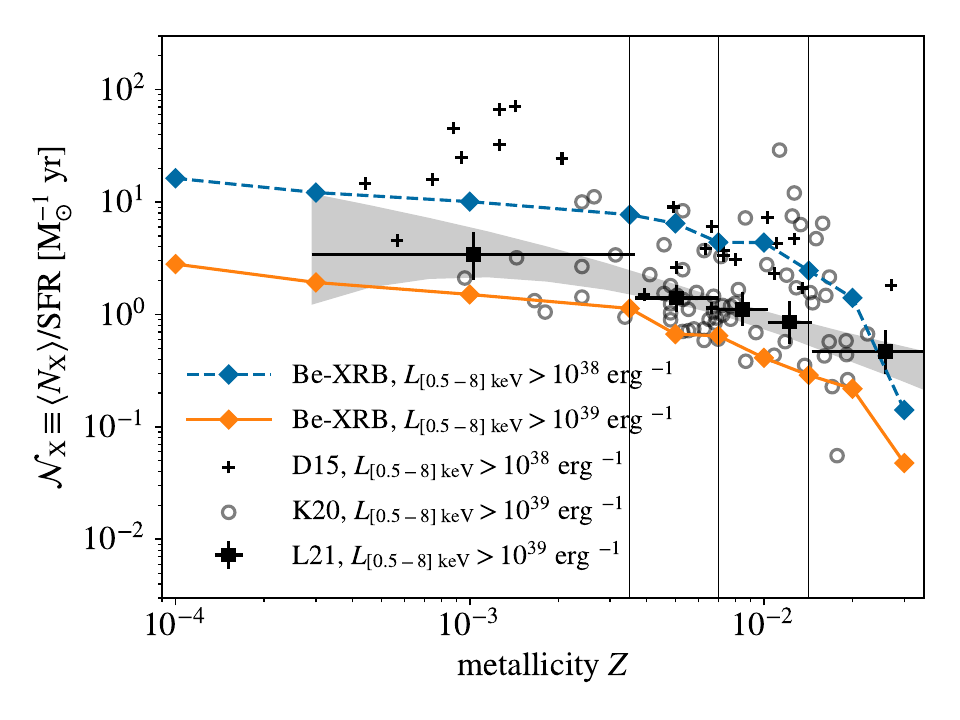}
    \vspace{-25pt}
    \caption{Number of (ultra-)luminous Be-XRBs (in outbursts) per unit SFR as a function of metallicity in the SR\_CS model with $f_{\rm corr}=0.5$. The BPS results for luminous ($L_{[0.5-8]~\rm keV}>10^{38}\ \rm erg\ s^{-1}$) and ultra-luminous ($L_{[0.5-8]~\rm keV}>10^{39}\ \rm erg\ s^{-1}$) sources are shown by the dashed and solid curves, respectively. For comparison, we show the observational data from \citet[D15, see their fig.~4]{Douna2015} for $L_{[0.5-8]~\rm keV}>10^{38}\ \rm erg\ s^{-1}$ with the crosses{, and those from \citet[K20, see their fig.~14]{Kovlakas2020} for $L_{[0.5-8]~\rm keV}>10^{39}\ \rm erg\ s^{-1}$ with circles, which include the data originally from \citet{Mapelli2010}.} 
    The uncertainties in these data are typically large ($\sim 0.5$~dex), as implied by their scatter around similar metallicities. The number counts for ultra-luminous sources ($L_{[0.5-8]~\rm keV}>10^{39}\ \rm erg\ s^{-1}$) from the observed HMXB sample in \citet[L21, see their table~3 and fig.~6]{Lehmer2021} are denoted by the squares with ($1\sigma$) error bars, and the shaded region denotes the 16-84\% confidence range obtained with mock populations of XRBs sampled from their best-fit model for $Z$-dependent X-ray luminosity functions. The thin vertical lines label the metallicities of the MW, LMC and SMC (from right to left).}
    \label{fig:nxsfr_fdcs}
\end{figure}

\begin{figure*}
    \centering
    \subfloat[$Z=10^{-4}$]{\includegraphics[width=0.7\columnwidth]{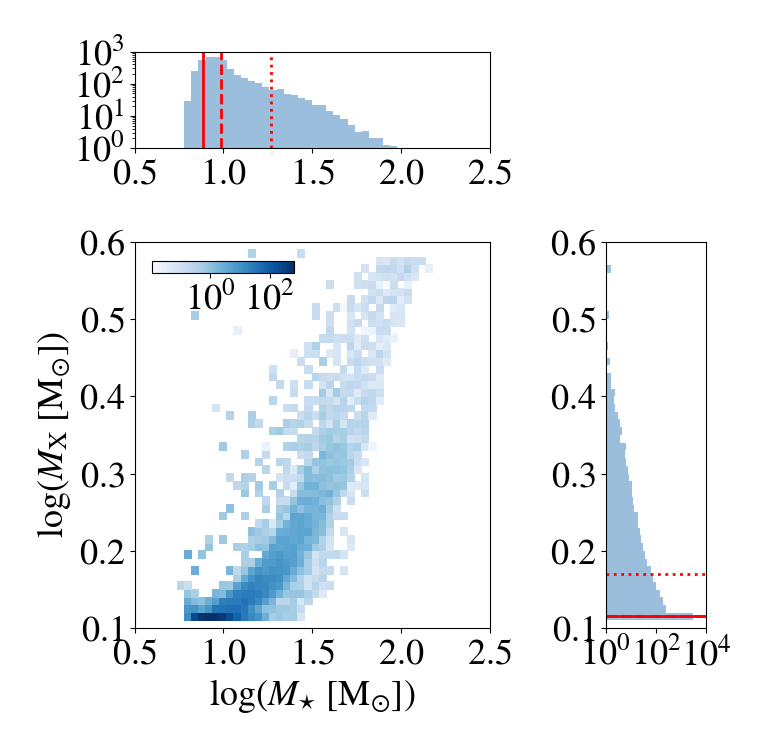}}
    \subfloat[$Z=0.0035$ (SMC)]{\includegraphics[width=0.7\columnwidth]{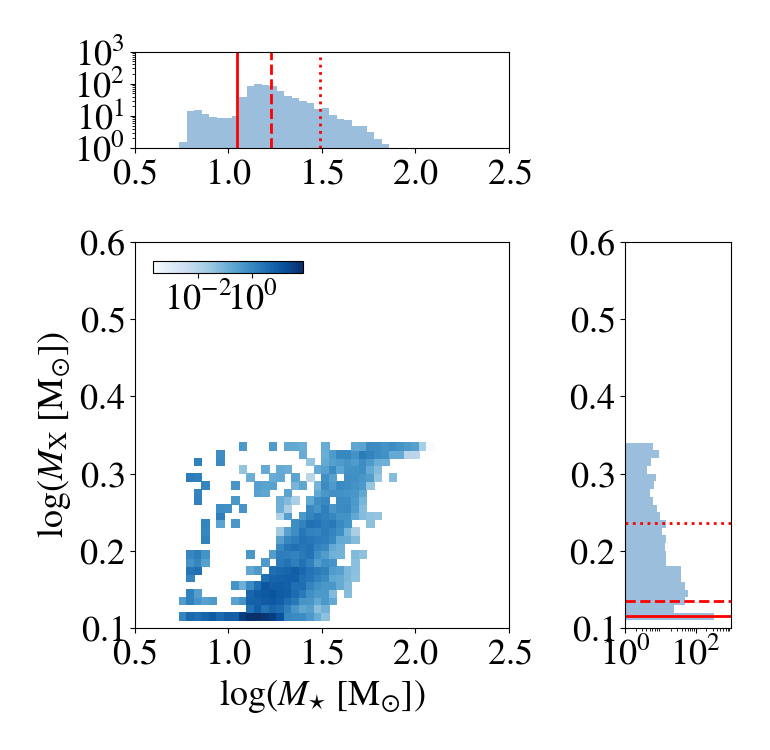}}
    \subfloat[$Z=0.0142$ (MW)]{\includegraphics[width=0.7\columnwidth]{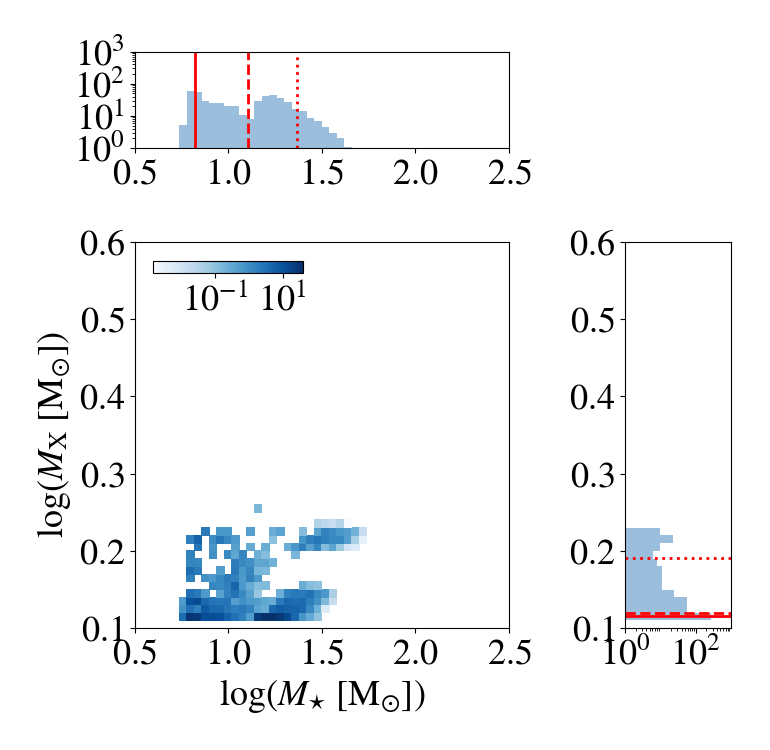}}\\
    \vspace{-17pt}
    \subfloat{\includegraphics[width=0.7\columnwidth]{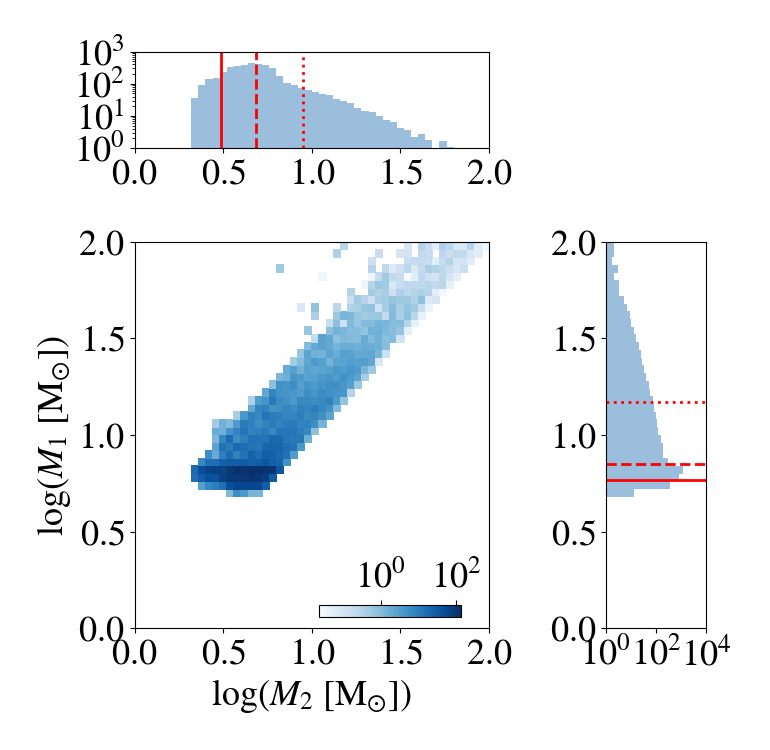}}
    \subfloat{\includegraphics[width=0.7\columnwidth]{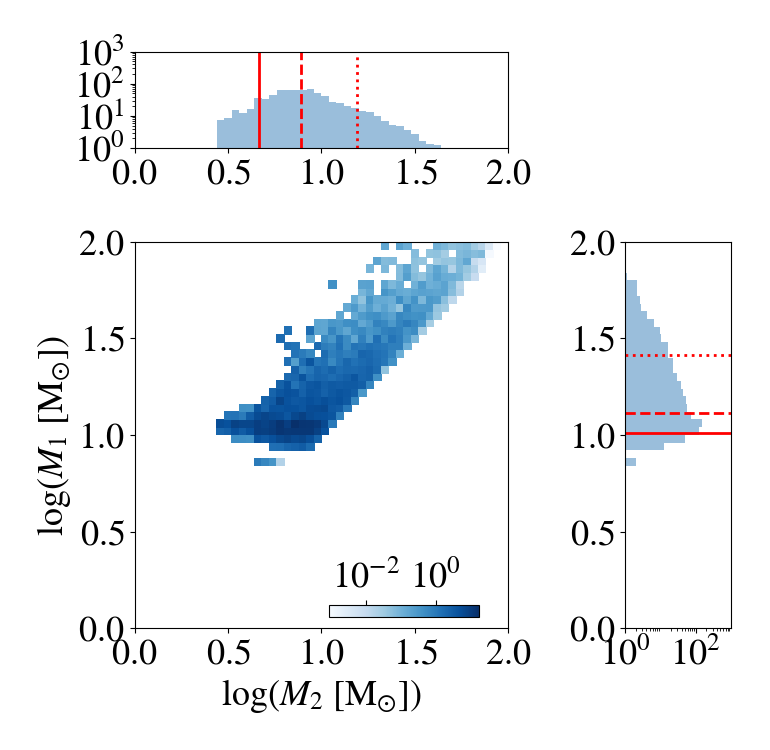}}
    \subfloat{\includegraphics[width=0.7\columnwidth]{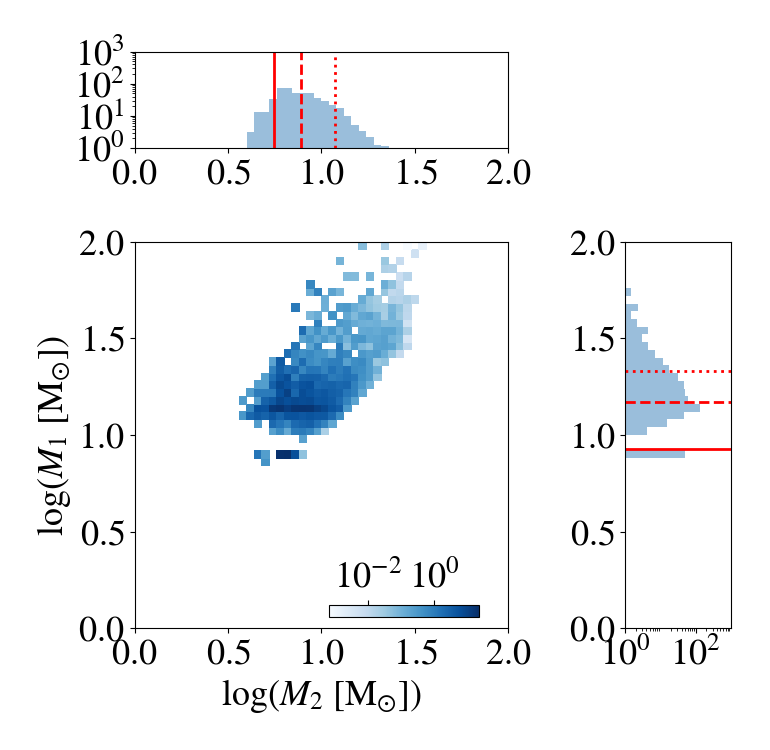}}\\
    \vspace{-17pt}
    \subfloat{\includegraphics[width=0.7\columnwidth]{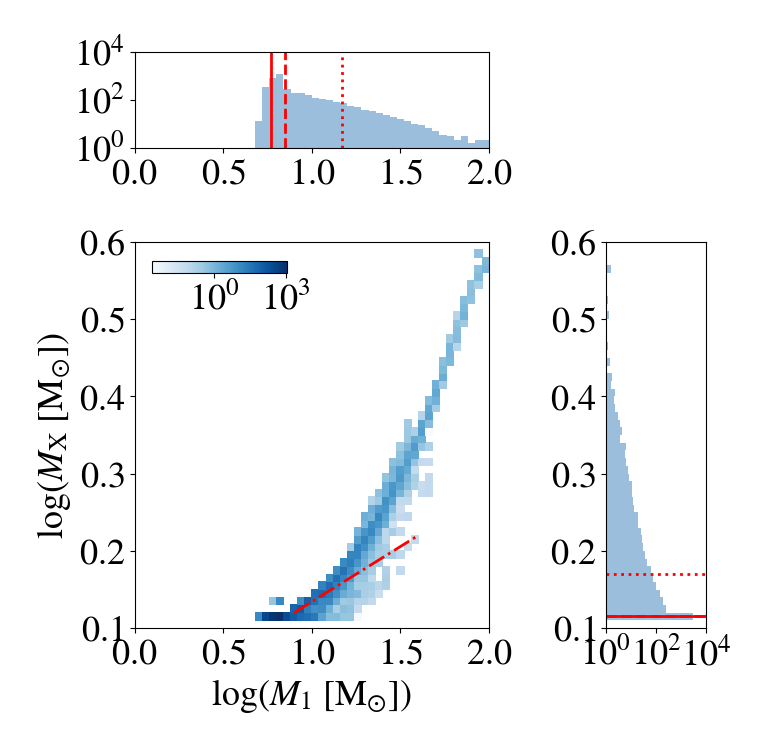}}
    \subfloat{\includegraphics[width=0.7\columnwidth]{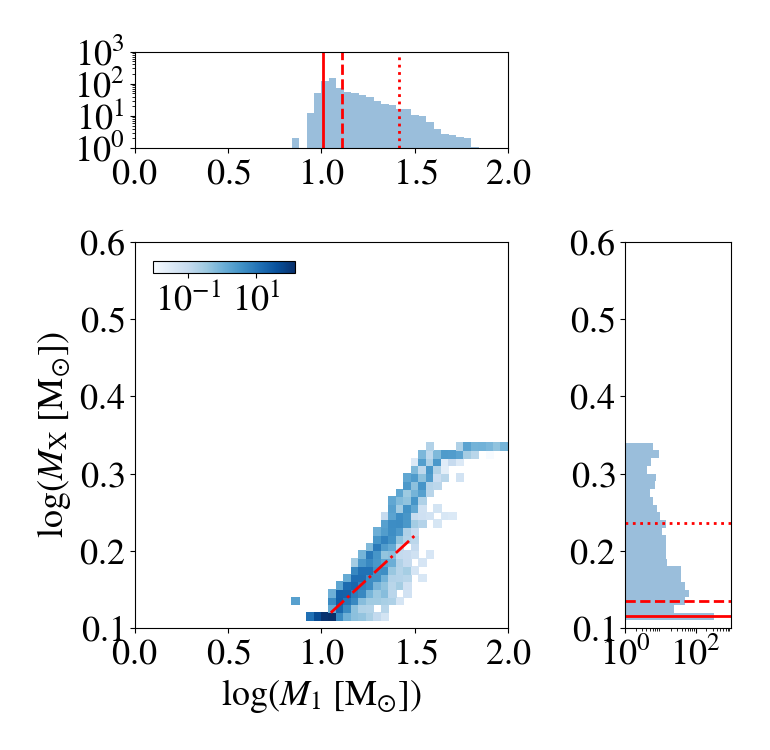}}
    \subfloat{\includegraphics[width=0.7\columnwidth]{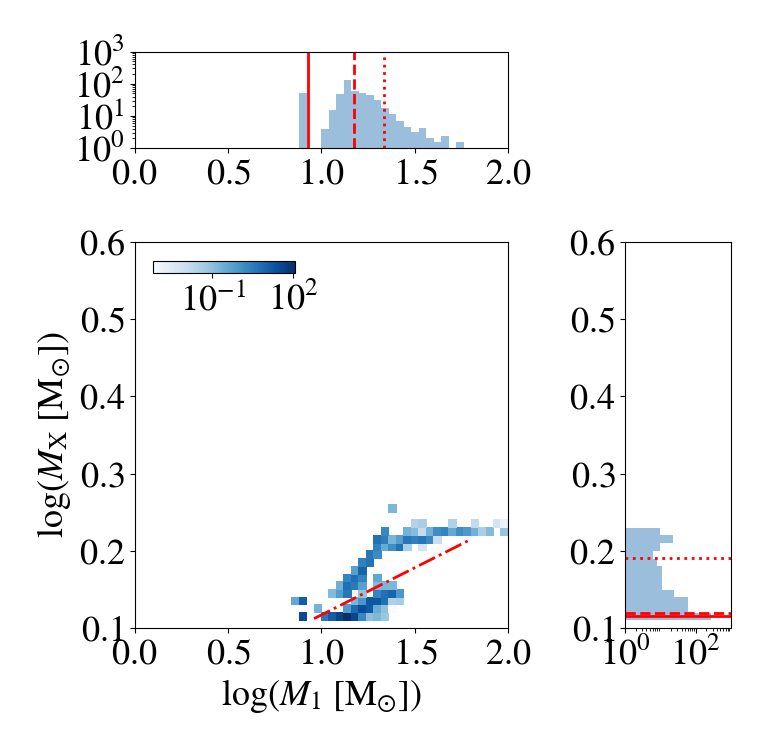}}
    \vspace{-8pt}
    \caption{Mass distributions of Be-XRBs (top) and their progenitors (middle), and the initial-remnant mass relation for primary stars of Be-XRB progenitors (bottom) in the SR\_CS model for $Z=10^{-4}$ (left), $Z=0.0035$ (middle) and $Z=0.0142$ (right). Here the (initial) primary star ($M_{1}$) is the progenitor of the compact object ($M_{\rm X}$), and the (initial) secondary star ($M_{2}$) is the progenitor of the O/Be star ($M_{\star}$). {Each Be-XRB $i$ is weighted by $\tau_{i}/M_{\rm tot}$, so that the number for each bin on these plots corresponds to the expected number of binaries in the Be-XRB phase per unit SFR ($\rm M_\odot\ yr^{-1}$). In the marginalized distributions, the solid, dashed and dotted lines mark the 10, 50 and 90 percentiles. The dash-dotted lines in the 2D maps of initial-remnant mass relation for primary stars (bottom) approximately divide the Be-XRB population into two groups with lower and higher $M_{\rm X}$ in which the primary stars experience different levels of mass loss (see the main text in Sec.~\ref{sec:properties}).}}% The data point areas are normalized within each scatter plot separately and should not be compared across different panels.}
    \label{fig:mass_dis}
\end{figure*}

In addition to the general Be-XRB population, we also derive the number of luminous and ultra-luminous sources with broad-band ($E\sim 0.5-8$~keV) X-ray luminosities $L_{[0.5-8]~\rm keV}>10^{38}\ \rm erg\ s^{-1}$ and $L_{[0.5-8]~\rm keV}>10^{39}\ \rm erg\ s^{-1}$ during outbursts, respectively, for which constraints from observations of HMXBs in nearby galaxies are available down to $Z\sim 0.0003$ {\citep{Mapelli2010,Douna2015,Kovlakas2020,Lehmer2021}. The results in our fiducial case are shown in Fig.~\ref{fig:nxsfr_fdcs}. For utra-luminous sources with $L_{[0.5-8]~\rm keV}>10^{39}\ \rm erg\ s^{-1}$, we find that the numbers of Be-XRBs predicted by our BPS runs are lower than those inferred from observations\footnote{When comparing our Be-XRB populations with observations, we ignore the effects of anisotropic emission which can be important for ULXs with NS accetors \citep{Wiktorowicz2019,Khan2022}. To the zeroth order, such effects can be absorbed into the duty cycle $f_{\rm duty}$.} of \textit{all} types of HMXBs \citep{Lehmer2021} by about 40\% at $Z\sim 0.0003-0.02$, which means that the simulated Be-XRBs can explain $\sim 60$\% of all ultra-luminous HMXBs, assuming that observations are complete. 
For instance, we have $\mathcal{N}_{\rm X}=0.29\ \rm M_{\odot}^{-1}\ yr$ at $Z=\rm Z_{\odot}$ for Be-XRBs, while the observed value for all types of HMXBs is $\mathcal{N}_{\rm X}=0.45_{-0.09}^{+0.06}\ \rm M_{\odot}^{-1}\ yr$ \citep{Kovlakas2020}. Interestingly, the predicted evolution with $Z$ of Be-XRBs follows a similar trend as that seen in observations of all types of HMXBs at $Z\sim 0.0003-0.02$.} %, implying potential dominance of Be-XRBs. }
%{The observed numbers from \citet{Lehmer2021} can be fully explained by our Be-XRB population at $Z\sim 0.0003-0.02$, if given a higher correction factor $f_{\rm corr}=0.8$ than that assumed by default ($f_{\rm corr}=0.5$), as shown in Appendix~\ref{apx:model}. %Fig.~\ref{fig:nxsfr_fdcs0.8}. 
{However, at $Z\gtrsim 0.02$ the decrease of $\mathcal{N}_{\rm X}$ with $Z$ is much stronger for our simulated Be-XRBs compared with the observed trend for all types of HMXBs, such that the predicted number for Be-XRBs becomes lower than the observed number for all types of HMXBs by a factor of a few at $Z=0.03$.} This is likely caused by strong stellar winds and poor statistics of the small number ($\sim 300$) of Be-XRBs identified from the BPS run for $Z=0.03$. 
If this feature is true, it means that Be-XRBs play much less important roles at $Z>0.02$, where wind-fed XRBs make up the majority of ultra-luminous sources \citep[e.g.,][]{Wiktorowicz2021}. 
The situation for sources with $L_{[0.5-8]~\rm keV}>10^{38}\ \rm erg\ s^{-1}$ \citep{Douna2015} is similar. Although not shown here for conciseness, we find similar trends for Be-XRBs with $L_{[0.5-8]~\rm keV}>10^{37}\ \rm erg\ s^{-1}$. In this case, we have $\mathcal{N}_{\rm X}\sim 10\ \rm M_{\odot}^{-1}\ yr$ at $Z\sim \rm Z_{\odot}$, consistent with observations of HMXBs in nearby galaxies \citep{Gilfanov2023}. Moreover, for Be-XRBs with $L_{[0.5-8]~\rm keV}\gtrsim 10^{35}\ \rm erg\ s^{-1}$, we predict $\mathcal{N}_{\rm X}\sim 60-130\ \rm M_{\odot}^{-1}\ yr$ at $Z\sim 0.0035-0.0142$, again below the observed rate $\sim 135\ \rm M_{\odot}^{-1}\ yr $ for all types of HMXBs in nearby star-forming galaxies \citep{Mineo2012,Gilfanov2023,Lazzarini2023}.

%It will be shown in Sec.~\ref{sec:lxsfr} below that 

%It will be shown in Appendix~\ref{apx:model} that under the same $f_{\rm corr}$ the other 3 models in Table~\ref{tab:model} with higher initial rotation rates and VDD densities predict slightly and significantly higher numbers of luminous Be-XRBs than the SR\_CS model for $Z<\rm Z_{\odot}$, respectively. 

\subsection{Masses and orbital parameters}\label{sec:properties}

For conciseness, in this section we only show the statistics of Be-XRBs for the SR\_CS model at three representative metallicities, $Z=10^{-4}$, $0.0035$ (SMC) and 0.0142 (MW), to illustrate the general trends. Each Be-XRB $i$ has a weight $\tau_{i}/M_{\rm tot}$, {so that the number for each bin in the plots of distributions (Fig.~\ref{fig:mass_dis} and \ref{fig:eloga}) corresponds to the expected number of binaries in the Be-XRB phase per unit SFR ($\rm M_\odot\ yr^{-1}$)}. %with a constant $\rm SFR=1\ M_{\odot}\ yr^{-1}$. 

Fig.~\ref{fig:mass_dis} shows the mass distributions of Be-XRBs and their progenitors as well as the initial-remnant mass relation for primary stars of Be-XRB progenitors. Here the (initial) primary star ($M_{1}$) is the progenitor of the compact object ($M_{\rm X}$), and the (initial) secondary star ($M_{2}$) is the progenitor of the O/Be star ($M_{\star}$). We find a positive correlation between the O/Be star mass and compact object mass, as well as between the progenitor masses, and that this correlation is stronger at lower $Z$. The reason is that more massive primary stars tend to have more massive remnants (i.e., higher $M_{\rm X}$) and also require more massive secondary stars to have stable mass transfer. 
%and meanwhile transfer more mass to the secondary stars, which also need to be more massive to have stable mass transfer. 
Both the typical primary and secondary masses increase with $Z$ which, combined with the bottom-heavy IMF, explains the decreasing formation efficiency of Be-XRBs at higher $Z$ (Fig.~\ref{fig:nxsfr}). This trend is caused by two effects: The formation of NSs and BHs requires more massive primary stars at higher $Z$ due to stronger stellar winds, and the secondary stars that are more compact at lower $Z$ are more easily spun up to become O/Be stars. These two effects further complement each other by the stability of mass transfer. 
Nevertheless, the high-mass tail of the progenitor mass distribution shrinks with increasing $Z$. This is caused by an effect that involves less massive stars given stronger winds at higher $Z$: Strong stellar winds from the most massive stars drive significant expansion of binary orbits so that mass transfer (which is usually required to make O/Be stars in the SR models) %\footnote{The trend of shrinking high-mass tail with increasing $Z$ is much weaker in the FR models where the secondary stars rotate rapidly from the beginning and do not need to be spun up by mass transfer to become O/Be stars.}) 
is suppressed, and the chance of forming close enough binaries to allow accretion from VDDs is also reduced. In fact, the trend of shrinking high-mass tail with increasing $Z$ is much weaker in the FR models where the secondary stars rotate rapidly from the beginning and do not need to be spun up by mass transfer to become O/Be stars.
%unstable mass transfer episodes are more likely to occur for massive stars that are larger at higher $Z$, which can significantly shrink binary orbits, leading to stellar mergers or Roche lobe overflow of the initial secondary stars (after the initial primary stars become compact objects).

\begin{figure}
    \centering
    \includegraphics[width=1\columnwidth]{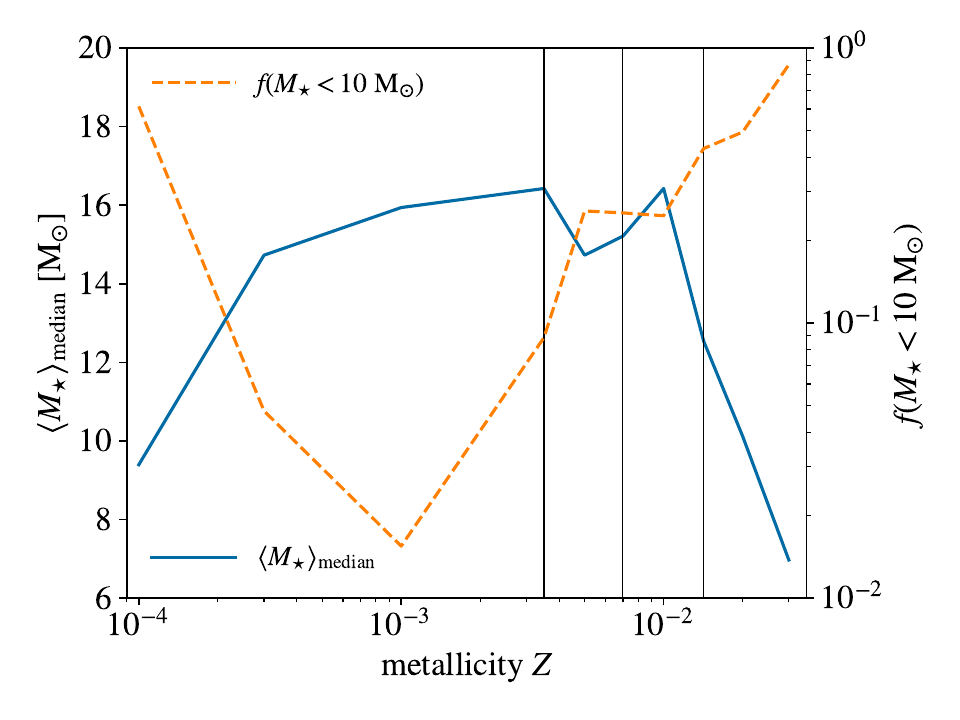}
    \vspace{-25pt}
    \caption{Median O/Be star mass (solid curve for the left axis) in Be-XRBs and fraction (dashed curve for the right axis) of Be-XRBs with low-mass ($<10\ \rm M_{\odot}$) O/Be stars in the SR\_CS model. The thin vertical lines label the metallicities of the MW, LMC and SMC (from right to left).}
    \label{fig:mbe}
\end{figure}

\begin{figure*}
    \centering
    \subfloat[$Z=10^{-4}$]{\includegraphics[width=0.7\columnwidth]{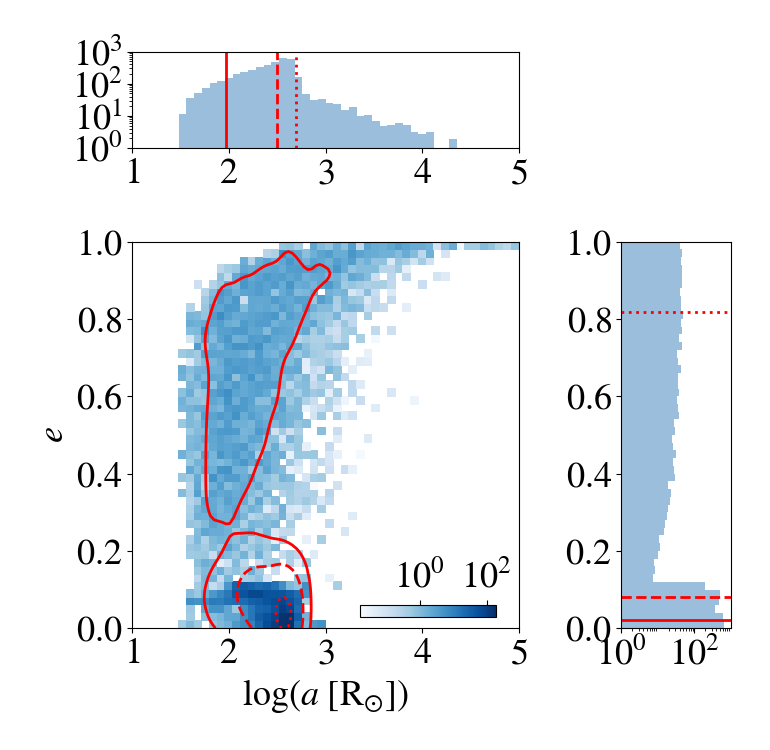}}
    \subfloat[$Z=0.0035$ (SMC)]{\includegraphics[width=0.7\columnwidth]{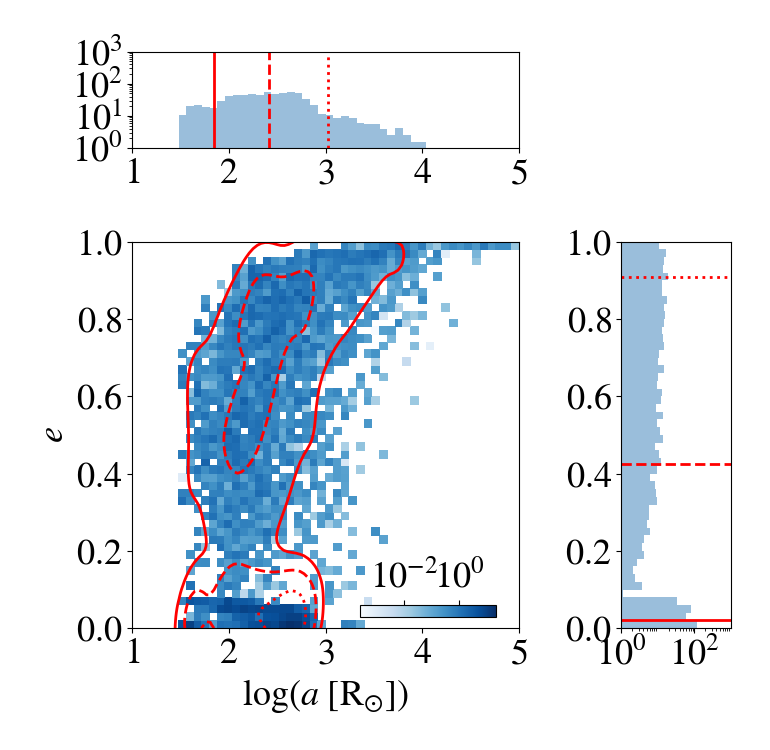}}
    \subfloat[$Z=0.0142$ (MW)]{\includegraphics[width=0.7\columnwidth]{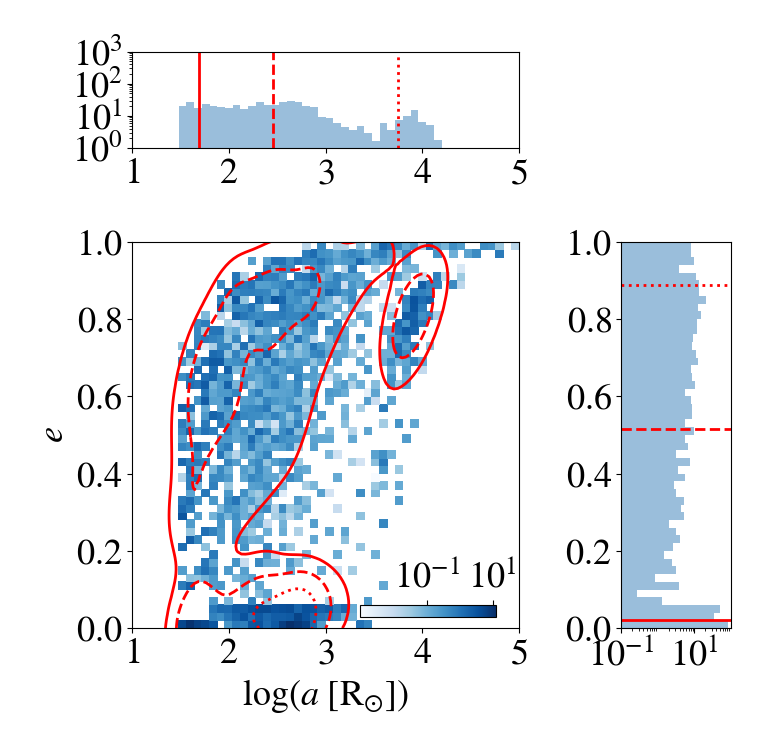}}
    \vspace{-8pt}
    \caption{Distribution of Be-XRB orbital parameters in the SR\_CS model for $Z=10^{-4}$ (left), $Z=0.0035$ (middle) and $Z=0.0142$ (right). {Each Be-XRB $i$ is weighted by $\tau_{i}/M_{\rm tot}$, so that the number for each bin here corresponds to the expected number of binaries in the Be-XRB phase per unit SFR ($\rm M_\odot\ yr^{-1}$). In the 2D maps of joint distribution, the solid, dashed and dotted contours enclose 90\%, 50\% and 10\% of systems from a Gaussian smoothed density field produced by the weighted data. In the marginalized distributions, the solid, dashed and dotted lines mark the 10, 50 and 90 percentiles.}}% The data point areas are normalized within each scatter plot separately and should not be compared across different panels.}
    \label{fig:eloga}
\end{figure*}

Because of stellar winds, the maximum compact object mass decreases with $Z$ from $M_{\rm X,\max}\simeq 4.3\ \rm M_{\odot}$ at $Z=10^{-4}$ to $M_{\rm X,\max}\simeq 1.6\ \rm M_{\odot}$ at $Z=0.03$. There are no BHs (with $M_{\rm X}>2.2\ \rm M_{\odot}$) at $Z\gtrsim 0.0035$, %consistent with previous BPS predictions and observations \citep{Zhang2004,Belczynski2009,Casares2014,Grudzinska2015,Brown2018,Zamanov2022}, 
and the fraction of BH systems is $\sim 1-3\%$ at lower $Z$. 
In general, our Be-XRBs are dominated by low-mass NSs with $M_{\rm X}\sim 1.3\ \rm M_{\odot}$. The mass distribution of O/Be stars shows complex $Z$ dependence. To demonstrate the general trends, 
we show the median O/Be star mass in Be-XRBs (solid curve) as well as the fraction of Be-XRBs hosting low-mass ($M_{\star}< 10\ \rm M_{\odot}$) O/Be stars (dashed curve) as a function of $Z$ in Fig.~\ref{fig:mbe}. The median O/Be star mass increases from $\sim 9\ \rm M_{\odot}$ at $Z=10^{-4}$ to $\sim 16\ \rm M_{\odot}$ at $Z=0.001$ and shows little evolution for $Z\sim 0.001-0.01$ before dropping rapidly for $Z>0.01$ down to $\sim 7\ \rm M_{\odot}$ at $Z=0.03$. 
The increasing trend at low $Z$ can be explained by the increase of secondary mass with $Z$ and higher mass transfer rates from more massive primary stars with larger radii (as $\dot{M}_{\rm acc,\max}\propto R_{\rm acc}$) 
at higher $Z$. For $Z\gtrsim 0.01$, formation of massive Be-XRBs from massive progenitors are increasingly suppressed by orbital expansion with stronger stellar winds at higher $Z$, which also reduce the masses of O/Be stars, so that $M_{\star}$ generally decreases with $Z$. 
The fraction of Be-XRBs with low-mass ($M_{\star}< 10\ \rm M_{\odot}$) O/Be stars (dashed curve in Fig.~\ref{fig:mbe}) decreases from $\sim 60$\% at $Z=10^{-4}$ to $\sim 1.5$\% at $Z=0.001$, and then increases quasi-monotonically with $Z$, reaching $\sim 87$\% at $Z=0.03$. This is consistent with the evolution of median O/Be star mass with $Z$ and can also be explained by the above arguments about secondary mass, mass transfer rate and stellar winds. %with the evolution of median O/Be star mass. %The corresponding O/Be star masses increase with $Z$, which drives the increase of median $M_{\star}$ with $Z$ at $Z\lesssim 0.007$. 

%\begin{comment}
%\end{comment}

We notice in Fig.~\ref{fig:mass_dis} that there are two groups of Be-XRBs in the $\log M_{\rm X}-\log M_{1}$ space with lower and higher $M_{\rm X}$ (approximately divided by the dash-dotted lines in the bottom panels of Fig.~\ref{fig:mass_dis}) that are more distinct at higher $Z$. These two groups are also related to the complex features of Be-XRB mass distributions in the $\log M_{\rm X}-\log M_{\star}$ space. The lower-$M_{\rm X}$ group %is more important at lower $Z$, which 
is made of binaries with relatively low-mass ($M_{1}\lesssim 25\ \rm M_{\odot}$, i.e., $\log(M_{1}\ [{\rm M_{\odot}}])\lesssim 1.4$) primary stars that always transfer significant mass to the secondary star. %and form NSs with $M_{\rm X}\sim 1.3\ \rm M_{\odot}$. 
These binaries also cluster around a primary mass that increases with $Z$. 
The higher-$M_{\rm X}$ group contains both low-mass and massive primary stars. %, and the fraction of low-mass primaries decreases with $Z$. 
The resulting O/Be star masses cover a broader range than those from the lower-$M_{\rm X}$ group. %Given the same primary mass, mass transfer is less significant in the higher-$M_{\rm X}$ group, leading to less massive O/Be stars, especially  $M_{\star}\lesssim 10\ \rm M_{\odot}$. 
We expect the primary stars in the lower-$M_{\rm X}$ group to completely lose their hydrogen envelope\footnote{Recently, a new sample of low-mass helium stars has been discovered in Magellanic Clouds, which are expected to originate from stars of initial masses $\sim 8-25\ \rm M_{\odot}$ that are stripped by binary interactions \citep{Drout2023,Gotberg2023}. These helium stars are similar to the NS progenitors of Be-XRBs in the low-$M_{\rm X}$ group generated by our BPS runs.} and even undergo mass transfer during the helium HG phase. 
In this way, most of them form low-mass NSs ($M_{\rm X}\sim 1.3\ \rm M_{\odot}$) via electron-capture SNe with no natal kicks. %The secondary stars then gain significant amounts of mass to become O/Be stars of $M_{\star}\sim 10-30\ \rm M_{\odot}$. 
We find that such systems account for about half of the Be-XRBs currently in the SMC (with each simulated Be-XRB re-weighted according to the star formation history of the SMC, see Appendix~\ref{apx:smc}), which is consistent with the results of \citet{Vinciguerra2020}. 
In contrast, the primary stars in the higher-$M_{\rm X}$ group keep a fraction of their hydrogen envelopes before collapse either due to higher initial masses, shorter lifetimes and/or less mass transfer in wider orbits. There is evidence of this scenario from observations of partially-stripped star + Be star binaries such as HR 6819 \citep{Frost2022} and SMCSGS-FS 69 \citep{Ramachandran2023}. In general, both groups produce O/Be stars of a broad mass range, and the mass distribution of O/Be stars becomes bi-polar at $Z\gtrsim 0.001$, which is more obvious for O/Be stars from the higher-$M_{\rm X}$ group. This likely results from the complex dependence of the mass loss/accretion rates on primary/secondary masses and orbital parameters in BSE models, which also correlate with the natal kicks and remnant masses of primary stars, as hinted by observations \citep{Zhao2023}. We defer a through discussion on this aspect to future work.

%The resulting O/Be star masses show a bi-polar distribution that is more obvious at higher $Z$, where the mass accretion rate of the secondary star strongly correlate 
%The difference between 
%(especially for low-mass primaries). %to form more massive remnants. 

Fig.~\ref{fig:eloga} shows the orbital parameter distribution of Be-XRBs. 
Similarly to the results in Sec.~\ref{sec:eps}, nearly-circular ($e\lesssim 0.1$) binaries with $a\sim 100-10^{3}\ \rm R_{\odot}$ make up a significant fraction of Be-XRBs, reaching $\sim 60\%$ at $Z=10^{-4}$. This is a natural consequence of binary interactions that tend to circularize binary orbits and our optimistic definition of the VDD boundary (Eq.~\ref{rcrit}) that allows faint objects with little accretion from beyond the tidal truncation radius to be counted as Be-XRBs (Sec.~\ref{sec:id}). These systems mostly belong to the aforementioned lower-$M_{\rm X}$ group and contain low-mass ($M_{\rm X}\sim 1.3\ \rm M_{\odot}$) NSs born in electron-capture SNe of helium stars with no natal kicks. Since strong mass transfer happens in their evolution histories, the O/Be stars are initially less massive and live longer than in the case of $e>0.1$. 
We also find that it is necessary to take into account such nearly-circular systems in order to explain the large population of Be-XRBs currently observed in the SMC, because the number of nearly-circular Be-XRBs as a function of time after a starburst has a strong peak at $\sim 30\ \rm Myr$ for $Z=Z_{\rm SMC}=0.0035$, and the SMC experienced a starburst just $\sim 20-40$~Myr ago \citep{Rubele2015}. This result is consistent with the finding in \citet{Linden2009} that Be-XRBs currently in the SMC primarily form through electron-capture SNe with low natal kicks. %, given the starburst $\sim 20-40$~Myr ago. 

If we ignore the nearly-circular binaries, our results are consistent with those in \citet[see their fig.~6]{Xing2021} who define the VDD boundary with the tidal truncation radius (Eq.~\ref{rtrunc}) such that nearly-circular systems are excluded. That is to say, wide ($a\gtrsim 300\ \rm R_{\odot}$) binaries with longer $a$ need to have higher $e$ to interact with the VDD at the pericenter. On the other hand, there is an upper limit of $e$ that increases with $a$ in close binaries ($a\lesssim 300\ \rm R_{\odot}$) to avoid RLO of the O/B star. Finally, we find that the distribution of $a$ is broader at higher $Z$, while the median is almost constant at $a\sim 300\ \rm R_{\odot}$. The increase of the fraction of wide Be-XRBs at higher $Z$ can be explained by the stronger winds that widen binary orbits more significantly. 
%the larger radii of stars that allow mass transfer in initial wider binaries to form Be-XRBs. 
The increasing relative importance of very close binaries ($a\lesssim 100\ \rm R_{\odot}$) at higher $Z$ corresponds to the decreasing importance of the lower-$M_{\rm X}$ group that produces a stronger peak around $a\sim 500\ \rm R_{\odot}$ at lower $Z$, and the trend will disappear if we exclude nearly-circular binaries.

%Implications on metallicity-dependent X-ray heating of the IGM from the simulated Be-XRB populations:
%\begin{itemize}
%(Specific) X-ray luminosity per unit star formation rate (SFR) $\mathcal{L}_{{\rm X},\nu}$: 
\subsection{X-ray outputs}\label{sec:lxsfr}
%\subsubsection{Evolution of X-ray luminosity after a starburst}

\begin{table*}
    \centering
    \caption{X-ray luminosity per unit SFR as a function of metallicity $Z$ for the 4 models defined in Table~\ref{tab:model} with $f_{\rm corr}=0.5$, in terms of $\log(\mathcal{L}_{\rm X}\equiv \langle L_{\rm X}\rangle/{\rm SFR}\ [\rm\ erg\ s^{-1}\ M_{\odot}^{-1}\ yr])$, in 4 energy bands: $0.1-2$~keV, $0.5-2$~keV, $2-10$~keV and $0.5-8$~keV, as well as for the bolometric luminosity.}
    \begin{tabular}{c|cccccccccc}
    \toprule
    Absolute metallicity $Z$ & 0.0001 & 0.0003 & 0.001 & 0.0035 & 0.005 & 0.007 & 0.01 & 0.0142 & 0.02 & 0.03 \\
    \toprule
    SR\_CS \\
    \midrule
$0.1-2$~keV & 39.73 & 39.57 & 39.48 & 39.35 & 39.22 & 39.09 & 39.00 & 38.78 & 38.63 & 37.67 \\
$0.5-2$~keV & 39.70 & 39.54 & 39.45 & 39.33 & 39.20 & 39.07 & 38.98 & 38.75 & 38.60 & 37.64 \\
$2-10$~keV & 40.07 & 39.90 & 39.81 & 39.70 & 39.59 & 39.47 & 39.37 & 39.16 & 39.01 & 38.07 \\
$0.5-8$~keV & 40.19 & 40.02 & 39.93 & 39.82 & 39.70 & 39.58 & 39.48 & 39.27 & 39.12 & 38.17 \\
Bolometric & 40.43 & 40.26 & 40.17 & 40.06 & 39.96 & 39.84 & 39.74 & 39.53 & 39.38 & 38.44 \\
    \toprule
    FR\_CS \\
    \midrule
$0.1-2$~keV & 39.78 & 39.62 & 39.55 & 39.40 & 39.36 & 39.23 & 39.05 & 38.96 & 38.61 & 37.65 \\
$0.5-2$~keV & 39.75 & 39.59 & 39.51 & 39.37 & 39.33 & 39.21 & 39.02 & 38.94 & 38.58 & 37.62 \\
$2-10$~keV & 40.12 & 39.94 & 39.87 & 39.74 & 39.70 & 39.59 & 39.41 & 39.32 & 39.00 & 38.10 \\
$0.5-8$~keV & 40.24 & 40.07 & 40.00 & 39.86 & 39.82 & 39.71 & 39.53 & 39.44 & 39.10 & 38.18 \\
Bolometric & 40.48 & 40.30 & 40.23 & 40.11 & 40.07 & 39.96 & 39.79 & 39.69 & 39.37 & 38.48 \\
    \toprule
    SR\_OP\\
    \midrule
$0.1-2$~keV & 40.33 & 40.17 & 40.05 & 39.94 & 39.77 & 39.52 & 39.17 & 38.78 & 38.63 & 37.67 \\
$0.5-2$~keV & 40.30 & 40.14 & 40.02 & 39.91 & 39.75 & 39.49 & 39.15 & 38.75 & 38.60 & 37.64 \\
$2-10$~keV & 40.63 & 40.44 & 40.33 & 40.22 & 40.08 & 39.85 & 39.52 & 39.16 & 39.01 & 38.07 \\
$0.5-8$~keV & 40.77 & 40.59 & 40.47 & 40.37 & 40.22 & 39.98 & 39.64 & 39.27 & 39.12 & 38.17 \\
Bolometric & 40.98 & 40.79 & 40.68 & 40.58 & 40.44 & 40.22 & 39.89 & 39.53 & 39.38 & 38.44 \\
    \toprule
    FR\_OP \\
    \midrule
$0.1-2$~keV & 40.40 & 40.26 & 40.17 & 40.05 & 39.88 & 39.64 & 39.29 & 38.96 & 38.61 & 37.65 \\
$0.5-2$~keV & 40.37 & 40.23 & 40.14 & 40.02 & 39.86 & 39.62 & 39.26 & 38.94 & 38.58 & 37.62 \\
$2-10$~keV & 40.68 & 40.53 & 40.44 & 40.32 & 40.17 & 39.95 & 39.63 & 39.32 & 39.00 & 38.10 \\
$0.5-8$~keV & 40.83 & 40.68 & 40.59 & 40.47 & 40.31 & 40.09 & 39.75 & 39.44 & 39.10 & 38.18 \\
Bolometric & 41.04 & 40.87 & 40.78 & 40.67 & 40.52 & 40.31 & 39.99 & 39.69 & 39.37 & 38.48 \\
    \bottomrule
    \end{tabular}
    \label{tab:lxsfr}
\end{table*}

\begin{comment}
\begin{figure*}
    \centering
    \subfloat[$Z=10^{-4}$]{\includegraphics[width=0.7\columnwidth]{spec0_sfr_bexrb_xt_Z1e-4.pdf}}
    \subfloat[$Z=0.0035$ (SMC)]{\includegraphics[width=0.7\columnwidth]{spec0_sfr_bexrb_xt_ZSMC.pdf}}
    \subfloat[$Z=0.0142$ (MW)]{\includegraphics[width=0.7\columnwidth]{spec0_sfr_bexrb_xt_Zsun.pdf}}
    \vspace{-10pt}
    \caption{SED of Be-XRBs per unit SFR from the SR\_CS model with $f_{\rm corr}=0.5$ at $Z=10^{-4}$ (left), $0.0035$ (middle) and 0.0142 (right). The total spectrum is shown with the thick solid curve. Contributions from Type I and II outbursts are denoted by the solid and dashed curves. Components of the LH, HS and SE states are also shown with the dash-dotted, dotted and long-dashed curves. In the $Z=10^{-4}$ case where BHs are present, we show the NS and BH contributions with the dash-dot-dotted and densely-dashed curves. The shaded regions show different levels of absorption by the ISM and IGM as defined in Fig.~\ref{fig:specmodel}.}
    \label{fig:spec_sfr}
\end{figure*}
\end{comment}

%Since Be-XRBs are short-lived objects, it is well justified to evaluate their X-ray outputs in terms of the X-ray luminosity per unit SFR for a long enough star formation timescale ($\tau$). 

\begin{figure}
    \centering
    \includegraphics[width=\columnwidth]{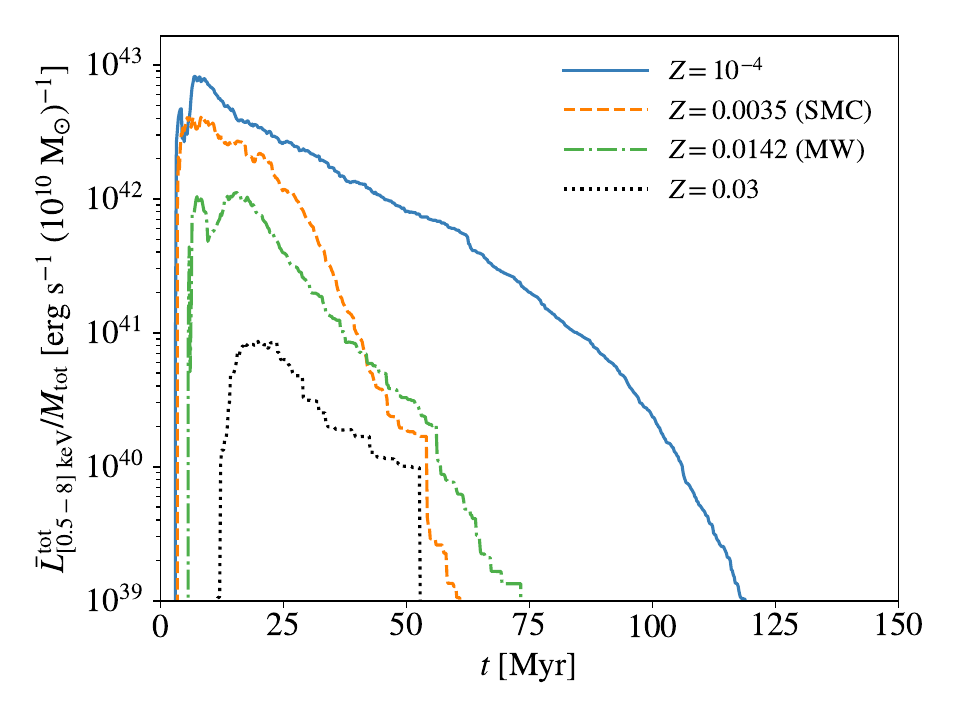}
    \vspace{-25pt}
    \caption{Evolution of X-ray luminosity in the $0.5-8$~keV band per unit stellar mass from Be-XRBs formed by an instantaneous starburst in the SR\_CS model with $f_{\rm corr}=0.5$, for $Z=10^{-4}$ (solid), 0.0035 (dashed), 0.0142 (dash-dotted) and 0.03 (dotted). Here $t=0$ corresponds to the moment of starburst.}% The later onset/peak at higher $Z$ is caused by stronger stellar winds that reduce the progenitor masses }
    \label{fig:lx_t}
\end{figure}

To characterize the X-ray outputs from the simulated Be-XRB populations, we start with the time evolution of total X-ray luminosity per unit stellar mass from an instantaneous starburst (at $t=0$):
\begin{align}
    \frac{\bar{L}^{\rm tot}_{\nu}(t)}{M_{\rm tot}}=\frac{\sum_{i}^{N}f_{{\rm duty},i}L_{\nu,i}\Theta(t-t_{i,\rm ini})\Theta(t_{i,\rm fin}-t)}{M_{\rm tot}}\ ,
\end{align}
where Be-XRB $i$ lives from $t_{i,\rm ini}$ to $t_{i,\rm fin}$ with a (specific) luminosity $L_{\nu,i}$ during outbursts for a certain energy (band)\footnote{The specific luminosity is defined as $L_{\nu}\equiv dL/d\nu$, while for a given energy band $\nu\in [\nu_{1},\nu_{2}]$, we have $L_{[\nu_{1},\nu_{2}]}\equiv \int_{\nu_{1}}^{\nu_{2}} L_{\nu}d\nu$.} and a duty cycle $f_{{\rm duty},i}$, and $\Theta$ is the Heaviside step function. Fig.~\ref{fig:lx_t} shows the results for the $0.5-8$~keV band from the SR\_CS model with $f_{\rm corr}=0.5$ at 4 representative metallicities. There is a delay of $\sim 3-10$~Myr between the starburst and the onset of X-ray emission from Be-XRBs, which reflects the evolutionary time for (the most) massive primary stars in Be-XRB progenitors to become compact objects. The total X-ray luminosity peaks at a few Myr after the starburst for $Z\lesssim 0.01$, while the peak (as well as the onset) of X-ray emission shifts to later stages at higher $Z$, up to $\sim 20$~Myr for $Z=0.03$. This is caused by the suppression of Be-XRBs from massive binaries by stellar winds that is more significant at higher $Z$ (see Sec.~\ref{sec:properties}) and longer lifetimes of (initially less massive) primary stars in Be-XRB progenitors with higher $Z$. The X-ray luminosity drops by at least 2 orders of magnitude within 100~Myr post the peak. The most metal-poor model with $Z=10^{-4}$ has a much slower drop compared with the other cases due to a higher fraction of Be-XRBs with low-mass Be stars from low-mass progenitors (see Figs.~\ref{fig:mass_dis} and \ref{fig:mbe}). 

\begin{figure}
    \centering
    %\subfloat[$Z=10^{-4}$]
    {\includegraphics[width=1\columnwidth]{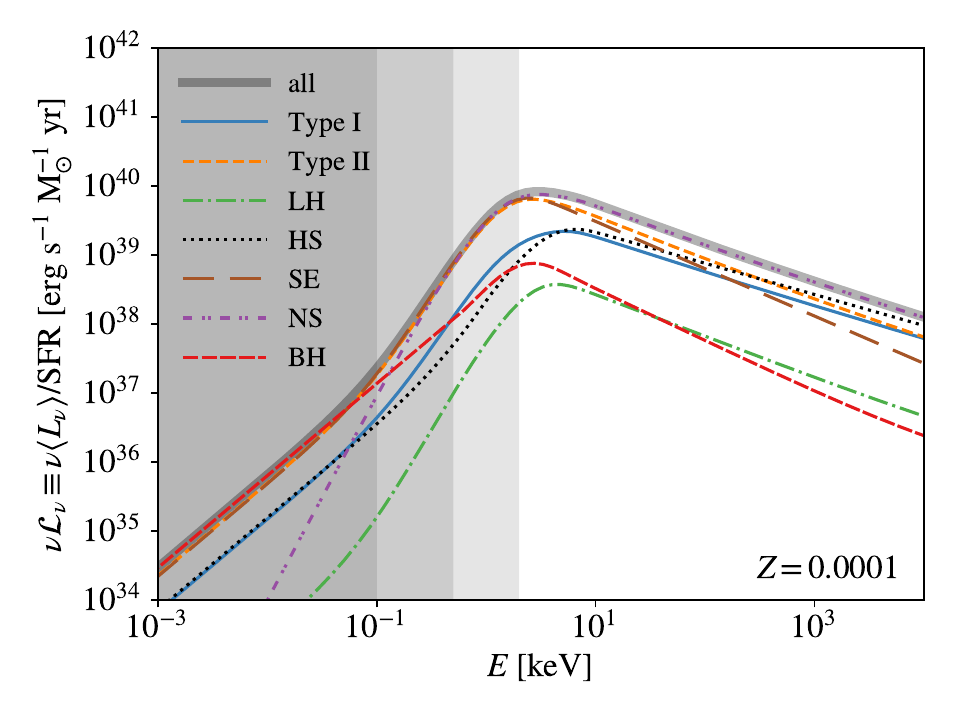}}\\
    \vspace{-20pt}
    %\subfloat[$Z=0.0035$ (SMC)]
    %{\includegraphics[width=1\columnwidth]{spec0_sfr_bexrb_xt_ZSMC.pdf}}\\
    %\vspace{-20pt}
    %\subfloat[$Z=0.0142$ (MW)]
    {\includegraphics[width=1\columnwidth]{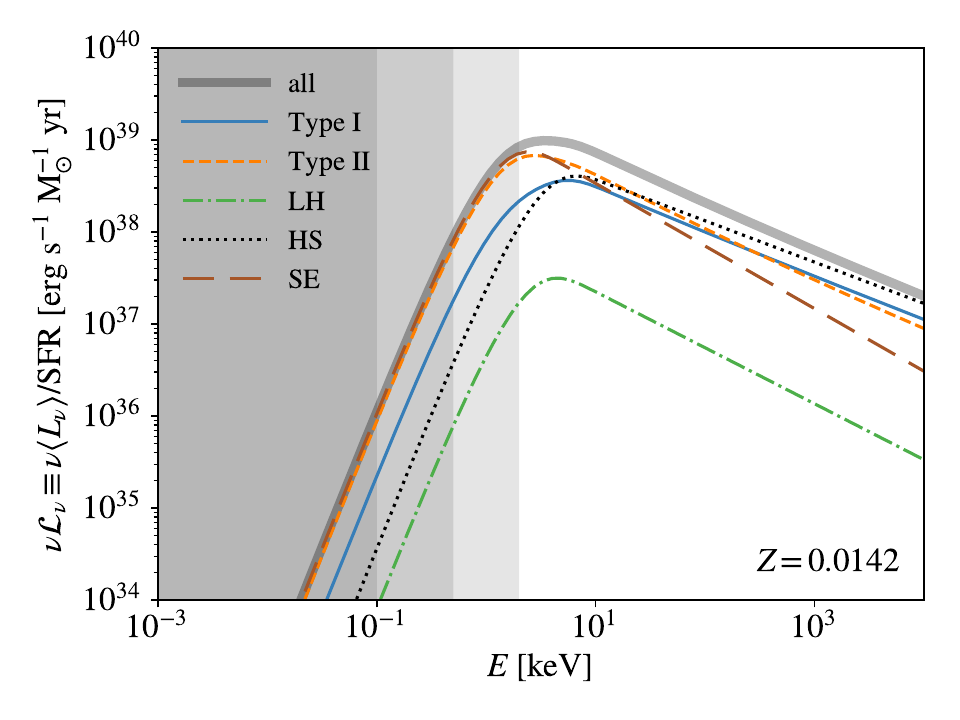}}
    \vspace{-25pt}
    \caption{SED of Be-XRBs per unit SFR from the SR\_CS model with $f_{\rm corr}=0.5$ at $Z=10^{-4}$ (top) %, $0.0035$ (middle) 
    and 0.0142 (bottom). The total spectrum is shown with the thick solid curve. Contributions from Type I and II outbursts are denoted by the solid and dashed curves. Components of the LH, HS and SE states are shown with the dash-dotted, dotted and long-dashed curves. In the $Z=10^{-4}$ case where BHs are present, we show the NS and BH contributions with the dash-dot-dotted and densely-dashed curves. The shaded regions show different levels of absorption as defined in Fig.~\ref{fig:specmodel}.}
    \label{fig:spec_sfr}
\end{figure}

Now, given the short-lived nature of Be-XRBs, their overall X-ray output can be well characterized by the (specific) X-ray luminosity per unit SFR:
\begin{align}
    \mathcal{L}_{\nu}&\equiv \frac{\langle L_{\nu}\rangle}{{\rm SFR}}%=\frac{1}{{\rm SFR}}\sum_{i}^{N}f_{{\rm duty},i}L_{\nu,i}\left(\frac{\tau_{i}{\rm SFR}}{M_{\rm tot}}\right)\notag\\
    %\left(\frac{M_{\rm tot}}{\tau_{\rm SF}}\right)^{-1}\notag\\
    =\sum_{i}^{N}\left(\frac{f_{{\rm duty},i}\tau_{i}}{M_{\rm tot}}\right)L_{\nu,i}\ ,\label{lxsfr}
\end{align}
where $\tau_{i}=t_{i,\rm fin}- t_{i,\rm ini}$ is the duration of the Be-XRB phase for binary $i$. 
Fig.~\ref{fig:spec_sfr} shows the full (intrinsic) SEDs in terms of $\nu\mathcal{L}_{\nu}$ for %3 representative metallicities, 
$Z=10^{-4}$ %, $0.0035$ (SMC) 
and 0.0142, in the SR\_CS model with $f_{\rm corr}=0.5$, where we also plot the contributions of different types of outbursts (Type I and II, see Sec.~\ref{sec:xray}), accretion regimes (LH, HS and SE, see Sec.~\ref{sec:spec}) and compact objects (NS and BH). In general, for the photon energy range $E\sim 0.5-8$~keV that typically contains $\sim 60\%$ of the integrated bolometric luminosity from a Be-XRB population, the SED is always dominated by NSs in the SE regime with $\eta>2$ mostly via Type~II outbursts, and the contribution of BHs remains below 10\%. This means that the spectral shape is almost independent of $Z$ in this energy range and the majority of X-ray emission comes from luminous ($\gtrsim 10^{38}\ \rm erg\ s^{-1}$) systems. 
In fact, SE systems contribute $\sim 55-68$\% of the total bolometric luminosity per unit SFR for $Z\in [10^{-4}, 0.03]$ with mildly increasing importance at lower $Z$. Similarly, Type~II outbursts make up $\sim 56-71$\% of the total X-ray output due to their luminous nature, even though the \textit{number} fraction of all (active) Type~II systems is (much) lower, $\sim 13-44\ (2-16$)\%. 
%The dominance of SE regime is more significant at lower $Z$ with more massive O/Be stars (see Fig.~\ref{fig:mass_dis}) that have denser VDDs (as well as in the OP models where VDDs are generally denser). 
For more energetic photons ($E\gtrsim 8$~keV), the contribution from NSs in the HS regime ($\eta\sim 0.05-2$) becomes important due to their harder spectra, especially at higher $Z$, while for $E\lesssim 0.1$~keV, BHs (mostly in the SE regime during Type~II outbursts) dominate the spectrum when they exist in Be-XRBs at $Z\lesssim 0.001$. 
%The dominance of Type~II outbursts results from their luminous nature, since the number fraction of Be-XRBs with Type~II outbursts is typically $\sim 40\%$ (and within the range of $12-58\%$ in all cases considered), which is also consistent with observations \citep{Cheng2014}. 

%When the intrinsic SED is known, one needs to further model the effect of ISM absorption by the host halo. 
%{\color{blue}(optional) We design a simple analytical model for high-$z$ haloes that captures the ISM absorption effect with a low-energy cutoff, as shown in Appendix~\ref{apx:ism}. We plan to combine our BPS results with this model in future work to explore the impact of Be-XRBs on the IGM.} 
Next, to better evaluate the metallicity dependence,  we calculate the X-ray luminosity per unit SFR in select energy bands for each simulated Be-XRB population. 
We start with two energy bands with very-soft ($0.1-2$~keV) and soft ($0.5-2$~keV) X-rays that are expected to escape the host haloes and interact with the IGM at $z\gtrsim 20$ and $z\sim 10$, respectively \citep{Das2017,Sartorio2023}. We further consider a hard band ($2-10$~keV) and a broad band ($0.5-8$~keV) 
%we calculate the integrated luminosity for very-soft ($ 0.1-2$~keV), soft ($0.5-2$~keV), hard ($2-10$~keV) and broad-band ($0.5-8$~keV) X-rays 
to make comparison with literature results (see below). %BPS results for other types of HMXBs (Sec.~\ref{sec:comp_bps} and observations of HMXBs in nearby galaxies (Sec.~\ref{sec:comp_obs}). 
Fig.~\ref{fig:lxsfr_Z} shows the X-ray luminosity per unit SFR as a function of $Z$ in these 4 bands for the SR\_CS model with $f_{\rm corr}=0.5$. The results for all the 4 models in Table~\ref{tab:model} are summarized in Table~\ref{tab:lxsfr}.

\subsubsection{Be-XRBs VS other types of HMXBs}\label{sec:comp_bps}

\begin{figure}
    \centering
    \includegraphics[width=\columnwidth]{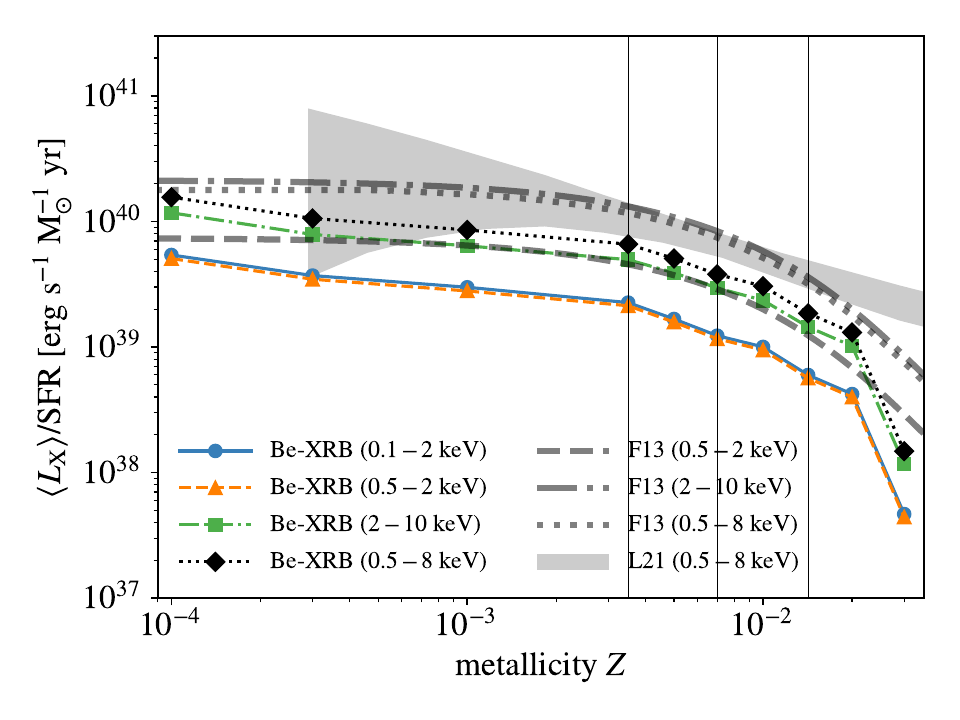}
    \vspace{-25pt}
    \caption{X-ray luminosity per unit SFR of Be-XRBs in the SR\_CS model with $f_{\rm corr}=0.5$, in the energy bands $E\sim 0.1-2$ (solid), $0.5-2$ (dashed), $2-10$ (dash-dotted) and $0.5-8$~keV (dotted). For comparison, we plot the best-fit models from BPS results for other types of HMXBs in \citet[F13]{Fragos2013} with the thick curves following the same line styles for the latter three bands. Here for the $0.5-2$~keV and $2-10$~keV bands, we use the fitting formulae and parameters in their eq.~3 and table 2, while the results for the $0.5-8$~keV band are taken from \citet[see their fig.~4]{Lehmer2021}. The observational results for the $0.5-8$~keV band from \citet[L21, see their fig.~4 and table~3]{Lehmer2021} are shown with the shaded region. The thin vertical lines label the metallicities of the MW, LMC and SMC (from right to left).}
    \label{fig:lxsfr_Z}
\end{figure}

We first compare our predictions for Be-XRBs with the BPS results for \textit{other} types of HMXBs powered by RLO and (spherical) stellar winds \citep[][]{Fragos2013}. %and observations for all types of HMXBs \citep[][]{Lehmer2021}. 
As shown in Fig.~\ref{fig:lxsfr_Z}, we find that (at the same $Z$) the X-ray luminosity (per unit SFR) from Be-XRBs in our SR\_CS model with $f_{\rm corr}=0.5$ is comparable (lower by up to a factor of 3) to that from other types of HMXBs predicted by \citet{Fragos2013} for $Z\sim 0.001-0.02$, where the evolution with $Z$ is also similar. This indicates that Be-XRBs can be as important as other types of HMXBs. Moreover, the striking similarity between the metallicity dependence of X-ray outputs in the two populations of HMXBs powered by distinct mechanisms (decretion disks of O/Be stars VS RLO and stellar winds) can be understood by the fact that their key properties (number counts and accretion rates of compact objects) are determined by the same binary stellar evolution processes (e.g., stellar winds, mass transfer and SN natal kicks) in this metallicity range. 
%between the $Z$-dependence X-ray outputs from HMXBs powered by quite different mechanisms. 

However, at $Z\lesssim 0.001$ and $Z\gtrsim 0.02$, our results for Be-XRBs show stronger evolution with $Z$. In \citet{Fragos2013}, the X-ray luminosity from other types of HMXBs is almost constant for $Z\lesssim 0.001$, while in our case, the X-ray luminosity from Be-XRBs increases by a factor of $\sim 2$ from $Z=0.001$ to $Z=10^{-4}$. %This difference may not be meaningful because the BPS runs in \citet{Fragos2013} only cover the metallicity range $Z\sim 0.001-0.025$, so that the weaker evolution at $Z\lesssim 0.001$ and $Z\gtrsim 0.02$ can be an artificial feature of their fitting formula (see their eq.~3). 
The rapid drop of X-ray luminosity at $Z>0.02$ in our case is caused by the stronger stellar winds at higher $Z$ that significantly reduce the number of (luminous) Be-XRBs (see Figs.~\ref{fig:nxsfr} and \ref{fig:nxsfr_fdcs}), such that the predicted X-ray output has large errors ($\sim 0.5$~dex) %and is likely underestimated 
due to the poor statistics of luminous Be-XRBs. {In fact, the X-ray output of the $Z=0.03$ model can be enhanced by a factor of $\sim 10$ when using the wind model in \citet{Hurley2002} that predicts weaker winds at high $Z$ than considered in our default wind prescription based on \citet{Schneider2018} and \citet{Sander2020}.  However, the effect of varying wind models on the total X-ray output is much weaker at lower metallicities (within factors of $\sim 3$ and $\sim 20$\% for $Z\lesssim 0.02$ and $Z\lesssim 0.005$, respectively).}
%which is also uncertain due to the small . 
%\footnote{The X-ray output of the $Z=0.03$ model can be enhanced by a factor of $\sim 10$ when using the wind model in \citet{Hurley2002} that produce weaker winds at high $Z$ than considered in our default wind prescription based on \citet{Schneider2018} and \citet{Sander2020}. { However, the effect of varying wind models on the total X-ray output is much weaker at lower metallicities (within a factor of $\sim 3$ and $\sim 20$\% for $Z\lesssim 0.02$ and $Z\lesssim 0.005$, respectively).}}

The difference between our results and those in \citet{Fragos2013} only varies slightly with the energy band considered, with the broad band ($0.5-8$~keV) having the smallest difference and the hard band ($2-10$~keV) the largest. This indicates that the overall SEDs of Be-XRB populations in our case are similar to those in \citet{Fragos2013} for other types of HMXBs. For instance, we have $L_{\rm [0.5-2]\ keV}/L_{\rm [2-10]\ keV}\simeq 0.41$, and this ratio is $\simeq 0.35$ in \citet{Fragos2013}. 
The small difference is caused by the different X-ray spectral models adopted in our work (Fig.~\ref{fig:spec_sfr}) and by \citet[see their fig.~1]{Fragos2013}. The latter predict slightly harder spectra with stronger X-ray emission at $E\gtrsim 2$~keV. 
The reason is that their spectral model \citep{Fragos2013xrb} only considers the LH and HS regimes and is calibrated to the BC factors derived from the early samples of Galactic XRBs in \citet{McClintock2006} and \citet{Wu2010}, while our model further includes the SE regime (which is dominant for Be-XRBs) and adopts the recent observational data for BC factors in \citet{Anastasopoulou2022}. 

% (see fig.~1 in \citet{Fragos2013} compared with our Fig.~\ref{fig:spec_sfr}).

\subsubsection{Comparison with HMXBs observed in nearby galaxies}\label{sec:comp_obs}

Next, we compare the overall X-ray outputs from our Be-XRB populations with observational results for HMXB populations in nearby galaxies \citep[][]{Douna2015,Lehmer2021}. Assuming that observations are complete, we find that {the simulated Be-XRBs in our fiducial model (SR\_CS with $f_{\rm corr}=0.5$)} can explain $\sim 60\%$ of the observed X-ray luminosity of (all types of) HMXBs in nearby galaxies \citep{Lehmer2021}, with a similar metallicity dependence (see the shaded region in Fig.~\ref{fig:lxsfr_Z}). The X-ray luminosity increases by about one order of magnitude from $Z\sim 0.02$ to $Z\lesssim 0.0003$. This is consistent with the results in Sec.~\ref{sec:eps} for number counts of ultra-luminous HMXBs with $L_{[0.5-8]~\rm keV}>10^{39}\ \rm erg\ s^{-1}$ (Fig.~\ref{fig:nxsfr_fdcs}). {%Since the total X-ray output from Be-XRBs is proportional to $f_{\rm corr}$, considering the lower limit $f_{\rm corr}=0.25$ allowed by the calibration of Be-XRB outburst luminosity (Sec.~\ref{sec:lx}), we conclude that our Be-XRB populations can produce at least $\sim 30\%$ of the total X-ray emission from observed HMXBs at $Z\sim 0.0003-0.02$. 
It is shown in Appendix~\ref{apx:model} %Fig.~\ref{fig:nxsfr_fdcs0.8} 
that both the observed number of (ultra-)luminous HMXBs and overall X-ray luminosity per unit SFR from \citet[for all types of HMXBs]{Lehmer2021} can be fully reproduced by our SR\_CS model of Be-XRBs at $Z\sim 0.0003-0.02$, if given a higher correction factor $f_{\rm corr}=0.8$ than that assumed by default ($f_{\rm corr}=0.5$). This is non-trivial because systems with $L_{[0.5-8]~\rm keV}>10^{39}\ \rm erg\ s^{-1}$ only account for $\sim 15-50$\% of the total X-ray output in our Be-XRB populations at $Z\sim 0.0003-0.02$ from the SR\_CS model with $f_{\rm corr}\sim 0.25-0.8$. Moreover, this indicates that $f_{\rm corr}\gtrsim 0.8$ is ruled out in our SR\_CS model, considering that other types of HMXBs also contribute to the observed values.} 
%In fact, we find that systems with $L_{[0.5-8]~\rm keV}>10^{39}\ \rm erg\ s^{-1}$ only account for $\sim 15-50$\% of the total X-ray output in our Be-XRB populations at $Z\sim 0.0003-0.02$ from the SR\_CS model with $f_{\rm corr}\sim 0.25-0.8$. It is therefore non-trivial that we obtain similar contributions from Be-XRBs in the number count of ultra-luminous HMXBs and total X-ray luminosity with respect to observations across 2 orders of magnitude in metallicity.} }%$21-36$\%
%that dominates the overall X-ray output at $Z\sim 0.0003-0.02$. 
%\sout{Considering the uncertainties in our results embodied by the correction factor $f_{\rm corr}\sim 0.25-1$, we conclude that Be-XRBs can contribute at least 30\% and up to 100\% of the X-ray emission of observed HMXBs at $Z\sim 0.0003-0.02$, and meanwhile make up a similar fraction of the observed ultra-luminous HMXB population.}

\begin{figure}
    \centering
    \includegraphics[width=\columnwidth]{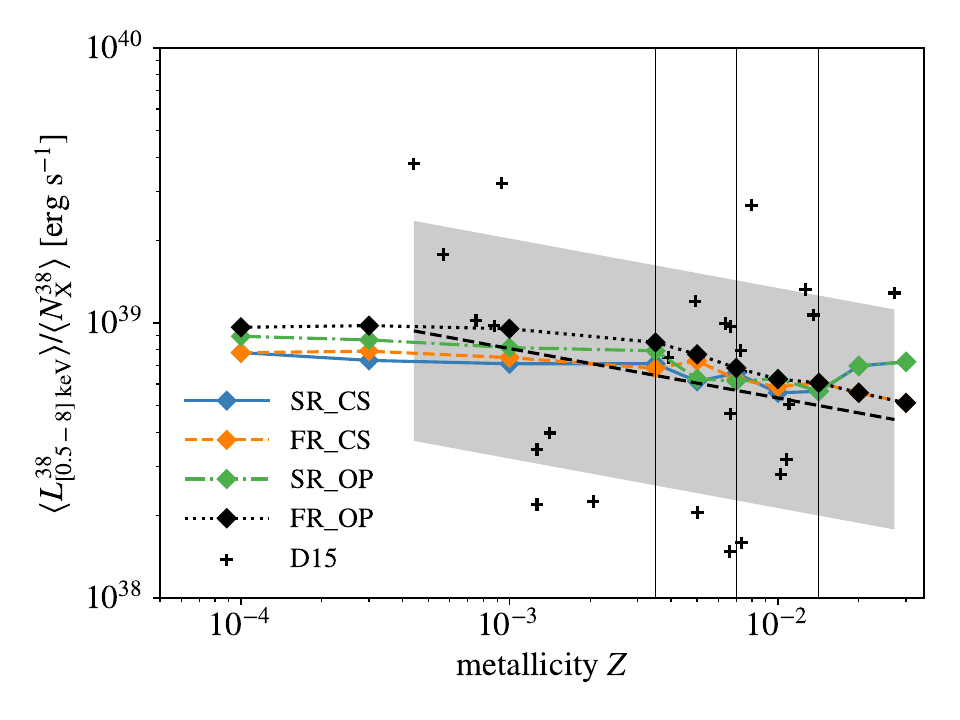}
    \vspace{-25pt}
    \caption{%{\color{blue}(optional)} 
    X-ray luminosity per luminous Be-XRBs with $L_{[0.5-8]~\rm keV}>10^{38}\ \rm erg\ s^{-1}$, %in the $0.5-8$~keV band, 
    for the SR\_CS (solid), FR\_CS (dashed), SR\_OP (dash-dotted) and FR\_OP (dotted) models with $f_{\rm corr}=0.5$. The observational data from nearby galaxies and the corresponding power-law fit from \citet[D15, see their fig.~5]{Douna2015} are shown with the crosses and dashed line surrounded by the shaded region. The thin vertical lines label the metallicities of the MW, LMC and SMC (from right to left).}
    \label{fig:l38}
\end{figure}

Finally, we calculate the X-ray luminosity per luminous Be-XRB with $L_{[0.5-8]~\rm keV}>10^{38}\ \rm erg\ s^{-1}$ in the $0.5-8$~keV band $\langle L_{\rm [0.5-8]\ keV}^{38}\rangle/\langle N_{\rm X}^{38}\rangle$, as shown in Fig.~\ref{fig:l38} for all the 4 models in Table~\ref{tab:model} assuming $f_{\rm corr}=0.5$. In general, we have $\langle L_{\rm [0.5-8]\ keV}^{38}\rangle/\langle N_{\rm X}^{38}\rangle\sim 0.5-1\times 10^{39}\ \rm erg\ s^{-1}$, insensitive to both metallicity and model parameters even with variations of $f_{\rm corr}$ in the range of $[0.25,1]$ (not shown for conciseness). Our results are consistent with observations of HMXBs in nearby galaxies for $Z\sim 0.0004-0.03$ \citep[see their fig.~5]{Douna2015}, which { further implies that the luminous X-ray sources in our Be-XRB populations are similar to those in observed HMXB populations}.

%and find that the is insensitive to metallicity for $Z\sim 0.0004-0.03$ in nearby galaxies \citep[see their fig.~5]{Douna2015}

\section{Caveats}\label{sec:dis}
To quantify more accurately the roles played by Be-XRBs, particularly in the ULX population {\citep[e.g.,][]{Kaaret2017,Kovlakas2020,Fabrika2021,Walton2022,King2023,Salvaggio2023,Tranin2023,Misra2023}}, and their imprints in the early Universe, more work needs to be done to overcome the following caveats of our study (with descending order of importance). 
%that should be taken into account in BPS of XRBs. 

\begin{enumerate}
    \item Although being more complete compared with previous studies, our model of the X-ray outbursts in Be-XRBs still relies on a phenomenological model based on simulations of steady-state VDDs with constant mass ejection rates (Eq.~\ref{mej}) and viscosity. However, in reality VDDs of O/Be stars, especially under the influence of companions, can be highly dynamical structures \citep{Carciofi2008,Haubois2012,Krticka2015,Panoglou2016,Vieira2017, Rimulo2018}, with variable mass ejection and disk dissipation/formation episodes of a few years \citep{Reig2011}. Such dynamical evolution of disk structures are particularly important for the most luminous Be-XRBs with non-periodic, strong Type~II outbursts (see Sec.~\ref{sec:fduty}). 
    In fact, according to our steady-state model, the majority ($\sim 60-70\%$) of X-ray emission from Be-XRBs are produced by the systems in which the compact object is able to accrete almost all ($\gtrsim 90$\%) of the materials ejected from the O/B star, so that significant variations of disk structures are expected to occur. 
    Additional uncertainties of our model may reside in the interpretation of simulation results and comparison with observations, as we do not model the detailed structure of accretion flows and geometry of X-ray emission, which are complex and still in debate, especially for ULXs {\citep[][]{Wiktorowicz2019,Fabrika2021,Mushtukov2022,King2023,Lasota2023}. For instance, taking into account the suppression of accretion by radiation-driven winds in the super-Eddington regime \citep{Shakura1973} reduces the X-ray outputs and number counts of ULXs from our Be-XRB populations by up to $\sim 60$\% and a factor of $\sim10$, respectively. }
    \item We do not fully explore the parameter space of (binary) stellar evolution. %relevant for Be-XRBs. 
    Although the mass-transfer efficiency in our BPS runs is calibrated to reproduce the number and range of orbital periods of Be-XRBs observed in the SMC, the agreement with observations is imperfect, and the other parameters (fixed in our case) governing the mass loss rate from stellar winds, stability of mass transfer, angular momentum loss, { spin-up of the accretor,} as well as natal kicks of SNe may affect Be-XRB properties and final X-ray outputs greatly {\citep[see, e.g.,][]{Linden2009,Zuo2014,Shao2014,Shao2020,Vinciguerra2020,Xing2021,Misra2022,Willcox2023}. %For instance, our model of SN natal kicks is only qualitatively consistent with recent parallax and proper motion measurements for young isolated radio pulsars and Galactic Be-XRBs \citep{Verbunt2017,Igoshev2021}, while the exact form of kick velocity distribution adopted in our BPS runs is different from that inferred from observations (for $Z=\rm Z_{\odot}$) and can be outdated, so that the total X-ray output can be overestimated by up to a factor of $\sim 3$ for $Z\lesssim 0.02$ 
    For instance, recent parallax and proper-motion measurements of young, isolated radio pulsars and Galactic Be-XRBs have placed new constraints on SN natal kicks. These support a kick-velocity distribution that is the sum of two Maxwellian distributions with $\sigma_{\rm kick}\sim 50$ and $340\ \rm km\ s^{-1}$, where the low-$\sigma_{\rm kick}$ component accounts for $20\pm 10$\% of all systems \citep{Verbunt2017,Igoshev2021}. Our model of natal kicks is qualitatively consistent with these results given the separate treatments of electron-capture SNe (with $\sigma_{\rm kick}=0$) and high-$\sigma_{\rm kick}$ cases (Type II and Ib/c with $\sigma_{\rm kick}=190$), although our $\sigma_{\rm kick}$ are smaller. We find by numerical experiments that compared with our default case, using the best-fit model of kick-velocity distribution in \citet{Igoshev2021} with generally stronger kicks will reduce the X-ray output from Be-XRBs by up to a factor of $\sim 3$ for $Z\lesssim 0.02$. However, with the kick-velocity distribution from \citet{Igoshev2021}, Be-XRBs on nearly circular ($e\lesssim 0.1$) orbits become extremely rare ($\lesssim 4$\%), and the predicted number of Be-XRBs in the SMC is lower by a factor of $\sim 2$ than that observed (see Appendix~\ref{apx:smc}). We defer a more detailed investigation into the effects of SN natal kicks on the X-ray output from Be-XRBs to future work. 
    The prescription of stellar winds is also important, especially for metal-rich stellar populations, causing up to one order of magnitude variations in the X-ray output at $Z=0.03$ (see Sec.~\ref{sec:comp_bps}).} %Using more up-to-date SN kick models may change our results by a factor of a few \citep{Vinciguerra2020}.}
    %Limited exploration/calibration of binary stellar evolution parameters
    \item In the construction of our binary populations, we do not consider the correlations between initial binary properties (i.e., binary fraction, distributions of masses and orbital parameters) nor their metallicity dependence, while there is evidence/hints of such correlations and evolution with metallicity from observations\footnote{In contrast to previous studies, recent observations of metal-poor "classical" dwarf spheroidal galaxies as MW satellites by \citet{Arroyo-Polonio2023} find no evidence of varying binary properties or their deviations from those at solar metallicity.} and simulations \citep[e.g.,][]{Gunawardhana2011,Marks2012,Duchene2013,Moe2017,Moe2019,Jerabkova2018,Lacchin2020,Chon2021,Chon2022,Tanvir2023,Rusakov2023}. %We also ignore the potential dependence of IMF and initial binary properties on metallicity and environments hinted by observations\footnote{There is also evidence that \citet{MariaArroyo-Polonio2023}} and simulations \citep[e.g.,][]{Gunawardhana2011,Marks2012,Jerabkova2018,Moe2019,Lacchin2020,Chon2021,Chon2022,Tanvir2023,Rusakov2023}. 
    Taking these effects into account may modify the predicted evolution of X-ray luminosity with metallicity and change our finding that the total X-ray output from Be-XRBs is insensitive to initial conditions.
    \item The effects of rotation on stellar evolution are not considered in our BPS runs, although it is shown by detailed stellar-evolution simulations \citep[e.g.,][]{Ekstrom2012,Georgy2013,Choi2017,Groh2019,Murphy2021} that fast rotation can impact the mass loss, timescales of evolution phases, stellar structure, nucleosynthesis and remnant masses in a complex and metallicity-dependent manner, particularly for initially fast-rotating stars, which can also change the properties of Be-XRBs that by definition contain fast-rotating O/B stars.  %In the lack of knowledge of the detailed mechanisms that connect the formation and properties of VDDs (i.e., the so-called `Be phenomenon') with stellar evolution processes are still unresolved \citep{Rivinius2013}  %Such effects may also be important in the modelling of Be-XRBs that contain fast-rotating O/Be stars. However, it is beyond the scope of this work to take into account these effects because the detailed mechanisms that connect the formation and properties of VDDs (i.e., the so-called `Be phenomenon') with stellar evolution processes are still unresolved \citep{Rivinius2013}.
    \item We use simple models calibrated to observations to derive the X-ray spectra of Be-XRBs, which only consider three regimes of accretion rates (low-hard, high-soft and super-Eddington) for two types of compact objects (NSs and BHs), while in reality, the X-ray spectra of Be-XRBs can have more complex dependence on compact object properties (e.g., masses, spins and magnetic fields) and accretion modes (e.g., thin disk and ADAF) that may also correlate with the stellar and orbital parameters as well as types of outbursts \citep{Martin2011,Okazaki2013,Cheng2014,Haberl2016,Xu2019,Franchini2021}. %Besides, since we ignored the beaming effects that can be important for luminous Be-XRBs with super-Eddington accretion, the X-ray luminosity 
\end{enumerate}

%\section{Summary}\label{sec:sumy}
\section{Summary and conclusions}\label{sec:sumy}
We improve population synthesis of Be-XRBs with a physically motivated model that combines recent hydrodynamic simulations of VDDs in Be-XRBs \citep{Brown2019} and VDD properties inferred from observations of classical Be stars \citep{Vieira2017,Rimulo2018}. Our model for the first time fully takes into account the dependence of X-ray outburst properties (i.e., strength and duty cycle) 
on stellar and orbital parameters of Be-XRBs, and also considers the classification of X-ray outbursts into the two conventional types in observations \citep[]{Reig2011,Rivinius2013}. Using the standard BSE models \citep{Hurley2002} in the BPS code \textsc{binary\_c}, we evolve large populations of randomly sampled massive binaries in the (absolute) metallicity range $Z\in [10^{-4},0.03]$ with initial conditions and BSE parameters calibrated to reproduce the population of observed Be-XRBs in the SMC (see Appendix~\ref{apx:smc}). %\citep[]{Coe2015,Rubele2015}. 
Finally, we apply an X-ray spectral model based on recent observations of HMXBs \citep{Anastasopoulou2022} to our simulated Be-XRBs to evaluate the {X-ray luminosity per unit SFR from Be-XRBs as a function of metallicity}. The effects of free model parameters for initial conditions %(orbital parameters and rotation rates) 
and VDD densities 
%(metallicity dependence and connections to peak accretion rates) 
are also explored. 
Our main findings are summarized as follows.

%Review of extra-galactic XRBs \citep{Gilfanov2023}, review of ULXs \citep{King2023}, statistics of ULXs and hyper-luminous X-ray sources \citep{Tranin2023}, variability of ULXs in the Cartwheel ring galaxy \citep{Salvaggio2023}, connection to compact object mergers \citep{Marchant2017,Fishbach2022}...

\begin{figure}
    \centering
    \includegraphics[width=\columnwidth]{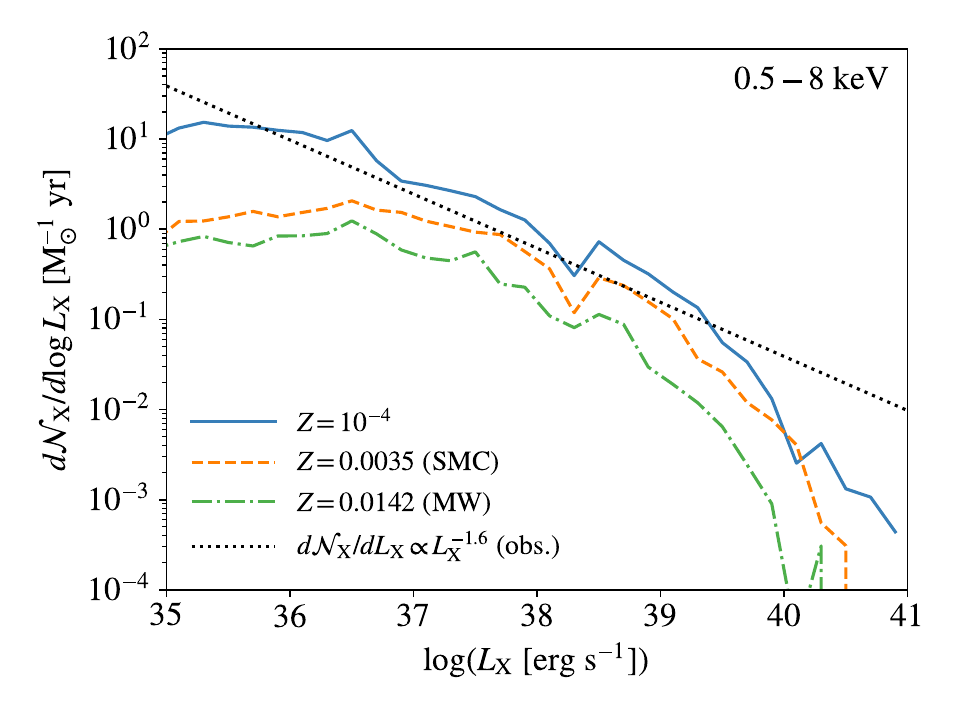}
    \vspace{-25pt}
    \caption{X-ray luminosity function (per unit SFR) of Be-XRBs in the 0.5-8 keV band predicted by our fiducial model (SR\_CS with $f_{\rm corr}=0.5$) for $Z=10^{-4}$ (solid), 0.0035 (dashed) and 0.0142 (dash-dotted). Here each Be-XRB is weighted by $f_{{\rm duty},i}\tau_{i}/M_{\rm tot}$. %like in the calculations of number counts and overall X-ray outputs (Eqs.~\ref{nxsfr} and \ref{lxsfr}). 
    The power-law fit ($d\mathcal{N}_{\rm X}/dL_{\rm X}\propto L_{\rm X}^{-1.6}$) to the average X-ray luminosity function of observed HMXBs in star-forming galaxies \citep[][see their fig.~5]{Mineo2012} is plotted with the dotted line for comparison. %(with arbitrary normalization). 
    The X-ray luminosity functions of our Be-XRBs are generally consistent with the observed power-law shape for $L_{\rm X}\sim 10^{36}-10^{39}\ \rm erg\ ^{-1}$, while the number fractions of systems in both the faint and luminous ends are lower for our Be-XRBs compared with the whole population of HMXBs in observations.}
    \label{fig:xlf}
\end{figure}

\begin{enumerate}
    \item {Be-XRBs can probably explain a non-negligible fraction of 
    %Be-XRBs can contribute a significant fraction ($\sim 60$\% in our fiducial case, up to $100$\% and at least $\sim 30$\% considering uncertainties) of 
    the total X-ray output from HMXBs observed in nearby galaxies \citep{Lehmer2021}. Assuming that current observations are complete, the X-ray luminosities per unit SFR from our simulated Be-XRB populations are at least $\sim 30\%$ of the observed values for \textit{all} types of HMXBs at $Z\sim 0.0003-0.02$. The minimum fraction $\sim 30\%$ is reached with the minimum correction factor $f_{\rm corr}=0.25$ allowed in our Be-XRB model calibrated to the observed relation between outburst X-ray luminosity and orbital period (Sec.~\ref{sec:lx}), and is subject to uncertainties from our imperfect modelling of VDDs, the limited parameter space of BPS explored, and the fact that our BPS simulations only consider Be-XRBs rather than all types of HMXBs (see Sec.~\ref{sec:dis}).} 
    In our fiducial model %(SR\_CS assuming $f_{\rm corr}=0.5$) 
    with no $Z$ dependence of VDD densities nor enhanced initial stellar rotation, the X-ray luminosity per unit SFR decreases by a factor of $\sim 8$ from $Z=0.0003$ to $Z=0.02$, which is mainly driven by the stronger stellar winds at higher $Z$ that reduce the number of binary stars able to become luminous Be-XRBs. {Interestingly, this trend is similar to that seen in observations of all types of HMXBs. }
    
    \item Our results for Be-XRBs are also similar to the $Z$-dependent X-ray outputs from other types of HMXBs powered by RLO and (spherical) stellar winds in previous BPS studies \citep[e.g.,][]{Fragos2013} for $Z\sim 0.001-0.02$ (Sec.~\ref{sec:comp_bps}). However, at $Z\lesssim 0.001$, the X-ray luminosity from Be-XRBs keeps increasing towards lower $Z$ (e.g., by a factor of $\sim 2$ from $Z=0.001$ to $Z=10^{-4}$), while there is almost no evolution in that from other types of HMXBs predicted by \citet{Fragos2013}. {This implies that Be-XRBs might play more important roles at lower metallicities, such that they can have strong impact on the thermal history and thus 21-cm signal from the IGM at Cosmic Dawn. More comprehensive BPS studies that consider both Be-XRBs and other types of HMXBs self-consistently are required to verify if this is indeed the case.} %The SEDs of Be-XRBs
    %compared
    \item Luminous ($\gtrsim 10^{38}\ \rm erg\ s^{-1}$) systems with non-periodic, major outbursts of Type~II (see Sec.~\ref{sec:fduty} for a detailed definition) and super-Eddington accretion (mostly on eccentric orbits with $e\gtrsim 0.4$), including ULXs ($\gtrsim 10^{39}\ \rm erg\ s^{-1}$), are important X-ray powerhouses in our Be-XRB populations across the metallicity range considered. In our fiducial model, $\sim 55-75\%$ of the total X-ray output of Be-XRBs comes from luminous sources with $L_{[0.5-8]\ \rm keV}>10^{38}\ \rm erg\ s^{-1}$, while ULXs with $L_{[0.5-8]\ \rm keV}>10^{39}\ \rm erg\ s^{-1}$ contribute $\sim 24-41\%$ of the total output. {Similarly to the case of total X-ray output, the simulated Be-XRBs can also account for non-negligible fractions of the number counts of luminous and ultra-luminous HMXBs in observations \citep{Mapelli2010,Douna2015,Kovlakas2020,Lehmer2021}, although more work needs to be done to fully evaluate the contributions of Be-XRBs.} Such luminous systems have been overlooked in previous BPS studies of Be-XRBs \citep[e.g.,][]{Zuo2014,Misra2022}, which use a simple empirical scaling law between outburst X-ray luminosity and orbital period \citep[see our Eq.~\ref{lxorig} and Fig.~\ref{fig:lx_p}]{Dai2006} that imposes a (likely artificial) cutoff at $\sim 10^{38}\ \rm erg\ s^{-1}$, while our improved Be-XRB model further considers the dependence of outburst accretion rate/luminosity on VDD density (and its scatter) as well as orbital eccentricity, so that the luminosity function of our Be-XRBs has a luminous tail up to $10^{41}\ \rm erg\ s^{-1}$, as shown in Fig.~\ref{fig:xlf}. 
    %Be-XRBs can contribute least 30\% and up to 100\% of the X-ray emission of HMXBs at $Z\sim 0.0003-0.02$, and meanwhile make up a similar fraction of ultra-luminous HMXB population\sout{Interestingly, we find that our Be-XRBs can account for similar fractions ($\sim 30-100$\%) of the number count of luminous and ultra-luminous HMXBs as their contribution to the total X-ray luminosity of HMXBs in observations.}
    \item The major uncertainty in the total X-ray output from Be-XRBs arises from the uncertainties in accretion rates and VDD properties rather than those in initial conditions when fixing the IMF (see Sec.~\ref{sec:ic}). The former can change the final X-ray output by a factor of a few, while the difference made by the latter is within {$\sim 50$\% (see Sec.~\ref{sec:eps} and Appendix~\ref{apx:model})}. The reason is that the X-ray luminosity of an accreting compact object is proportional to the accretion rate, which is then proportional to the VDD density (or mass ejection rate of the O/B star) to the first order, and there are still large uncertainties in these quantities from idealized simulations and limited observations, especially at low metallicity. On the other hand, a key feature of our BPS simulations is that in most binaries that can potentially become Be-XRBs the secondary star will be spun up to become an O/Be star (via stable mass transfer) regardless of its initial rotation rate, which is consistent with the scenario that most (young) O/Be stars are produced by binary interactions as predicted/hinted by theoretical models and observations {\citep{Shao2014,Klement2019,Bodensteiner2020,DorigoJones2020,Hastings2020,Hastings2021,Dallas2022,Dodd2023,Wang2023}}. 
\end{enumerate}

In general, our BPS results highlight {the possibility that Be-XRBs constitute an important component in the population of HMXBs, making significant contributions to the overall X-ray output and number count of ULXs}. 

Indeed, robust modelling of XRBs with BPS is a challenging task {(see Sec.~\ref{sec:dis})} because of large uncertainties in binary stellar evolution, particularly for Be-XRBs due to their transient nature and our lack of understanding on the detailed mechanisms that drive the formation and evolution of VDDs around fast-rotating massive stars 
\citep[for candidate scenarios, see, e.g.,][]{Granada2013,Hastings2020,Zhao2020,Cranmer2009,Rogers2013,Lee2014,Ressler2021}. Nevertheless, such theoretical efforts are worthwhile considering the important roles played by XRBs in the Epoch of Reionization and Cosmic Dawn \citep[e.g.,][]{Fialkov2014,Fialkov2014rich,Pacucci2014}, as well as the wealth of high-$z$ observational data to come in the next decades for the 21-cm signal by radio telescopes such as HERA \citep{DeBoer2017}, SARAS \citep{Singh2022}, REACH \citep{deLeraAcedo2022}, NenuFAR \citep{Mertens2021}{, MIST \citep{Monsalve2023}} and SKA \citep{Koopmans2015} which will place constraints on high-$z$ XRBs and the underlying star/galaxy/structure formation processes. 

The present paper is a small step that points out the importance of Be-XRBs. In future work, we plan to model Be-XRBs (with possible improvements of the aforementioned caveats) and other types of XRBs in one BPS framework to predict their impact on the 21-cm signal as well as cosmic X-ray background and derive constraints from existing observational data \citep[e.g.,][]{Lehmer2012,Moretti2012,Cappelluti2017,Fialkov2017,Bowman2018,Bevins2023,Lazare2023,Rossland2023}. We will also extend the metallicity range down to the extremely metal-poor ($Z\lesssim 10^{-6}$) regime of {Pop~III stars}\footnote{The properties of {Pop~III} binaries are still highly uncertain in the lack of direct observations. Recent radiative hydrodynamic simulations \citep[][]{Sugimura2020,Sugimura2023,Park2023,Park2022} { found that outward migrations of Pop~III protostars and their circumstellar disks by accretion of gas with high angular momentum are common. This implies that close binaries of massive Pop~III stars are likely rare.
%have found that close binaries are likely rare for the first stars (due to outward migrations of protostars and their circumstellar disks by accretion of gas with high angular momentum). This implies that 
If this is true, the formation efficiency and X-ray outputs of XRBs from Pop~III stars will be much lower than those of XRBs from metal-rich stars (considered in this paper) that are dominated by close binaries \citep{Liu2021binary}. Nevertheless, these simulations are still limited by resolution (with sink radii $\sim 100$~AU) and lack of magnetic fields, whose results are yet to be validated by future simulations and observations.}}, which are expected to be promising progenitors of Be-XRBs \citep{Sartorio2023} given their compact and fast-rotating\footnote{According to hydrodynamic simulations of primordial star formation, Pop~III stars tend to be born as fast ($W_{0}\sim 0.5-1$) rotators \citep[][]{Stacy2011,Stacy2013,Hirano2018}, which is also supported by stellar archaeology observations {\citep[e.g.,][]{Chiappini2006,Chiappini2011,Chiappini2013,Maeder2015,Choplin2017,Choplin2019,Jeena2023}}.} nature that favors stable mass transfer \citep{Inayoshi2017} and formation of O/Be stars. Another interesting topic to investigate is the connection between (Be-)XRBs and compact object mergers \citep[see, e.g.,][]{Marchant2017,Mondal2020,Fishbach2022,Kotko2023,Liotine2023} that may foster the synergy between 21-cm and gravitational wave observations to better constrain binary stellar evolution in high-$z$ galaxies\footnote{A significant fraction ($\sim 60-85\%$) of compact object mergers {(mainly from binary NSs)}, in our BPS runs for $Z\sim 10^{-4}-0.02$ have progenitors that undergo a Be-XRB phase. The fraction becomes smaller but still non-negligible, $\sim 10-20\%$, when we exclude Be-XRBs on nearly-circular ($e\lesssim 0.1$) orbits. %that are usually faint and difficult to detect. 
Larger samples of binaries are required to study the properties of such compact object mergers with Be-XRB progenitors. Here we only comment on our preliminary results of the number fraction, {which imply a strong connection between Be-XRBs and binary NS mergers. The reason is that the main formation channels of them both involve stable mass transfer during the HG phase, as also seen in previous BPS studies \citep{Vigna-Gomez2018,Vinciguerra2020}.}}.

%Pop~III stars tend to be born as fast rotators \citep[$W_{0}\sim 0.5-1$,][]{Stacy2011,Stacy2013} unless braked by strong magnetic fields \citep[]{Hirano2018}, which is supported by stellar archaeology observations \citep{Chiappini2006,Chiappini2013,Maeder2015,Choplin2019,K2023}, and their compactness favors stable mass transfer \citep{Inayoshi2017}

%\begin{itemize}
    %\item Dependence on BPS parameters 
    %\item Including all types of HMXBs in one BPS framework to fully evaluate their relative importance in the overall X-ray output and X-ray luminosity function
    %\item Pop~III stars
    %\item Impact on the 21-cm signal
    %\item Connection between Be-XRBs and compact object mergers
%\end{itemize}

\begin{comment}
\begin{figure}
    \centering
    \includegraphics[width=\columnwidth]{LXSFR_ZCS1.pdf}
    \vspace{-25pt}
    \caption{{\color{blue}(optional)} Same as Fig.~\ref{fig:lxsfr_Z} but comparing the results for the $0.1-2$~keV (solid) and $2-10$~keV (dash-dotted) bands from our SR\_CS model assuming $f_{\rm corr}=0.5$ with those of XRBs (excluding Be-XRBs) from metal-free ($Z=0$) stars in the low-mass (LM, orange), Fiducial (Fid, red) and high-mass (HM, black) models of \citet[S23, see their table~2 and fig.~6]{Sartorio2023}. }
    \label{fig:lxsfr_Z_pop3}
\end{figure}
\end{comment}

%\subsection{Implications on the 21-cm signal}

%\subsection{Comparison with previous studies}

%\section{Summary}\label{sec:sum} 

\section*{Acknowledgements}
The authors would like to thank the anonymous referee for the insightful comments and David D. Hendriks for his help with \href{https://gitlab.com/binary_c/binary_c-python}{\textsc{binary\_c-python}}. BL and AF are supported by the Royal Society University Research Fellowship. NS gratefully acknowledges the support of the Research Foundation - Flanders (FWO Vlaanderen) grant 1290123N. RGI acknowledges funding by the STFC consolidated grants ST/L003910/1
and ST/R000603/1, and thanks the BRIDGCE network. This work excessively used the public packages \texttt{numpy} \citep{vanderWalt2011}, \texttt{matplotlib} \citep{Hunter2007} and \texttt{scipy} \citep{jones2001scipy}. The authors wish to express their gratitude to the developers of these packages and to those who maintain them.
%The authors acknowledge the Texas Advanced Computing Center (TACC) for providing HPC resources under XSEDE allocation TG-AST120024.
%This work was supported by National Science Foundation (NSF) grant AST-1413501. %The authors acknowledge the Texas Advanced Computing Center (TACC) for providing HPC resources under XSEDE allocation TG-AST120024.
%Support for this work was provided by NASA through the NASA Hubble Fellowship grant HST-HF2-51418.001-A awarded by the Space Telescope Science Institute, which is operated by the Association of Universities for Research in  Astronomy, Inc., for NASA, under contract NAS5-26555. 

\section*{Data availability}
The data of Be-XRBs and \texttt{python} scripts used to manage BPS runs and conduct post-processing will be shared on reasonable request to the corresponding authors. The BPS code \textsc{binary\_c} (including the Be-XRB module) is available at \url{https://gitlab.com/binary_c/binary_c/-/tree/development-BeXRB}. The corresponding \texttt{python} interface \textsc{binary\_c-python} is available at \url{https://gitlab.com/binary_c/binary_c-python/-/tree/feature/source_file_sampling}.

\bibliographystyle{mnras}
\bibliography{ref}

\begin{thebibliography}{}
\makeatletter
\relax
\def\mn@urlcharsother{\let\do\@makeother \do\$\do\&\do\#\do\^\do\_\do\%\do\~}
\def\mn@doi{\begingroup\mn@urlcharsother \@ifnextchar [ {\mn@doi@}
  {\mn@doi@[]}}
\def\mn@doi@[#1]#2{\def\@tempa{#1}\ifx\@tempa\@empty \href
  {http://dx.doi.org/#2} {doi:#2}\else \href {http://dx.doi.org/#2} {#1}\fi
  \endgroup}
\def\mn@eprint#1#2{\mn@eprint@#1:#2::\@nil}
\def\mn@eprint@arXiv#1{\href {http://arxiv.org/abs/#1} {{\tt arXiv:#1}}}
\def\mn@eprint@dblp#1{\href {http://dblp.uni-trier.de/rec/bibtex/#1.xml}
  {dblp:#1}}
\def\mn@eprint@#1:#2:#3:#4\@nil{\def\@tempa {#1}\def\@tempb {#2}\def\@tempc
  {#3}\ifx \@tempc \@empty \let \@tempc \@tempb \let \@tempb \@tempa \fi \ifx
  \@tempb \@empty \def\@tempb {arXiv}\fi \@ifundefined
  {mn@eprint@\@tempb}{\@tempb:\@tempc}{\expandafter \expandafter \csname
  mn@eprint@\@tempb\endcsname \expandafter{\@tempc}}}

\bibitem[\protect\citeauthoryear{{Abdusalam}, {Ablimit}, {Hashim}, {L{\"u}},
  {Mardini}  \& {Wang}}{{Abdusalam} et~al.}{2020}]{Abdusalam2020}
{Abdusalam} K.,  {Ablimit} I.,  {Hashim} P.,  {L{\"u}} G.~L.,  {Mardini} M.~K.,
    {Wang} Z.~J.,  2020, \mn@doi [\apj] {10.3847/1538-4357/abb5a8}, \href
  {https://ui.adsabs.harvard.edu/abs/2020ApJ...902..125A} {902, 125}

\bibitem[\protect\citeauthoryear{{Acharya}, {Cyr}  \& {Chluba}}{{Acharya}
  et~al.}{2023}]{Acharya2023}
{Acharya} S.~K.,  {Cyr} B.,   {Chluba} J.,  2023, \mn@doi [\mnras]
  {10.1093/mnras/stad1540}, \href
  {https://ui.adsabs.harvard.edu/abs/2023MNRAS.523.1908A} {523, 1908}

\bibitem[\protect\citeauthoryear{{Aird}, {Coil}  \& {Georgakakis}}{{Aird}
  et~al.}{2017}]{Aird2017}
{Aird} J.,  {Coil} A.~L.,   {Georgakakis} A.,  2017, \mn@doi [\mnras]
  {10.1093/mnras/stw2932}, \href
  {https://ui.adsabs.harvard.edu/abs/2017MNRAS.465.3390A} {465, 3390}

\bibitem[\protect\citeauthoryear{{Allende Prieto}, {Lambert}  \&
  {Asplund}}{{Allende Prieto} et~al.}{2001}]{AllendePrieto2001}
{Allende Prieto} C.,  {Lambert} D.~L.,   {Asplund} M.,  2001, \mn@doi [\apjl]
  {10.1086/322874}, \href
  {https://ui.adsabs.harvard.edu/abs/2001ApJ...556L..63A} {556, L63}

\bibitem[\protect\citeauthoryear{{Anastasopoulou}, {Zezas}, {Steiner}  \&
  {Reig}}{{Anastasopoulou} et~al.}{2022}]{Anastasopoulou2022}
{Anastasopoulou} K.,  {Zezas} A.,  {Steiner} J.~F.,   {Reig} P.,  2022, \mn@doi
  [\mnras] {10.1093/mnras/stac940}, \href
  {https://ui.adsabs.harvard.edu/abs/2022MNRAS.513.1400A} {513, 1400}

\bibitem[\protect\citeauthoryear{{Antoniou} \& {Zezas}}{{Antoniou} \&
  {Zezas}}{2016}]{Antoniou2016}
{Antoniou} V.,  {Zezas} A.,  2016, \mn@doi [\mnras] {10.1093/mnras/stw167},
  \href {https://ui.adsabs.harvard.edu/abs/2016MNRAS.459..528A} {459, 528}

\bibitem[\protect\citeauthoryear{{Antoniou}, {Hatzidimitriou}, {Zezas}  \&
  {Reig}}{{Antoniou} et~al.}{2009}]{Antoniou2009}
{Antoniou} V.,  {Hatzidimitriou} D.,  {Zezas} A.,   {Reig} P.,  2009, \mn@doi
  [\apj] {10.1088/0004-637X/707/2/1080}, \href
  {https://ui.adsabs.harvard.edu/abs/2009ApJ...707.1080A} {707, 1080}

\bibitem[\protect\citeauthoryear{{Antoniou}, {Zezas}, {Hatzidimitriou}  \&
  {Kalogera}}{{Antoniou} et~al.}{2010}]{Antoniou2010}
{Antoniou} V.,  {Zezas} A.,  {Hatzidimitriou} D.,   {Kalogera} V.,  2010,
  \mn@doi [\apjl] {10.1088/2041-8205/716/2/L140}, \href
  {https://ui.adsabs.harvard.edu/abs/2010ApJ...716L.140A} {716, L140}

\bibitem[\protect\citeauthoryear{{Antoniou} et~al.,}{{Antoniou}
  et~al.}{2019}]{Antoniou2019}
{Antoniou} V.,  et~al., 2019, \mn@doi [\apj] {10.3847/1538-4357/ab4a7a}, \href
  {https://ui.adsabs.harvard.edu/abs/2019ApJ...887...20A} {887, 20}

\bibitem[\protect\citeauthoryear{{Arcos}, {Jones}, {Sigut}, {Kanaan}  \&
  {Cur{\'e}}}{{Arcos} et~al.}{2017}]{Arcos2017}
{Arcos} C.,  {Jones} C.~E.,  {Sigut} T.~A.~A.,  {Kanaan} S.,   {Cur{\'e}} M.,
  2017, \mn@doi [\apj] {10.3847/1538-4357/aa6f5f}, \href
  {https://ui.adsabs.harvard.edu/abs/2017ApJ...842...48A} {842, 48}

\bibitem[\protect\citeauthoryear{{Arroyo-Polonio}, {Battaglia}, {Thomas},
  {Irwin}, {McConnachie}  \& {Tolstoy}}{{Arroyo-Polonio}
  et~al.}{2023}]{Arroyo-Polonio2023}
{Arroyo-Polonio} J.~M.,  {Battaglia} G.,  {Thomas} G.~F.,  {Irwin} M.~J.,
  {McConnachie} A.~W.,   {Tolstoy} E.,  2023, \mn@doi [\aap]
  {10.1051/0004-6361/202346843}, \href
  {https://ui.adsabs.harvard.edu/abs/2023A&A...677A..95A} {677, A95}

\bibitem[\protect\citeauthoryear{{Artale}, {Tissera}  \& {Pellizza}}{{Artale}
  et~al.}{2015}]{Artale2015}
{Artale} M.~C.,  {Tissera} P.~B.,   {Pellizza} L.~J.,  2015, \mn@doi [\mnras]
  {10.1093/mnras/stv218}, \href
  {https://ui.adsabs.harvard.edu/abs/2015MNRAS.448.3071A} {448, 3071}

\bibitem[\protect\citeauthoryear{{Asplund}, {Grevesse}, {Sauval}, {Allende
  Prieto}  \& {Kiselman}}{{Asplund} et~al.}{2004}]{Asplund2004}
{Asplund} M.,  {Grevesse} N.,  {Sauval} A.~J.,  {Allende Prieto} C.,
  {Kiselman} D.,  2004, \mn@doi [\aap] {10.1051/0004-6361:20034328}, \href
  {https://ui.adsabs.harvard.edu/abs/2004A&A...417..751A} {417, 751}

\bibitem[\protect\citeauthoryear{{Asplund}, {Grevesse}, {Sauval}  \&
  {Scott}}{{Asplund} et~al.}{2009}]{Asplund2009}
{Asplund} M.,  {Grevesse} N.,  {Sauval} A.~J.,   {Scott} P.,  2009, \mn@doi
  [\araa] {10.1146/annurev.astro.46.060407.145222}, \href
  {https://ui.adsabs.harvard.edu/abs/2009ARA&A..47..481A} {47, 481}

\bibitem[\protect\citeauthoryear{{Barkana}}{{Barkana}}{2018}]{Barkana2018}
{Barkana} R.,  2018, \mn@doi [\nat] {10.1038/nature25791}, \href
  {https://ui.adsabs.harvard.edu/abs/2018Natur.555...71B} {555, 71}

\bibitem[\protect\citeauthoryear{{Bastian} et~al.,}{{Bastian}
  et~al.}{2017}]{Bastian2017}
{Bastian} N.,  et~al., 2017, \mn@doi [\mnras] {10.1093/mnras/stw3042}, \href
  {https://ui.adsabs.harvard.edu/abs/2017MNRAS.465.4795B} {465, 4795}

\bibitem[\protect\citeauthoryear{{Basu-Zych} et~al.,}{{Basu-Zych}
  et~al.}{2013}]{Basu-Zych2013}
{Basu-Zych} A.~R.,  et~al., 2013, \mn@doi [\apj] {10.1088/0004-637X/762/1/45},
  \href {https://ui.adsabs.harvard.edu/abs/2013ApJ...762...45B} {762, 45}

\bibitem[\protect\citeauthoryear{{Bavera} et~al.,}{{Bavera}
  et~al.}{2020}]{Bavera2020}
{Bavera} S.~S.,  et~al., 2020, \mn@doi [\aap] {10.1051/0004-6361/201936204},
  \href {https://ui.adsabs.harvard.edu/abs/2020A&A...635A..97B} {635, A97}

\bibitem[\protect\citeauthoryear{{Belczynski} \& {Ziolkowski}}{{Belczynski} \&
  {Ziolkowski}}{2009}]{Belczynski2009}
{Belczynski} K.,  {Ziolkowski} J.,  2009, \mn@doi [\apj]
  {10.1088/0004-637X/707/2/870}, \href
  {https://ui.adsabs.harvard.edu/abs/2009ApJ...707..870B} {707, 870}

\bibitem[\protect\citeauthoryear{{Bera}, {Samui}  \& {Datta}}{{Bera}
  et~al.}{2023}]{Bera2023}
{Bera} A.,  {Samui} S.,   {Datta} K.~K.,  2023, \mn@doi [\mnras]
  {10.1093/mnras/stac3814}, \href
  {https://ui.adsabs.harvard.edu/abs/2023MNRAS.519.4869B} {519, 4869}

\bibitem[\protect\citeauthoryear{{Bevins}, {Heimersheim}, {Abril-Cabezas},
  {Fialkov}, {de Lera Acedo}, {Handley}, {Singh}  \& {Barkana}}{{Bevins}
  et~al.}{2023}]{Bevins2023}
{Bevins} H. T.~J.,  {Heimersheim} S.,  {Abril-Cabezas} I.,  {Fialkov} A.,  {de
  Lera Acedo} E.,  {Handley} W.,  {Singh} S.,   {Barkana} R.,  2023, \mn@doi
  [\mnras] {10.1093/mnras/stad3194}, \href
  {https://ui.adsabs.harvard.edu/abs/2023MNRAS.tmp.3075B} {}

\bibitem[\protect\citeauthoryear{{Bodensteiner}, {Shenar}  \&
  {Sana}}{{Bodensteiner} et~al.}{2020}]{Bodensteiner2020}
{Bodensteiner} J.,  {Shenar} T.,   {Sana} H.,  2020, \mn@doi [\aap]
  {10.1051/0004-6361/202037640}, \href
  {https://ui.adsabs.harvard.edu/abs/2020A&A...641A..42B} {641, A42}

\bibitem[\protect\citeauthoryear{{Bowman}, {Rogers}, {Monsalve}, {Mozdzen}  \&
  {Mahesh}}{{Bowman} et~al.}{2018}]{Bowman2018}
{Bowman} J.~D.,  {Rogers} A. E.~E.,  {Monsalve} R.~A.,  {Mozdzen} T.~J.,
  {Mahesh} N.,  2018, \mn@doi [\nat] {10.1038/nature25792}, \href
  {https://ui.adsabs.harvard.edu/abs/2018Natur.555...67B} {555, 67}

\bibitem[\protect\citeauthoryear{{Brorby}, {Kaaret}, {Prestwich}  \&
  {Mirabel}}{{Brorby} et~al.}{2016}]{Brorby2016}
{Brorby} M.,  {Kaaret} P.,  {Prestwich} A.,   {Mirabel} I.~F.,  2016, \mn@doi
  [\mnras] {10.1093/mnras/stw284}, \href
  {https://ui.adsabs.harvard.edu/abs/2016MNRAS.457.4081B} {457, 4081}

\bibitem[\protect\citeauthoryear{{Brown}, {Ho}, {Coe}  \& {Okazaki}}{{Brown}
  et~al.}{2018}]{Brown2018}
{Brown} R.~O.,  {Ho} W.~C.~G.,  {Coe} M.~J.,   {Okazaki} A.~T.,  2018, \mn@doi
  [\mnras] {10.1093/mnras/sty973}, \href
  {https://ui.adsabs.harvard.edu/abs/2018MNRAS.477.4810B} {477, 4810}

\bibitem[\protect\citeauthoryear{{Brown}, {Coe}, {Ho}  \& {Okazaki}}{{Brown}
  et~al.}{2019}]{Brown2019}
{Brown} R.~O.,  {Coe} M.~J.,  {Ho} W.~C.~G.,   {Okazaki} A.~T.,  2019, \mn@doi
  [\mnras] {10.1093/mnras/stz1757}, \href
  {https://ui.adsabs.harvard.edu/abs/2019MNRAS.488..387B} {488, 387}

\bibitem[\protect\citeauthoryear{{Cappelluti} et~al.,}{{Cappelluti}
  et~al.}{2017}]{Cappelluti2017}
{Cappelluti} N.,  et~al., 2017, \mn@doi [\apj] {10.3847/1538-4357/aa5ea4},
  \href {https://ui.adsabs.harvard.edu/abs/2017ApJ...837...19C} {837, 19}

\bibitem[\protect\citeauthoryear{{Carciofi} \& {Bjorkman}}{{Carciofi} \&
  {Bjorkman}}{2006}]{Carciofi2006}
{Carciofi} A.~C.,  {Bjorkman} J.~E.,  2006, \mn@doi [\apj]
  {10.1086/49948310.48550/arXiv.astro-ph/0511228}, \href
  {https://ui.adsabs.harvard.edu/abs/2006ApJ...639.1081C} {639, 1081}

\bibitem[\protect\citeauthoryear{{Carciofi} \& {Bjorkman}}{{Carciofi} \&
  {Bjorkman}}{2008}]{Carciofi2008}
{Carciofi} A.~C.,  {Bjorkman} J.~E.,  2008, \mn@doi [\apj] {10.1086/589875},
  \href {https://ui.adsabs.harvard.edu/abs/2008ApJ...684.1374C} {684, 1374}

\bibitem[\protect\citeauthoryear{{Casares}, {Negueruela}, {Rib{\'o}}, {Ribas},
  {Paredes}, {Herrero}  \& {Sim{\'o}n-D{\'\i}az}}{{Casares}
  et~al.}{2014}]{Casares2014}
{Casares} J.,  {Negueruela} I.,  {Rib{\'o}} M.,  {Ribas} I.,  {Paredes} J.~M.,
  {Herrero} A.,   {Sim{\'o}n-D{\'\i}az} S.,  2014, \mn@doi [\nat]
  {10.1038/nature12916}, \href
  {https://ui.adsabs.harvard.edu/abs/2014Natur.505..378C} {505, 378}

\bibitem[\protect\citeauthoryear{Chashkina, Lipunova, Abolmasov  \&
  Poutanen}{Chashkina et~al.}{2019}]{chashkina_super-eddington_2019}
Chashkina A.,  Lipunova G.,  Abolmasov P.,   Poutanen J.,  2019, \mn@doi [\aap]
  {10.1051/0004-6361/201834414}, 626, A18

\bibitem[\protect\citeauthoryear{{Chatterjee}, {Dayal}, {Choudhury}  \&
  {Schneider}}{{Chatterjee} et~al.}{2020}]{Chatterjee2020}
{Chatterjee} A.,  {Dayal} P.,  {Choudhury} T.~R.,   {Schneider} R.,  2020,
  \mn@doi [\mnras] {10.1093/mnras/staa1609}, \href
  {https://ui.adsabs.harvard.edu/abs/2020MNRAS.496.1445C} {496, 1445}

\bibitem[\protect\citeauthoryear{{Cheng}, {Shao}  \& {Li}}{{Cheng}
  et~al.}{2014}]{Cheng2014}
{Cheng} Z.~Q.,  {Shao} Y.,   {Li} X.~D.,  2014, \mn@doi [\apj]
  {10.1088/0004-637X/786/2/128}, \href
  {https://ui.adsabs.harvard.edu/abs/2014ApJ...786..128C} {786, 128}

\bibitem[\protect\citeauthoryear{{Chiappini}}{{Chiappini}}{2013}]{Chiappini2013}
{Chiappini} C.,  2013, \mn@doi [Astronomische Nachrichten]
  {10.1002/asna.201311902}, \href
  {https://ui.adsabs.harvard.edu/abs/2013AN....334..595C} {334, 595}

\bibitem[\protect\citeauthoryear{{Chiappini}, {Hirschi}, {Meynet},
  {Ekstr{\"o}m}, {Maeder}  \& {Matteucci}}{{Chiappini}
  et~al.}{2006}]{Chiappini2006}
{Chiappini} C.,  {Hirschi} R.,  {Meynet} G.,  {Ekstr{\"o}m} S.,  {Maeder} A.,
  {Matteucci} F.,  2006, \mn@doi [\aap] {10.1051/0004-6361:20064866}, \href
  {https://ui.adsabs.harvard.edu/abs/2006A&A...449L..27C} {449, L27}

\bibitem[\protect\citeauthoryear{{Chiappini}, {Frischknecht}, {Meynet},
  {Hirschi}, {Barbuy}, {Pignatari}, {Decressin}  \& {Maeder}}{{Chiappini}
  et~al.}{2011}]{Chiappini2011}
{Chiappini} C.,  {Frischknecht} U.,  {Meynet} G.,  {Hirschi} R.,  {Barbuy} B.,
  {Pignatari} M.,  {Decressin} T.,   {Maeder} A.,  2011, \mn@doi [\nat]
  {10.1038/nature10000}, \href
  {https://ui.adsabs.harvard.edu/abs/2011Natur.472..454C} {472, 454}

\bibitem[\protect\citeauthoryear{{Choi}, {Conroy}  \& {Byler}}{{Choi}
  et~al.}{2017}]{Choi2017}
{Choi} J.,  {Conroy} C.,   {Byler} N.,  2017, \mn@doi [\apj]
  {10.3847/1538-4357/aa679f}, \href
  {https://ui.adsabs.harvard.edu/abs/2017ApJ...838..159C} {838, 159}

\bibitem[\protect\citeauthoryear{{Chon}, {Omukai}  \& {Schneider}}{{Chon}
  et~al.}{2021}]{Chon2021}
{Chon} S.,  {Omukai} K.,   {Schneider} R.,  2021, \mn@doi [\mnras]
  {10.1093/mnras/stab2497}, \href
  {https://ui.adsabs.harvard.edu/abs/2021MNRAS.508.4175C} {508, 4175}

\bibitem[\protect\citeauthoryear{{Chon}, {Ono}, {Omukai}  \&
  {Schneider}}{{Chon} et~al.}{2022}]{Chon2022}
{Chon} S.,  {Ono} H.,  {Omukai} K.,   {Schneider} R.,  2022, \mn@doi [\mnras]
  {10.1093/mnras/stac1549}, \href
  {https://ui.adsabs.harvard.edu/abs/2022MNRAS.514.4639C} {514, 4639}

\bibitem[\protect\citeauthoryear{{Choplin}, {Hirschi}, {Meynet}  \&
  {Ekstr{\"o}m}}{{Choplin} et~al.}{2017}]{Choplin2017}
{Choplin} A.,  {Hirschi} R.,  {Meynet} G.,   {Ekstr{\"o}m} S.,  2017, \mn@doi
  [\aap] {10.1051/0004-6361/201731948}, \href
  {https://ui.adsabs.harvard.edu/abs/2017A&A...607L...3C} {607, L3}

\bibitem[\protect\citeauthoryear{{Choplin}, {Tominaga}  \&
  {Ishigaki}}{{Choplin} et~al.}{2019}]{Choplin2019}
{Choplin} A.,  {Tominaga} N.,   {Ishigaki} M.~N.,  2019, \mn@doi [\aap]
  {10.1051/0004-6361/201936187}, \href
  {https://ui.adsabs.harvard.edu/abs/2019A&A...632A..62C} {632, A62}

\bibitem[\protect\citeauthoryear{{Coe} \& {Kirk}}{{Coe} \&
  {Kirk}}{2015}]{Coe2015}
{Coe} M.~J.,  {Kirk} J.,  2015, \mn@doi [\mnras] {10.1093/mnras/stv1283}, \href
  {https://ui.adsabs.harvard.edu/abs/2015MNRAS.452..969C} {452, 969}

\bibitem[\protect\citeauthoryear{{Cranmer}}{{Cranmer}}{2009}]{Cranmer2009}
{Cranmer} S.~R.,  2009, \mn@doi [\apj] {10.1088/0004-637X/701/1/396}, \href
  {https://ui.adsabs.harvard.edu/abs/2009ApJ...701..396C} {701, 396}

\bibitem[\protect\citeauthoryear{{Cyr}, {Jones}, {Panoglou}, {Carciofi}  \&
  {Okazaki}}{{Cyr} et~al.}{2017}]{Cyr2017}
{Cyr} I.~H.,  {Jones} C.~E.,  {Panoglou} D.,  {Carciofi} A.~C.,   {Okazaki}
  A.~T.,  2017, \mn@doi [\mnras] {10.1093/mnras/stx1427}, \href
  {https://ui.adsabs.harvard.edu/abs/2017MNRAS.471..596C} {471, 596}

\bibitem[\protect\citeauthoryear{{Dai}, {Liu}  \& {Li}}{{Dai}
  et~al.}{2006}]{Dai2006}
{Dai} H.-L.,  {Liu} X.-W.,   {Li} X.-D.,  2006, \mn@doi [\apj]
  {10.1086/508735}, \href
  {https://ui.adsabs.harvard.edu/abs/2006ApJ...653.1410D} {653, 1410}

\bibitem[\protect\citeauthoryear{{Dallas}, {Oey}  \& {Castro}}{{Dallas}
  et~al.}{2022}]{Dallas2022}
{Dallas} M.~M.,  {Oey} M.~S.,   {Castro} N.,  2022, \mn@doi [\apj]
  {10.3847/1538-4357/ac8988}, \href
  {https://ui.adsabs.harvard.edu/abs/2022ApJ...936..112D} {936, 112}

\bibitem[\protect\citeauthoryear{{Das}, {Mesinger}, {Pallottini}, {Ferrara}  \&
  {Wise}}{{Das} et~al.}{2017}]{Das2017}
{Das} A.,  {Mesinger} A.,  {Pallottini} A.,  {Ferrara} A.,   {Wise} J.~H.,
  2017, \mn@doi [\mnras] {10.1093/mnras/stx943}, \href
  {https://ui.adsabs.harvard.edu/abs/2017MNRAS.469.1166D} {469, 1166}

\bibitem[\protect\citeauthoryear{{Davies}, {Kudritzki}, {Gazak}, {Plez},
  {Bergemann}, {Evans}  \& {Patrick}}{{Davies} et~al.}{2015}]{Davies2015}
{Davies} B.,  {Kudritzki} R.-P.,  {Gazak} Z.,  {Plez} B.,  {Bergemann} M.,
  {Evans} C.,   {Patrick} L.,  2015, \mn@doi [\apj]
  {10.1088/0004-637X/806/1/21}, \href
  {https://ui.adsabs.harvard.edu/abs/2015ApJ...806...21D} {806, 21}

\bibitem[\protect\citeauthoryear{{DeBoer} et~al.,}{{DeBoer}
  et~al.}{2017}]{DeBoer2017}
{DeBoer} D.~R.,  et~al., 2017, \mn@doi [\pasp]
  {10.1088/1538-3873/129/974/045001}, \href
  {https://ui.adsabs.harvard.edu/abs/2017PASP..129d5001D} {129, 045001}

\bibitem[\protect\citeauthoryear{{Dodd}, {Oudmaijer}, {Radley}, {Vioque}  \&
  {Frost}}{{Dodd} et~al.}{2023}]{Dodd2023}
{Dodd} J.~M.,  {Oudmaijer} R.~D.,  {Radley} I.~C.,  {Vioque} M.,   {Frost}
  A.~J.,  2023, \mn@doi [\mnras] {10.1093/mnras/stad3105}, \href
  {https://ui.adsabs.harvard.edu/abs/2023MNRAS.tmp.2984D} {}

\bibitem[\protect\citeauthoryear{{Dorigo Jones}, {Oey}, {Paggeot}, {Castro}  \&
  {Moe}}{{Dorigo Jones} et~al.}{2020}]{DorigoJones2020}
{Dorigo Jones} J.,  {Oey} M.~S.,  {Paggeot} K.,  {Castro} N.,   {Moe} M.,
  2020, \mn@doi [\apj] {10.3847/1538-4357/abbc6b}, \href
  {https://ui.adsabs.harvard.edu/abs/2020ApJ...903...43D} {903, 43}

\bibitem[\protect\citeauthoryear{{Douna}, {Pellizza}, {Mirabel}  \&
  {Pedrosa}}{{Douna} et~al.}{2015}]{Douna2015}
{Douna} V.~M.,  {Pellizza} L.~J.,  {Mirabel} I.~F.,   {Pedrosa} S.~E.,  2015,
  \mn@doi [\aap] {10.1051/0004-6361/201525617}, \href
  {https://ui.adsabs.harvard.edu/abs/2015A&A...579A..44D} {579, A44}

\bibitem[\protect\citeauthoryear{{Drout}, {G{\"o}tberg}, {Ludwig}, {Groh}, {de
  Mink}, {O'Grady}  \& {Smith}}{{Drout} et~al.}{2023}]{Drout2023}
{Drout} M.~R.,  {G{\"o}tberg} Y.,  {Ludwig} B.~A.,  {Groh} J.~H.,  {de Mink}
  S.~E.,  {O'Grady} A.~J.~G.,   {Smith} N.,  2023, \mn@doi [arXiv e-prints]
  {10.48550/arXiv.2307.00061}, \href
  {https://ui.adsabs.harvard.edu/abs/2023arXiv230700061D} {p. arXiv:2307.00061}

\bibitem[\protect\citeauthoryear{{Duch{\^e}ne} \& {Kraus}}{{Duch{\^e}ne} \&
  {Kraus}}{2013}]{Duchene2013}
{Duch{\^e}ne} G.,  {Kraus} A.,  2013, \mn@doi [\araa]
  {10.1146/annurev-astro-081710-102602}, \href
  {https://ui.adsabs.harvard.edu/abs/2013ARA&A..51..269D} {51, 269}

\bibitem[\protect\citeauthoryear{{Eggleton}}{{Eggleton}}{1983}]{Eggleton1983}
{Eggleton} P.~P.,  1983, \mn@doi [\apj] {10.1086/160960}, \href
  {https://ui.adsabs.harvard.edu/abs/1983ApJ...268..368E} {268, 368}

\bibitem[\protect\citeauthoryear{{Eide}, {Graziani}, {Ciardi}, {Feng},
  {Kakiichi}  \& {Di Matteo}}{{Eide} et~al.}{2018}]{Eide2018}
{Eide} M.~B.,  {Graziani} L.,  {Ciardi} B.,  {Feng} Y.,  {Kakiichi} K.,   {Di
  Matteo} T.,  2018, \mn@doi [\mnras] {10.1093/mnras/sty272}, \href
  {https://ui.adsabs.harvard.edu/abs/2018MNRAS.476.1174E} {476, 1174}

\bibitem[\protect\citeauthoryear{{Ekstr{\"o}m}, {Meynet}, {Maeder}  \&
  {Barblan}}{{Ekstr{\"o}m} et~al.}{2008}]{Ekstrom2008}
{Ekstr{\"o}m} S.,  {Meynet} G.,  {Maeder} A.,   {Barblan} F.,  2008, \mn@doi
  [\aap] {10.1051/0004-6361:20078095}, \href
  {https://ui.adsabs.harvard.edu/abs/2008A&A...478..467E} {478, 467}

\bibitem[\protect\citeauthoryear{{Ekstr{\"o}m} et~al.,}{{Ekstr{\"o}m}
  et~al.}{2012}]{Ekstrom2012}
{Ekstr{\"o}m} S.,  et~al., 2012, \mn@doi [\aap] {10.1051/0004-6361/201117751},
  \href {https://ui.adsabs.harvard.edu/abs/2012A&A...537A.146E} {537, A146}

\bibitem[\protect\citeauthoryear{{Ewall-Wice}, {Chang}, {Lazio}, {Dor{\'e}},
  {Seiffert}  \& {Monsalve}}{{Ewall-Wice} et~al.}{2018}]{Ewall-Wice2018}
{Ewall-Wice} A.,  {Chang} T.~C.,  {Lazio} J.,  {Dor{\'e}} O.,  {Seiffert} M.,
  {Monsalve} R.~A.,  2018, \mn@doi [\apj] {10.3847/1538-4357/aae51d}, \href
  {https://ui.adsabs.harvard.edu/abs/2018ApJ...868...63E} {868, 63}

\bibitem[\protect\citeauthoryear{{Fabrika}, {Atapin}, {Vinokurov}  \&
  {Sholukhova}}{{Fabrika} et~al.}{2021}]{Fabrika2021}
{Fabrika} S.~N.,  {Atapin} K.~E.,  {Vinokurov} A.~S.,   {Sholukhova} O.~N.,
  2021, \mn@doi [Astrophysical Bulletin] {10.1134/S1990341321010077}, \href
  {https://ui.adsabs.harvard.edu/abs/2021AstBu..76....6F} {76, 6}

\bibitem[\protect\citeauthoryear{{Farmer}, {Renzo}, {de Mink}, {Marchant}  \&
  {Justham}}{{Farmer} et~al.}{2019}]{Farmer2019}
{Farmer} R.,  {Renzo} M.,  {de Mink} S.~E.,  {Marchant} P.,   {Justham} S.,
  2019, \mn@doi [\apj] {10.3847/1538-4357/ab518b}, \href
  {https://ui.adsabs.harvard.edu/abs/2019ApJ...887...53F} {887, 53}

\bibitem[\protect\citeauthoryear{{Fialkov} \& {Barkana}}{{Fialkov} \&
  {Barkana}}{2014}]{Fialkov2014rich}
{Fialkov} A.,  {Barkana} R.,  2014, \mn@doi [\mnras] {10.1093/mnras/stu1744},
  \href {https://ui.adsabs.harvard.edu/abs/2014MNRAS.445..213F} {445, 213}

\bibitem[\protect\citeauthoryear{{Fialkov}, {Barkana}, {Visbal},
  {Tseliakhovich}  \& {Hirata}}{{Fialkov} et~al.}{2013}]{Fialkov2013}
{Fialkov} A.,  {Barkana} R.,  {Visbal} E.,  {Tseliakhovich} D.,   {Hirata}
  C.~M.,  2013, \mn@doi [\mnras] {10.1093/mnras/stt650}, \href
  {https://ui.adsabs.harvard.edu/abs/2013MNRAS.432.2909F} {432, 2909}

\bibitem[\protect\citeauthoryear{{Fialkov}, {Barkana}  \& {Visbal}}{{Fialkov}
  et~al.}{2014}]{Fialkov2014}
{Fialkov} A.,  {Barkana} R.,   {Visbal} E.,  2014, \mn@doi [\nat]
  {10.1038/nature12999}, \href
  {https://ui.adsabs.harvard.edu/abs/2014Natur.506..197F} {506, 197}

\bibitem[\protect\citeauthoryear{{Fialkov}, {Cohen}, {Barkana}  \&
  {Silk}}{{Fialkov} et~al.}{2017}]{Fialkov2017}
{Fialkov} A.,  {Cohen} A.,  {Barkana} R.,   {Silk} J.,  2017, \mn@doi [\mnras]
  {10.1093/mnras/stw2540}, \href
  {https://ui.adsabs.harvard.edu/abs/2017MNRAS.464.3498F} {464, 3498}

\bibitem[\protect\citeauthoryear{{Fishbach} \& {Kalogera}}{{Fishbach} \&
  {Kalogera}}{2022}]{Fishbach2022}
{Fishbach} M.,  {Kalogera} V.,  2022, \mn@doi [\apjl]
  {10.3847/2041-8213/ac64a5}, \href
  {https://ui.adsabs.harvard.edu/abs/2022ApJ...929L..26F} {929, L26}

\bibitem[\protect\citeauthoryear{{Fornasini} et~al.,}{{Fornasini}
  et~al.}{2019}]{Fornasini2019}
{Fornasini} F.~M.,  et~al., 2019, \mn@doi [\apj] {10.3847/1538-4357/ab4653},
  \href {https://ui.adsabs.harvard.edu/abs/2019ApJ...885...65F} {885, 65}

\bibitem[\protect\citeauthoryear{{Fornasini}, {Civano}  \& {Suh}}{{Fornasini}
  et~al.}{2020}]{Fornasini2020}
{Fornasini} F.~M.,  {Civano} F.,   {Suh} H.,  2020, \mn@doi [\mnras]
  {10.1093/mnras/staa1211}, \href
  {https://ui.adsabs.harvard.edu/abs/2020MNRAS.495..771F} {495, 771}

\bibitem[\protect\citeauthoryear{{Fornasini}, {Antoniou}  \&
  {Dubus}}{{Fornasini} et~al.}{2023}]{Fornasini2023}
{Fornasini} F.~M.,  {Antoniou} V.,   {Dubus} G.,  2023, \mn@doi [arXiv
  e-prints] {10.48550/arXiv.2308.02645}, \href
  {https://ui.adsabs.harvard.edu/abs/2023arXiv230802645F} {p. arXiv:2308.02645}

\bibitem[\protect\citeauthoryear{{Fortin}, {Garc{\'\i}a}, {Simaz Bunzel}  \&
  {Chaty}}{{Fortin} et~al.}{2023}]{Fortin2023}
{Fortin} F.,  {Garc{\'\i}a} F.,  {Simaz Bunzel} A.,   {Chaty} S.,  2023,
  \mn@doi [\aap] {10.1051/0004-6361/202245236}, \href
  {https://ui.adsabs.harvard.edu/abs/2023A&A...671A.149F} {671, A149}

\bibitem[\protect\citeauthoryear{{Fragos} et~al.,}{{Fragos}
  et~al.}{2013a}]{Fragos2013xrb}
{Fragos} T.,  et~al., 2013a, \mn@doi [\apj] {10.1088/0004-637X/764/1/41}, \href
  {https://ui.adsabs.harvard.edu/abs/2013ApJ...764...41F} {764, 41}

\bibitem[\protect\citeauthoryear{{Fragos}, {Lehmer}, {Naoz}, {Zezas}  \&
  {Basu-Zych}}{{Fragos} et~al.}{2013b}]{Fragos2013}
{Fragos} T.,  {Lehmer} B.~D.,  {Naoz} S.,  {Zezas} A.,   {Basu-Zych} A.,
  2013b, \mn@doi [\apjl] {10.1088/2041-8205/776/2/L31}, \href
  {https://ui.adsabs.harvard.edu/abs/2013ApJ...776L..31F} {776, L31}

\bibitem[\protect\citeauthoryear{{Franchini} \& {Martin}}{{Franchini} \&
  {Martin}}{2021}]{Franchini2021}
{Franchini} A.,  {Martin} R.~G.,  2021, \mn@doi [\apjl]
  {10.3847/2041-8213/ac4029}, \href
  {https://ui.adsabs.harvard.edu/abs/2021ApJ...923L..18F} {923, L18}

\bibitem[\protect\citeauthoryear{{Frost} et~al.,}{{Frost}
  et~al.}{2022}]{Frost2022}
{Frost} A.~J.,  et~al., 2022, \mn@doi [\aap] {10.1051/0004-6361/202143004},
  \href {https://ui.adsabs.harvard.edu/abs/2022A&A...659L...3F} {659, L3}

\bibitem[\protect\citeauthoryear{{Georgy} et~al.,}{{Georgy}
  et~al.}{2013}]{Georgy2013}
{Georgy} C.,  et~al., 2013, \mn@doi [\aap] {10.1051/0004-6361/201322178}, \href
  {https://ui.adsabs.harvard.edu/abs/2013A&A...558A.103G} {558, A103}

\bibitem[\protect\citeauthoryear{{Gessey-Jones} et~al.,}{{Gessey-Jones}
  et~al.}{2022}]{Gessey-Jones2022}
{Gessey-Jones} T.,  et~al., 2022, \mn@doi [\mnras] {10.1093/mnras/stac2049},
  \href {https://ui.adsabs.harvard.edu/abs/2022MNRAS.516..841G} {516, 841}

\bibitem[\protect\citeauthoryear{{Gessey-Jones}, {Fialkov}, {de Lera Acedo},
  {Handley}  \& {Barkana}}{{Gessey-Jones} et~al.}{2023}]{Gessey-Jones2023}
{Gessey-Jones} T.,  {Fialkov} A.,  {de Lera Acedo} E.,  {Handley} W.~J.,
  {Barkana} R.,  2023, \mn@doi [\mnras] {10.1093/mnras/stad3014}, \href
  {https://ui.adsabs.harvard.edu/abs/2023MNRAS.526.4262G} {526, 4262}

\bibitem[\protect\citeauthoryear{{Ghara}, {Mellema}  \& {Zaroubi}}{{Ghara}
  et~al.}{2022}]{Ghara2022}
{Ghara} R.,  {Mellema} G.,   {Zaroubi} S.,  2022, \mn@doi [\jcap]
  {10.1088/1475-7516/2022/03/055}, \href
  {https://ui.adsabs.harvard.edu/abs/2022JCAP...03..055G} {2022, 055}

\bibitem[\protect\citeauthoryear{{Gilfanov}, {Fabbiano}, {Lehmer}  \&
  {Zezas}}{{Gilfanov} et~al.}{2022}]{Gilfanov2023}
{Gilfanov} M.,  {Fabbiano} G.,  {Lehmer} B.,   {Zezas} A.,  2022, in , Handbook
  of X-ray and Gamma-ray Astrophysics.
p.~105, \mn@doi{10.1007/978-981-16-4544-0_108-1}

\bibitem[\protect\citeauthoryear{{Gotberg} et~al.,}{{Gotberg}
  et~al.}{2023}]{Gotberg2023}
{Gotberg} Y.,  et~al., 2023, \mn@doi [arXiv e-prints]
  {10.48550/arXiv.2307.00074}, \href
  {https://ui.adsabs.harvard.edu/abs/2023arXiv230700074G} {p. arXiv:2307.00074}

\bibitem[\protect\citeauthoryear{{Granada}, {Ekstr{\"o}m}, {Georgy},
  {Krti{\v{c}}ka}, {Owocki}, {Meynet}  \& {Maeder}}{{Granada}
  et~al.}{2013}]{Granada2013}
{Granada} A.,  {Ekstr{\"o}m} S.,  {Georgy} C.,  {Krti{\v{c}}ka} J.,  {Owocki}
  S.,  {Meynet} G.,   {Maeder} A.,  2013, \mn@doi [\aap]
  {10.1051/0004-6361/201220559}, \href
  {https://ui.adsabs.harvard.edu/abs/2013A&A...553A..25G} {553, A25}

\bibitem[\protect\citeauthoryear{{Groh} et~al.,}{{Groh}
  et~al.}{2019}]{Groh2019}
{Groh} J.~H.,  et~al., 2019, \mn@doi [\aap] {10.1051/0004-6361/201833720},
  \href {https://ui.adsabs.harvard.edu/abs/2019A&A...627A..24G} {627, A24}

\bibitem[\protect\citeauthoryear{{Grudzinska} et~al.,}{{Grudzinska}
  et~al.}{2015}]{Grudzinska2015}
{Grudzinska} M.,  et~al., 2015, \mn@doi [\mnras] {10.1093/mnras/stv1419}, \href
  {https://ui.adsabs.harvard.edu/abs/2015MNRAS.452.2773G} {452, 2773}

\bibitem[\protect\citeauthoryear{{Gunawardhana} et~al.,}{{Gunawardhana}
  et~al.}{2011}]{Gunawardhana2011}
{Gunawardhana} M.~L.~P.,  et~al., 2011, \mn@doi [\mnras]
  {10.1111/j.1365-2966.2011.18800.x}, \href
  {https://ui.adsabs.harvard.edu/abs/2011MNRAS.415.1647G} {415, 1647}

\bibitem[\protect\citeauthoryear{{Haberl} \& {Sturm}}{{Haberl} \&
  {Sturm}}{2016}]{Haberl2016}
{Haberl} F.,  {Sturm} R.,  2016, \mn@doi [\aap] {10.1051/0004-6361/201527326},
  \href {https://ui.adsabs.harvard.edu/abs/2016A&A...586A..81H} {586, A81}

\bibitem[\protect\citeauthoryear{{Hansen} \& {Phinney}}{{Hansen} \&
  {Phinney}}{1997}]{Hansen1997}
{Hansen} B. M.~S.,  {Phinney} E.~S.,  1997, \mn@doi [\mnras]
  {10.1093/mnras/291.3.569}, \href
  {https://ui.adsabs.harvard.edu/abs/1997MNRAS.291..569H} {291, 569}

\bibitem[\protect\citeauthoryear{{Hassan} et~al.,}{{Hassan}
  et~al.}{2023}]{Hassan2023}
{Hassan} S.,  et~al., 2023, \mn@doi [arXiv e-prints]
  {10.48550/arXiv.2305.02703}, \href
  {https://ui.adsabs.harvard.edu/abs/2023arXiv230502703H} {p. arXiv:2305.02703}

\bibitem[\protect\citeauthoryear{{Hastings}, {Wang}  \& {Langer}}{{Hastings}
  et~al.}{2020}]{Hastings2020}
{Hastings} B.,  {Wang} C.,   {Langer} N.,  2020, \mn@doi [\aap]
  {10.1051/0004-6361/201937018}, \href
  {https://ui.adsabs.harvard.edu/abs/2020A&A...633A.165H} {633, A165}

\bibitem[\protect\citeauthoryear{{Hastings}, {Langer}, {Wang}, {Schootemeijer}
  \& {Milone}}{{Hastings} et~al.}{2021}]{Hastings2021}
{Hastings} B.,  {Langer} N.,  {Wang} C.,  {Schootemeijer} A.,   {Milone} A.~P.,
   2021, \mn@doi [\aap] {10.1051/0004-6361/202141269}, \href
  {https://ui.adsabs.harvard.edu/abs/2021A&A...653A.144H} {653, A144}

\bibitem[\protect\citeauthoryear{{Haubois}, {Carciofi}, {Rivinius}, {Okazaki}
  \& {Bjorkman}}{{Haubois} et~al.}{2012}]{Haubois2012}
{Haubois} X.,  {Carciofi} A.~C.,  {Rivinius} T.,  {Okazaki} A.~T.,   {Bjorkman}
  J.~E.,  2012, \mn@doi [\apj] {10.1088/0004-637X/756/2/156}, \href
  {https://ui.adsabs.harvard.edu/abs/2012ApJ...756..156H} {756, 156}

\bibitem[\protect\citeauthoryear{Hendriks \& Izzard}{Hendriks \&
  Izzard}{2023a}]{Hendriks2023}
Hendriks D.~D.,  Izzard R.~G.,  2023a, \mn@doi [Journal of Open Source
  Software] {10.21105/joss.04642}, 8, 4642

\bibitem[\protect\citeauthoryear{{Hendriks} \& {Izzard}}{{Hendriks} \&
  {Izzard}}{2023b}]{Hendriks2023ms}
{Hendriks} D.~D.,  {Izzard} R.~G.,  2023b, \mn@doi [\mnras]
  {10.1093/mnras/stad2077}, \href
  {https://ui.adsabs.harvard.edu/abs/2023MNRAS.524.4315H} {524, 4315}

\bibitem[\protect\citeauthoryear{{Hendriks}, {van Son}, {Renzo}, {Izzard}  \&
  {Farmer}}{{Hendriks} et~al.}{2023}]{Hendriks2023gw}
{Hendriks} D.~D.,  {van Son} L.~A.~C.,  {Renzo} M.,  {Izzard} R.~G.,   {Farmer}
  R.,  2023, \mn@doi [\mnras] {10.1093/mnras/stad2857}, \href
  {https://ui.adsabs.harvard.edu/abs/2023MNRAS.526.4130H} {526, 4130}

\bibitem[\protect\citeauthoryear{{Hirano} \& {Bromm}}{{Hirano} \&
  {Bromm}}{2018}]{Hirano2018}
{Hirano} S.,  {Bromm} V.,  2018, \mn@doi [\mnras] {10.1093/mnras/sty487}, \href
  {https://ui.adsabs.harvard.edu/abs/2018MNRAS.476.3964H} {476, 3964}

\bibitem[\protect\citeauthoryear{{Hohle}, {Neuh{\"a}user}  \& {Schutz}}{{Hohle}
  et~al.}{2010}]{Hohle2010}
{Hohle} M.~M.,  {Neuh{\"a}user} R.,   {Schutz} B.~F.,  2010, \mn@doi
  [Astronomische Nachrichten] {10.1002/asna.200911355}, \href
  {https://ui.adsabs.harvard.edu/abs/2010AN....331..349H} {331, 349}

\bibitem[\protect\citeauthoryear{{Huang}}{{Huang}}{1963}]{Huang1963}
{Huang} S.-S.,  1963, \mn@doi [\apj] {10.1086/147648}, \href
  {https://ui.adsabs.harvard.edu/abs/1963ApJ...138..342H} {138, 342}

\bibitem[\protect\citeauthoryear{{Hummel}, {Stacy}, {Jeon}, {Oliveri}  \&
  {Bromm}}{{Hummel} et~al.}{2015}]{Hummel2015}
{Hummel} J.~A.,  {Stacy} A.,  {Jeon} M.,  {Oliveri} A.,   {Bromm} V.,  2015,
  \mn@doi [\mnras] {10.1093/mnras/stv1902}, \href
  {https://ui.adsabs.harvard.edu/abs/2015MNRAS.453.4136H} {453, 4136}

\bibitem[\protect\citeauthoryear{{Hummel}, {Stacy}  \& {Bromm}}{{Hummel}
  et~al.}{2016}]{Hummel2016}
{Hummel} J.~A.,  {Stacy} A.,   {Bromm} V.,  2016, \mn@doi [\mnras]
  {10.1093/mnras/stw1127}, \href
  {https://ui.adsabs.harvard.edu/abs/2016MNRAS.460.2432H} {460, 2432}

\bibitem[\protect\citeauthoryear{{Hunter}}{{Hunter}}{2007}]{Hunter2007}
{Hunter} J.~D.,  2007, \mn@doi [Computing in Science and Engineering]
  {10.1109/MCSE.2007.55}, \href
  {https://ui.adsabs.harvard.edu/abs/2007CSE.....9...90H} {9, 90}

\bibitem[\protect\citeauthoryear{{Hurley}, {Pols}  \& {Tout}}{{Hurley}
  et~al.}{2000}]{Hurley2000}
{Hurley} J.~R.,  {Pols} O.~R.,   {Tout} C.~A.,  2000, \mn@doi [\mnras]
  {10.1046/j.1365-8711.2000.03426.x}, \href
  {https://ui.adsabs.harvard.edu/abs/2000MNRAS.315..543H} {315, 543}

\bibitem[\protect\citeauthoryear{{Hurley}, {Tout}  \& {Pols}}{{Hurley}
  et~al.}{2002}]{Hurley2002}
{Hurley} J.~R.,  {Tout} C.~A.,   {Pols} O.~R.,  2002, \mn@doi [\mnras]
  {10.1046/j.1365-8711.2002.05038.x}, \href
  {https://ui.adsabs.harvard.edu/abs/2002MNRAS.329..897H} {329, 897}

\bibitem[\protect\citeauthoryear{{Igoshev}, {Chruslinska}, {Dorozsmai}  \&
  {Toonen}}{{Igoshev} et~al.}{2021}]{Igoshev2021}
{Igoshev} A.~P.,  {Chruslinska} M.,  {Dorozsmai} A.,   {Toonen} S.,  2021,
  \mn@doi [\mnras] {10.1093/mnras/stab2734}, \href
  {https://ui.adsabs.harvard.edu/abs/2021MNRAS.508.3345I} {508, 3345}

\bibitem[\protect\citeauthoryear{{Inayoshi}, {Hirai}, {Kinugawa}  \&
  {Hotokezaka}}{{Inayoshi} et~al.}{2017}]{Inayoshi2017}
{Inayoshi} K.,  {Hirai} R.,  {Kinugawa} T.,   {Hotokezaka} K.,  2017, \mn@doi
  [\mnras] {10.1093/mnras/stx757}, \href
  {https://ui.adsabs.harvard.edu/abs/2017MNRAS.468.5020I} {468, 5020}

\bibitem[\protect\citeauthoryear{{Izzard} \& {Halabi}}{{Izzard} \&
  {Halabi}}{2018}]{Izzard2018}
{Izzard} R.~G.,  {Halabi} G.~M.,  2018, \mn@doi [arXiv e-prints]
  {10.48550/arXiv.1808.06883}, \href
  {https://ui.adsabs.harvard.edu/abs/2018arXiv180806883I} {p. arXiv:1808.06883}

\bibitem[\protect\citeauthoryear{{Izzard} \& {Jermyn}}{{Izzard} \&
  {Jermyn}}{2023}]{Izzard2023}
{Izzard} R.~G.,  {Jermyn} A.~S.,  2023, \mn@doi [\mnras]
  {10.1093/mnras/stac2899}, \href
  {https://ui.adsabs.harvard.edu/abs/2023MNRAS.521...35I} {521, 35}

\bibitem[\protect\citeauthoryear{{Izzard}, {Tout}, {Karakas}  \&
  {Pols}}{{Izzard} et~al.}{2004}]{Izzard2004}
{Izzard} R.~G.,  {Tout} C.~A.,  {Karakas} A.~I.,   {Pols} O.~R.,  2004, \mn@doi
  [\mnras] {10.1111/j.1365-2966.2004.07446.x}, \href
  {https://ui.adsabs.harvard.edu/abs/2004MNRAS.350..407I} {350, 407}

\bibitem[\protect\citeauthoryear{{Izzard}, {Dray}, {Karakas}, {Lugaro}  \&
  {Tout}}{{Izzard} et~al.}{2006}]{Izzard2006}
{Izzard} R.~G.,  {Dray} L.~M.,  {Karakas} A.~I.,  {Lugaro} M.,   {Tout} C.~A.,
  2006, \mn@doi [\aap] {10.1051/0004-6361:20066129}, \href
  {https://ui.adsabs.harvard.edu/abs/2006A&A...460..565I} {460, 565}

\bibitem[\protect\citeauthoryear{{Izzard}, {Glebbeek}, {Stancliffe}  \&
  {Pols}}{{Izzard} et~al.}{2009}]{Izzard2009}
{Izzard} R.~G.,  {Glebbeek} E.,  {Stancliffe} R.~J.,   {Pols} O.~R.,  2009,
  \mn@doi [\aap] {10.1051/0004-6361/200912827}, \href
  {https://ui.adsabs.harvard.edu/abs/2009A&A...508.1359I} {508, 1359}

\bibitem[\protect\citeauthoryear{Izzard, Preece, Jofre, Halabi, Masseron  \&
  Tout}{Izzard et~al.}{2017}]{Izzard2017}
Izzard R.~G.,  Preece H.,  Jofre P.,  Halabi G.~M.,  Masseron T.,   Tout C.~A.,
   2017, \mn@doi [\mnras] {10.1093/mnras/stx2355}, 473, 2984

\bibitem[\protect\citeauthoryear{{Janssens}, {Shenar}, {Degenaar},
  {Bodensteiner}, {Sana}, {Audenaert}  \& {Frost}}{{Janssens}
  et~al.}{2023}]{Janssens2023}
{Janssens} S.,  {Shenar} T.,  {Degenaar} N.,  {Bodensteiner} J.,  {Sana} H.,
  {Audenaert} J.,   {Frost} A.~J.,  2023, \mn@doi [\aap]
  {10.1051/0004-6361/202347318}, \href
  {https://ui.adsabs.harvard.edu/abs/2023A&A...677L...9J} {677, L9}

\bibitem[\protect\citeauthoryear{{Jeena}, {Banerjee}, {Chiaki}  \&
  {Heger}}{{Jeena} et~al.}{2023}]{Jeena2023}
{Jeena} S.~K.,  {Banerjee} P.,  {Chiaki} G.,   {Heger} A.,  2023, \mn@doi
  [\mnras] {10.1093/mnras/stad3028}, \href
  {https://ui.adsabs.harvard.edu/abs/2023MNRAS.526.4467J} {526, 4467}

\bibitem[\protect\citeauthoryear{{Jeon}, {Pawlik}, {Greif}, {Glover}, {Bromm},
  {Milosavljevi{\'c}}  \& {Klessen}}{{Jeon} et~al.}{2012}]{Jeon2012}
{Jeon} M.,  {Pawlik} A.~H.,  {Greif} T.~H.,  {Glover} S. C.~O.,  {Bromm} V.,
  {Milosavljevi{\'c}} M.,   {Klessen} R.~S.,  2012, \mn@doi [\apj]
  {10.1088/0004-637X/754/1/34}, \href
  {https://ui.adsabs.harvard.edu/abs/2012ApJ...754...34J} {754, 34}

\bibitem[\protect\citeauthoryear{{Je{\v{r}}{\'a}bkov{\'a}}, {Hasani Zonoozi},
  {Kroupa}, {Beccari}, {Yan}, {Vazdekis}  \& {Zhang}}{{Je{\v{r}}{\'a}bkov{\'a}}
  et~al.}{2018}]{Jerabkova2018}
{Je{\v{r}}{\'a}bkov{\'a}} T.,  {Hasani Zonoozi} A.,  {Kroupa} P.,  {Beccari}
  G.,  {Yan} Z.,  {Vazdekis} A.,   {Zhang} Z.~Y.,  2018, \mn@doi [\aap]
  {10.1051/0004-6361/201833055}, \href
  {https://ui.adsabs.harvard.edu/abs/2018A&A...620A..39J} {620, A39}

\bibitem[\protect\citeauthoryear{{Johnson}, {Dalla Vecchia}  \&
  {Khochfar}}{{Johnson} et~al.}{2013}]{Johnson2013}
{Johnson} J.~L.,  {Dalla Vecchia} C.,   {Khochfar} S.,  2013, \mn@doi [\mnras]
  {10.1093/mnras/sts011}, \href
  {https://ui.adsabs.harvard.edu/abs/2013MNRAS.428.1857J} {428, 1857}

\bibitem[\protect\citeauthoryear{Jones, Oliphant, Peterson  et~al.}{Jones
  et~al.}{2001}]{jones2001scipy}
Jones E.,  Oliphant T.,  Peterson P.,   et~al., 2001

\bibitem[\protect\citeauthoryear{{Kaaret}, {Feng}  \& {Roberts}}{{Kaaret}
  et~al.}{2017}]{Kaaret2017}
{Kaaret} P.,  {Feng} H.,   {Roberts} T.~P.,  2017, \mn@doi [\araa]
  {10.1146/annurev-astro-091916-055259}, \href
  {https://ui.adsabs.harvard.edu/abs/2017ARA&A..55..303K} {55, 303}

\bibitem[\protect\citeauthoryear{{Kamran}, {Ghara}, {Majumdar}, {Mellema},
  {Bharadwaj}, {Pritchard}, {Mondal}  \& {Iliev}}{{Kamran}
  et~al.}{2022}]{Kamran2022}
{Kamran} M.,  {Ghara} R.,  {Majumdar} S.,  {Mellema} G.,  {Bharadwaj} S.,
  {Pritchard} J.~R.,  {Mondal} R.,   {Iliev} I.~T.,  2022, \mn@doi [\jcap]
  {10.1088/1475-7516/2022/11/001}, \href
  {https://ui.adsabs.harvard.edu/abs/2022JCAP...11..001K} {2022, 001}

\bibitem[\protect\citeauthoryear{{Karino}}{{Karino}}{2022}]{Karino2022}
{Karino} S.,  2022, \mn@doi [\mnras] {10.1093/mnras/stac1334}, \href
  {https://ui.adsabs.harvard.edu/abs/2022MNRAS.514..191K} {514, 191}

\bibitem[\protect\citeauthoryear{{Kaur}, {Qin}, {Mesinger}, {Pallottini},
  {Fragos}  \& {Basu-Zych}}{{Kaur} et~al.}{2022}]{Kaur2022}
{Kaur} H.~D.,  {Qin} Y.,  {Mesinger} A.,  {Pallottini} A.,  {Fragos} T.,
  {Basu-Zych} A.,  2022, \mn@doi [\mnras] {10.1093/mnras/stac1226}, \href
  {https://ui.adsabs.harvard.edu/abs/2022MNRAS.513.5097K} {513, 5097}

\bibitem[\protect\citeauthoryear{{Kennea}, {Coe}, {Evans}, {Townsend},
  {Campbell}  \& {Udalski}}{{Kennea} et~al.}{2021}]{Kennea2021}
{Kennea} J.~A.,  {Coe} M.~J.,  {Evans} P.~A.,  {Townsend} L.~J.,  {Campbell}
  Z.~A.,   {Udalski} A.,  2021, \mn@doi [\mnras] {10.1093/mnras/stab2632},
  \href {https://ui.adsabs.harvard.edu/abs/2021MNRAS.508..781K} {508, 781}

\bibitem[\protect\citeauthoryear{{Khan}, {Middleton}, {Wiktorowicz}, {Dauser},
  {Roberts}  \& {Wilms}}{{Khan} et~al.}{2022}]{Khan2022}
{Khan} N.,  {Middleton} M.~J.,  {Wiktorowicz} G.,  {Dauser} T.,  {Roberts}
  T.~P.,   {Wilms} J.,  2022, \mn@doi [\mnras] {10.1093/mnras/stab3049}, \href
  {https://ui.adsabs.harvard.edu/abs/2022MNRAS.509.2493K} {509, 2493}

\bibitem[\protect\citeauthoryear{{Khokhlov} et~al.,}{{Khokhlov}
  et~al.}{2018}]{Khokhlov2018}
{Khokhlov} S.~A.,  et~al., 2018, \mn@doi [\apj] {10.3847/1538-4357/aab49d},
  \href {https://ui.adsabs.harvard.edu/abs/2018ApJ...856..158K} {856, 158}

\bibitem[\protect\citeauthoryear{{King}}{{King}}{2009}]{King2009}
{King} A.~R.,  2009, \mn@doi [\mnras] {10.1111/j.1745-3933.2008.00594.x}, \href
  {https://ui.adsabs.harvard.edu/abs/2009MNRAS.393L..41K} {393, L41}

\bibitem[\protect\citeauthoryear{{King}, {Davies}, {Ward}, {Fabbiano}  \&
  {Elvis}}{{King} et~al.}{2001}]{King2001}
{King} A.~R.,  {Davies} M.~B.,  {Ward} M.~J.,  {Fabbiano} G.,   {Elvis} M.,
  2001, \mn@doi [\apjl] {10.1086/320343}, \href
  {https://ui.adsabs.harvard.edu/abs/2001ApJ...552L.109K} {552, L109}

\bibitem[\protect\citeauthoryear{{King}, {Lasota}  \& {Middleton}}{{King}
  et~al.}{2023}]{King2023}
{King} A.,  {Lasota} J.-P.,   {Middleton} M.,  2023, \mn@doi [\nar]
  {10.1016/j.newar.2022.101672}, \href
  {https://ui.adsabs.harvard.edu/abs/2023NewAR..9601672K} {96, 101672}

\bibitem[\protect\citeauthoryear{{Klement} et~al.,}{{Klement}
  et~al.}{2017}]{Klement2017}
{Klement} R.,  et~al., 2017, \mn@doi [\aap] {10.1051/0004-6361/201629932},
  \href {https://ui.adsabs.harvard.edu/abs/2017A&A...601A..74K} {601, A74}

\bibitem[\protect\citeauthoryear{{Klement} et~al.,}{{Klement}
  et~al.}{2019}]{Klement2019}
{Klement} R.,  et~al., 2019, \mn@doi [\apj] {10.3847/1538-4357/ab48e7}, \href
  {https://ui.adsabs.harvard.edu/abs/2019ApJ...885..147K} {885, 147}

\bibitem[\protect\citeauthoryear{{Koopmans} et~al.,}{{Koopmans}
  et~al.}{2015}]{Koopmans2015}
{Koopmans} L.,  et~al., 2015, in Advancing Astrophysics with the Square
  Kilometre Array (AASKA14). p.~1 (\mn@eprint {arXiv} {1505.07568}),
  \mn@doi{10.22323/1.215.0001}

\bibitem[\protect\citeauthoryear{{Kotko} \& {Belczynski}}{{Kotko} \&
  {Belczynski}}{2023}]{Kotko2023}
{Kotko} I.,  {Belczynski} K.,  2023, \mn@doi [arXiv e-prints]
  {10.48550/arXiv.2305.08640}, \href
  {https://ui.adsabs.harvard.edu/abs/2023arXiv230508640K} {p. arXiv:2305.08640}

\bibitem[\protect\citeauthoryear{{Kovlakas}, {Zezas}, {Andrews}, {Basu-Zych},
  {Fragos}, {Hornschemeier}, {Lehmer}  \& {Ptak}}{{Kovlakas}
  et~al.}{2020}]{Kovlakas2020}
{Kovlakas} K.,  {Zezas} A.,  {Andrews} J.~J.,  {Basu-Zych} A.,  {Fragos} T.,
  {Hornschemeier} A.,  {Lehmer} B.,   {Ptak} A.,  2020, \mn@doi [\mnras]
  {10.1093/mnras/staa2481}, \href
  {https://ui.adsabs.harvard.edu/abs/2020MNRAS.498.4790K} {498, 4790}

\bibitem[\protect\citeauthoryear{{Kovlakas}, {Fragos}, {Schaerer}  \&
  {Mesinger}}{{Kovlakas} et~al.}{2022}]{Kovlakas2022}
{Kovlakas} K.,  {Fragos} T.,  {Schaerer} D.,   {Mesinger} A.,  2022, \mn@doi
  [\aap] {10.1051/0004-6361/202244252}, \href
  {https://ui.adsabs.harvard.edu/abs/2022A&A...665A..28K} {665, A28}

\bibitem[\protect\citeauthoryear{{Kretschmar} et~al.,}{{Kretschmar}
  et~al.}{2019}]{Kretschmar2019}
{Kretschmar} P.,  et~al., 2019, \mn@doi [\nar] {10.1016/j.newar.2020.101546},
  \href {https://ui.adsabs.harvard.edu/abs/2019NewAR..8601546K} {86, 101546}

\bibitem[\protect\citeauthoryear{{Kroupa}}{{Kroupa}}{1995}]{Kroupa1995}
{Kroupa} P.,  1995, \mn@doi [\mnras] {10.1093/mnras/277.4.1507}, \href
  {https://ui.adsabs.harvard.edu/abs/1995MNRAS.277.1507K} {277, 1507}

\bibitem[\protect\citeauthoryear{{Kroupa}}{{Kroupa}}{2001}]{Kroupa2001}
{Kroupa} P.,  2001, \mn@doi [\mnras] {10.1046/j.1365-8711.2001.04022.x}, \href
  {https://ui.adsabs.harvard.edu/abs/2001MNRAS.322..231K} {322, 231}

\bibitem[\protect\citeauthoryear{{Krti{\v{c}}ka}, {Owocki}  \&
  {Meynet}}{{Krti{\v{c}}ka} et~al.}{2011}]{Krticka2011}
{Krti{\v{c}}ka} J.,  {Owocki} S.~P.,   {Meynet} G.,  2011, \mn@doi [\aap]
  {10.1051/0004-6361/20101595110.48550/arXiv.1101.1732}, \href
  {https://ui.adsabs.harvard.edu/abs/2011A&A...527A..84K} {527, A84}

\bibitem[\protect\citeauthoryear{{Krti{\v{c}}ka}, {Kurf{\"u}rst}  \&
  {Krti{\v{c}}kov{\'a}}}{{Krti{\v{c}}ka} et~al.}{2015}]{Krticka2015}
{Krti{\v{c}}ka} J.,  {Kurf{\"u}rst} P.,   {Krti{\v{c}}kov{\'a}} I.,  2015,
  \mn@doi [\aap] {10.1051/0004-6361/201424867}, \href
  {https://ui.adsabs.harvard.edu/abs/2015A&A...573A..20K} {573, A20}

\bibitem[\protect\citeauthoryear{{Kulkarni}, {Visbal}  \& {Bryan}}{{Kulkarni}
  et~al.}{2021}]{Kulkarni2021}
{Kulkarni} M.,  {Visbal} E.,   {Bryan} G.~L.,  2021, \mn@doi [\apj]
  {10.3847/1538-4357/ac08a3}, \href
  {https://ui.adsabs.harvard.edu/abs/2021ApJ...917...40K} {917, 40}

\bibitem[\protect\citeauthoryear{{Kuranov}, {Postnov}  \&
  {Yungelson}}{{Kuranov} et~al.}{2020}]{Kuranov2020}
{Kuranov} A.~G.,  {Postnov} K.~A.,   {Yungelson} L.~R.,  2020, \mn@doi
  [Astronomy Letters] {10.1134/S1063773720100084}, \href
  {https://ui.adsabs.harvard.edu/abs/2020AstL...46..658K} {46, 658}

\bibitem[\protect\citeauthoryear{{Lacchin}, {Matteucci}, {Vincenzo}  \&
  {Palla}}{{Lacchin} et~al.}{2020}]{Lacchin2020}
{Lacchin} E.,  {Matteucci} F.,  {Vincenzo} F.,   {Palla} M.,  2020, \mn@doi
  [\mnras] {10.1093/mnras/staa585}, \href
  {https://ui.adsabs.harvard.edu/abs/2020MNRAS.495.3276L} {495, 3276}

\bibitem[\protect\citeauthoryear{{Lang}}{{Lang}}{1992}]{Lang1992}
{Lang} K.~R.,  1992, {Astrophysical Data}.
Springer-Verlag

\bibitem[\protect\citeauthoryear{{Lasota} \& {King}}{{Lasota} \&
  {King}}{2023}]{Lasota2023}
{Lasota} J.-P.,  {King} A.,  2023, \mn@doi [\mnras] {10.1093/mnras/stad2926},
  \href {https://ui.adsabs.harvard.edu/abs/2023MNRAS.526.2506L} {526, 2506}

\bibitem[\protect\citeauthoryear{{Lazare}, {Sarkar}  \& {Kovetz}}{{Lazare}
  et~al.}{2023}]{Lazare2023}
{Lazare} H.,  {Sarkar} D.,   {Kovetz} E.~D.,  2023, \mn@doi [arXiv e-prints]
  {10.48550/arXiv.2307.15577}, \href
  {https://ui.adsabs.harvard.edu/abs/2023arXiv230715577L} {p. arXiv:2307.15577}

\bibitem[\protect\citeauthoryear{{Lazzarini} et~al.,}{{Lazzarini}
  et~al.}{2023}]{Lazzarini2023}
{Lazzarini} M.,  et~al., 2023, \mn@doi [\apj] {10.3847/1538-4357/acdbc8}, \href
  {https://ui.adsabs.harvard.edu/abs/2023ApJ...952..114L} {952, 114}

\bibitem[\protect\citeauthoryear{{Lee}, {Neiner}  \& {Mathis}}{{Lee}
  et~al.}{2014}]{Lee2014}
{Lee} U.,  {Neiner} C.,   {Mathis} S.,  2014, \mn@doi [\mnras]
  {10.1093/mnras/stu1256}, \href
  {https://ui.adsabs.harvard.edu/abs/2014MNRAS.443.1515L} {443, 1515}

\bibitem[\protect\citeauthoryear{{Lehmer} et~al.,}{{Lehmer}
  et~al.}{2012}]{Lehmer2012}
{Lehmer} B.~D.,  et~al., 2012, \mn@doi [\apj] {10.1088/0004-637X/752/1/46},
  \href {https://ui.adsabs.harvard.edu/abs/2012ApJ...752...46L} {752, 46}

\bibitem[\protect\citeauthoryear{{Lehmer} et~al.,}{{Lehmer}
  et~al.}{2016}]{Lehmer2016}
{Lehmer} B.~D.,  et~al., 2016, \mn@doi [\apj] {10.3847/0004-637X/825/1/7},
  \href {https://ui.adsabs.harvard.edu/abs/2016ApJ...825....7L} {825, 7}

\bibitem[\protect\citeauthoryear{{Lehmer} et~al.,}{{Lehmer}
  et~al.}{2019}]{Lehmer2019}
{Lehmer} B.~D.,  et~al., 2019, \mn@doi [\apjs] {10.3847/1538-4365/ab22a8},
  \href {https://ui.adsabs.harvard.edu/abs/2019ApJS..243....3L} {243, 3}

\bibitem[\protect\citeauthoryear{{Lehmer} et~al.,}{{Lehmer}
  et~al.}{2021}]{Lehmer2021}
{Lehmer} B.~D.,  et~al., 2021, \mn@doi [\apj] {10.3847/1538-4357/abcec1}, \href
  {https://ui.adsabs.harvard.edu/abs/2021ApJ...907...17L} {907, 17}

\bibitem[\protect\citeauthoryear{{Lehmer}, {Eufrasio}, {Basu-Zych}, {Garofali},
  {Gilbertson}, {Mesinger}  \& {Yukita}}{{Lehmer} et~al.}{2022}]{Lehmer2022}
{Lehmer} B.~D.,  {Eufrasio} R.~T.,  {Basu-Zych} A.,  {Garofali} K.,
  {Gilbertson} W.,  {Mesinger} A.,   {Yukita} M.,  2022, \mn@doi [\apj]
  {10.3847/1538-4357/ac63a7}, \href
  {https://ui.adsabs.harvard.edu/abs/2022ApJ...930..135L} {930, 135}

\bibitem[\protect\citeauthoryear{{Lewis}, {Pillepich}, {Nelson}, {Klessen}  \&
  {Glover}}{{Lewis} et~al.}{2023}]{Lewis2023}
{Lewis} J. S.~W.,  {Pillepich} A.,  {Nelson} D.,  {Klessen} R.~S.,   {Glover}
  S. C.~O.,  2023, \mn@doi [arXiv e-prints] {10.48550/arXiv.2305.09721}, \href
  {https://ui.adsabs.harvard.edu/abs/2023arXiv230509721L} {p. arXiv:2305.09721}

\bibitem[\protect\citeauthoryear{{Linden}, {Sepinsky}, {Kalogera}  \&
  {Belczynski}}{{Linden} et~al.}{2009}]{Linden2009}
{Linden} T.,  {Sepinsky} J.~F.,  {Kalogera} V.,   {Belczynski} K.,  2009,
  \mn@doi [\apj] {10.1088/0004-637X/699/2/1573}, \href
  {https://ui.adsabs.harvard.edu/abs/2009ApJ...699.1573L} {699, 1573}

\bibitem[\protect\citeauthoryear{{Linden}, {Kalogera}, {Sepinsky}, {Prestwich},
  {Zezas}  \& {Gallagher}}{{Linden} et~al.}{2010}]{Linden2010}
{Linden} T.,  {Kalogera} V.,  {Sepinsky} J.~F.,  {Prestwich} A.,  {Zezas} A.,
  {Gallagher} J.~S.,  2010, \mn@doi [\apj] {10.1088/0004-637X/725/2/1984},
  \href {https://ui.adsabs.harvard.edu/abs/2010ApJ...725.1984L} {725, 1984}

\bibitem[\protect\citeauthoryear{{Liotine}, {Zevin}, {Berry}, {Doctor}  \&
  {Kalogera}}{{Liotine} et~al.}{2023}]{Liotine2023}
{Liotine} C.,  {Zevin} M.,  {Berry} C. P.~L.,  {Doctor} Z.,   {Kalogera} V.,
  2023, \mn@doi [\apj] {10.3847/1538-4357/acb8b2}, \href
  {https://ui.adsabs.harvard.edu/abs/2023ApJ...946....4L} {946, 4}

\bibitem[\protect\citeauthoryear{{Liu} \& {Bromm}}{{Liu} \&
  {Bromm}}{2020}]{Liu2020did}
{Liu} B.,  {Bromm} V.,  2020, \mn@doi [\mnras] {10.1093/mnras/staa2143}, \href
  {https://ui.adsabs.harvard.edu/abs/2020MNRAS.497.2839L} {497, 2839}

\bibitem[\protect\citeauthoryear{{Liu}, {Schauer}  \& {Bromm}}{{Liu}
  et~al.}{2019}]{Liu2019}
{Liu} B.,  {Schauer} A. T.~P.,   {Bromm} V.,  2019, \mn@doi [\mnras]
  {10.1093/mnras/stz1583}, \href
  {https://ui.adsabs.harvard.edu/abs/2019MNRAS.487.4711L} {487, 4711}

\bibitem[\protect\citeauthoryear{{Liu}, {Meynet}  \& {Bromm}}{{Liu}
  et~al.}{2021}]{Liu2021binary}
{Liu} B.,  {Meynet} G.,   {Bromm} V.,  2021, \mn@doi [\mnras]
  {10.1093/mnras/staa3671}, \href
  {https://ui.adsabs.harvard.edu/abs/2021MNRAS.501..643L} {501, 643}

\bibitem[\protect\citeauthoryear{{Lodato} \& {Price}}{{Lodato} \&
  {Price}}{2010}]{Lodato2010}
{Lodato} G.,  {Price} D.~J.,  2010, \mn@doi [\mnras]
  {10.1111/j.1365-2966.2010.16526.x}, \href
  {https://ui.adsabs.harvard.edu/abs/2010MNRAS.405.1212L} {405, 1212}

\bibitem[\protect\citeauthoryear{{Ma}, {Ciardi}, {Eide}  \& {Helgason}}{{Ma}
  et~al.}{2018}]{Ma2018}
{Ma} Q.,  {Ciardi} B.,  {Eide} M.~B.,   {Helgason} K.,  2018, \mn@doi [\mnras]
  {10.1093/mnras/sty1806}, \href
  {https://ui.adsabs.harvard.edu/abs/2018MNRAS.480...26M} {480, 26}

\bibitem[\protect\citeauthoryear{{Ma}, {Ghara}, {Ciardi}, {Iliev}, {Koopmans},
  {Mellema}, {Mondal}  \& {Zaroubi}}{{Ma} et~al.}{2023}]{Ma202321}
{Ma} Q.-B.,  {Ghara} R.,  {Ciardi} B.,  {Iliev} I.~T.,  {Koopmans} L. V.~E.,
  {Mellema} G.,  {Mondal} R.,   {Zaroubi} S.,  2023, \mn@doi [\mnras]
  {10.1093/mnras/stad1203}, \href
  {https://ui.adsabs.harvard.edu/abs/2023MNRAS.522.3284M} {522, 3284}

\bibitem[\protect\citeauthoryear{{Madau}}{{Madau}}{2018}]{Madau2018}
{Madau} P.,  2018, \mn@doi [\mnras] {10.1093/mnrasl/sly125}, \href
  {https://ui.adsabs.harvard.edu/abs/2018MNRAS.480L..43M} {480, L43}

\bibitem[\protect\citeauthoryear{{Madau} \& {Fragos}}{{Madau} \&
  {Fragos}}{2017}]{Madau2017}
{Madau} P.,  {Fragos} T.,  2017, \mn@doi [\apj] {10.3847/1538-4357/aa6af9},
  \href {https://ui.adsabs.harvard.edu/abs/2017ApJ...840...39M} {840, 39}

\bibitem[\protect\citeauthoryear{{Maeder}, {Meynet}  \& {Chiappini}}{{Maeder}
  et~al.}{2015}]{Maeder2015}
{Maeder} A.,  {Meynet} G.,   {Chiappini} C.,  2015, \mn@doi [\aap]
  {10.1051/0004-6361/201424153}, \href
  {https://ui.adsabs.harvard.edu/abs/2015A&A...576A..56M} {576, A56}

\bibitem[\protect\citeauthoryear{{Magg} et~al.,}{{Magg}
  et~al.}{2022}]{Magg2022tr}
{Magg} M.,  et~al., 2022, \mn@doi [\mnras] {10.1093/mnras/stac1664}, \href
  {https://ui.adsabs.harvard.edu/abs/2022MNRAS.514.4433M} {514, 4433}

\bibitem[\protect\citeauthoryear{{Mapelli}, {Ripamonti}, {Zampieri}, {Colpi}
  \& {Bressan}}{{Mapelli} et~al.}{2010}]{Mapelli2010}
{Mapelli} M.,  {Ripamonti} E.,  {Zampieri} L.,  {Colpi} M.,   {Bressan} A.,
  2010, \mn@doi [\mnras] {10.1111/j.1365-2966.2010.17048.x}, \href
  {https://ui.adsabs.harvard.edu/abs/2010MNRAS.408..234M} {408, 234}

\bibitem[\protect\citeauthoryear{{Maravelias}, {Zezas}, {Antoniou}  \&
  {Hatzidimitriou}}{{Maravelias} et~al.}{2014}]{Maravelias2014}
{Maravelias} G.,  {Zezas} A.,  {Antoniou} V.,   {Hatzidimitriou} D.,  2014,
  \mn@doi [\mnras] {10.1093/mnras/stt2302}, \href
  {https://ui.adsabs.harvard.edu/abs/2014MNRAS.438.2005M} {438, 2005}

\bibitem[\protect\citeauthoryear{{Marchant}, {Langer}, {Podsiadlowski},
  {Tauris}, {de Mink}, {Mandel}  \& {Moriya}}{{Marchant}
  et~al.}{2017}]{Marchant2017}
{Marchant} P.,  {Langer} N.,  {Podsiadlowski} P.,  {Tauris} T.~M.,  {de Mink}
  S.,  {Mandel} I.,   {Moriya} T.~J.,  2017, \mn@doi [\aap]
  {10.1051/0004-6361/201630188}, \href
  {https://ui.adsabs.harvard.edu/abs/2017A&A...604A..55M} {604, A55}

\bibitem[\protect\citeauthoryear{{Marks}, {Kroupa}, {Dabringhausen}  \&
  {Pawlowski}}{{Marks} et~al.}{2012}]{Marks2012}
{Marks} M.,  {Kroupa} P.,  {Dabringhausen} J.,   {Pawlowski} M.~S.,  2012,
  \mn@doi [\mnras] {10.1111/j.1365-2966.2012.20767.x}, \href
  {https://ui.adsabs.harvard.edu/abs/2012MNRAS.422.2246M} {422, 2246}

\bibitem[\protect\citeauthoryear{{Martin}, {Pringle}, {Tout}  \&
  {Lubow}}{{Martin} et~al.}{2011}]{Martin2011}
{Martin} R.~G.,  {Pringle} J.~E.,  {Tout} C.~A.,   {Lubow} S.~H.,  2011,
  \mn@doi [\mnras] {10.1111/j.1365-2966.2011.19231.x}, \href
  {https://ui.adsabs.harvard.edu/abs/2011MNRAS.416.2827M} {416, 2827}

\bibitem[\protect\citeauthoryear{{McClintock} \& {Remillard}}{{McClintock} \&
  {Remillard}}{2006}]{McClintock2006}
{McClintock} J.~E.,  {Remillard} R.~A.,  2006, in , Vol.~39, Compact stellar
  X-ray sources.
pp 157--213

\bibitem[\protect\citeauthoryear{{Mertens}, {Semelin}  \& {Koopmans}}{{Mertens}
  et~al.}{2021}]{Mertens2021}
{Mertens} F.~G.,  {Semelin} B.,   {Koopmans} L.~V.~E.,  2021, in {Siebert} A.,
  et~al., eds, SF2A-2021: Proceedings of the Annual meeting of the French
  Society of Astronomy and Astrophysics. pp 211--214 (\mn@eprint {arXiv}
  {2109.10055}), \mn@doi{10.48550/arXiv.2109.10055}

\bibitem[\protect\citeauthoryear{{Mineo}, {Gilfanov}  \& {Sunyaev}}{{Mineo}
  et~al.}{2012}]{Mineo2012}
{Mineo} S.,  {Gilfanov} M.,   {Sunyaev} R.,  2012, \mn@doi [\mnras]
  {10.1111/j.1365-2966.2011.19862.x}, \href
  {https://ui.adsabs.harvard.edu/abs/2012MNRAS.419.2095M} {419, 2095}

\bibitem[\protect\citeauthoryear{{Mirocha} \& {Furlanetto}}{{Mirocha} \&
  {Furlanetto}}{2019}]{Mirocha2019}
{Mirocha} J.,  {Furlanetto} S.~R.,  2019, \mn@doi [\mnras]
  {10.1093/mnras/sty3260}, \href
  {https://ui.adsabs.harvard.edu/\#abs/2019MNRAS.483.1980M} {483, 1980}

\bibitem[\protect\citeauthoryear{{Mirouh}, {Hendriks}, {Dykes}, {Moe}  \&
  {Izzard}}{{Mirouh} et~al.}{2023}]{Mirouh2023}
{Mirouh} G.~M.,  {Hendriks} D.~D.,  {Dykes} S.,  {Moe} M.,   {Izzard} R.~G.,
  2023, \mn@doi [\mnras] {10.1093/mnras/stad2048}, \href
  {https://ui.adsabs.harvard.edu/abs/2023MNRAS.524.3978M} {524, 3978}

\bibitem[\protect\citeauthoryear{{Misra}, {Fragos}, {Tauris}, {Zapartas}  \&
  {Aguilera-Dena}}{{Misra} et~al.}{2020}]{Misra2020}
{Misra} D.,  {Fragos} T.,  {Tauris} T.~M.,  {Zapartas} E.,   {Aguilera-Dena}
  D.~R.,  2020, \mn@doi [\aap] {10.1051/0004-6361/202038070}, \href
  {https://ui.adsabs.harvard.edu/abs/2020A&A...642A.174M} {642, A174}

\bibitem[\protect\citeauthoryear{{Misra} et~al.,}{{Misra}
  et~al.}{2023a}]{Misra2023}
{Misra} D.,  et~al., 2023a, \mn@doi [arXiv e-prints]
  {10.48550/arXiv.2309.15904}, \href
  {https://ui.adsabs.harvard.edu/abs/2023arXiv230915904M} {p. arXiv:2309.15904}

\bibitem[\protect\citeauthoryear{{Misra} et~al.,}{{Misra}
  et~al.}{2023b}]{Misra2022}
{Misra} D.,  et~al., 2023b, \mn@doi [\aap] {10.1051/0004-6361/202244929}, \href
  {https://ui.adsabs.harvard.edu/abs/2023A&A...672A..99M} {672, A99}

\bibitem[\protect\citeauthoryear{{Moe} \& {Di Stefano}}{{Moe} \& {Di
  Stefano}}{2017}]{Moe2017}
{Moe} M.,  {Di Stefano} R.,  2017, \mn@doi [\apjs] {10.3847/1538-4365/aa6fb6},
  \href {https://ui.adsabs.harvard.edu/abs/2017ApJS..230...15M} {230, 15}

\bibitem[\protect\citeauthoryear{{Moe}, {Kratter}  \& {Badenes}}{{Moe}
  et~al.}{2019}]{Moe2019}
{Moe} M.,  {Kratter} K.~M.,   {Badenes} C.,  2019, \mn@doi [\apj]
  {10.3847/1538-4357/ab0d88}, \href
  {https://ui.adsabs.harvard.edu/abs/2019ApJ...875...61M} {875, 61}

\bibitem[\protect\citeauthoryear{{Mondal} \& {Barkana}}{{Mondal} \&
  {Barkana}}{2023}]{Mondal2023}
{Mondal} R.,  {Barkana} R.,  2023, \mn@doi [Nature Astronomy]
  {10.1038/s41550-023-02057-y}, \href
  {https://ui.adsabs.harvard.edu/abs/2023NatAs...7.1025M} {7, 1025}

\bibitem[\protect\citeauthoryear{{Mondal}, {Belczy{\'n}ski}, {Wiktorowicz},
  {Lasota}  \& {King}}{{Mondal} et~al.}{2020}]{Mondal2020}
{Mondal} S.,  {Belczy{\'n}ski} K.,  {Wiktorowicz} G.,  {Lasota} J.-P.,   {King}
  A.~R.,  2020, \mn@doi [\mnras] {10.1093/mnras/stz3227}, \href
  {https://ui.adsabs.harvard.edu/abs/2020MNRAS.491.2747M} {491, 2747}

\bibitem[\protect\citeauthoryear{{Mondal}, {Barkana}  \& {Fialkov}}{{Mondal}
  et~al.}{2023}]{Mondal2023dm}
{Mondal} R.,  {Barkana} R.,   {Fialkov} A.,  2023, \mn@doi [arXiv e-prints]
  {10.48550/arXiv.2310.15530}, \href
  {https://ui.adsabs.harvard.edu/abs/2023arXiv231015530M} {p. arXiv:2310.15530}

\bibitem[\protect\citeauthoryear{{Monsalve} et~al.,}{{Monsalve}
  et~al.}{2023}]{Monsalve2023}
{Monsalve} R.~A.,  et~al., 2023, \mn@doi [arXiv e-prints]
  {10.48550/arXiv.2309.02996}, \href
  {https://ui.adsabs.harvard.edu/abs/2023arXiv230902996M} {p. arXiv:2309.02996}

\bibitem[\protect\citeauthoryear{{Montero-Camacho}, {Zhang}  \&
  {Mao}}{{Montero-Camacho} et~al.}{2023}]{Montero-Camacho2023}
{Montero-Camacho} P.,  {Zhang} Y.,   {Mao} Y.,  2023, \mn@doi [arXiv e-prints]
  {10.48550/arXiv.2307.10598}, \href
  {https://ui.adsabs.harvard.edu/abs/2023arXiv230710598M} {p. arXiv:2307.10598}

\bibitem[\protect\citeauthoryear{{Moretti}, {Vattakunnel}, {Tozzi},
  {Salvaterra}, {Severgnini}, {Fugazza}, {Haardt}  \& {Gilli}}{{Moretti}
  et~al.}{2012}]{Moretti2012}
{Moretti} A.,  {Vattakunnel} S.,  {Tozzi} P.,  {Salvaterra} R.,  {Severgnini}
  P.,  {Fugazza} D.,  {Haardt} F.,   {Gilli} R.,  2012, \mn@doi [\aap]
  {10.1051/0004-6361/201219921}, \href
  {https://ui.adsabs.harvard.edu/abs/2012A&A...548A..87M} {548, A87}

\bibitem[\protect\citeauthoryear{{Mu{\~n}oz}, {Qin}, {Mesinger}, {Murray},
  {Greig}  \& {Mason}}{{Mu{\~n}oz} et~al.}{2022}]{Munoz2022}
{Mu{\~n}oz} J.~B.,  {Qin} Y.,  {Mesinger} A.,  {Murray} S.~G.,  {Greig} B.,
  {Mason} C.,  2022, \mn@doi [\mnras] {10.1093/mnras/stac185}, \href
  {https://ui.adsabs.harvard.edu/abs/2022MNRAS.511.3657M} {511, 3657}

\bibitem[\protect\citeauthoryear{{Munar-Adrover}, {Paredes}, {Rib{\'o}},
  {Iwasawa}, {Zabalza}  \& {Casares}}{{Munar-Adrover}
  et~al.}{2014}]{Munar-Adrover2014}
{Munar-Adrover} P.,  {Paredes} J.~M.,  {Rib{\'o}} M.,  {Iwasawa} K.,  {Zabalza}
  V.,   {Casares} J.,  2014, \mn@doi [\apjl] {10.1088/2041-8205/786/2/L11},
  \href {https://ui.adsabs.harvard.edu/abs/2014ApJ...786L..11M} {786, L11}

\bibitem[\protect\citeauthoryear{{Murphy} et~al.,}{{Murphy}
  et~al.}{2021}]{Murphy2021}
{Murphy} L.~J.,  et~al., 2021, \mn@doi [\mnras] {10.1093/mnras/staa3803}, \href
  {https://ui.adsabs.harvard.edu/abs/2021MNRAS.501.2745M} {501, 2745}

\bibitem[\protect\citeauthoryear{{Mushtukov} \& {Portegies Zwart}}{{Mushtukov}
  \& {Portegies Zwart}}{2023}]{Mushtukov2023}
{Mushtukov} A.~A.,  {Portegies Zwart} S.,  2023, \mn@doi [\mnras]
  {10.1093/mnras/stac3431}, \href
  {https://ui.adsabs.harvard.edu/abs/2023MNRAS.518.5457M} {518, 5457}

\bibitem[\protect\citeauthoryear{{Mushtukov} \& {Tsygankov}}{{Mushtukov} \&
  {Tsygankov}}{2022}]{Mushtukov2022}
{Mushtukov} A.,  {Tsygankov} S.,  2022, \mn@doi [arXiv e-prints]
  {10.48550/arXiv.2204.14185}, \href
  {https://ui.adsabs.harvard.edu/abs/2022arXiv220414185M} {p. arXiv:2204.14185}

\bibitem[\protect\citeauthoryear{{Nixon} \& {Pringle}}{{Nixon} \&
  {Pringle}}{2020}]{Nixon2020}
{Nixon} C.~J.,  {Pringle} J.~E.,  2020, \mn@doi [\apjl]
  {10.3847/2041-8213/abd17e}, \href
  {https://ui.adsabs.harvard.edu/abs/2020ApJ...905L..29N} {905, L29}

\bibitem[\protect\citeauthoryear{{Ogilvie}}{{Ogilvie}}{1999}]{Ogilvie1999}
{Ogilvie} G.~I.,  1999, \mn@doi [\mnras] {10.1046/j.1365-8711.1999.02340.x},
  \href {https://ui.adsabs.harvard.edu/abs/1999MNRAS.304..557O} {304, 557}

\bibitem[\protect\citeauthoryear{{Okazaki}}{{Okazaki}}{2001}]{Okazaki2001}
{Okazaki} A.~T.,  2001, \mn@doi [\pasj] {10.1093/pasj/53.1.119}, \href
  {https://ui.adsabs.harvard.edu/abs/2001PASJ...53..119O} {53, 119}

\bibitem[\protect\citeauthoryear{{Okazaki} \& {Negueruela}}{{Okazaki} \&
  {Negueruela}}{2001}]{Okazaki2001bexrb}
{Okazaki} A.~T.,  {Negueruela} I.,  2001, \mn@doi [\aap]
  {10.1051/0004-6361:20011083}, \href
  {https://ui.adsabs.harvard.edu/abs/2001A&A...377..161O} {377, 161}

\bibitem[\protect\citeauthoryear{{Okazaki}, {Bate}, {Ogilvie}  \&
  {Pringle}}{{Okazaki} et~al.}{2002}]{Okazaki2002}
{Okazaki} A.~T.,  {Bate} M.~R.,  {Ogilvie} G.~I.,   {Pringle} J.~E.,  2002,
  \mn@doi [\mnras] {10.1046/j.1365-8711.2002.05960.x}, \href
  {https://ui.adsabs.harvard.edu/abs/2002MNRAS.337..967O} {337, 967}

\bibitem[\protect\citeauthoryear{{Okazaki}, {Hayasaki}  \&
  {Moritani}}{{Okazaki} et~al.}{2013}]{Okazaki2013}
{Okazaki} A.~T.,  {Hayasaki} K.,   {Moritani} Y.,  2013, \mn@doi [\pasj]
  {10.1093/pasj/65.2.41}, \href
  {https://ui.adsabs.harvard.edu/abs/2013PASJ...65...41O} {65, 41}

\bibitem[\protect\citeauthoryear{{Pacucci}, {Mesinger}, {Mineo}  \&
  {Ferrara}}{{Pacucci} et~al.}{2014}]{Pacucci2014}
{Pacucci} F.,  {Mesinger} A.,  {Mineo} S.,   {Ferrara} A.,  2014, \mn@doi
  [\mnras] {10.1093/mnras/stu1240}, \href
  {https://ui.adsabs.harvard.edu/abs/2014MNRAS.443..678P} {443, 678}

\bibitem[\protect\citeauthoryear{{Paczy{\'n}ski} \&
  {Sienkiewicz}}{{Paczy{\'n}ski} \& {Sienkiewicz}}{1972}]{Paczynski1972}
{Paczy{\'n}ski} B.,  {Sienkiewicz} R.,  1972, \actaa, \href
  {https://ui.adsabs.harvard.edu/abs/1972AcA....22...73P} {22, 73}

\bibitem[\protect\citeauthoryear{{Pallottini}, {Ferrara}, {Gallerani},
  {Salvadori}  \& {D'Odorico}}{{Pallottini} et~al.}{2014}]{Pallottini2014}
{Pallottini} A.,  {Ferrara} A.,  {Gallerani} S.,  {Salvadori} S.,   {D'Odorico}
  V.,  2014, \mn@doi [\mnras] {10.1093/mnras/stu451}, \href
  {https://ui.adsabs.harvard.edu/abs/2014MNRAS.440.2498P} {440, 2498}

\bibitem[\protect\citeauthoryear{{Panoglou}, {Carciofi}, {Vieira}, {Cyr},
  {Jones}, {Okazaki}  \& {Rivinius}}{{Panoglou} et~al.}{2016}]{Panoglou2016}
{Panoglou} D.,  {Carciofi} A.~C.,  {Vieira} R.~G.,  {Cyr} I.~H.,  {Jones}
  C.~E.,  {Okazaki} A.~T.,   {Rivinius} T.,  2016, \mn@doi [\mnras]
  {10.1093/mnras/stw1508}, \href
  {https://ui.adsabs.harvard.edu/abs/2016MNRAS.461.2616P} {461, 2616}

\bibitem[\protect\citeauthoryear{{Panoglou}, {Faes}, {Carciofi}, {Okazaki},
  {Baade}, {Rivinius}  \& {Borges Fernandes}}{{Panoglou}
  et~al.}{2018}]{Panoglou2018}
{Panoglou} D.,  {Faes} D.~M.,  {Carciofi} A.~C.,  {Okazaki} A.~T.,  {Baade} D.,
   {Rivinius} T.,   {Borges Fernandes} M.,  2018, \mn@doi [\mnras]
  {10.1093/mnras/stx2497}, \href
  {https://ui.adsabs.harvard.edu/abs/2018MNRAS.473.3039P} {473, 3039}

\bibitem[\protect\citeauthoryear{{Park}, {Ricotti}  \& {Sugimura}}{{Park}
  et~al.}{2021a}]{Park2021I}
{Park} J.,  {Ricotti} M.,   {Sugimura} K.,  2021a, \mn@doi [\mnras]
  {10.1093/mnras/stab2999}, \href
  {https://ui.adsabs.harvard.edu/abs/2021MNRAS.508.6176P} {508, 6176}

\bibitem[\protect\citeauthoryear{{Park}, {Ricotti}  \& {Sugimura}}{{Park}
  et~al.}{2021b}]{Park2021II}
{Park} J.,  {Ricotti} M.,   {Sugimura} K.,  2021b, \mn@doi [\mnras]
  {10.1093/mnras/stab3000}, \href
  {https://ui.adsabs.harvard.edu/abs/2021MNRAS.508.6193P} {508, 6193}

\bibitem[\protect\citeauthoryear{{Park}, {Ricotti}  \& {Sugimura}}{{Park}
  et~al.}{2023a}]{Park2023}
{Park} J.,  {Ricotti} M.,   {Sugimura} K.,  2023a, \mn@doi [arXiv e-prints]
  {10.48550/arXiv.2307.14562}, \href
  {https://ui.adsabs.harvard.edu/abs/2023arXiv230714562P} {p. arXiv:2307.14562}

\bibitem[\protect\citeauthoryear{{Park}, {Ricotti}  \& {Sugimura}}{{Park}
  et~al.}{2023b}]{Park2022}
{Park} J.,  {Ricotti} M.,   {Sugimura} K.,  2023b, \mn@doi [\mnras]
  {10.1093/mnras/stad895}, \href
  {https://ui.adsabs.harvard.edu/abs/2023MNRAS.521.5334P} {521, 5334}

\bibitem[\protect\citeauthoryear{{Paxton} et~al.,}{{Paxton}
  et~al.}{2018}]{Paxton2018}
{Paxton} B.,  et~al., 2018, \mn@doi [\apjs] {10.3847/1538-4365/aaa5a8}, \href
  {https://ui.adsabs.harvard.edu/abs/2018ApJS..234...34P} {234, 34}

\bibitem[\protect\citeauthoryear{{Paxton} et~al.,}{{Paxton}
  et~al.}{2019}]{Paxton2019}
{Paxton} B.,  et~al., 2019, \mn@doi [\apjs] {10.3847/1538-4365/ab2241}, \href
  {https://ui.adsabs.harvard.edu/abs/2019ApJS..243...10P} {243, 10}

\bibitem[\protect\citeauthoryear{Pradhan, Paul, Bozzo, Maitra  \& Paul}{Pradhan
  et~al.}{2021}]{pradhan_comprehensive_2021}
Pradhan P.,  Paul B.,  Bozzo E.,  Maitra C.,   Paul B.~C.,  2021, \mn@doi
  [\mnras] {10.1093/mnras/stab024}, 502, 1163

\bibitem[\protect\citeauthoryear{{Prestwich}, {Tsantaki}, {Zezas}, {Jackson},
  {Roberts}, {Foltz}, {Linden}  \& {Kalogera}}{{Prestwich}
  et~al.}{2013}]{Prestwich2013}
{Prestwich} A.~H.,  {Tsantaki} M.,  {Zezas} A.,  {Jackson} F.,  {Roberts}
  T.~P.,  {Foltz} R.,  {Linden} T.,   {Kalogera} V.,  2013, \mn@doi [\apj]
  {10.1088/0004-637X/769/2/92}, \href
  {https://ui.adsabs.harvard.edu/abs/2013ApJ...769...92P} {769, 92}

\bibitem[\protect\citeauthoryear{Pringle}{Pringle}{1981}]{Pringle1987}
Pringle J.~E.,  1981, \mn@doi [\araa] {10.1146/annurev.aa.19.090181.001033},
  19, 137

\bibitem[\protect\citeauthoryear{{Pritchard} \& {Furlanetto}}{{Pritchard} \&
  {Furlanetto}}{2007}]{Pritchard2007}
{Pritchard} J.~R.,  {Furlanetto} S.~R.,  2007, \mn@doi [\mnras]
  {10.1111/j.1365-2966.2007.11519.x}, \href
  {https://ui.adsabs.harvard.edu/abs/2007MNRAS.376.1680P} {376, 1680}

\bibitem[\protect\citeauthoryear{Qiao \& Liu}{Qiao \&
  Liu}{2020}]{qiao_systematic_2020}
Qiao E.,  Liu B.~F.,  2020, \mn@doi [\mnras] {10.1093/mnras/stz3510}, 492, 615

\bibitem[\protect\citeauthoryear{{Qin}, {Mesinger}, {Park}, {Greig}  \&
  {Mu{\~n}oz}}{{Qin} et~al.}{2020}]{Qin2020}
{Qin} Y.,  {Mesinger} A.,  {Park} J.,  {Greig} B.,   {Mu{\~n}oz} J.~B.,  2020,
  \mn@doi [\mnras] {10.1093/mnras/staa1131}, \href
  {https://ui.adsabs.harvard.edu/abs/2020MNRAS.495..123Q} {495, 123}

\bibitem[\protect\citeauthoryear{{Raguzova} \& {Popov}}{{Raguzova} \&
  {Popov}}{2005}]{Raguzova2005}
{Raguzova} N.~V.,  {Popov} S.~B.,  2005, \mn@doi [Astronomical and
  Astrophysical Transactions] {10.1080/10556790500497311}, \href
  {https://ui.adsabs.harvard.edu/abs/2005A&AT...24..151R} {24, 151}

\bibitem[\protect\citeauthoryear{{Ramachandran}, {Klencki}, {Sander}, {Pauli},
  {Shenar}, {Oskinova}  \& {Hamann}}{{Ramachandran}
  et~al.}{2023}]{Ramachandran2023}
{Ramachandran} V.,  {Klencki} J.,  {Sander} A.~A.~C.,  {Pauli} D.,  {Shenar}
  T.,  {Oskinova} L.~M.,   {Hamann} W.~R.,  2023, \mn@doi [\aap]
  {10.1051/0004-6361/202346818}, \href
  {https://ui.adsabs.harvard.edu/abs/2023A&A...674L..12R} {674, L12}

\bibitem[\protect\citeauthoryear{{Reig}}{{Reig}}{2011}]{Reig2011}
{Reig} P.,  2011, \mn@doi [\apss] {10.1007/s10509-010-0575-8}, \href
  {https://ui.adsabs.harvard.edu/abs/2011Ap&SS.332....1R} {332, 1}

\bibitem[\protect\citeauthoryear{{Reis}, {Fialkov}  \& {Barkana}}{{Reis}
  et~al.}{2021}]{Reis2021}
{Reis} I.,  {Fialkov} A.,   {Barkana} R.,  2021, \mn@doi [\mnras]
  {10.1093/mnras/stab2089}, \href
  {https://ui.adsabs.harvard.edu/abs/2021MNRAS.506.5479R} {506, 5479}

\bibitem[\protect\citeauthoryear{{Ressler}}{{Ressler}}{2021}]{Ressler2021}
{Ressler} S.~M.,  2021, \mn@doi [\mnras] {10.1093/mnras/stab2880}, \href
  {https://ui.adsabs.harvard.edu/abs/2021MNRAS.508.4887R} {508, 4887}

\bibitem[\protect\citeauthoryear{{Riccio} et~al.,}{{Riccio}
  et~al.}{2023}]{Riccio2023}
{Riccio} G.,  et~al., 2023, \mn@doi [\aap] {10.1051/0004-6361/202346857}, \href
  {https://ui.adsabs.harvard.edu/abs/2023A&A...678A.164R} {678, A164}

\bibitem[\protect\citeauthoryear{{Richardson} et~al.,}{{Richardson}
  et~al.}{2023}]{Richardson2023}
{Richardson} N.~D.,  et~al., 2023, \mn@doi [\nat] {10.1038/s41586-022-05618-9},
  \href {https://ui.adsabs.harvard.edu/abs/2023Natur.614...45R} {614, 45}

\bibitem[\protect\citeauthoryear{{Ricotti}}{{Ricotti}}{2016}]{Ricotti2016}
{Ricotti} M.,  2016, \mn@doi [\mnras] {10.1093/mnras/stw1672}, \href
  {https://ui.adsabs.harvard.edu/abs/2016MNRAS.462..601R} {462, 601}

\bibitem[\protect\citeauthoryear{{Riley} et~al.,}{{Riley}
  et~al.}{2022}]{Riley2022}
{Riley} J.,  et~al., 2022, \mn@doi [\apjs] {10.3847/1538-4365/ac416c}, \href
  {https://ui.adsabs.harvard.edu/abs/2022ApJS..258...34R} {258, 34}

\bibitem[\protect\citeauthoryear{{R{\'\i}mulo} et~al.,}{{R{\'\i}mulo}
  et~al.}{2018}]{Rimulo2018}
{R{\'\i}mulo} L.~R.,  et~al., 2018, \mn@doi [\mnras] {10.1093/mnras/sty431},
  \href {https://ui.adsabs.harvard.edu/abs/2018MNRAS.476.3555R} {476, 3555}

\bibitem[\protect\citeauthoryear{{Rivinius}}{{Rivinius}}{2019}]{Rivinius2019}
{Rivinius} T.,  2019, \mn@doi [IAU Symposium] {10.1017/S1743921318008207},
  \href {https://ui.adsabs.harvard.edu/abs/2019IAUS..346..105R} {346, 105}

\bibitem[\protect\citeauthoryear{{Rivinius}, {Carciofi}  \&
  {Martayan}}{{Rivinius} et~al.}{2013}]{Rivinius2013}
{Rivinius} T.,  {Carciofi} A.~C.,   {Martayan} C.,  2013, \mn@doi [\aapr]
  {10.1007/s00159-013-0069-0}, \href
  {https://ui.adsabs.harvard.edu/abs/2013A&ARv..21...69R} {21, 69}

\bibitem[\protect\citeauthoryear{{Rogers}, {Lin}, {McElwaine}  \&
  {Lau}}{{Rogers} et~al.}{2013}]{Rogers2013}
{Rogers} T.~M.,  {Lin} D.~N.~C.,  {McElwaine} J.~N.,   {Lau} H.~H.~B.,  2013,
  \mn@doi [\apj] {10.1088/0004-637X/772/1/21}, \href
  {https://ui.adsabs.harvard.edu/abs/2013ApJ...772...21R} {772, 21}

\bibitem[\protect\citeauthoryear{{Rossland} et~al.,}{{Rossland}
  et~al.}{2023}]{Rossland2023}
{Rossland} S.,  et~al., 2023, \mn@doi [\aj] {10.3847/1538-3881/acd0ae}, \href
  {https://ui.adsabs.harvard.edu/abs/2023AJ....166...20R} {166, 20}

\bibitem[\protect\citeauthoryear{{Rubele} et~al.,}{{Rubele}
  et~al.}{2015}]{Rubele2015}
{Rubele} S.,  et~al., 2015, \mn@doi [\mnras] {10.1093/mnras/stv141}, \href
  {https://ui.adsabs.harvard.edu/abs/2015MNRAS.449..639R} {449, 639}

\bibitem[\protect\citeauthoryear{{Rusakov}, {Steinhardt}  \&
  {Sneppen}}{{Rusakov} et~al.}{2023}]{Rusakov2023}
{Rusakov} V.,  {Steinhardt} C.~L.,   {Sneppen} A.,  2023, \mn@doi [\apjs]
  {10.3847/1538-4365/acdde3}, \href
  {https://ui.adsabs.harvard.edu/abs/2023ApJS..268...10R} {268, 10}

\bibitem[\protect\citeauthoryear{{Safranek-Shrader}, {Agarwal}, {Federrath},
  {Dubey}, {Milosavljevi{\'c}}  \& {Bromm}}{{Safranek-Shrader}
  et~al.}{2012}]{Safranek-Shrader2012}
{Safranek-Shrader} C.,  {Agarwal} M.,  {Federrath} C.,  {Dubey} A.,
  {Milosavljevi{\'c}} M.,   {Bromm} V.,  2012, \mn@doi [\mnras]
  {10.1111/j.1365-2966.2012.21852.x}, \href
  {https://ui.adsabs.harvard.edu/abs/2012MNRAS.426.1159S} {426, 1159}

\bibitem[\protect\citeauthoryear{{Salvaggio}, {Wolter}, {Belfiore}  \&
  {Colpi}}{{Salvaggio} et~al.}{2023}]{Salvaggio2023}
{Salvaggio} C.,  {Wolter} A.,  {Belfiore} A.,   {Colpi} M.,  2023, \mn@doi
  [\mnras] {10.1093/mnras/stad943}, \href
  {https://ui.adsabs.harvard.edu/abs/2023MNRAS.522.1377S} {522, 1377}

\bibitem[\protect\citeauthoryear{{Sana} et~al.,}{{Sana}
  et~al.}{2012}]{Sana2012}
{Sana} H.,  et~al., 2012, \mn@doi [Science] {10.1126/science.1223344}, \href
  {https://ui.adsabs.harvard.edu/abs/2012Sci...337..444S} {337, 444}

\bibitem[\protect\citeauthoryear{{Sander} \& {Vink}}{{Sander} \&
  {Vink}}{2020}]{Sander2020}
{Sander} A. A.~C.,  {Vink} J.~S.,  2020, \mn@doi [\mnras]
  {10.1093/mnras/staa2712}, \href
  {https://ui.adsabs.harvard.edu/abs/2020MNRAS.499..873S} {499, 873}

\bibitem[\protect\citeauthoryear{{Sartorio} et~al.,}{{Sartorio}
  et~al.}{2023}]{Sartorio2023}
{Sartorio} N.~S.,  et~al., 2023, \mn@doi [\mnras] {10.1093/mnras/stad697},
  \href {https://ui.adsabs.harvard.edu/abs/2023MNRAS.tmp..702S} {}

\bibitem[\protect\citeauthoryear{{Schauer}, {Liu}  \& {Bromm}}{{Schauer}
  et~al.}{2019}]{Schauer2019}
{Schauer} A. T.~P.,  {Liu} B.,   {Bromm} V.,  2019, \mn@doi [\apjl]
  {10.3847/2041-8213/ab1e51}, \href
  {https://ui.adsabs.harvard.edu/abs/2019ApJ...877L...5S} {877, L5}

\bibitem[\protect\citeauthoryear{{Schauer}, {Glover}, {Klessen}  \&
  {Clark}}{{Schauer} et~al.}{2021}]{Schauer2021}
{Schauer} A. T.~P.,  {Glover} S. C.~O.,  {Klessen} R.~S.,   {Clark} P.,  2021,
  \mn@doi [\mnras] {10.1093/mnras/stab1953}, \href
  {https://ui.adsabs.harvard.edu/abs/2021MNRAS.507.1775S} {507, 1775}

\bibitem[\protect\citeauthoryear{{Schmidtke}, {Cowley}  \&
  {Udalski}}{{Schmidtke} et~al.}{2013}]{Schmidtke2013}
{Schmidtke} P.~C.,  {Cowley} A.~P.,   {Udalski} A.,  2013, \mn@doi [\mnras]
  {10.1093/mnras/stt159}, \href
  {https://ui.adsabs.harvard.edu/abs/2013MNRAS.431..252S} {431, 252}

\bibitem[\protect\citeauthoryear{{Schneider} et~al.,}{{Schneider}
  et~al.}{2018}]{Schneider2018}
{Schneider} F.~R.~N.,  et~al., 2018, \mn@doi [Science]
  {10.1126/science.aan0106}, \href
  {https://ui.adsabs.harvard.edu/abs/2018Sci...359...69S} {359, 69}

\bibitem[\protect\citeauthoryear{{Schootemeijer}, {Lennon}, {Garcia}, {Langer},
  {Hastings}  \& {Sch{\"u}rmann}}{{Schootemeijer}
  et~al.}{2022}]{Schootemeijer2022}
{Schootemeijer} A.,  {Lennon} D.~J.,  {Garcia} M.,  {Langer} N.,  {Hastings}
  B.,   {Sch{\"u}rmann} C.,  2022, \mn@doi [\aap]
  {10.1051/0004-6361/202244730}, \href
  {https://ui.adsabs.harvard.edu/abs/2022A&A...667A.100S} {667, A100}

\bibitem[\protect\citeauthoryear{{Sguera}, {Sidoli}, {Bird}  \& {La
  Palombara}}{{Sguera} et~al.}{2023}]{Sguera2023}
{Sguera} V.,  {Sidoli} L.,  {Bird} A.~J.,   {La Palombara} N.,  2023, \mn@doi
  [\mnras] {10.1093/mnras/stad1494}, \href
  {https://ui.adsabs.harvard.edu/abs/2023MNRAS.523.1192S} {523, 1192}

\bibitem[\protect\citeauthoryear{{Shakura} \& {Sunyaev}}{{Shakura} \&
  {Sunyaev}}{1973}]{Shakura1973}
{Shakura} N.~I.,  {Sunyaev} R.~A.,  1973, \aap, \href
  {https://ui.adsabs.harvard.edu/abs/1973A&A....24..337S} {24, 337}

\bibitem[\protect\citeauthoryear{{Shao} \& {Li}}{{Shao} \&
  {Li}}{2014}]{Shao2014}
{Shao} Y.,  {Li} X.-D.,  2014, \mn@doi [\apj] {10.1088/0004-637X/796/1/37},
  \href {https://ui.adsabs.harvard.edu/abs/2014ApJ...796...37S} {796, 37}

\bibitem[\protect\citeauthoryear{{Shao} \& {Li}}{{Shao} \&
  {Li}}{2020}]{Shao2020}
{Shao} Y.,  {Li} X.-D.,  2020, \mn@doi [\apj] {10.3847/1538-4357/aba118}, \href
  {https://ui.adsabs.harvard.edu/abs/2020ApJ...898..143S} {898, 143}

\bibitem[\protect\citeauthoryear{{Shao}, {Li}  \& {Dai}}{{Shao}
  et~al.}{2019}]{Shao2019}
{Shao} Y.,  {Li} X.-D.,   {Dai} Z.-G.,  2019, \mn@doi [\apj]
  {10.3847/1538-4357/ab4d50}, \href
  {https://ui.adsabs.harvard.edu/abs/2019ApJ...886..118S} {886, 118}

\bibitem[\protect\citeauthoryear{{Shao}, {Xu}, {Wang}, {Yang}, {Li}, {Zhang}
  \& {Chen}}{{Shao} et~al.}{2023}]{Shao2023}
{Shao} Y.,  {Xu} Y.,  {Wang} Y.,  {Yang} W.,  {Li} R.,  {Zhang} X.,   {Chen}
  X.,  2023, \mn@doi [Nature Astronomy] {10.1038/s41550-023-02024-7}, \href
  {https://ui.adsabs.harvard.edu/abs/2023NatAs.tmp..144S} {}

\bibitem[\protect\citeauthoryear{{Sidoli} \& {Paizis}}{{Sidoli} \&
  {Paizis}}{2018}]{Sidoli2018}
{Sidoli} L.,  {Paizis} A.,  2018, \mn@doi [\mnras] {10.1093/mnras/sty2428},
  \href {https://ui.adsabs.harvard.edu/abs/2018MNRAS.481.2779S} {481, 2779}

\bibitem[\protect\citeauthoryear{{Singh} et~al.,}{{Singh}
  et~al.}{2022}]{Singh2022}
{Singh} S.,  et~al., 2022, \mn@doi [Nature Astronomy]
  {10.1038/s41550-022-01610-5}, \href
  {https://ui.adsabs.harvard.edu/abs/2022NatAs...6..607S} {6, 607}

\bibitem[\protect\citeauthoryear{{Soberman}, {Phinney}  \& {van den
  Heuvel}}{{Soberman} et~al.}{1997}]{Soberman1997}
{Soberman} G.~E.,  {Phinney} E.~S.,   {van den Heuvel} E.~P.~J.,  1997, \mn@doi
  [\aap] {10.48550/arXiv.astro-ph/9703016}, \href
  {https://ui.adsabs.harvard.edu/abs/1997A&A...327..620S} {327, 620}

\bibitem[\protect\citeauthoryear{Sokolova-Lapa et~al.,}{Sokolova-Lapa
  et~al.}{2021}]{sokolova-lapa_x-ray_2021}
Sokolova-Lapa E.,  et~al., 2021, \mn@doi [\aap] {10.1051/0004-6361/202040228},
  651, A12

\bibitem[\protect\citeauthoryear{{Stacy} \& {Bromm}}{{Stacy} \&
  {Bromm}}{2007}]{Stacy2007}
{Stacy} A.,  {Bromm} V.,  2007, \mn@doi [\mnras]
  {10.1111/j.1365-2966.2007.12247.x}, \href
  {https://ui.adsabs.harvard.edu/abs/2007MNRAS.382..229S} {382, 229}

\bibitem[\protect\citeauthoryear{{Stacy}, {Bromm}  \& {Loeb}}{{Stacy}
  et~al.}{2011}]{Stacy2011}
{Stacy} A.,  {Bromm} V.,   {Loeb} A.,  2011, \mn@doi [\mnras]
  {10.1111/j.1365-2966.2010.18152.x}, \href
  {https://ui.adsabs.harvard.edu/abs/2011MNRAS.413..543S} {413, 543}

\bibitem[\protect\citeauthoryear{{Stacy}, {Greif}, {Klessen}, {Bromm}  \&
  {Loeb}}{{Stacy} et~al.}{2013}]{Stacy2013}
{Stacy} A.,  {Greif} T.~H.,  {Klessen} R.~S.,  {Bromm} V.,   {Loeb} A.,  2013,
  \mn@doi [\mnras] {10.1093/mnras/stt264}, \href
  {https://ui.adsabs.harvard.edu/abs/2013MNRAS.431.1470S} {431, 1470}

\bibitem[\protect\citeauthoryear{{Steiner}, {Narayan}, {McClintock}  \&
  {Ebisawa}}{{Steiner} et~al.}{2009}]{Steiner2009}
{Steiner} J.~F.,  {Narayan} R.,  {McClintock} J.~E.,   {Ebisawa} K.,  2009,
  \mn@doi [\pasp] {10.1086/648535}, \href
  {https://ui.adsabs.harvard.edu/abs/2009PASP..121.1279S} {121, 1279}

\bibitem[\protect\citeauthoryear{{Suffak}, {Jones}  \& {Carciofi}}{{Suffak}
  et~al.}{2022}]{Suffak2022}
{Suffak} M.,  {Jones} C.~E.,   {Carciofi} A.~C.,  2022, \mn@doi [\mnras]
  {10.1093/mnras/stab3024}, \href
  {https://ui.adsabs.harvard.edu/abs/2022MNRAS.509..931S} {509, 931}

\bibitem[\protect\citeauthoryear{{Sugimura}, {Matsumoto}, {Hosokawa}, {Hirano}
  \& {Omukai}}{{Sugimura} et~al.}{2020}]{Sugimura2020}
{Sugimura} K.,  {Matsumoto} T.,  {Hosokawa} T.,  {Hirano} S.,   {Omukai} K.,
  2020, \mn@doi [\apjl] {10.3847/2041-8213/ab7d37}, \href
  {https://ui.adsabs.harvard.edu/abs/2020ApJ...892L..14S} {892, L14}

\bibitem[\protect\citeauthoryear{{Sugimura}, {Matsumoto}, {Hosokawa}, {Hirano}
  \& {Omukai}}{{Sugimura} et~al.}{2023}]{Sugimura2023}
{Sugimura} K.,  {Matsumoto} T.,  {Hosokawa} T.,  {Hirano} S.,   {Omukai} K.,
  2023, \mn@doi [arXiv e-prints] {10.48550/arXiv.2307.15108}, \href
  {https://ui.adsabs.harvard.edu/abs/2023arXiv230715108S} {p. arXiv:2307.15108}

\bibitem[\protect\citeauthoryear{{Takhistov}, {Lu}, {Gelmini}, {Hayashi},
  {Inoue}  \& {Kusenko}}{{Takhistov} et~al.}{2022}]{Takhistov2022}
{Takhistov} V.,  {Lu} P.,  {Gelmini} G.~B.,  {Hayashi} K.,  {Inoue} Y.,
  {Kusenko} A.,  2022, \mn@doi [\jcap] {10.1088/1475-7516/2022/03/017}, \href
  {https://ui.adsabs.harvard.edu/abs/2022JCAP...03..017T} {2022, 017}

\bibitem[\protect\citeauthoryear{{Tanvir} \& {Krumholz}}{{Tanvir} \&
  {Krumholz}}{2023}]{Tanvir2023}
{Tanvir} T.~S.,  {Krumholz} M.~R.,  2023, \mn@doi [arXiv e-prints]
  {10.48550/arXiv.2305.20039}, \href
  {https://ui.adsabs.harvard.edu/abs/2023arXiv230520039T} {p. arXiv:2305.20039}

\bibitem[\protect\citeauthoryear{{Tranin}, {Webb}  \& {Godet}}{{Tranin}
  et~al.}{2023}]{Tranin2023}
{Tranin} H.,  {Webb} N.,   {Godet} O.,  2023, \mn@doi [arXiv e-prints]
  {10.48550/arXiv.2304.11216}, \href
  {https://ui.adsabs.harvard.edu/abs/2023arXiv230411216T} {p. arXiv:2304.11216}

\bibitem[\protect\citeauthoryear{{Ucci} et~al.,}{{Ucci}
  et~al.}{2023}]{Ucci2023}
{Ucci} G.,  et~al., 2023, \mn@doi [\mnras] {10.1093/mnras/stac2654}, \href
  {https://ui.adsabs.harvard.edu/abs/2023MNRAS.518.3557U} {518, 3557}

\bibitem[\protect\citeauthoryear{{Ventura}, {Trinca}, {Schneider}, {Graziani},
  {Valiante}  \& {Wyithe}}{{Ventura} et~al.}{2023}]{Ventura2023}
{Ventura} E.~M.,  {Trinca} A.,  {Schneider} R.,  {Graziani} L.,  {Valiante} R.,
    {Wyithe} J. S.~B.,  2023, \mn@doi [\mnras] {10.1093/mnras/stad237}, \href
  {https://ui.adsabs.harvard.edu/abs/2023MNRAS.tmp..252V} {}

\bibitem[\protect\citeauthoryear{{Verbunt}, {Igoshev}  \& {Cator}}{{Verbunt}
  et~al.}{2017}]{Verbunt2017}
{Verbunt} F.,  {Igoshev} A.,   {Cator} E.,  2017, \mn@doi [\aap]
  {10.1051/0004-6361/201731518}, \href
  {https://ui.adsabs.harvard.edu/abs/2017A&A...608A..57V} {608, A57}

\bibitem[\protect\citeauthoryear{{Vieira}, {Carciofi}, {Bjorkman}, {Rivinius},
  {Baade}  \& {R{\'\i}mulo}}{{Vieira} et~al.}{2017}]{Vieira2017}
{Vieira} R.~G.,  {Carciofi} A.~C.,  {Bjorkman} J.~E.,  {Rivinius} T.,  {Baade}
  D.,   {R{\'\i}mulo} L.~R.,  2017, \mn@doi [\mnras] {10.1093/mnras/stw2542},
  \href {https://ui.adsabs.harvard.edu/abs/2017MNRAS.464.3071V} {464, 3071}

\bibitem[\protect\citeauthoryear{{Vigna-G{\'o}mez} et~al.,}{{Vigna-G{\'o}mez}
  et~al.}{2018}]{Vigna-Gomez2018}
{Vigna-G{\'o}mez} A.,  et~al., 2018, \mn@doi [\mnras] {10.1093/mnras/sty2463},
  \href {https://ui.adsabs.harvard.edu/abs/2018MNRAS.481.4009V} {481, 4009}

\bibitem[\protect\citeauthoryear{{Vinciguerra} et~al.,}{{Vinciguerra}
  et~al.}{2020}]{Vinciguerra2020}
{Vinciguerra} S.,  et~al., 2020, \mn@doi [\mnras] {10.1093/mnras/staa2177},
  \href {https://ui.adsabs.harvard.edu/abs/2020MNRAS.498.4705V} {498, 4705}

\bibitem[\protect\citeauthoryear{{Walter}, {Lutovinov}, {Bozzo}  \&
  {Tsygankov}}{{Walter} et~al.}{2015}]{Walter2015}
{Walter} R.,  {Lutovinov} A.~A.,  {Bozzo} E.,   {Tsygankov} S.~S.,  2015,
  \mn@doi [\aapr] {10.1007/s00159-015-0082-6}, \href
  {https://ui.adsabs.harvard.edu/abs/2015A&ARv..23....2W} {23, 2}

\bibitem[\protect\citeauthoryear{{Walton}, {Mackenzie}, {Gully}, {Patel},
  {Roberts}, {Earnshaw}  \& {Mateos}}{{Walton} et~al.}{2022}]{Walton2022}
{Walton} D.~J.,  {Mackenzie} A.~D.~A.,  {Gully} H.,  {Patel} N.~R.,  {Roberts}
  T.~P.,  {Earnshaw} H.~P.,   {Mateos} S.,  2022, \mn@doi [\mnras]
  {10.1093/mnras/stab3001}, \href
  {https://ui.adsabs.harvard.edu/abs/2022MNRAS.509.1587W} {509, 1587}

\bibitem[\protect\citeauthoryear{{Wang} et~al.,}{{Wang}
  et~al.}{2023}]{Wang2023}
{Wang} C.,  et~al., 2023, \mn@doi [\aap] {10.1051/0004-6361/202245413}, \href
  {https://ui.adsabs.harvard.edu/abs/2023A&A...670A..43W} {670, A43}

\bibitem[\protect\citeauthoryear{{Wiktorowicz}, {Sobolewska}, {Lasota}  \&
  {Belczynski}}{{Wiktorowicz} et~al.}{2017}]{Wiktorowicz2017}
{Wiktorowicz} G.,  {Sobolewska} M.,  {Lasota} J.-P.,   {Belczynski} K.,  2017,
  \mn@doi [\apj] {10.3847/1538-4357/aa821d}, \href
  {https://ui.adsabs.harvard.edu/abs/2017ApJ...846...17W} {846, 17}

\bibitem[\protect\citeauthoryear{{Wiktorowicz}, {Lasota}, {Middleton}  \&
  {Belczynski}}{{Wiktorowicz} et~al.}{2019}]{Wiktorowicz2019}
{Wiktorowicz} G.,  {Lasota} J.-P.,  {Middleton} M.,   {Belczynski} K.,  2019,
  \mn@doi [\apj] {10.3847/1538-4357/ab0f27}, \href
  {https://ui.adsabs.harvard.edu/abs/2019ApJ...875...53W} {875, 53}

\bibitem[\protect\citeauthoryear{{Wiktorowicz}, {Lasota}, {Belczynski}, {Lu},
  {Liu}  \& {I{\l}kiewicz}}{{Wiktorowicz} et~al.}{2021}]{Wiktorowicz2021}
{Wiktorowicz} G.,  {Lasota} J.-P.,  {Belczynski} K.,  {Lu} Y.,  {Liu} J.,
  {I{\l}kiewicz} K.,  2021, \mn@doi [\apj] {10.3847/1538-4357/ac0cf7}, \href
  {https://ui.adsabs.harvard.edu/abs/2021ApJ...918...60W} {918, 60}

\bibitem[\protect\citeauthoryear{{Willcox}, {MacLeod}, {Mandel}  \&
  {Hirai}}{{Willcox} et~al.}{2023}]{Willcox2023}
{Willcox} R.,  {MacLeod} M.,  {Mandel} I.,   {Hirai} R.,  2023, \mn@doi [arXiv
  e-prints] {10.48550/arXiv.2308.06666}, \href
  {https://ui.adsabs.harvard.edu/abs/2023arXiv230806666W} {p. arXiv:2308.06666}

\bibitem[\protect\citeauthoryear{{Wise}, {Turk}, {Norman}  \& {Abel}}{{Wise}
  et~al.}{2012}]{Wise2012}
{Wise} J.~H.,  {Turk} M.~J.,  {Norman} M.~L.,   {Abel} T.,  2012, \mn@doi
  [\apj] {10.1088/0004-637X/745/1/50}, \href
  {https://ui.adsabs.harvard.edu/abs/2012ApJ...745...50W} {745, 50}

\bibitem[\protect\citeauthoryear{{Wood}, {Bjorkman}  \& {Bjorkman}}{{Wood}
  et~al.}{1997}]{Wood1997}
{Wood} K.,  {Bjorkman} K.~S.,   {Bjorkman} J.~E.,  1997, \mn@doi [\apj]
  {10.1086/303747}, \href
  {https://ui.adsabs.harvard.edu/abs/1997ApJ...477..926W} {477, 926}

\bibitem[\protect\citeauthoryear{{Wu}, {Yu}, {Li}, {Maccarone}  \& {Li}}{{Wu}
  et~al.}{2010}]{Wu2010}
{Wu} Y.~X.,  {Yu} W.,  {Li} T.~P.,  {Maccarone} T.~J.,   {Li} X.~D.,  2010,
  \mn@doi [\apj] {10.1088/0004-637X/718/2/620}, \href
  {https://ui.adsabs.harvard.edu/abs/2010ApJ...718..620W} {718, 620}

\bibitem[\protect\citeauthoryear{{Xing} \& {Li}}{{Xing} \&
  {Li}}{2021}]{Xing2021}
{Xing} Z.-P.,  {Li} X.-D.,  2021, \mn@doi [\apj] {10.3847/1538-4357/ac16e1},
  \href {https://ui.adsabs.harvard.edu/abs/2021ApJ...920...67X} {920, 67}

\bibitem[\protect\citeauthoryear{{Xu} \& {Li}}{{Xu} \& {Li}}{2019}]{Xu2019}
{Xu} X.-T.,  {Li} X.-D.,  2019, \mn@doi [\apj] {10.3847/1538-4357/aafee0},
  \href {https://ui.adsabs.harvard.edu/abs/2019ApJ...872..102X} {872, 102}

\bibitem[\protect\citeauthoryear{{Xu}, {Wise}  \& {Norman}}{{Xu}
  et~al.}{2013}]{Xu2013}
{Xu} H.,  {Wise} J.~H.,   {Norman} M.~L.,  2013, \mn@doi [\apj]
  {10.1088/0004-637X/773/2/83}, \href
  {https://ui.adsabs.harvard.edu/abs/2013ApJ...773...83X} {773, 83}

\bibitem[\protect\citeauthoryear{{Yang}}{{Yang}}{2021}]{Yang2021}
{Yang} Y.,  2021, \mn@doi [\mnras] {10.1093/mnras/stab2966}, \href
  {https://ui.adsabs.harvard.edu/abs/2021MNRAS.508.5709Y} {508, 5709}

\bibitem[\protect\citeauthoryear{Yang, Laycock, Christodoulou, Fingerman, Coe
  \& Drake}{Yang et~al.}{2017}]{yang_comprehensive_2017}
Yang J.,  Laycock S. G.~T.,  Christodoulou D.~M.,  Fingerman S.,  Coe M.~J.,
  Drake J.~J.,  2017, \mn@doi [\apj] {10.3847/1538-4357/aa6898}, 839, 119

\bibitem[\protect\citeauthoryear{{Yang}, {Li}  \& {Li}}{{Yang}
  et~al.}{2023}]{Yang2023}
{Yang} Y.,  {Li} X.,   {Li} G.,  2023, \mn@doi [\prd]
  {10.1103/PhysRevD.107.103501}, \href
  {https://ui.adsabs.harvard.edu/abs/2023PhRvD.107j3501Y} {107, 103501}

\bibitem[\protect\citeauthoryear{{Yates}, {Hendriks}, {Vijayan}, {Izzard},
  {Thomas}  \& {Das}}{{Yates} et~al.}{2023}]{Yates2023}
{Yates} R.~M.,  {Hendriks} D.,  {Vijayan} A.~P.,  {Izzard} R.~G.,  {Thomas}
  P.~A.,   {Das} P.,  2023, \mn@doi [arXiv e-prints]
  {10.48550/arXiv.2310.15218}, \href
  {https://ui.adsabs.harvard.edu/abs/2023arXiv231015218Y} {p. arXiv:2310.15218}

\bibitem[\protect\citeauthoryear{{Zamanov} et~al.,}{{Zamanov}
  et~al.}{2022}]{Zamanov2022}
{Zamanov} R.~K.,  et~al., 2022, \mn@doi [Astronomische Nachrichten]
  {10.1002/asna.20224019}, \href
  {https://ui.adsabs.harvard.edu/abs/2022AN....34324019Z} {343, e24019}

\bibitem[\protect\citeauthoryear{{Zhang}, {Li}  \& {Wang}}{{Zhang}
  et~al.}{2004}]{Zhang2004}
{Zhang} F.,  {Li} X.~D.,   {Wang} Z.~R.,  2004, \mn@doi [\apj]
  {10.1086/381540}, \href
  {https://ui.adsabs.harvard.edu/abs/2004ApJ...603..663Z} {603, 663}

\bibitem[\protect\citeauthoryear{{Zhao} \& {Fuller}}{{Zhao} \&
  {Fuller}}{2020}]{Zhao2020}
{Zhao} X.,  {Fuller} J.,  2020, \mn@doi [\mnras] {10.1093/mnras/staa1097},
  \href {https://ui.adsabs.harvard.edu/abs/2020MNRAS.495..249Z} {495, 249}

\bibitem[\protect\citeauthoryear{{Zhao}, {Gandhi}, {Brown}, {Knigge},
  {Charles}, {Maccarone}  \& {Nuchvanichakul}}{{Zhao} et~al.}{2023}]{Zhao2023}
{Zhao} Y.,  {Gandhi} P.,  {Brown} C.~D.,  {Knigge} C.,  {Charles} P.~A.,
  {Maccarone} T.~J.,   {Nuchvanichakul} P.,  2023, \mn@doi [\mnras]
  {10.1093/mnras/stad2226}, \href
  {https://ui.adsabs.harvard.edu/abs/2023MNRAS.tmp.2163Z} {}

\bibitem[\protect\citeauthoryear{{Zuo}, {Li}  \& {Gu}}{{Zuo}
  et~al.}{2014}]{Zuo2014}
{Zuo} Z.-Y.,  {Li} X.-D.,   {Gu} Q.-S.,  2014, \mn@doi [\mnras]
  {10.1093/mnras/stt1918}, \href
  {https://ui.adsabs.harvard.edu/abs/2014MNRAS.437.1187Z} {437, 1187}

\bibitem[\protect\citeauthoryear{{de Lera Acedo} et~al.,}{{de Lera Acedo}
  et~al.}{2022}]{deLeraAcedo2022}
{de Lera Acedo} E.,  et~al., 2022, \mn@doi [Nature Astronomy]
  {10.1038/s41550-022-01709-9}, \href
  {https://ui.adsabs.harvard.edu/abs/2022NatAs...6..984D} {6, 984}

\bibitem[\protect\citeauthoryear{{van der Walt}, {Colbert}  \&
  {Varoquaux}}{{van der Walt} et~al.}{2011}]{vanderWalt2011}
{van der Walt} S.,  {Colbert} S.~C.,   {Varoquaux} G.,  2011, \mn@doi
  [Computing in Science and Engineering] {10.1109/MCSE.2011.37}, \href
  {https://ui.adsabs.harvard.edu/abs/2011CSE....13b..22V} {13, 22}

\makeatother
\end{thebibliography}

\appendix

\section{Be-XRBs in the SMC}\label{apx:smc}

\begin{figure}
    \centering
    \includegraphics[width=\columnwidth]{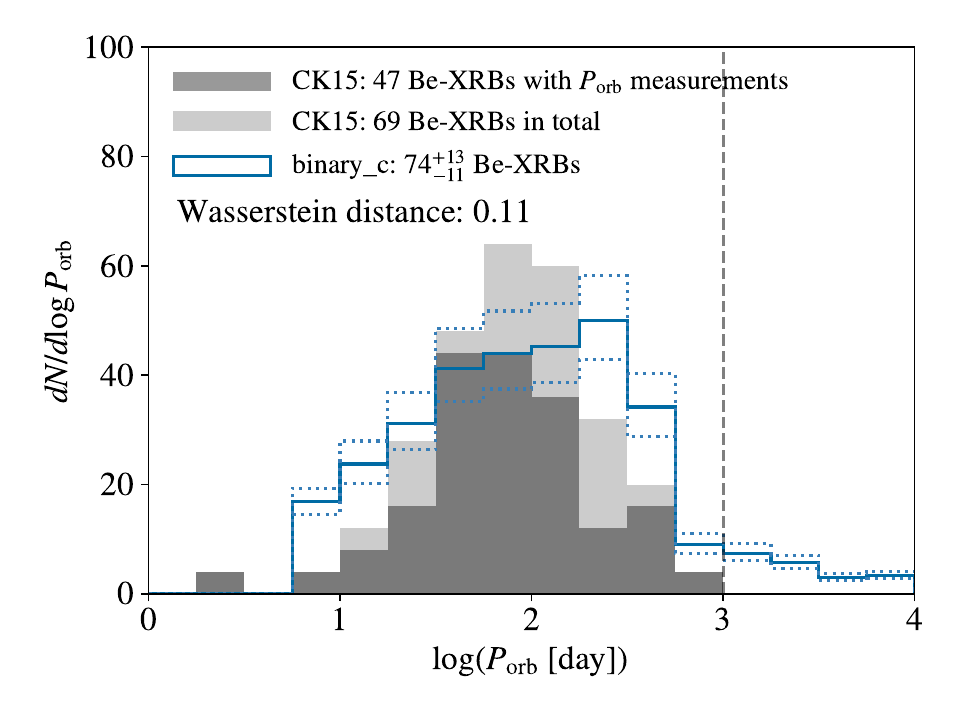}
    \vspace{-25pt}
    \caption{Orbital period distribution of Be-XRBs in the SMC predicted by the SR\_CS model (contours) with the default initial binary statistics, compared with the observed population (histograms) from \citet[][CK15]{Coe2015}. The solid contour shows the most likely case, enclosed by the dotted contours that reflect the uncertainties of the SMC star formation history \citep{Rubele2015}. The darker histograms show the distribution for the 47 Be-XRBs with measurements of $P_{\rm orb}$, while the lighter histograms also include the 22 Be-XRBs with $P_{\rm orb}$ inferred from the empirical scaling law $\log(P_{\rm orb}\ [{\rm day}])=0.4329\log(P_{\rm s}\ [{\rm s}])+ 1.043$ \citep{Vinciguerra2020} given the spin period $P_{\rm s}$. The SR\_CS model predicts $74_{-11}^{+13}$ Be-XRBs with $P_{\rm orb}<10^{3}$~days (dashed vertical line) in the SMC, consistent with the number 69 in observations. The predicted and observed distributions at $P_{\rm orb}<10^{3}$~days agree well with a Wasserstein distance of 0.11.}
    \label{fig:smc_fd}
\end{figure}

\begin{figure}
    \centering
    \includegraphics[width=\columnwidth]{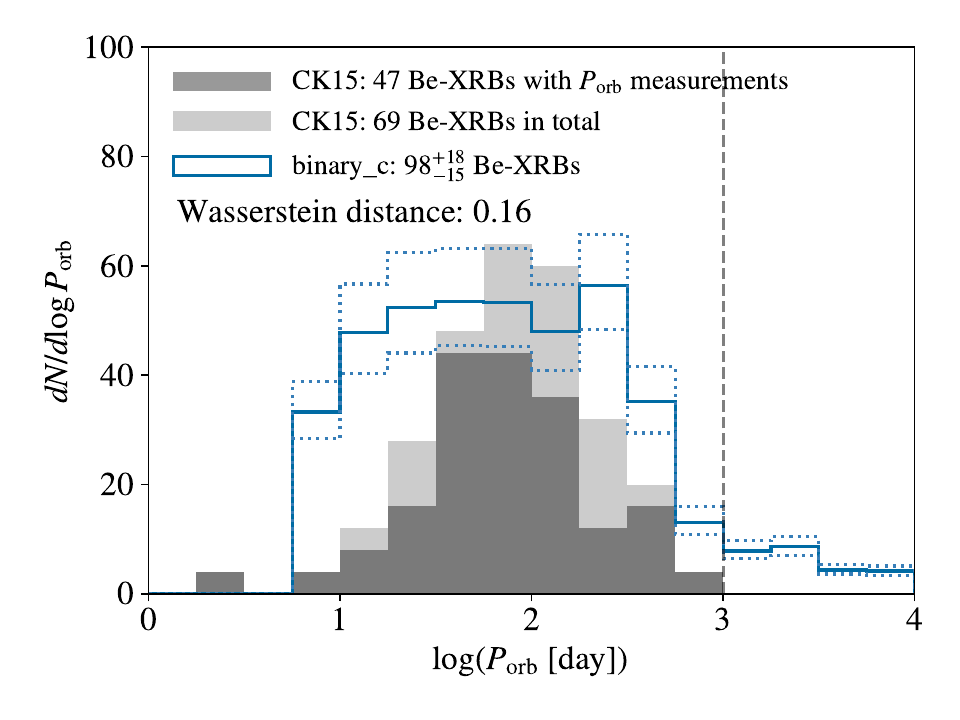}
    \vspace{-25pt}
    \caption{Same as Fig.~\ref{fig:smc_fd} but for the FR\_CS model with enhanced initial rotation (see Table~\ref{tab:model}). The FR\_CS model predicts more Be-XRBs, $N_{\rm BeXRB}= 98_{-15}^{+18}$, compared with the SR\_CS model. However, the over-predication of low-period systems in the FR\_CS model is more significant, leading to a larger Wasserstein distance of 0.16.}
    \label{fig:smc_fr}
\end{figure}

\begin{figure}
    \centering
    \includegraphics[width=\columnwidth]{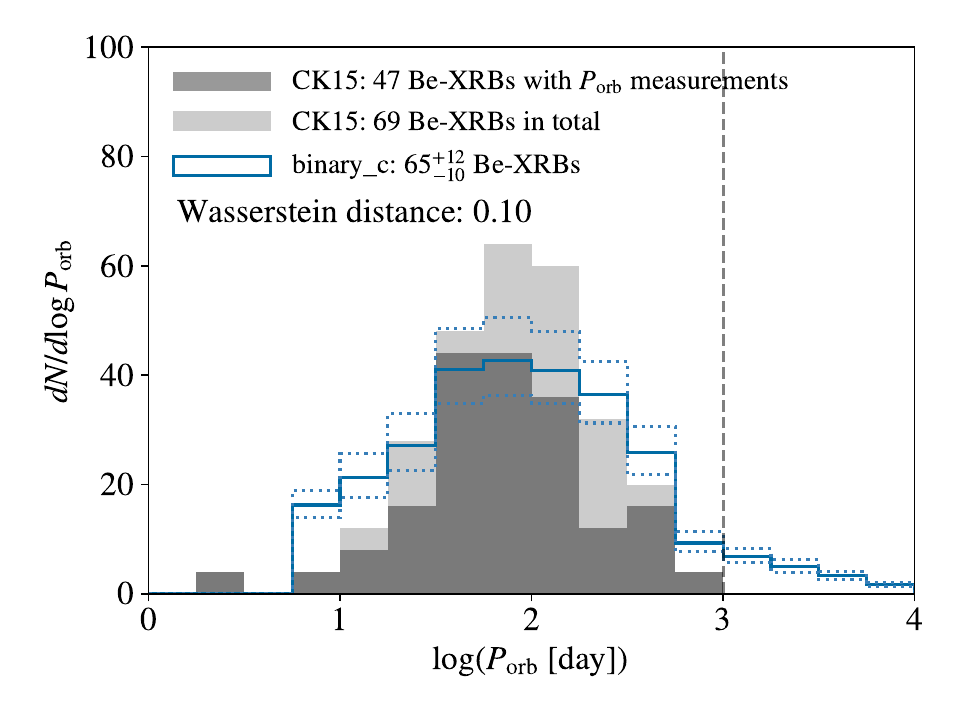}
    \vspace{-25pt}
    \caption{%{\color{blue}(optional)} 
    Same as Fig.~\ref{fig:smc_fd} but for the SR\_CS model with the hybrid initial orbital period distribution (Eqs.~\ref{fporb}-\ref{fporb_sana}). Now we have $N_{\rm BeXRB}= 65_{-10}^{+12}$, lower than the number $74_{-11}^{+13}$ in the default case, but still consistent with observations. The observed $P_{\rm orb}$ distribution is better reproduced with a slightly smaller Wasserstein distance of 0.1 than in the default case.}
    \label{fig:smc_porb}
\end{figure}

As mentioned in Sec.~\ref{sec:bps}, the BSE parameters and initial conditions of our BPS runs are chosen to reproduce the observed population of Be-XRBs in the SMC \citep{Coe2015} at the metallicity $Z_{\rm SMC}=0.0035$ \citep{Davies2015}. The agreements with observations are evaluated in two aspects: (1) The total number $N_{\rm BeXRB}$ of binaries currently in the Be-XRB phase with orbital periods $P_{\rm orb}<10^{3}$~days\footnote{Here we only count the binaries with $P_{\rm orb}<10^{3}$~days because all the observed Be-XRBs reside in this range and observations are more likely to be incomplete for long-period binaries.} is larger than 69, given the 69 Be-XRBs already found in the SMC \citep{Coe2015}. (2) The orbital period distribution of these binaries should cover $P_{\rm orb}\sim 10-10^{3}$~days with a peak around 100 days \citep{Coe2015,Vinciguerra2020}.

The BPS results (i.e., $N$ Be-XRBs from a single-age stellar population of a total mass $M_{\rm tot}$) are convolved with the star formation history of the SMC \citep[see their fig.~16]{Rubele2015} to calculate $N_{\rm BeXRB}$ as
\begin{align}
    N_{\rm BeXRB}=\sum_{i}^{N}w_{i}\ ,\quad w_{i}=\int_{t_{i,\rm ini}}^{t_{i,\rm fin}}[{\rm SFR}(t)/M_{\rm tot}]dt\ ,\label{nxrb_smc}
    %\int_{0}^{t}[L_{\nu,\rm tot}(t-t')/M_{\rm tot}]{{\rm SFR}}(t')dt'\ ,\\
    %L_{\nu,\rm tot}(t)&=\sum_{i}^{N}f_{{\rm duty},i}L_{\nu,i}\Theta(t-t_{i,\rm ini})\Theta(t_{i,\rm fin}-t)\ ,\label{lxtot}
\end{align}
where ${\rm SFR}(t)$ is the star formation history as a function of \textit{look-back} time $t$, and $t_{i,\rm ini\ (fin)}$ marks the beginning (end) of the Be-XRB phase of binary $i$. Similarly, the orbital period distribution is derived by applying the weight $w_{i}$ in Eq.~\ref{nxrb_smc} to each Be-XRB. %We expect the number count $N_{\rm BeXRB}$ obtained in this way to have relative errors $\sim 20-50$\% corresponding to the uncertainties of the SMC star formation history \citep[see their fig.~16]{Rubele2015} within the past 50~Myr, during which period formed the progenitors of most Be-XRBs alive at present (Fig.~\ref{fig:lx_t}). 

The results for the SR\_CS model (see Table~\ref{tab:model}) with the default initial binary statistics are shown in Fig.~\ref{fig:smc_fd}, which are almost the same as those for the SR\_OP model because our Be-XRB routine (Sec.~\ref{sec:bexrb}) does not affect binary stellar evolution. 
The number count requirement is satisfied with $N_{\rm BeXRB}= 74_{-11}^{+13}$, very close to the observed number 69, implying that the observed sample of Be-XRBs in the SMC is nearly complete. Here the uncertainties in $N_{\rm BeXRB}$ arise from the errors in ${\rm SFR}(t)$. The observed orbital period distribution is well reproduced by the SR\_CS model at $P_{\rm orb}<10^{3}$~days with a Wasserstein distance of 0.11. Our results are also consistent with those from the BPS study by \citet[see their fig.~3 and table~2]{Vinciguerra2020}\footnote{\citet{Vinciguerra2020} consider all Be-XRB \textit{candidates} made of a MS B star and a compact object with no modelling of the VDD. Therefore, they can produce much higher numbers in some cases compared with our results with stricter criteria for Be-XRBs (see Sec.~\ref{sec:id}).}. Nevertheless, the predicted distribution is broader and peaks at $\sim 300$~days compared with the observed distribution with a narrower peak around 100~days. The over-prediction of long-period ($P_{\rm orb}\gtrsim 200$~days) systems may be explained by incompleteness in observations but the discrepancy at $P_{\rm orb}\sim 10-100$~days is difficult to reconcile, which indicates that our binary stellar evolution and Be-XRB models are still imperfect.

For comparison, we show the results for the FR\_CS model in Fig.~\ref{fig:smc_fr} (with the default initial conditions), which are also almost the same as those for the FR\_OP model. Here we have $N_{\rm BeXRB}= 98_{-15}^{+18}$, allowing more space for the potential incompleteness of observations. With optimistic star formation rates of the SMC, the predicted number of Be-XRBs is larger than the observed number in almost every bin of $P_{\rm orb}$. %(see the upper dotted contour in Fig.~\ref{fig:smc_fr}). 
However, the discrepancy in the shape of the orbital period distribution is larger with a Wasserstein distance of 0.16. In particular, the number of low-period ($P_{\rm orb}\sim 10$~days) systems is over-predicted by up to a factor of $\sim 10$.

%{\color{blue}(optional)} 
Finally, we find that the orbital period distribution of Be-XRBs in the SMC is sensitive to the initial binary properties. When using the hybrid initial orbital period distribution (Eqs.~\ref{fporb}-\ref{fporb_sana}) rather than the default log-flat initial separation distribution, for the SR\_CS model we obtain $N_{\rm BeXRB}= 65_{-10}^{+12}$, lower than the number $74_{-11}^{+13}$ in the default case, but still consistent with observations. The predicted $P_{\rm orb}$ distribution now also peaks around 100~days as in observations, achieving a slightly smaller Wasserstein distance of 0.1, %with respect to the observed distribution, 
as shown in Fig.~\ref{fig:smc_porb}. The results for the FR models are similar to those under the default initial conditions (Fig.~\ref{fig:smc_fr}) and not shown.

%\newpage

\section{Additional results from alternative models}\label{apx:model}
\begin{figure}
    \centering
    \includegraphics[width=\columnwidth]{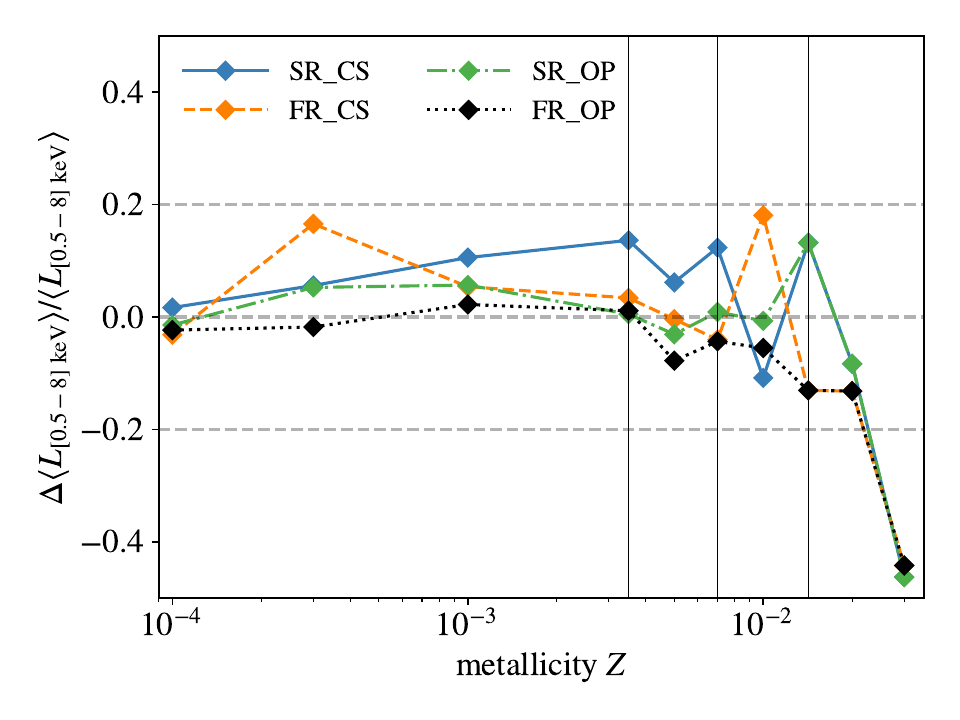}
    \vspace{-25pt}
    \caption{%{\color{blue}(optional)} 
    Relative difference between the X-ray luminosities per unit SFR in the $0.5-8$~keV band with $f_{\rm corr}=0.5$ from the two initial condition models defined in Sec.~\ref{sec:ic}. In the y axis, the denominator is the result from the default model with a log-flat initial separation distribution, while the nominator is the result from the hybrid initial orbital period distribution (Eqs.~\ref{fporb}-\ref{fporb_sana}) minus that from the default model. The thin vertical lines label the metallicities of the MW, LMC and SMC (from right to left)}
    \label{fig:dlx_sfr}
\end{figure}

To illustrate the weak dependence of our results on the distribution of initial binary orbital parameters, Fig.~\ref{fig:dlx_sfr} shows the relative difference between the X-ray outputs in the $0.5-8$~keV band from the two initial condition models considered in our study, i.e., the default model with a log-flat distribution of initial separations and the alternative model with an observation-based hybrid initial orbital period distribution (Sec.~\ref{sec:ic}). In general, the relative difference is small, i.e., within $20\%$ for $Z\lesssim 0.02$ and up to $\sim 50\%$ for the $Z=0.03$ case (with poor statistics). Therefore, in the rest of this appendix, we only show the results from the default initial condition model.
%For all the 4 models of initial rotation and VDD density parameters (Table~\ref{tab:model}), the relative difference between the X-ray outputs from the two initial condition models is within $20\%$ for $Z\lesssim 0.02$ and up to $\sim 50\%$ for the $Z=0.03$ case (with poor statistics), as shown in Fig.~\ref{fig:dlx_sfr} for the $0.5-8$~keV band.

\begin{figure}
    \centering
    \includegraphics[width=\columnwidth]{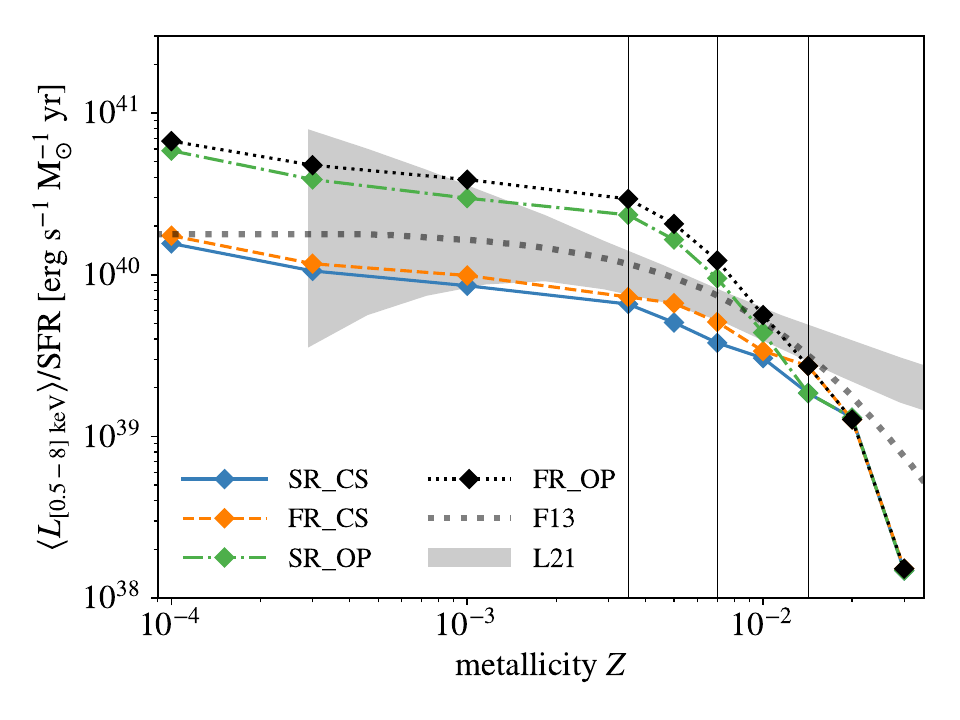}
    \vspace{-25pt}
    \caption{Same as Fig.~\ref{fig:lxsfr_Z} but comparing the results for the $0.5-8$~keV band from the SR\_CS (solid), FR\_CS (dashed), SR\_OP (dash-dotted) and FR\_OP (dotted) models with $f_{\rm corr}=0.5$.} %We also show the power-law fit of observational data from \citet[B16]{Brorby2016} with the thick long-dashed line, which can be regarded as optimistic extrapolation of the strong $Z$ evolution in observations at $Z\gtrsim 0.001$ to lower metallicities.}% The thin vertical lines label the metallicities of the MW, LMC and SMC (from right to left).}
    \label{fig:lxsfr_Z_all}
\end{figure}

\begin{figure}
    \centering
    \includegraphics[width=\columnwidth]{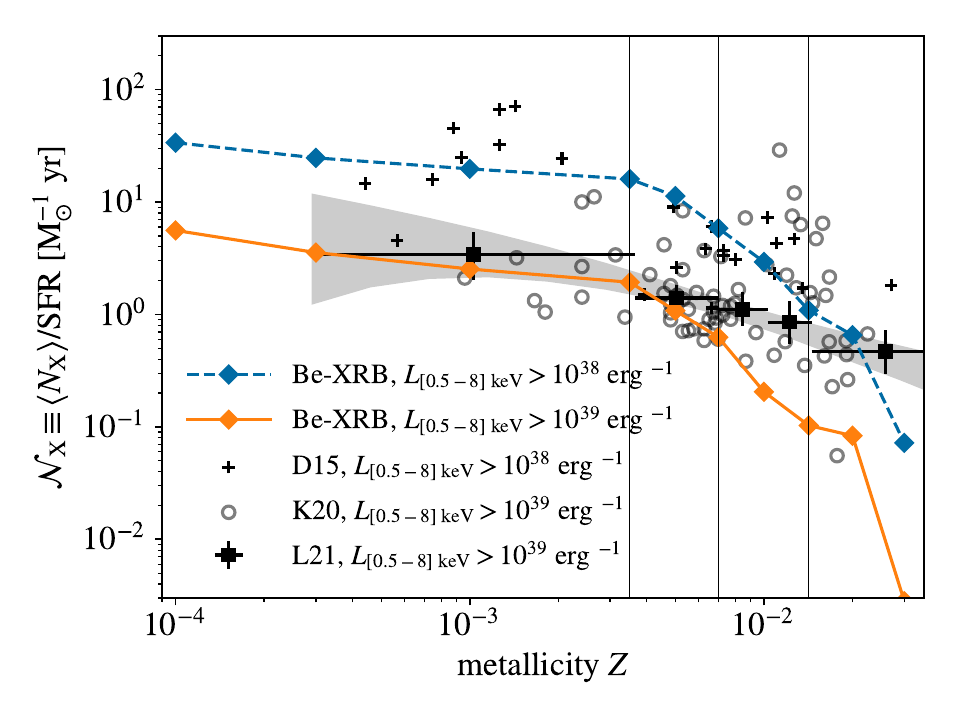}\\
    \vspace{-20pt}
    \includegraphics[width=\columnwidth]{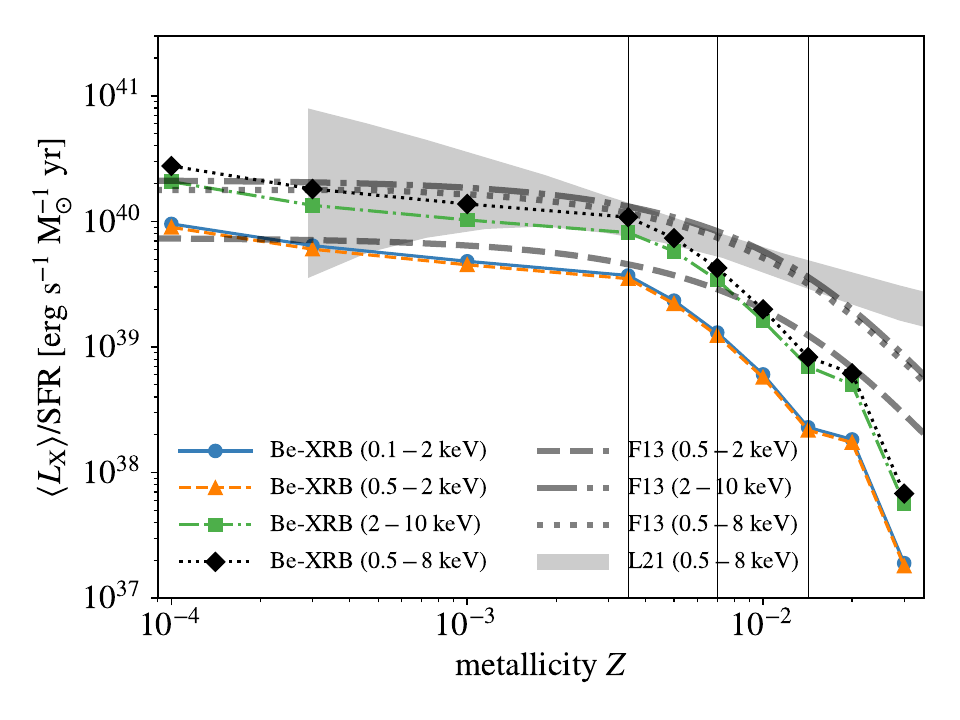}
    \vspace{-25pt}
    \caption{Same as Figs.~\ref{fig:nxsfr_fdcs} (top) and \ref{fig:lxsfr_Z} (bottom) but for the SR\_OP model with a lower correction factor $f_{\rm corr}=0.25$.}
    \label{fig:nxsfr_fdop0.25}
\end{figure}

\begin{figure}
    \centering
    \includegraphics[width=\columnwidth]{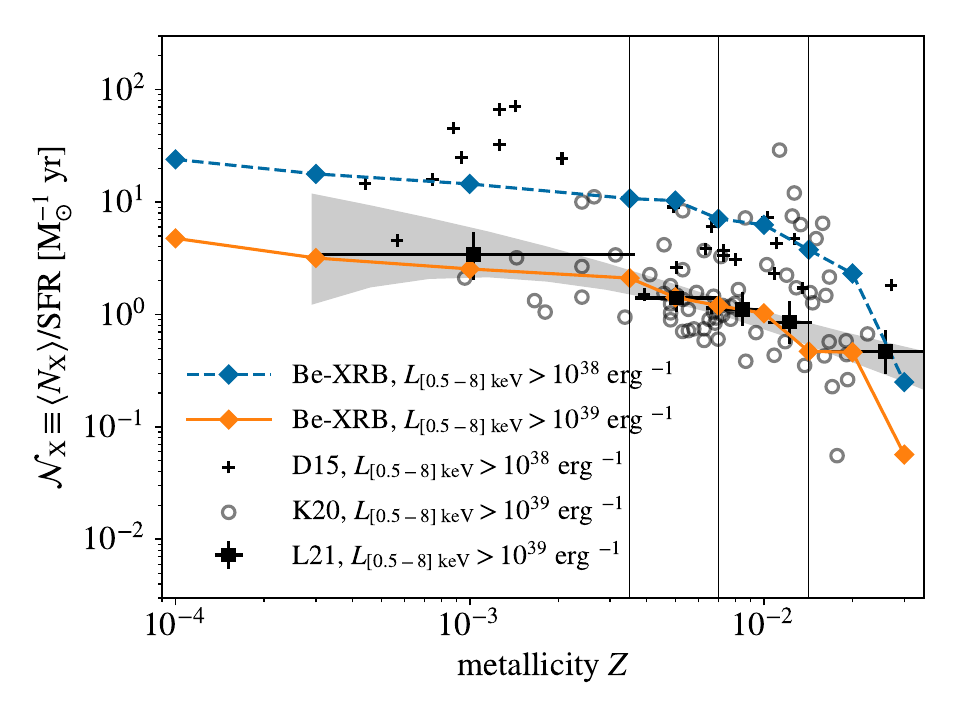}\\
    \vspace{-20pt}
    \includegraphics[width=\columnwidth]{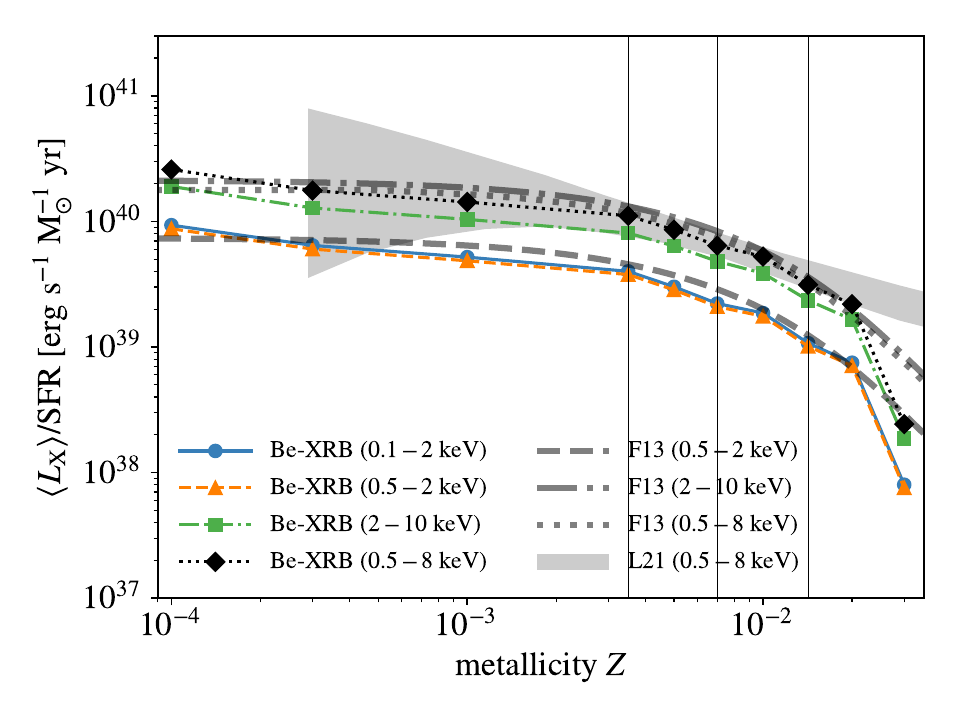}
    \vspace{-25pt}
    \caption{Same as Figs.~\ref{fig:nxsfr_fdcs} (top) and \ref{fig:lxsfr_Z} (bottom) but with a higher correction factor $f_{\rm corr}=0.8$.}
    \label{fig:nxsfr_fdcs0.8}
\end{figure}

To understand the impact of model parameters on the total X-ray output from Be-XRBs, %on X-ray outputs from Be-XRBs, 
in Fig.~\ref{fig:lxsfr_Z_all} we compare the X-ray luminosity per unit SFR in the $0.5-8$~keV band for all the 4 models in Table~\ref{tab:model} with $f_{\rm corr}=0.5$. {We find that models with higher initial rotation rates (FR) and VDD densities (OP) produce slightly ($\lesssim 50\%$) and significantly (up to a factor of $\sim 4$) higher X-ray luminosities than in the fiducial case SR\_CS, respectively. In the OP models with $f_{\rm corr}=0.5$, the luminosity from Be-XRBs is higher than that observed for all types of HMXBs \citep{Lehmer2021} by up to a factor of $\sim2$. This means that the OP models can only be consistent with observations with $f_{\rm corr}\lesssim 0.25$. Indeed, it is shown in \ref{fig:nxsfr_fdop0.25} that the X-ray luminosity and number of (ultra-)luminous HMXBs per unit SFR from Be-XRBs in the SR\_OP model reach the observed values for all types of HMXBs given $f_{\rm corr}=0.25$. Since $f_{\rm corr}=0.25$ is already the lower limit allowed by the calibration of accretion rate/luminosity for Be-XRBs (Sec.~\ref{sec:lx}), and other types of HMXBs also contribute to the observed luminosity/number, we conclude that the OP models are strongly disfavoured. Similarly, we find that $f_{\rm corr}\gtrsim 0.8$ is ruled out for the CS models, as illustrated in Fig.~\ref{fig:nxsfr_fdcs0.8} where the observed X-ray luminosity and number of (ultra-)luminous HMXBs per unit SFR for all types of HMXBs from \citet{Lehmer2021} are reproduced solely by Be-XRBs in the SR\_CS model with $f_{\rm corr}=0.8$.}
%However, the luminosity can be reduced to the observed range by a lower correction factor $f_{\rm corr}=0.25$.
%{\color{blue}(optional)} 
%Fig.~\ref{fig:nxsfr_alt} further shows the number of (ultra-)luminous Be-XRBs per unit SFR as a function of metallicity in the FR\_CS, SR\_OP and FR\_OP models in comparison with the results of the SR\_CS model in Fig.~\ref{fig:nxsfr_fdcs} assuming $f_{\rm corr}=0.5$. 
%Figs.~\ref{fig:nxsfr_fdcs0.8} and \ref{fig:nxsfr_fdop0.25} further show the fine-tuned results from the SR\_CS model with $f_{\rm corr}=0.8$ and the SR\_OP model with $f_{\rm corr}=0.25$ that are able to reproduce the total X-ray luminosity from HMXBs as well as the number of ultra-luminous HMXBs observed in nearby galaxies \citep{Lehmer2021} at $Z\lesssim 0.02$ and $Z\lesssim 0.0035$, respectively. In the SR\_OP model with $f_{\rm corr}=0.25$, the observed numbers for HMXBs with $L_{[0.5-8]~\rm keV}>10^{38}\ \rm erg\ s^{-1}$ \citep{Douna2015} are even better reproduced (see the bottom panel of Fig.~\ref{fig:nxsfr_fdop0.25} in comparison with Fig.~\ref{fig:nxsfr_fdcs}).}

%\clearpage
%\newpage

\bsp	% typesetting comment
\label{lastpage}

%%%%%%%%%%%%%%%%%%%%%%%%%%%%%%%%%%%%%%%%%%%%%%%%%%
\end{document}